\documentclass[preprint,12pt]{elsarticle}

\usepackage{mathbbol}
\usepackage{amssymb} 
\usepackage{bbm}
\usepackage{amsmath}
\usepackage[obeyspaces,spaces]{url}
\usepackage{etoolbox}
\AtBeginEnvironment{pmatrix}{\setlength{\arraycolsep}{2pt}}
\usepackage{makecell}
\usepackage{siunitx}
\usepackage[labelfont=bf]{caption}
\usepackage[labelformat=simple]{subcaption}
\usepackage{slashed}
\usepackage{graphicx}
\usepackage{grffile}
\usepackage{boldline}
\usepackage{multirow}
\usepackage{hyperref}

\usepackage{xcolor}
\usepackage{color}
\pagecolor{white}
\usepackage{listings}
\allowdisplaybreaks

\usepackage{pstricks}
\usepackage{color}

\lstdefinestyle{mystyle}{
    language = Mathematica,
    frame=single,
    rulesepcolor=\color{gray},
    backgroundcolor=\color{white},
    commentstyle=\color{lightgray},
    keywordstyle=\color{darkgray},
    numberstyle=\tiny\color{gray},
    stringstyle=\color{gray},
    basicstyle=\ttfamily\footnotesize,
    breakatwhitespace=false,
    breaklines=true,
    captionpos=b,
    keepspaces=true,
    numbers=left,
    numbersep=5pt,
    showspaces=false,
    showstringspaces=false,
    showtabs=false,
    tabsize=2
}
\newcounter{bla}

\journal{ }

\usepackage{etoolbox}
\makeatletter
\patchcmd{\ps@pprintTitle}
  {Preprint submitted to }
  {\normalfont{CERN-TH-2021-074, PSI-PR-21-08, ZU-TH 19/21}}
  {}{}
\makeatother

\begin{document}

\begin{frontmatter}



\title{Complete Lagrangian and Set of Feynman Rules for Scalar Leptoquarks}


\author[a,b,c]{Andreas Crivellin}
\author[d,e,f]{Luc Schnell\corref{author}}

\cortext[author] {Corresponding author.\\\textit{E-mail address:} schnell@mpp.mpg.de}
\address[a]{CERN Theory Division, CH--1211 Geneva 23, Switzerland}
\address[b]{Physik-Institut, Universit\"at Z\"urich,
	Winterthurerstrasse 190, CH-8057 Z\"urich, Switzerland}
\address[c]{Paul Scherrer Institut, CH--5232 Villigen PSI, Switzerland}
\address[d]{Laboratoire de Physique Th\'eorique et Hautes \'Energies, LPTHE, Sorbonne Universit\'e, CNRS, 4 place Jussieu, FR-75252 Paris Cedex 05, France}
\address[e]{Departement Physik, ETH Zürich, Otto-Stern-Weg 1, CH-8093 Zürich, Switzerland}
\address[f]{D\'epartement de Physique, \'Ecole Polytechnique, Route de Saclay, FR-91128 Palaiseau Cedex, France}

\begin{abstract}
Leptoquarks (LQs) have attracted increasing attention within recent years, mainly since they can explain the flavor anomalies found in $R(D^{(*)})$, $b \rightarrow s \ell^+ \ell^-$ transitions and the anomalous magnetic moment of the muon. In this article, we lay the groundwork for further automated analyses by presenting the complete Lagrangian and the corresponding set of Feynman rules for scalar leptoquarks. This means we consider the five representations $\Phi_1, \Phi_{\tilde1}, \Phi_2, \Phi_{\tilde2}$ and $\Phi_3$ and include the triple and quartic self-interactions, as well as couplings to the Standard Model (SM) fermions, gauge bosons and the Higgs. The calculations are performed using FeynRules and all model files are publicly available online at \url{https://gitlab.com/lucschnell/SLQrules}.
\end{abstract}


\end{frontmatter}
\clearpage

\noindent{\bf PROGRAM SUMMARY}

\begin{small}
\noindent
{\em Program Title:} SLQrules \\
{\em CPC Library link to program files:} (to be added by Technical Editor) \\
{\em Developer's repository link:} \url{https://gitlab.com/lucschnell/SLQrules} \\
{\em Code Ocean capsule:} (to be added by Technical Editor)\\
{\em Licensing provisions:} CC By 4.0 \\
{\em Programming language:} Mathematica, FeynRules \\
{\em Nature of problem:} In order to explain the deviations from SM predictions in $R(D^{(*)})$, $b \to \ell^+\ell^-$ transitions and the muon AMM jointly, models involving multiple LQ representations are necessary. This significantly increases the number of possible interactions and creates the need for computational tools that allow for studying the phenomenology of these LQ models in an automated manner. While model files exist for each LQ representation individually~[1], we are not aware of any publicly available model files that combine all scalar LQ representations as well as their self-interactions. \\
{\em Solution method:} We implemented the complete scalar LQ Lagrangian in a FeynRules~[2] model file and provide the corresponding MOD and UFO model files. These can be imported directly in FeynArts~[3] and MadGraph~[4] to obtain a versatile toolbox for the study of scalar LQs.  \\

\end{small}

\clearpage

\tableofcontents
\clearpage

\section{Introduction}
\label{sec:introduction}

Leptoquarks (LQs) are hypothetical beyond the Standard Model (BSM) particles that were first proposed in the context of Grand Unified Theories~\cite{Pati:1974yy, Georgi:1974sy, Georgi:1974yf, Fritzsch:1974nn}. What makes them stand out, and defines them, are their direct couplings to leptons and quarks (i.e. they convert a quark into a lepton and vice versa).  LQs were first systematically classified in Ref.~\cite{Buchmuller:1986zs}, where ten possible LQ representations under the Standard Model (SM) gauge group were found, of which five are scalar (spin 0) and five are vector (spin 1) particles. While they have received varying degrees of attention since then, LQs have undergone a renaissance in recent years. This can be attributed to the emergence of observed flavor anomalies, i.e. the deviations from the SM predictions measured in several flavour observables. In particular,  $R(D^{(*)})$~\cite{Lees:2012xj,Lees:2013uzd,Aaij:2015yra,Aaij:2017deq,Aaij:2017uff,Abdesselam:2019dgh}, $b \rightarrow s \ell^+ \ell^-$ observables~\cite{CMS:2014xfa,Aaij:2015oid,Abdesselam:2016llu,Aaij:2017vbb,Aaij:2019wad,Aaij:2020nrf,Aaij:2021vac} and the muon anomalous magnetic moment $a_\mu$~\cite{Bennett:2006fi,Abi:2021gix} deviate from their SM predictions by $>3\sigma$~\cite{Amhis:2016xyh,Murgui:2019czp,Shi:2019gxi,Blanke:2019qrx,Kumbhakar:2019avh}, $>5\sigma$~\cite{Capdevila:2017bsm, Altmannshofer:2017yso,Alguero:2019ptt,Alok:2019ufo,Ciuchini:2019usw,Aebischer:2019mlg, Arbey:2019duh,Kumar:2019nfv} and $>4.2\sigma$~\cite{Aoyama:2020ynm}, respectively. Recently, the authors of Ref.~\cite{Bobeth:2021lya} unveiled a further $\approx 4 \sigma$ tension with the SM prediction in the forward-backward asymmetry of $\bar{B} \to D^{*} \ell \nu$. While these hints for new physics are by themselves interesting, it is intriguing to note that they all fall into a common pattern, namely lepton flavour universality violation.

Models with LQs can account for $b\to s\ell^+\ell^-$ data~\cite{Alonso:2015sja, Calibbi:2015kma, Hiller:2016kry, Bhattacharya:2016mcc, Buttazzo:2017ixm, Barbieri:2015yvd, Barbieri:2016las, Calibbi:2017qbu, Crivellin:2017dsk, Bordone:2018nbg, Kumar:2018kmr, Crivellin:2018yvo, Crivellin:2019szf, Cornella:2019hct, Bordone:2019uzc, Bernigaud:2019bfy,Aebischer:2018acj,Fuentes-Martin:2019ign,Popov:2019tyc,Fajfer:2015ycq,Blanke:2018sro,deMedeirosVarzielas:2019lgb,Varzielas:2015iva,Crivellin:2019dwb,Saad:2020ihm,Saad:2020ucl,Gherardi:2020qhc,DaRold:2020bib}, $R(D^{(*)})$~\cite{Alonso:2015sja, Calibbi:2015kma, Fajfer:2015ycq, Bhattacharya:2016mcc, Buttazzo:2017ixm, Barbieri:2015yvd, Barbieri:2016las, Calibbi:2017qbu, Bordone:2017bld, Bordone:2018nbg, Kumar:2018kmr, Biswas:2018snp, Crivellin:2018yvo, Blanke:2018sro, Heeck:2018ntp,deMedeirosVarzielas:2019lgb, Cornella:2019hct, Bordone:2019uzc,Sahoo:2015wya, Chen:2016dip, Dey:2017ede, Becirevic:2017jtw, Chauhan:2017ndd, Becirevic:2018afm, Popov:2019tyc,Fajfer:2012jt, Deshpande:2012rr, Freytsis:2015qca, Bauer:2015knc, Li:2016vvp, Zhu:2016xdg, Popov:2016fzr, Deshpand:2016cpw, Becirevic:2016oho, Cai:2017wry, Altmannshofer:2017poe, Kamali:2018fhr, Mandal:2018kau, Azatov:2018knx, Wei:2018vmk, Angelescu:2018tyl, Kim:2018oih, Aydemir:2019ynb, Crivellin:2019qnh, Yan:2019hpm,Crivellin:2017zlb, Marzocca:2018wcf, Bigaran:2019bqv,Crivellin:2019dwb,Saad:2020ihm,Dev:2020qet,Saad:2020ucl,Altmannshofer:2020axr,Fuentes-Martin:2020bnh,Gherardi:2020qhc,DaRold:2020bib} and/or $a_\mu$~\cite{Bauer:2015knc,Djouadi:1989md, Chakraverty:2001yg,Cheung:2001ip,Popov:2016fzr,Chen:2016dip,Biggio:2016wyy,Davidson:1993qk,Couture:1995he,Mahanta:2001yc,Queiroz:2014pra,ColuccioLeskow:2016dox,Chen:2017hir,Das:2016vkr,Crivellin:2017zlb,Cai:2017wry,Crivellin:2018qmi,Kowalska:2018ulj,Dorsner:2019itg,Crivellin:2019dwb,DelleRose:2020qak,Saad:2020ihm,Bigaran:2020jil,Dorsner:2020aaz,Fuentes-Martin:2020bnh,Gherardi:2020qhc,Babu:2020hun,Crivellin:2020tsz}, making them prime NP candidates. Also the deviation in the $\bar{B} \to D^{*} \ell \nu$ forward-backward asymmetry can be explained with LQs, as was pointed out by Ref.~\cite{Carvunis:2021dss}. As a result, LQs have been studied in direct searches at LHC~\cite{Kramer:1997hh,Kramer:2004df,Faroughy:2016osc,Greljo:2017vvb, Blumlein:1996qp, Dorsner:2017ufx, Cerri:2018ypt, Bandyopadhyay:2018syt, Hiller:2018wbv,Faber:2018afz,Schmaltz:2018nls,Chandak:2019iwj,Allanach:2019zfr, Buonocore:2020erb,Haisch:2020xjd,Borschensky:2020hot,Crivellin:2021egp}, leptonic observables~\cite{Crivellin:2020mjs} and oblique electroweak (EW) parameters, Higgs couplings to gauge bosons~\cite{Keith:1997fv,Dorsner:2016wpm,Bhaskar:2020kdr,Zhang:2019jwp,Gherardi:2020det,Crivellin:2020ukd} and a wide range of low energy precision probes~\cite{Crivellin:2021egp, Shanker:1981mj,Shanker:1982nd,Leurer:1993em,Leurer:1993qx,Davidson:1993qk, Grossman:2019bzp,Seng:2020wjq,Belfatto:2019swo,Coutinho:2019aiy,Crivellin:2020lzu,Capdevila:2020rrl,Crivellin:2020ebi,Kirk:2020wdk,Alok:2020jod,Crivellin:2020oup,Crivellin:2020klg,Crivellin:2021njn,Belfatto:2021jhf,Branco:2021vhs, Bobeth:2017ecx,Dorsner:2019vgp,Mandal:2019gff}. 

Many models that are now being proposed to explain the flavour anomalies jointly contain multiple scalar leptoquark representations~\cite{Crivellin:2019dwb, Saad:2020ihm, Crivellin:2017zlb, Gherardi:2020qhc, Bigaran:2019bqv, Greljo:2021xmg, DaRold:2020bib, Babu:2020hun, Bigaran:2020jil, Gargalionis:2019drk, Saad:2020ucl, Chen:2016dip}. In addition to couplings of LQs to the SM fermions and gauge bosons, also interactions with the particles of the Higgs sector and between the LQs themselves are present in such a setup. In this article, we lay the groundwork for future automated LQ analyses by presenting the complete set of Feynman rules for all five scalar LQs. We include the $\text{LQ-LQ-Higgs(-Higgs)}$, LQ-LQ-LQ(-Higgs) and LQ-LQ-LQ-LQ couplings, as well as all couplings to the SM fermions and gauge bosons. The FeynRules~\cite{Alloul:2013bka} model files as well as the corresponding  MOD and UFO files are publicly available online at \url{https://gitlab.com/lucschnell/SLQrules} and can be used for automatized LQ phenomenology using FeynArts~\cite{Hahn:2000kx} or MadGraph~\cite{Alwall:2014hca}. \\

In Sections~\ref{sec:SMLagrangian} and \ref{sec:LQLagrangian}, we present our conventions by defining the complete Lagrangian for the SM extended with all five scalar LQ representations. Sec.~\ref{sec:methods} gives an overview on the computational tools and in Sec.~\ref{sec:SMInteractions}, the Feynman rules for the SM interactions are listed, while the ones for the scalar LQs are given in Sec.~\ref{sec:LQInteractions}. A detailed description of our treatment of charge-conjugate fermions can be found in~\ref{sec:appendix}.

\section{SM Lagrangian}
\label{sec:SMLagrangian}
The SM fermions and the Higgs field transform under the $SU(3)_c \times SU(2)_L \times U(1)_Y$ gauge group as
\begin{equation}
\renewcommand{\arraystretch}{1.1}
    \centering
    \begin{tabular}{c | c c c c c c}
         & $Q^\prime_i$ & $u^\prime_i$ & $d^\prime_i$ & $L^\prime_i$ & $\ell^\prime_i$ & $H$\\ \hline
    	$SU(3)_c$ & 3 & 3 & 3 & 1 & 1 & 1 \\
    	$SU(2)_L$ & 2 & 1 & 1 & 2 & 1 & 2 \\
    	$U(1)_Y$ & $\frac{1}{6}$ & $\frac{2}{3}$ & -$\frac{1}{3}$ & -$\frac{1}{2}$ & -1 & $\frac{1}{2}$
    \end{tabular}\,.
\end{equation}
The prime indicates that these are the weak eigenstates. With this convention, the electric charge $Q$ can be calculated via the Gell-Mann-Nishijima formula
\begin{equation}
\label{eq:Gell-Mann-Nishijima}
    Q = I_3 + Y \,,
\end{equation}
where $Y$ is the $U(1)_Y$ hypercharge and $I_3$ the third component of the weak isospin. The left-handed quarks are components of the $SU(2)_L$-doublet $Q^\prime$ and the right-handed ones correspond to the singlets $u^\prime, d^\prime$
\begin{equation}
    Q^\prime_i = \begin{pmatrix}
            u^\prime_{i, L}\\ 
            d^\prime_{i, L}
        \end{pmatrix}, 
    u^\prime_i = \left(u^\prime_{i, R}\right) \text{ and } 
    d^\prime_i = \left(d^\prime_{i, R}\right)\,,
\end{equation}
where $i = 1,2,3$ is the flavor index. Analogous conventions hold for the leptons
\begin{equation}
    L^\prime_i = \begin{pmatrix}
            \nu^\prime_{i, L}\\ 
            \ell^\prime_{i, L}
        \end{pmatrix} \text{ and } 
    \ell^\prime_i = \left(\ell^\prime_{i, R}\right) \,.
\end{equation}
Note that right-handed neutrinos are not considered in this work. The SM Lagrangian then takes the form
\begin{align}
    \begin{aligned}
        \mathcal{L}^{\text{SM}}&= - \frac{1}{4} B^{\mu \nu} B_{\mu\nu} 
        - \frac{1}{4} W^{I, \mu \nu} W_{I, \mu\nu}
        - \frac{1}{4} G^{\alpha, \mu \nu} G_{\alpha, \mu\nu} \\
        &\phantom{12} + i \left(\bar{Q}_i^{\prime \intercal} \slashed{D} Q_i^\prime\right)   + i \left(\bar{L}_i^{\prime \intercal} \slashed{D} L_i^\prime\right)  + i \bar{u}_i^\prime \slashed{D} u_i^\prime  + i \bar{d}_i^\prime \slashed{D} d_i^\prime + i \bar{\ell}_i^\prime \slashed{D} \ell_i^\prime  \\
        &\phantom{12} +\left(\left(D^\mu H\right)^{\dagger} D_\mu H\right) + \mu_H^2 \left(H^\dagger H \right) - \lambda_H \left(H^\dagger H \right)^2 \\
        &\phantom{12} - \left(Y^d_{ij} \left(\bar{Q}^{\prime \intercal}_i H\right) d^\prime_j 
        +Y^\ell_{ij} \left(\bar{L}^{\prime \intercal}_i H\right) \ell_j^\prime  
        + Y^u_{ij} \left( \bar{Q}^{\prime \intercal}_i \tilde{H} \right) u_j^\prime + \text{ h.c.}\right) \\
        &\phantom{12} + \mathcal{L}_{\text{gauge-fixing}} \,.
        \label{eq:SM_lagrangian}
    \end{aligned}
\end{align}
We use brackets to indicate the $SU(2)_L$ contractions and denote the transpose of a $SU(2)_L$ multiplet $\Phi$ as $\Phi^\intercal$. $B^{\mu\nu}, W^{I, \mu \nu}$ (with $I \in \{1,2,3\}$) and $G^{\alpha, \mu \nu}$ (with $\alpha \in \{1, \dots ,8\}$) are the field strength tensors
\begin{equation}
    \begin{aligned}
    B_{\mu \nu} &= \partial_\mu B_\nu - \partial_\nu B_\mu \,, \\
    W^{I}_{\mu \nu} &= \partial_\mu W^I_\nu - \partial_\nu W^I_\mu + g f^{IJK} W_\mu^J W_\nu^K \,, \\
    G^{\alpha}_{\mu \nu} &= \partial_\mu G^\alpha_\nu - \partial_\nu G^\alpha_\mu + g f^{\alpha \beta \gamma} W_\mu^\beta W_\nu^\gamma \,,
    \end{aligned}
\end{equation} 
where $B_\mu, W^I_\mu, G^\alpha_\mu$ are the gauge fields corresponding to $U(1)_{Y}$, $SU(2)_{L}$ and $SU(3)_{c}$, respectively. $f^{IJK}$, $f^{\alpha \beta \gamma}$ denote the structure constants of $SU(2)_L$, $SU(3)_c$. $Y^d$, $Y^\ell$, $Y^u$ are the Yukawa couplings and $\tilde{H} \equiv i\sigma_2 H^\dagger$, where $\sigma_2$ is the second Pauli matrix. In our conventions, the Pauli matrices are given by
\begin{equation}
\sigma_1 = \begin{pmatrix} 0 & 1 \\ 1 & 0 \end{pmatrix}, \, \sigma_2 = \begin{pmatrix} 0 & -i \\ i & 0 \end{pmatrix} ,\, \sigma_3 = \begin{pmatrix} 1 & 0 \\ 0 & -1 \end{pmatrix}\,.
\end{equation}
The covariant derivative is defined as
\begin{equation}
D_\mu \phi = \partial_\mu \phi - ig_1 Y B_\mu \phi - ig_2  \tau_{I} W^{I}_\mu \phi - ig_s T_{\alpha} G^{\alpha}_\mu \phi \,,
\end{equation}
with $I \in \{ 1,2,3 \}$, $\alpha \in \{1, \dots, 8\}$. $\tau_{I} = \frac{\sigma_I}{2}, T_{\alpha} = \frac{\lambda_\alpha}{2}$ refer to the generators of the fundamental $SU(2)_L$, $SU(3)_c$ representations here. The gauge-fixing part $\mathcal{L}_{\text{gauge-fixing}}$ of the Lagrangian is described in Eq.~(\ref{eq:gaugefixing}). \\

Having acquired a non-zero vacuum expectation value (VEV) $v \approx 246$ GeV after spontaneous symmetry breaking, the Higgs doublet $H$ can be expanded in modes around $v$
\begin{equation}
H = \begin{pmatrix}
-i\varphi^+ \\
\frac{v + h + i\varphi_Z}{\sqrt{2}}\\
\end{pmatrix}\,,
\end{equation}
where $\varphi^+$ and $\varphi_Z$ are the Goldstone bosons and $h$ the physical Higgs field with $m_h \approx $ \SI{125}{\giga \electronvolt}. Expressed in terms of the couplings $\mu_H, \lambda_H$ (see Eq.~(\ref{eq:SM_lagrangian})), the parameters $v, m_h$ read
\begin{equation}
    \begin{aligned}
    m_h &= \sqrt{2} \mu_H \,, \\
    v &= \dfrac{\mu_H}{\sqrt{\lambda_H}} \,. 
    \end{aligned}
\end{equation}
The non-vanishing VEV leads to mixing of the $U(1)_Y$ gauge boson with the neutral component of the $SU(2)_L$ ones
\begin{align}
    \begin{aligned}
        \begin{pmatrix}
            A_\mu\\ 
            Z_\mu
        \end{pmatrix} &= 
        \begin{pmatrix}
            \cos \theta_w & \sin \theta_w \\
            -\sin \theta_w & \cos \theta_w
        \end{pmatrix}
                \begin{pmatrix}
            B_\mu\\ 
            W^3_\mu
        \end{pmatrix}\,, \\
     \end{aligned}
     \label{eq:EWMixing1}
\end{align}
where $\theta_w$ is the Weinberg angle. The remaining two $SU(2)_L$ gauge bosons form electric charge eigenstates
\begin{align}
    \begin{aligned}
        W^\pm_\mu &= \frac{1}{\sqrt{2}} \left( W^1_\mu \mp i W^2_\mu \right) \,. \label{eq:EWMixing2}
        \end{aligned}
\end{align}
In the following, $\cos \theta_w, \sin \theta_w$ will be abbreviated as $c_w, s_w$. Instead of using the coupling constants $g_1, g_2$ directly, we will state the Feynman rules using the electric charge $e$. The conversion back to $g_1, g_2$ is easily carried out using 
\begin{align}
    \begin{aligned}
        e = c_w   g_1 \,, \\
        e = s_w   g_2\,.
    \end{aligned}
\end{align}
The masses acquired by the gauge bosons in Eqs.~(\ref{eq:EWMixing1}) and (\ref{eq:EWMixing2}) via spontaneous symmetry breaking are
\begin{equation}
    \begin{aligned}
        m_A &= 0 \,, \\
        m_W &= \frac{v}{2}\frac{e}{s_w} \,, \\
        m_Z &= \frac{v}{2}\frac{e}{s_w c_w} \,,
    \end{aligned}
    \label{eq:bosonMasses}
\end{equation}
from which it becomes clear that $A_\mu$ corresponds to the photon. As stated above, $Y^d, Y^u, Y^\ell$ denote the Yukawa couplings of the Higgs doublet for the down-type quarks, up-type quarks and leptons, respectively. After spontaneous symmetry breaking, they yield non-diagonal mass terms in the weak basis
\begin{equation}
    \begin{aligned}
        m^u_{ij} \bar{u}_i^\prime u_j^\prime= \frac{v}{\sqrt{2}}  Y^u_{ij}\ \bar{u}^\prime_{i} u^\prime_{j}\,, \\
        m^d_{ij} \bar{d}_i^\prime d_j^\prime= \frac{v}{\sqrt{2}}  Y^d_{ij}\ \bar{d}^\prime_{i} d^\prime_{j}\,, \\
        m^\ell_{ij} \bar{\ell}_i^\prime \ell_j^\prime= \frac{v}{\sqrt{2}}  Y^\ell_{ij}\ \bar{\ell}^\prime_{i} \ell^\prime_{j}\,. 
    \end{aligned}
\end{equation}
These can be diagonalized by rotating the fermion fields
\begin{equation}
\begin{aligned}
u_{i, L}^\prime &\equiv U^{uL}_{ij} u_{j, L} \,, & u_{i, R}^\prime &\equiv U^{uR}_{ij} u_{j, R}   \,, \\
d_{i, L}^\prime &\equiv U^{dL}_{ij} d_{j, L} \,, & d_{i, R}^\prime &\equiv U^{dR}_{ij} d_{j, R}  \,, \\
\ell_{i, L}^\prime &\equiv U^{\ell L}_{ij} \ell_{j, L}  \,, & \ell_{i, R}^\prime &\equiv U^{\ell R}_{ij} \ell_{j, R}\,,
\end{aligned}
\label{eq:fermionRotation}
\end{equation}
where $U^{uL}, U^{uR}, U^{dL}, U^{dR}$, $U^{\ell L}, U^{\ell R}$ are unitary matrices, $u^\prime, d^\prime, \ell^\prime$ the weak eigenstates and $u, d, \ell$ the mass eigenstates. The Yukawa couplings in this new basis 
\begin{equation}
    \begin{aligned}
    \hat{Y}^u &\equiv \left(U^{uL}\right)^\dagger Y^u U^{uR} \,, \\
    \hat{Y}^d &\equiv \left(U^{dL}\right)^\dagger Y^d U^{dR} \,,\\
    \hat{Y}^\ell &\equiv \left(U^{\ell L}\right)^\dagger Y^\ell U^{\ell R} \,,\\
    \end{aligned}
    \label{eq:YukawaDiagonalization}
\end{equation}
are diagonal and are related to the fermion masses by
\begin{equation}
    \begin{aligned}
    \frac{v}{\sqrt{2}} \hat{Y}^u &= \text{diag}(m_{u_1}, m_{u_2},m_{u_3}) \,, \\
    \frac{v}{\sqrt{2}} \hat{Y}^d &= \text{diag}(m_{d_1}, m_{d_2},m_{d_3}) \,, \\
    \frac{v}{\sqrt{2}} \hat{Y}^\ell &= \text{diag}(m_{\ell_1}, m_{\ell_2},m_{\ell_3}) \,.\\
    \end{aligned}
    \label{eq:fermionMasses}
\end{equation}
The unitary matrices in Eq.~(\ref{eq:fermionRotation}) can for the most part be absorbed into redefinitions of the fermion fields, the only remaining physical matrix being in the weak charged current for quarks, the CKM matrix $V$
\begin{equation}
\begin{aligned}
    \mathcal{L}^{SM} &\supset \frac{e}{\sqrt{2} s_w} W_\mu^+ \bar{u}^\prime_{i,L} \gamma^\mu d^\prime_{i,L} + \text{ h.c.} \\
    &= \frac{e}{\sqrt{2} s_w} W_\mu^+ \bar{u}_{k,L} \gamma^\mu \underbrace{\left(U^{uL} \right)_{ki}^\dagger \left(U^{dL}\right)_{il}}_{V_{kl}} d_{l,L} + \text{ h.c.} \,.
\end{aligned}
\end{equation}
Without loss of generality, we can therefore assume that the relations between the weak eigenstates $u^\prime, d^\prime, \ell^\prime, \nu^\prime$ and the mass eigenstates $u,d, \ell, \nu$ take the form
\begin{equation}
    \begin{aligned}
        u^\prime_{i, L} &= u_{i, L} \,, &u^\prime_{i, R} &= u_{i, R} \,, \\
        d^\prime_{i, L} &= V_{ij} d_{j, L} \,, &d^\prime_{i, R} &= d_{i, R} \,, \\
        \ell^\prime_{i, L} &= \ell_{i, L} \,, &\ell^\prime_{i, R} &= \ell_{i, R} \,, \\
        \nu^\prime_{i, L} &= \nu_{i, L} \,, & &
    \end{aligned}
    \label{eq:massEigenstates}
\end{equation}
meaning that CKM elements only appear in couplings involving left-handed down-type quarks (corresponding to the up basis). Note that while this choice does not affect the SM Feynman rules, it becomes relevant in LQ interactions with SM fermions. The gauge freedoms of the SM gauge bosons are given in Ref.~\cite{Romao:2012pq} (our conventions correspond to $\eta_Y, \eta_Z, \eta_\theta, \eta_G = 1$ and $\eta, \eta_s, \eta_e, \eta^\prime = -1$). They are fixed using the Fadeev-Popov prescription~\cite{Faddeev:1967fc}. Working in the Feynman gauge, we add the following gauge-fixing terms, after spontaneous symmetry breaking and expressed in terms of the physical fields $A_\mu, Z_\mu, W_\mu^\pm, G_\mu^\alpha$ and their corresponding ghost fields $c_A, c_Z, c_{W^\pm}, c_{G^\alpha}$, to the Lagrangian
\begin{equation}
\begin{aligned}
 \mathcal{L}_{\text{gauge-fixing}} &= -\frac{1}{2} \left( \partial^\mu A_\mu \right)^2 - \frac{1}{2}\left(\partial^\mu Z_\mu - m_Z \varphi_Z \right)^2 \\
 &\phantom{12} - \left(\partial^\mu W_\mu^- + i m_W \varphi^- \right)\left(\partial^\mu W_\mu^+ - i m_W \varphi^+ \right) \\
 &\phantom{12}- \frac{1}{2}\left(\partial^\mu G_\mu^\alpha \right)^2 \\
&\phantom{12} +\sum_{B \in \{A,Z,W^\pm \}} \bigg( \bar{c}_{W^+} \frac{\partial(\delta F_{W^+})}{\partial \alpha_B} + \bar{c}_{W^-} \frac{\partial(\delta F_{W^+})}{\partial \alpha_B}  \\
&\phantom{12 + \sum_{B \in \{A,Z,W^\pm \}}}+ \bar{c}_Z \frac{\partial(\delta F_Z)}{\partial \alpha_B} + \bar{c}_A \frac{\partial(\delta F_A)}{\partial \alpha_B}\bigg)c_B \\
&\phantom{12} + \sum_{\alpha, \beta = 1}^{8} \bar{c}_{G^\alpha} \frac{\partial(\delta F_{G^\alpha})}{\partial \alpha_{G^\beta}} c_{G^\beta}\,.
 \end{aligned}
 \label{eq:gaugefixing}
\end{equation}
The explicit expressions for $\delta F_{W^+}, \delta F_Z, \delta F_A, \delta F_{G^\alpha}$ are given in Ref.~\cite{Romao:2012pq}.

\clearpage
\section{LQ Lagrangian}
\label{sec:LQLagrangian}
The five possible scalar LQ representations $\Phi_1, \Phi_{\tilde{1}}, \Phi_2, \Phi_{\tilde{2}}, \Phi_3$ transform under the $SU(3)_c \times SU(2)_L \times U(1)_Y$ gauge group as~\cite{Buchmuller:1986zs} 
\begin{equation}
\renewcommand{\arraystretch}{1.1}
    \centering
    \begin{tabular}{c | c c c c c}
         & $\Phi_1,$ & $\Phi_{\tilde{1}}$ & $\Phi_2$ & $ \Phi_{\tilde{2}}$ & $\Phi_3$\\ \hline
    	$SU(3)_c$ & 3 & 3 & 3 & 3 & 3 \\
    	$SU(2)_L$ & 1 & 1 & 2 & 2 & 3 \\
    	$U(1)_Y$ & $-\frac{1}{3}$ & $-\frac{4}{3}$ & $\frac{7}{6}$ & $\frac{1}{6}$ & $-\frac{1}{3}$
    \end{tabular}\,.
\end{equation}
Note that we adopt the convention that all scalar LQs transform in the fundamental $SU(3)_c$ representation. Expanding the $SU(2)_L$ multiplets into their components, the following electric charge eigenstates are found
\renewcommand{\arraystretch}{1.5}
\begin{align}
    \begin{aligned}
        \Phi_1 = \left( \Phi_1^{-1/3}\right), \ 
        \Phi_{\tilde{1}} = \left( \Phi_{\tilde{1}}^{-4/3}\right), \ 
        \Phi_2 = \begin{pmatrix} \Phi^{+5/3}_2 \\ \Phi^{+2/3}_2 \end{pmatrix},\ \\
        \renewcommand{\arraystretch}{1.5}
        \Phi_{\tilde{2}} = \begin{pmatrix} \Phi^{+2/3}_{\tilde{2}} \\ \Phi^{-1/3}_{\tilde{2}} \end{pmatrix},\ 
        \sigma \cdot \Phi_{3} = \begin{pmatrix}
            \Phi_3^{-1/3} & \sqrt{2} \Phi_3^{+2/3} \\
            \sqrt{2} \Phi_3^{-4/3} & -\Phi_3^{-1/3}
        \end{pmatrix}\,. 
    \end{aligned}
\end{align}
We decompose the scalar LQ Lagrangian as
\begin{equation}
    \mathcal{L}^{\text{LQ}} = \mathcal{L}_{2 \Phi} + \mathcal{L}_{\text{kin}} +
    \mathcal{L}_{\text{f}}  + \mathcal{L}_{3 \Phi} + \mathcal{L}_{4 \Phi} \,,
\end{equation}
where $\mathcal{L}_{2 \Phi}$ contains the LQ masses and the LQ-LQ-Higgs(-Higgs) couplings, $\mathcal{L}_{\text{kin}}$  the kinetic terms and the couplings to the SM gauge bosons, $\mathcal{L}_{\text{f}}$ the couplings to the SM fermions, $\mathcal{L}_{3 \Phi}$ the LQ-LQ-LQ(-Higgs) couplings and finally $\mathcal{L}_{4 \Phi}$ the LQ-LQ-LQ-LQ interaction terms. The individual parts of the Lagrangians are discussed in more detail below.

\subsection{LQ Masses and Higgs Interactions ($\mathcal{L}_{2 \Phi}$)}
\label{sec:LQ-Higgs}
The LQ masses and the LQ-LQ-Higgs(-Higgs) interactions are implemented by the Lagrangian \cite{Crivellin:2020ukd, Hirsch:1996qy}
\begin{equation}
    \begin{aligned}
        \mathcal{L}_{2\Phi} 
        = &- \sum_{a = 1}^3\big(m_{a}^2+Y_{a} \left(H^{\dagger}H\right)\big)\big(\Phi_a^{\dagger}\Phi_a\big)
        - \sum_{a = 1}^2 \big(m_{\tilde a}^2+Y_{\tilde a} \left(H^{\dagger}H\right) \big)\big(\Phi_{\tilde a}^{\dagger}\Phi_{\tilde a}\big) \\
&- Y_{22} \big(H^\intercal i\sigma_{2} \Phi_2\big)^\dagger \big(H^\intercal i\sigma_{2} \Phi_2\big)
        - Y_{\tilde{2}\tilde{2}} \big(H^\intercal i\sigma_{2} \Phi_{\tilde{2}} \big)^\dagger \big(H^\intercal i\sigma_{2} \Phi_{\tilde{2}} \big) \\
        &- iY_{33} \epsilon^{IJK} \left(H^\dagger \sigma^I H \right) \Phi_3^{J\dagger} \Phi_3^K\\
         \Big[&- A_{1\tilde{2}}\big( \Phi_{\tilde{2}}^\dagger H \big)\Phi_1 + A_{\tilde{2}3}\big(\Phi_{\tilde{2}}^\dagger \left(\sigma \cdot \Phi_3\right) H \big) + Y_{2\tilde{2}}\big(\Phi_{2}^\dagger H \big)\big(H^\intercal i\sigma_{2}\Phi_{\tilde{2}}\big)\\
        &+Y_{\tilde{1}3}\big(H^\intercal i\sigma_{2}\left(\sigma\cdot\Phi_{3}\right)^\dagger H\big)\Phi_{\tilde{1}} + Y_{13} \big(H^{\dagger}\left(\sigma\cdot\Phi_{3} \right)H \big)\Phi_{1}^{\dagger} + \text{h.c.}\Big] \,.
    \end{aligned}
    \label{eq:LQ_mixing}
\end{equation}
The color indices are omitted since they just involve trivial contractions. For example, the term with $Y_{22}$ in Eq.~(\ref{eq:LQ_mixing}) would be $- Y_{22} \left(H^\intercal i\sigma_{2} \Phi_{2, c_1} \right)^\dagger $ $ \left(H^\intercal i\sigma_{2} \Phi_{2, c_1} \right)$ in full notation. We use the convention that couplings with mass dimension one (zero) are denoted by $A$ ($Y$). \\

The terms in Eq.~(\ref{eq:LQ_mixing}) lead to mixing among the LQ eigenstates of the same electric charge after EW symmetry breaking. It is therefore convenient to collect them in the electric charge eigenstate vectors
\begin{equation}
    \begin{aligned}
        \Phi^{-1/3} &\equiv 
        \renewcommand{\arraystretch}{1.5}
                    \begin{pmatrix}
                        \Phi_1^{-1/3} \\ 
                        \Phi^{-1/3}_{\tilde{2}} \\
                        \Phi_3^{-1/3}
                    \end{pmatrix} \,, 
        &\Phi^{+2/3} &\equiv 
        \renewcommand{\arraystretch}{1.5}
                    \begin{pmatrix}
                        \Phi_2^{+2/3} \\ 
                        \Phi^{+2/3}_{\tilde{2}} \\
                        \Phi_3^{+2/3}
                    \end{pmatrix} \,, \\
        \Phi^{-4/3} &\equiv 
        \renewcommand{\arraystretch}{1.5}
                    \begin{pmatrix}
                        \Phi_{\tilde{1}}^{-4/3} \\ 
                        \Phi^{-4/3}_{3}
                    \end{pmatrix} \,,
       &\Phi^{+5/3} &\equiv 
       \renewcommand{\arraystretch}{1.5}
                    \begin{pmatrix}
                        \Phi_2^{+5/3}
                    \end{pmatrix}\,. 
    \end{aligned}
\label{eq:ChargeVectors}
\end{equation}
After spontaneous symmetry breaking, non-diagonal mass matrices $\mathbbm{M}^q$ in
\begin{equation}
    \mathcal{L}_{2\Phi} \supset - \sum_{q \in Q} \Phi^q{}^\dagger \mathbbm{M}^q \Phi^q
\end{equation} 
are generated with $Q = \left \{ -\frac{4}{3}, -\frac{1}{3}, +\frac{2}{3}, +\frac{5}{3} \right \}$ and~\cite{Crivellin:2020ukd, Hirsch:1996qy, AristizabalSierra:2007nf}
\begin{align}
    \begin{aligned}
        \mathbbm{M}^{-1/3} &=     \begin{pmatrix}
                                    			m_1^2 + \frac{v^2}{2} Y_1 & \frac{v}{\sqrt{2}} A_{1\tilde{2}}^*  & \frac{v^2}{2} Y_{13} \\ 
                                    			\frac{v}{\sqrt{2}} A_{1\tilde{2}} & m_{\tilde{2}}^2 + \frac{v^2}{2}Y_{\tilde{2}} & \frac{v}{\sqrt{2}} A_{\tilde{2}3} \\ 
                                    			\frac{v^2}{2} Y_{13}^* & \frac{v}{\sqrt{2}} A^*_{\tilde{2}3} & m_3^2 + \frac{v^2}{2} Y_3
                                			\end{pmatrix} \,, \\
        \mathbbm{M}^{+2/3} &=    \begin{pmatrix}
                                    			m_2^2 + \frac{v^2}{2} Y_2 & \frac{v^2}{2} Y_{2\tilde{2}} & 0 \\ 
                                    			\frac{v^2}{2} Y_{2\tilde{2}}^* & m_{\tilde{2}}^2 + \frac{v^2}{2} \left( Y_{\tilde{2}} + Y_{\tilde{2}\tilde{2}} \right) & -v A_{\tilde{2}3} \\ 
                                    			0 & - v A_{\tilde{2}3}^* & m_3^2 + \frac{v^2}{2} \left(Y_3 + Y_{33} \right)
                                			\end{pmatrix}  \,, \\
        \mathbbm{M}^{-4/3} &=    \begin{pmatrix}
                                   			m_{\tilde{1}}^2 + \frac{v^2}{2} Y_{\tilde{1}} & \frac{v^2}{\sqrt{2}}Y_{\tilde{1}3}^*\\ 
                                    			\frac{v^2}{\sqrt{2}} Y_{\tilde{1}3} & m_3^2 + \frac{v^2}{2}\left(Y_3 - Y_{33}\right) \\ 
                                			\end{pmatrix}   \,, \\
        \mathbbm{M}^{+5/3} &= m_2^2 + \frac{v^2}{2} \left(Y_2 + Y_{22} \right) \,.
    \end{aligned}
\end{align}
They can be diagonalized by rotating the LQs with unitary matrices
\begin{equation}
    W^q \Phi^q \equiv \hat{\Phi}^q \,,
\end{equation}
in order to arrive at the physical basis with digonal mass matrices. $W^q$ can be calculated perturbatively to arbitrary order in $v$~\cite{Crivellin:2020ukd, Rosiek:2015jua}. At $\mathcal{O}(v^2)$ one finds
\begin{align}
    \begin{aligned}
        W^{-1/3}& \approx \begin{pmatrix}
        1\!-\!\frac{v^2|A_{1\tilde 2}|^2}{4 (m_{1\tilde{2}}^2)^2}& \frac{v A_{1\tilde 2}^{*}}{\sqrt{2} m_{1\tilde{2}}^2} & \frac{v^2(Y_{13}m_{1 \tilde{2}}^2+A_{1\tilde 2}^{*}A_{\tilde 23})}{2 m_{13}^2 m_{1\tilde{2}}^2}\\
        \frac{-v A_{1\tilde 2}}{\sqrt{2} m_{1\tilde{2}}^2} & 1\!-\!\frac{v^2}{4}\left(\!\frac{|A_{1\tilde 2}|^2}{(m_{1\tilde{2}}^2 )^2}\!+\!\frac{|A_{\tilde 23}|^2}{(m_{3 \tilde{2}}^2)^2}\!\right)&
        \frac{-v A_{\tilde 23}}{\sqrt{2} m_{3\tilde{2}}^2}\\
        \frac{-v^2(Y_{13}^{*}(m_{3\tilde{2}}^2)+A_{1\tilde 2}A_{\tilde 23}^{*})}{2 m_{13}^2 m_{3\tilde{2}}^2}&\frac{v A_{\tilde 23}^{*}}{\sqrt{2} m_{3\tilde{2}}^2}& 1\!-\!\frac{v^2 |A_{\tilde 23}|^2}{4 (m_{3\tilde{2}}^2)^2}
        \end{pmatrix}\,,\\ \rule{0pt}{3em}
W^{+2/3}& \approx \begin{pmatrix}
        1& \frac{v^2 Y_{2\tilde 2}}{2 m_{2 \tilde{2}}^2} & 0\\
        \frac{-v^2 Y_{2\tilde 2}^{*}}{2 m_{2\tilde{2}}^2} & 1\!-\!\frac{v^{2}|A_{\tilde 23}|^2}{2 (m_{3 \tilde{2}}^2 )^2} & \frac{-vA_{\tilde 23}}{m_{\tilde{2}3}^2}\\
        0 & \frac{vA_{\tilde 23}^{*}}{m_{\tilde{2}3}^2} & 1-\frac{v^2|A_{\tilde 23}|^2}{2 (m_{3 \tilde{2}}^2 )^2 }
        \end{pmatrix}\,, \,W^{-4/3}\approx\begin{pmatrix}
        1 & \frac{v^2 Y_{\tilde 13}^{*}}{\sqrt{2}m_{\tilde{1}3}^2}\\
        \frac{-v^2 Y_{\tilde 13}}{\sqrt{2} m_{\tilde{1}3}^2}& 1
        \end{pmatrix}\,, \\
W^{+5/3} &= 1 \,,
    \end{aligned}
    \label{eq:Wmatrices}
\end{align}
where $m^2_{ab} \equiv m_a^2 - m_b^2$ for $a,b \in \left \{1, \tilde{1}, 2, \tilde{2}, 3 \right \}$. 
The diagonal mass matrices 
\begin{equation}
 \hat{\mathbbm{M}}^q \equiv W^q \mathbbm{M}^q \left( W^q \right)^\dagger 
\end{equation}
then read~\cite{Crivellin:2020ukd}
\begin{equation}
\begin{aligned}
\hat{\mathbbm{M}}^{-1/3}&\approx \text{diag}\Bigg(m_{1}^2+\frac{v^{2}}{2}\left(\!Y_{1}\!-\!\frac{|A_{1\tilde 2}|^2}{m_{\tilde{2}1}^2}\!\right) , \, m_{\tilde{2}}^2+\frac{v^{2}}{2}\left(\!Y_{\tilde{2}}\!+\!\frac{|A_{1\tilde 2}|^2}{m_{\tilde{2}1}^2}\!+\!\frac{|A_{\tilde 23}|^2}{m_{\tilde{2}3}^2}\!\right),\, \\ 
 & \phantom{\approx \text{diag}\Bigg(\,} m_{3}^2+\frac{v^{2}}{2}\left(\!Y_{3}\!-\!\frac{|A_{\tilde 23}|^2}{m_{\tilde{2}3}^2}\right)\Bigg)\,, \\
\hat{\mathbbm{M}}^{+2/3} &\approx \text{diag}\Bigg(m_{2}^2+\frac{v^{2}}{2}Y_{2} , \, m_{\tilde{2}}^2 +\frac{v^{2}}{2}\left(Y_{\tilde{2}}+\!Y_{\tilde{2}\tilde{2}}\!+\!\frac{2|A_{\tilde 23}|^2}{m_{\tilde{2}3}^2}\! \right),\, \\ 
 & \phantom{\approx \text{diag}\Bigg(\,} m_{3}^2+\frac{v^{2}}{2}\left(\!Y_{3} + Y_{33} \!-\!\frac{2|A_{\tilde 23}|^2}{m_{\tilde{2}3}^2}\!\right)
\Bigg)\,, \\
\hat{\mathbbm{M}}^{-4/3}&\approx \text{diag}\left(m_{\tilde{1}}^2+\frac{v^{2}}{2}Y_{\tilde{1}}, \, m_{3}^2+\frac{v^{2}}{2}\left(Y_{3} - Y_{33}\right)\right)\,, \\
\hat{\mathbbm{M}}^{+5/3}&=m_{2}^2+\frac{v^2}{2}\left(Y_{2}+Y_{22}\right)\,,
\end{aligned}
\label{eq:LQmassMatrices}
\end{equation}
up to order $v^2$. \\

The interactions of LQs with the Higgs field $h$ are determined by
\begin{align}
    \begin{aligned}
    \mathcal{L}_{2\Phi} &\supset \phantom{+} \sum_{q \in Q} \left( \Phi^{q\dagger} \mathbb{\Gamma}^q \Phi^q \right)h + \frac{1}{2} \sum_{q \in Q}  \left( \Phi^{q\dagger} \mathbb{\Lambda}^q \Phi^q \right)hh \\
    & = \phantom{+} \sum_{q \in Q} \left( \hat{\Phi}^{q\dagger} \hat{\mathbb{\Gamma}}^q \hat{\Phi}^q \right)h + \frac{1}{2} \sum_{q \in Q} \left( \hat{\Phi}^{q\dagger} \hat{\mathbb{\Lambda}}^q \hat{\Phi}^q \right)hh \,, \\
    \end{aligned}
\end{align}
with
\begin{equation*}
\begin{aligned}
        \mathbb{\Gamma}^{-1/3} &= 
                -v
                \begin{pmatrix}
                   Y_1  & \frac{A_{1\tilde{2}}^{*}}{\sqrt{2}v} & Y_{13}\\ 
                   \frac{A_{1\tilde{2}}}{\sqrt{2}v}  & Y_{\tilde{2}} & \frac{A_{\tilde{2}3}}{\sqrt{2}v}\\ 
                    Y_{13}^{*} & \frac{A^{*}_{\tilde{2}3}}{\sqrt{2}v} & Y_{3}
                \end{pmatrix}\,, 
         &\mathbb{\Gamma}^{+2/3} &= 
                \renewcommand{\arraystretch}{1.5}
                -v
                \begin{pmatrix}
                   Y_2  & Y_{2\tilde{2}} & 0\\ 
                  Y^{*}_{2\tilde{2}}  & Y_{\tilde{2}} + Y_{\tilde2 \tilde2} & -\frac{A_{23}}{v}\\ 
                    0 & -\frac{A^{*}_{23}}{v} & Y_3 + Y_{33}
                \end{pmatrix}\,, \\ \rule{0pt}{3em}
\end{aligned}
\end{equation*}
\begin{equation}
\begin{aligned}
        \mathbb{\Gamma}^{-4/3} &= 
                        \renewcommand{\arraystretch}{1.5}
                        -v
                        \begin{pmatrix}
                          Y_{\tilde{1}}  & \sqrt{2} Y^{*}_{\tilde{1}3} \\ 
                          \sqrt{2} Y_{\tilde{1}3}  & Y_3 - Y_{33} \\
                        \end{pmatrix}\,, 
        &\mathbb{\Gamma}^{+5/3} &= -v \Big(Y_2 + Y_{22} \Big) \,, \\
        \mathbb{\Lambda}^{-1/3} &= 
                -\phantom{v}\begin{pmatrix}
                   Y_1  & 0 & Y_{13}\\ 
                   0  & Y_{\tilde{2}} & 0\\ 
                    Y_{13}^{*} & 0 & Y_3
                \end{pmatrix} \,,
        &\mathbb{\Lambda}^{+2/3} &=
                -\phantom{v} \begin{pmatrix}
                   Y_2 & Y_{2\tilde{2}} & 0\\ 
                  Y^{*}_{2\tilde{2}}  & Y_{\tilde{2}} + Y_{\tilde2 \tilde2} & 0\\ 
                    0 & 0 & Y_3 + Y_{33}
                \end{pmatrix} \,, \\ \rule{0pt}{3em}
        \mathbb{\Lambda}^{-4/3} &=
                -\phantom{v}\begin{pmatrix}
                   Y_{\tilde{1}}  & \sqrt{2} Y^{*}_{\tilde{1}3} \\ 
                  \sqrt{2} Y_{\tilde{1}3}  & Y_3 - Y_{33} \\
                \end{pmatrix}\,, 
        &\mathbb{\Lambda}^{+5/3} &= -\phantom{v}\Big(Y_2 + Y_{22}\Big)\,, \\
    \end{aligned}
\end{equation}
and
\begin{align}
	\begin{aligned}
		\hat{\mathbb{\Gamma}}^q &= W^q \mathbb{\Gamma}^q \left(W^{q}\right)^\dagger \text{ for } q \in Q \,, \\
		\hat{\mathbb{\Lambda}}^q &= W^q \mathbb{\Lambda}^q \left(W^{q}\right)^\dagger \text{ for } q \in Q \,.
	\end{aligned}
\end{align} \\
The explicit expressions for these coupling matrices up to $\mathcal{O}(v^2)$ are given in Ref.~\cite{Crivellin:2020ukd}. \\

When listing the Feynman rules for interactions involving LQs of the same charge, we will have to relate the LQ label $a \in \left \{ 1, \tilde{1}, 2, \tilde{2}, 3 \right \}$ to the position in $\Phi^q$. For this we use the coefficients
\begin{equation}
    n_a^q \text{ with } q \in \left \{-\frac{4}{3},-\frac{1}{3}, +\frac{2}{3}, +\frac{5}{3} \right \} \text{ and } a \in \{1, \tilde{1}, 2, \tilde{2}, 3 \}.
    \label{eq:nqi}
\end{equation}
Explicitly, we have
\renewcommand{\arraystretch}{1.5}
\begin{center}
    \begin{tabular}{ r|c c c c c c|}
     $n_a^q$ & $a = 1$ & $a = \tilde{1}$ & $a = 2$ & $a = \tilde{2}$ & $a = 3$ \\
    \hline
       $q = -\frac{1}{3}$ & 1 & 0 & 0 & 2 & 3 \\
       $q = +\frac{2}{3}$ & 0 & 0 & 1 & 2 & 3 \\
       $q = -\frac{4}{3}$ & 0 & 1 & 0 & 0 & 2 \\
       $q = +\frac{5}{3}$ & 0 & 0 & 1 & 0 & 0 \\
    \end{tabular}\,.
\end{center}
Whenever $q$ is clear from the context, we write $n^q_a$ as $n_a$ to simplify the notation.

\subsection{Kinetic Terms and Interactions with SM Gauge Bosons ($\mathcal{L}_{\text{kin}}$)}
\label{sec:LQkin}
The kinetic terms for the LQs as well as their couplings to the SM gauge bosons are presented in this section. The same conventions as for the SM are used. For the three-dimensional adjoint representation of $SU(2)_L$, the $I^{\text{th}}$ generator ($I \in \{1,2,3\}$) has the $3\times3$ matrix form  $(\tilde{\epsilon}_I)_{JK} \equiv i \epsilon_{IJK}$, where $\epsilon_{IJK}$ is the three-dimensional Levi-Civita tensor.  We use the convention $\epsilon_{123} = +1$.
\begin{equation}
    \renewcommand{\arraystretch}{1.5}
    \centering
    \begin{tabular}{c | c c}
          & $\mathcal{L}_{\text{kin}}$ & $D_\mu$ \\ \hline \rule{0pt}{2em}
    	
    	$\Phi_1$ &  $\left(\left( D_\mu \Phi_1 \right)^{\dagger}D^\mu \Phi_1 \right)$ & $ \partial_\mu + \frac{i}{3}g_1 B_\mu - i g_s \frac{\lambda_\alpha}{2} G^\alpha_\mu$\\ \rule{0pt}{2em}

    	$\Phi_{\tilde{1}}$ & $\left(\left( D_\mu \Phi_{\tilde{1}} \right)^{\dagger}D^\mu \Phi_{\tilde{1}} \right)$ & $ \partial_\mu + \frac{4i}{3}g_1  B_\mu - i g_s \frac{\lambda_\alpha}{2} G^\alpha_\mu $\\ \rule{0pt}{2em}

    	$\Phi_{2}$ &  $\left(\left( D_\mu \Phi_{2} \right)^{\dagger}D^\mu \Phi_{2} \right)$ & $ \partial_\mu - \frac{7i}{6}g_1 B_\mu - i g_2 \frac{\sigma_I}{2} W_\mu^{I} - i g_s \frac{\lambda_\alpha}{2} G^\alpha_\mu $\\ \rule{0pt}{2em}

    	$\Phi_{\tilde{2}}$  & $\left(\left( D_\mu \Phi_{\tilde{2}} \right)^{\dagger}D^\mu \Phi_{\tilde{2}}\right)$ & $ \partial_\mu - \frac{i}{6}g_1 B_\mu - i g_2 \frac{\sigma_I}{2} W_\mu^{I} - i g_s \frac{\lambda_\alpha}{2} G^\alpha_\mu$\\ \rule{0pt}{2em}

    	$\Phi_{3}$ &  $\left(\left( D_\mu \Phi_{3} \right)^{\dagger}D^\mu \Phi_{3}\right)$ & $ \partial_\mu + \frac{i}{3}g_1 B_\mu - i g_2 \tilde{\epsilon}_I W_\mu^{I} - i g_s \frac{\lambda_\alpha}{2} G^\alpha_\mu$\\
    \end{tabular}
\end{equation}
By definition, the $\Phi^q$ couple to the photon with the electric charge $q$. The couplings to the electroweak SM gauge bosons are given by 
\begin{align}
    \begin{aligned}
        \mathcal{L}_\text{kin} &\supset \phantom{+} i\dfrac{e}{s_w c_w} \sum_{q \in Q}\left(\Phi^{q\dagger}\mathbbm{Z}^{q} \overset{\leftrightarrow}{\partial^\mu} \Phi^{q} \right) Z_{\mu} \\ 
        &\phantom{12}+ i\dfrac{e}{s_w} \sum_{q \in Q\backslash \left\{+\frac{5}{3}\right\}} \left( \left(\Phi^{q+1}\right)^\dagger \mathbbm{W}_{q}^{q+1} \overset{\leftrightarrow}{\partial^\mu} \Phi^q\right)W^+_\mu \,,
    \end{aligned}
\end{align}
where $\Phi^\dagger \overset{\leftrightarrow}{\partial^\mu} \Phi \equiv \Phi^\dagger \partial^\mu \Phi - \left(\partial^\mu \Phi^\dagger\right) \Phi$, $Q = \left\{ -\frac{4}{3}, -\frac{1}{3}, +\frac{2}{3}, +\frac{5}{3} \right\}$ and
\begin{equation*}
    \begin{aligned}
        \mathbbm{Z}^{-1/3} &= 
                \renewcommand{\arraystretch}{1.0}
                \begin{pmatrix}
                    \dfrac{1}{3}s_w^2 & 0 & 0 \\ 
                    0 & -\dfrac{1}{2} + \dfrac{1}{3}s_w^2 & 0\\ 
                    0 & 0 & \dfrac{1}{3}s_w^2
                \end{pmatrix} \,,
        &\mathbbm{Z}^{-4/3} &= 
                        \renewcommand{\arraystretch}{1.0}
                        \begin{pmatrix}
                            \dfrac{4}{3}s_w^2 & 0\\ 
                            0 & -1 + \dfrac{4}{3}s_w^2
                        \end{pmatrix} \,, \\
        \mathbbm{Z}^{+2/3} &= 
                \renewcommand{\arraystretch}{1.0}
                \begin{pmatrix}
                    -\dfrac{1}{2} - \dfrac{2}{3}s_w^2 & 0 & 0 \\
                    0 & \dfrac{1}{2} - \dfrac{2}{3}s_w^2 & 0\\ 0 & 0 & 1 - \dfrac{2}{3}s_w^2
                \end{pmatrix} \,,
        &\mathbbm{Z}^{+5/3} &= 
                        \left(\dfrac{1}{2} - \dfrac{5}{3} s_w^2\right)\,,
  \end{aligned}
\end{equation*}
\begin{equation}
  \begin{aligned}
        \mathbbm{W}^{+2/3}_{-1/3} &=
                \renewcommand{\arraystretch}{1.0}
                \begin{pmatrix}
                    0 & 0 & 0 \\ 
                    0 & 1 & 0\\ 
                    0 & 0 & - \sqrt{2}
                \end{pmatrix} \,, &
        \mathbbm{W}^{-1/3}_{-4/3} &=
                \renewcommand{\arraystretch}{1.0}
                \begin{pmatrix}
                    0 & 0  \\ 
                    0 & 0 \\
                    0 & \sqrt{2}
                \end{pmatrix} \,, &
        \mathbbm{W}^{+5/3}_{+2/3} &= 
                \renewcommand{\arraystretch}{1.0}
                \begin{pmatrix}
                   1 & 0 & 0 \\
                \end{pmatrix} \,.
    \end{aligned}
    \label{eq:LQchargeMatrices}
\end{equation}
Such interactions were first considered in Ref.~\cite{Blumlein:1992ej}. Expressed in the mass basis for the LQs, they take the form
\begin{align}
    \begin{aligned}
        \mathcal{L}_\text{kin} &\supset +i\dfrac{e}{s_w c_w} \sum_{q \in Q}\left(\hat{\Phi}^{q\dagger}\hat{\mathbbm{Z}}^{q} \overset{\leftrightarrow}{\partial^\mu} \hat{\Phi}^{q} \right) Z_{\mu} \\ 
        &\phantom{12}+ i\dfrac{e}{s_w} \sum_{q \in Q\backslash\{+\frac{5}{3}\}} \left( \left(\hat{\Phi}^{q+1}\right)^\dagger \hat{\mathbbm{W}}_{q}^{q+1} \overset{\leftrightarrow}{\partial^\mu} \hat{\Phi}^q\right)W^+_\mu \,.
    \end{aligned}
\end{align}
with
\begin{align}
	\begin{aligned}
		\hat{\mathbbm{Z}}^q &\equiv W^q \mathbbm{Z}^q \left(W^{q}\right)^\dagger \text{ for } q \in Q \,, \\
		\hat{\mathbbm{W}}_q^{q+1} &\equiv W^{q+1} \mathbbm{W}_q^{q+1} \left(W^{q}\right)^\dagger \text{ for } q \in Q \backslash \left \{ +\frac{5}{3} \right \}\,. \\
	\end{aligned}
\end{align}
Explicitly, the matrices $\hat{\mathbbm{Z}}^q, \hat{\mathbbm{W}}_q^{q+1}$ are given by
\begin{align}
    \begin{aligned}
        \hat{\mathbbm{Z}}^{-1/3}& \approx \begin{pmatrix}
        \frac{1}{3}s_w^2 - \frac{v^2 \left|A_{1\tilde{2}} \right|^2}{4 \left(m^2_{1\tilde{2}}\right)^2} & -\frac{v A_{1\tilde{2}}^*}{2\sqrt{2} m_{1\tilde{2}}^2}& -\frac{v^2 A_{\tilde{2}3} A^*_{1 \tilde{2}}}{4m^2_{1\tilde{2}} m_{3\tilde{2}}^2} \\
        -\frac{v A_{1\tilde{2}}}{2 \sqrt{2} m_{1\tilde{2}}^2} & -\frac{1}{2} + \frac{1}{3} s_w^2 + \frac{v^2}{4}\left(\frac{\left | A_{1 \tilde{2}}\right |^2}{\left(m_{1\tilde{2}}^2\right)^2} +  \frac{\left | A_{\tilde{2}3}\right |^2}{\left(m_{3\tilde{2}}^2\right)^2} \right)& - \frac{v A_{\tilde{2}3}}{2 \sqrt{2} m_{3 \tilde{2}}^2} \\
        -\frac{v^2 A_{1\tilde{2}} A_{\tilde{2}3}^*}{4 m_{1 \tilde{2}}^2  m_{3\tilde{2}}^2} & - \frac{v A_{\tilde{2}3}^*}{2 \sqrt{2} m_{3 \tilde{2}}^2}& \frac{1}{3}s_w^2 - \frac{v^2 \left| A_{\tilde{2}3} \right|^2}{4 \left(m_{3 \tilde{2}}^2 \right)^2} \\
        \end{pmatrix}\,,\\ \rule{0pt}{3em}
\hat{\mathbbm{Z}}^{+2/3}& \approx \begin{pmatrix}
        -\frac{1}{2} - \frac{2}{3}s_w^2 & \frac{v^2Y_{2\tilde{2}}}{2 m_{2 \tilde{2}}^2} & 0 \\
        \frac{v^2Y^*_{2\tilde{2}}}{2 m_{2 \tilde{2}}^2} & \frac{1}{2}-\frac{2}{3}s_w^2 + \frac{ v^2 \left| A_{\tilde{2}3}\right|^2}{2\left(m_{\tilde{2}3}^2\right)^2} & -\frac{v A_{\tilde{2}3}}{2 m_{\tilde{2}3}^2} \\
        0 & -\frac{v A_{\tilde{2}3}^*}{2 m_{\tilde{2}3}^2} & 1 - \frac{2}{3} s_w^2 -  \frac{v^2 \left| A_{\tilde{2}3}\right|^2}{2 \left(m_{\tilde{2}3}^2 \right)^2} \\
        \end{pmatrix}\,, \\ \rule{0pt}{1em}
        \hat{\mathbbm{Z}}^{-4/3}&\approx\begin{pmatrix}
        \frac{4}{3}s_w^2 & -\frac{v^2 Y_{\tilde{1}3}^*}{\sqrt{2} m_{\tilde{1}3}^2}\\
        -\frac{v^2 Y_{\tilde{1}3}}{\sqrt{2}m_{\tilde{1}3}^2} & -1 + \frac{4}{3}s_w^2\\
        \end{pmatrix}\,, ~\hat{\mathbbm{Z}}^{+5/3} = \left(\frac{1}{2} - \frac{5}{3}s_w^2 \right)\,, \\
        \hat{\mathbbm{W}}^{+2/3}_{-1/3}& \approx \renewcommand*{\arraystretch}{2.0}  \begin{pmatrix}
	0 & \frac{v^2 Y_{2 \tilde{2}}}{2 m^2_{2 \tilde{2}}} & 0 \\
	\frac{v A_{1 \tilde{2}}}{\sqrt{2} m_{1 \tilde{2}}^2}& 1+ \frac{v^2}{4} \left( \frac{\left | A_{\tilde{2} 3} \right|^2}{\left(m_{\tilde{2} 3}^2\right)^2} -\frac{\left | A_{1\tilde{2}} \right|^2}{\left(m_{1\tilde{2}}^2\right)^2} \right)&  \frac{v A_{\tilde{2}3}}{\sqrt{2} m_{\tilde{2}3}^2}\\
	\frac{v^2}{\sqrt{2} m_{13}^2} \left(\frac{A_{1\tilde{2}}A_{\tilde{2}3}^*}{m^2_{\tilde{2}3}} - Y_{13}^* \right)&  0 & -\sqrt{2} + \frac{v^2\left| A_{\tilde{2}3}\right|^2}{2\sqrt{2} \left(m_{3 \tilde{2}}^2\right)^2}\\
        \end{pmatrix}\,,\\
        \hat{\mathbbm{W}}^{-1/3}_{-4/3}&\approx\begin{pmatrix}
        0 & \frac{v^2}{\sqrt{2} m_{13}^2} \left(Y_{13} + \frac{A_{\tilde{2}3}A_{1 \tilde{2}}^*}{m_{1\tilde{2}}^2} \right) \\
        0 & \frac{vA_{\tilde{2}3}}{m_{\tilde{2}3}^2}\\
        \frac{v^2 Y_{\tilde{1}3}}{m_{\tilde{1}3}^2} & \sqrt{2} - \frac{v^2 \left|A_{\tilde{2}3} \right |^2}{2 \sqrt{2} \left(m_{\tilde{2}3}^2 \right)^2}\end{pmatrix}\,, ~
\hat{\mathbbm{W}}^{+5/3}_{+2/3} \approx \begin{pmatrix}
       1 & -\frac{v^2 Y_{2\tilde{2}}}{2 m_{2\tilde{2}}^2} & 0 \end{pmatrix}\,,
    \end{aligned}
\end{align}
up to $\mathcal{O}(v^2)$, where again $m^2_{ab} \equiv m_a^2 - m_b^2$ for $a,b \in \left \{1, \tilde{1}, 2, \tilde{2}, 3 \right \}$.  \\

\clearpage
\subsection{Interactions with SM Fermions ($\mathcal{L}_{\text{f}}$)}
The interactions between the five scalar LQs and the SM fermions are listed below. $Y_{a}^{AB}{}$ with $a \in \{1, \tilde1, 2, \tilde2, 3 \}$ and $A,B \in \{L, R\}$ are arbitrary complex $3 \times 3$ matrices coupling LQs to a quark and a lepton. $Y_a^{Q,AA}$ with $a \in \left \{ 1, \tilde{1}, 2, \tilde{2}, 3 \right \}$ and $A \in \left \{ L, R\right \}$ couple LQs to two quarks. $Y_1^{Q, LL}$ is symmetric in flavour space (i.e. $Y_{1, ij}^{Q, LL}$ = $Y_{1, ji}^{Q, LL}$ ), $Y_{\tilde{1}}^{Q, RR}$ and $Y_{3}^{Q, LL}$ are anti-symmetric and $Y_{1}^{Q, RR}$ is again an arbitrary complex matrix. We omit the color indices whenever they just involve trivial contractions. 
\begin{equation}
    \renewcommand{\arraystretch}{1.5}
    \centering
    \begin{tabular}{c | c} 
         & $\mathcal{L}_{f}$ \\ \hline \rule{0pt}{2em}
         
    	\multirow{3}{*}{$\Phi_1$} & $Y_{1, ij}^{RR}\, \bar{u}^{\prime c}_i\ell^\prime_{j}\Phi_{1}^{\dagger} +Y_{1, ij}^{LL}\, \left(\bar{Q}_{i}^{\prime c \intercal} i\sigma_{2}L^\prime_{j}\right)\Phi_{1}^{\dagger}+\text{h.c.}$ \\ 
	\multirow{3}{*} &  $Y_{1, ij}^{Q, LL}\, \left(\bar{Q}^{\prime c \intercal}_{i, c_1} i\sigma_2 Q^\prime_{j, c_2}\right) \Phi_{1, c_3} \epsilon^{c_1 c_2 c_3} +\text{h.c.}$ \\ 
	\multirow{3}{*} &  $Y_{1, ij}^{Q, RR} \bar{u}^{\prime c}_{i, c1} d^\prime_{j, c_2} \Phi_{1, c_3} \epsilon^{c_1 c_2 c_3} +\text{h.c.}$ \\ \rule{0pt}{2em}

    	\multirow{2}{*}{$\Phi_{\tilde{1}}$} & $Y^{RR}_{\tilde{1}, ij}\, \bar{d}^{\prime c}_{i}\ell^\prime_{j}\Phi_{\tilde{1}}^{\dagger} +\text{h.c.}$ \\
	\multirow{2}{*} & $Y^{Q, RR}_{\tilde{1}, ij}\, \bar{u}^{\prime c}_{i, c_1} u^\prime_{j, c_2}\Phi_{\tilde{1}, c_3} \epsilon^{c_1 c_2 c_3} +\text{h.c.}$ \\ \rule{0pt}{2em}

    	$\Phi_{2}$  & $Y_{2, ij}^{RL}\, \left(\Phi_{2}^\intercal \bar{u}^\prime_i i\sigma_2 L^\prime_{j} \right)+Y_{2, ij}^{LR}\, \left(\bar{Q}^{\prime \intercal}_i \ell^\prime_j\Phi_{2}\right)+\text{h.c.}$\\ \rule{0pt}{2em}

    	$\Phi_{\tilde{2}}$ & $Y_{\tilde{2}, ij}^{RL}\, \left(\Phi_{\tilde{2}}^\intercal \bar{d}^\prime_{i} i\sigma_2 L^\prime_{j}\right)+\text{h.c.}$ \\ \rule{0pt}{2em}
	
	\multirow{2}{*}{$\Phi_{3}$}  & $Y_{3, ij}^{LL}\, \left(\bar{Q}^{\prime c \intercal}_{i} i\sigma_{2}\left(\sigma\cdot\Phi_{3}\right)^{\dagger}L^\prime_{j}\right)+\text{h.c.}$ \\
	\multirow{2}{*} &  $Y^{Q, LL}_{3, ij}\, \left(\bar{Q}^{\prime c \intercal}_{i, c_1} i\sigma_2 \left(\sigma \cdot \Phi_{3, c_3}\right)Q^\prime_{j, c_2}\right) \epsilon^{c_1 c_2 c_3} +\text{h.c.}$ \\
    \end{tabular}
    \label{eq:LQ_yukawa}
\end{equation}
Again, we use brackets to indicate the $SU(2)_L$ contractions. Note that we stated the Lagrangian above before EW symmetry breaking using the weak eigenstates of the fermions (indicated by the prime). When going to the mass eigenbasis (after EW symmetry breaking), CKM matrix elements enter interactions involving left-handed down-type quarks according to Eq.~(\ref{eq:massEigenstates}). 
%
The charge-conjugate of a fermion field $\Psi$ is denoted as $\Psi^c$, where
\begin{equation}
\begin{aligned}
	\Psi^c &= C \bar{\Psi}^\intercal \,, \\
	\bar{\Psi}^c &= - \Psi^\intercal C^{-1}\,,
\end{aligned}
\label{eq:Cmatrix}
\end{equation}
with $C$ the charge conjugation matrix. A detailed description of our treatment of charge-conjugate SM fermions is given in \ref{sec:appendix}.

\subsection{Triple LQ Interactions ($\mathcal{L}_{3 \Phi}$)}
Let us now turn to the LQ-LQ-LQ(-Higgs) interactions. In order to get a $SU(3)_c$ singlet, either three LQs or three anti-LQs have to be combined, since 
\begin{align}
    \begin{aligned}
    3 \otimes 3 \otimes 3 &= 1 \oplus 8 \oplus 8 \oplus 10 \\
    \overline{3} \otimes \overline{3} \otimes \overline{3} &= 1 \oplus 8 \oplus 8 \oplus \overline{10}
    \end{aligned}
\end{align}
are the only tensor products of three SU(3) (anti-)triplets $3, \overline{3}$ that contain a singlet. Regarding the hypercharge $Y$, one finds the possible combinations listed in Table~\ref{tab:triple}. 
\begin{table}[t]
    \renewcommand{\arraystretch}{1.2}
    \begin{center}
        \begin{tabular}{rr|c c c c c c c }
            & & \multicolumn{7}{c}{LQ Fields (and their $Y$)}\\
            & & $\Phi_1$ & $\Phi_{\tilde{1}}$ & $\Phi_2$ & $\Phi_{\tilde{2}}$ & $\Phi_{3}$ & $H$ & $H^\dagger$ \\ [0.5ex] 
            & & $-\frac{1}{3}$ & $-\frac{4}{3}$ & $\frac{7}{6}$ & $\frac{1}{6}$ & $-\frac{1}{3}$ & $\frac{1}{2}$ & $-\frac{1}{2}$ \\ [0.7ex]
            \cline{1-9}
            \multicolumn{1}{ c  }{\multirow{12}{*}{ \rotatebox[origin=c]{90}{Interaction Terms}} } & \multicolumn{1}{ r|  }{$Y_{112}$ } & 2 & 0 & 1 & 0 & 0 & 0 & 1 \\
            \multicolumn{1}{ c  }{\multirow{12}{*}} & \multicolumn{1}{ r|  }{$Y_{11\tilde2}$ } & 2 & 0 & 0 & 1 & 0 & 1 & 0 \\
            \multicolumn{1}{ c  }{\multirow{12}{*}} & \multicolumn{1}{ r|  }{$Y_{1\tilde12}$ } & 1 & 1 & 1 & 0 & 0 & 1 & 0 \\
            \multicolumn{1}{ c  }{\multirow{12}{*}} & \multicolumn{1}{ r|  }{$Y_{123}$ } & 1 & 0 & 1 & 0 & 1 & 0 & 1 \\
            \multicolumn{1}{ c  }{\multirow{12}{*}} & \multicolumn{1}{ r|  }{$A_{1\tilde2\tilde2}$ } & 1 & 0 & 0 & 2 & 0 & 0 & 0 \\
            \multicolumn{1}{ c  }{\multirow{12}{*}} & \multicolumn{1}{ r|  }{$Y_{1\tilde23}$ } & 1 & 0 & 0 & 1 & 1 & 1 & 0 \\
            \multicolumn{1}{ c  }{\multirow{12}{*}} & \multicolumn{1}{ r|  }{$A_{\tilde12\tilde2}$ } & 0 & 1 & 1 & 1 & 0 & 0 & 0 \\
            \multicolumn{1}{ c  }{\multirow{12}{*}} & \multicolumn{1}{ r|  }{$Y_{\tilde123}$ } & 0 & 1 & 1 & 0 & 1 & 1 & 0 \\
            \multicolumn{1}{ c  }{\multirow{12}{*}} & \multicolumn{1}{ r|  }{$Y_{233}$ } & 0 & 0 & 1 & 0 & 2 & 0 & 1 \\
            \multicolumn{1}{ c  }{\multirow{12}{*}} & \multicolumn{1}{ r|  }{$Y_{\tilde2 \tilde2 \tilde2}$ } & 0 & 0 & 0 & 3 & 0 & 0 & 1 \\
            \multicolumn{1}{ c  }{\multirow{12}{*}} & \multicolumn{1}{ r|  }{$A_{\tilde2 \tilde2 3}$ } & 0 & 0 & 0 & 2 & 1 & 0 & 0 \\
            \multicolumn{1}{ c  }{\multirow{12}{*}} & \multicolumn{1}{ r|  }{$Y_{\tilde2 33}$ } & 0 & 0 & 0 & 1 & 2 & 1 & 0 \\
        \end{tabular}
    \end{center}
\caption{Collection of all interaction terms with three LQ fields that result in a vanishing total weak hypercharge $Y$. The numbers in the table indicate the number of the corresponding LQ fields interacting in the vertex. The same combinations with all fields replaced by their anti-fields correspond to the Hermitian conjugates and are also valid solutions.  }
\label{tab:triple}
\end{table}
However, not all of these combinations can be implemented. Since the Lagrangian is symmetric under the exchange of identical bosons and the color terms are totally anti-symmetric, the $SU(2)_L$ part needs to be anti-symmetric for identical bosons as well. This is not satisfied by $Y_{112}, Y_{11\tilde{2}}, Y_{\tilde{2} \tilde{2} \tilde{2}}, A_{\tilde{2}\tilde{2}3}$ in Table~\ref{tab:triple}. Hence, the Lagrangian is composed of the terms~\cite{Kovalenko:2002eh, Klapdor-Kleingrothaus:2002rvk, Arnold:2012sd, Hambye:2017qix}
\begin{align}
    \begin{aligned}
        \mathcal{L}_{3 \Phi} 
        &= \phantom{12} A_{1\tilde{2}\tilde{2}} \ \Phi_{1, c_1} \left(\Phi_{\tilde{2}, c_2}^\intercal i\sigma_2 \Phi_{\tilde{2}, c_3}\right)
        + A_{\tilde{1}2\tilde{2}} \ \Phi_{\tilde{1}, c_1} \left(\Phi_{2, c_2}^\intercal i\sigma_2 \Phi_{\tilde{2}, c_3}\right) \\
        &\phantom{12}+ Y_{1 \tilde{1}2} \ \Phi_{1, c_1} \Phi_{\tilde{1}, c_2} \left( \Phi_{2, c_3}^\intercal i \sigma_2 H \right)
        + Y_{123} \ \Phi_{1, c_1} \left(H^\dagger \left( \sigma \cdot \Phi_{3, c_3} \right) \Phi_{2, c_2}\right) \\
        &\phantom{12}+ Y_{1\tilde{2}3} \ \Phi_{1, c_1} \left(\Phi_{\tilde{2}, c_2}^\intercal i \sigma_2 \left( \sigma \cdot \Phi_{3, c_3} \right) H\right)
        + Y_{\tilde{1}23} \ \Phi_{\tilde{1}, c_1} \left(\Phi_{2, c_2}^\intercal i \sigma_2 \left( \sigma \cdot \Phi_{3, c_3} \right) H\right)  \\
        &\phantom{12}+ Y_{233} \ \left(H^\dagger \sigma^I \Phi_{2, c_1}\right)\left( \Phi^J_{3, c_2} i \epsilon^{IJK} \Phi^K_{3, c_3}\right) \\
        &\phantom{12}+  Y_{\tilde{2}33} \ \left(\Phi_{\tilde{2}, c_1}^\intercal i \sigma_2 \sigma^I H\right) \left( \Phi^J_{3, c_2} i\epsilon^{IJK}\Phi^K_{3, c_3} \right) \\
        &\phantom{12}+ \text{h.c.} \,.& 
        \label{eq:LQ_triplet_interactions}
    \end{aligned}
\end{align}
The Levi-Civita tensors $\epsilon^{c_1 c_2 c_3}$ are omitted, such that i.e.~the first term would read $A_{1\tilde{2}\tilde{2}} \epsilon^{c_1c_2c_3} \ \Phi_{1, c_1} \left(\Phi_{\tilde{2}, c_2}^\intercal i\sigma_2 \Phi_{\tilde{2}, c_3}\right)$ in the full notation. 

\subsection{Quartic LQ Interactions ($\mathcal{L}_{4 \Phi}$)}
To get four-LQ interaction terms in the Lagrangian, one needs to combine two LQs and two anti-LQs, since
\begin{equation}
3 \otimes \overline{3} \otimes 3 \otimes \overline{3} = 1 \oplus 1 \oplus 8 \oplus 8 \oplus 8 \oplus 8 \oplus 10 \oplus \overline{10} \oplus 27
\label{eq:SU(3)_singlets}
\end{equation}
is the only tensor product of four $SU(3)_{c}$ fundamental representations $\left(3, \overline{3}\right)$ that contains a singlet. Trivial combinations of LQ fields that contribute to the four-LQ Lagrangian are of the type
\begin{equation}
\begin{aligned}
Y^{(1)}_{a} \left(\Phi^\dagger_{a, c_1} \Phi_{a, c_1}\right) \left( \Phi^\dagger_{a, c_2} \Phi_{a, c_2} \right)\,, \\
Y^{(1)}_{ab} \left(\Phi^\dagger_{a, c_1} \Phi_{a, c_1}\right) \left( \Phi^\dagger_{b, c_2} \Phi_{b, c_2} \right)\,,
\end{aligned}
\label{eq:obviouscombination}
\end{equation}
where $a \neq b \in \left\{1,\tilde{1}, 2, \tilde{2}, 3 \right\}$. All additional combinations that result in a vanishing total weak hypercharge $Y$ are given in Tab.~\ref{tab:quartic}. 
\begin{table}[t]
    \renewcommand{\arraystretch}{1.2}
    \begin{center}
        \begin{tabular}{rr|c c c c c }
            & & \multicolumn{5}{c}{LQ Fields (and their $Y$)}\\
            & & $\Phi_1,\Phi_1^\dagger$  & $\Phi_{\tilde{1}},\Phi_{\tilde{1}}^\dagger$ & $\Phi_2,\Phi_2^\dagger$ & $\Phi_{\tilde{2}},\Phi_{\tilde{2}}^\dagger$ & $\Phi_3,\Phi_3^\dagger$ \\ [0.5ex] 
            & & $-\frac{1}{3}, \frac{1}{3}$ & $-\frac{4}{3}, \frac{4}{3}$ & $\frac{7}{6}, -\frac{7}{6}$ & $\frac{1}{6}, -\frac{1}{6}$& $-\frac{1}{3}, \frac{1}{3}$ \\ [0.7ex]
            \cline{1-7}
            \multicolumn{1}{ c  }{\multirow{8}{*}{ \rotatebox[origin=c]{90}{Interaction Terms}} } & \multicolumn{1}{r|}{$Y_{1113}$ } & $1,2$ & $0,0$ & $0,0$ & $0,0$ & $1,0$ \\
            \multicolumn{1}{ c  }{\multirow{8}{*}} & \multicolumn{1}{ r|  }{$Y_{1\tilde{1}\tilde{1}3}$ } & $0,1$ & $1,1$ & $0,0$ & $0,0$ & $1,0$ \\
            \multicolumn{1}{ c  }{\multirow{8}{*}} & \multicolumn{1}{ r|  }{$Y_{1\tilde{1}\tilde{2}2}$ } & $0,1$ & $1,0$ & $1,0$ & $0,1$ & $0,0$ \\
            \multicolumn{1}{ c  }{\multirow{8}{*}} & \multicolumn{1}{ r|  }{$Y_{1223}$ } & $0,1$ & $0,0$ & $1,1$ & $0,0$ & $1,0$ \\
            \multicolumn{1}{ c  }{\multirow{8}{*}} & \multicolumn{1}{ r|  }{$Y_{1\tilde{2}\tilde{2}3}$ } & $0,1$ & $0,0$ & $0,0$ & $1,1$ & $1,0$ \\
            \multicolumn{1}{ c  }{\multirow{8}{*}} & \multicolumn{1}{ r|  }{$Y_{1313}$ } & $0,2$ & $0,0$ & $0,0$ & $0,0$ & $2,0$ \\
            \multicolumn{1}{ c  }{\multirow{8}{*}} & \multicolumn{1}{ r|  }{$Y_{1333}$ } & $0,1$ & $0,0$ & $0,0$ & $0,0$ & $2,1$ \\
            \multicolumn{1}{ c  }{\multirow{8}{*}} & \multicolumn{1}{ r|  }{$Y_{\tilde{1}\tilde{2}23}$ } & $0,0$ & $0,1$ & $0,1$ & $1,0$ & $1,0$ \\
        \end{tabular}
    \end{center}
\caption{Collection of all interaction terms with four LQ fields that result in a vanishing total weak hypercharge $Y$. We do not include the trivial combinations stated in Eq.~(\ref{eq:obviouscombination}). The numbers in the table indicate the number of the corresponding LQ fields in the interaction term. The same combinations with all fields replaced by their anti-fields (i.e. with the pairs listed in the table permuted) correspond to the Hermitian conjugates and are also valid solutions. }
\label{tab:quartic}
\end{table}
Taking into account $SU(2)_L$ invariance, the terms corresponding to $Y_{1113}$ and $Y_{1\tilde{1}\tilde{1}3}$ cannot be implemented, but besides $Y^{(1)}_{2}, Y^{(1)}_{\tilde{2}}, Y^{(1)}_{3}, Y^{(1)}_{2\tilde{2}}, Y^{(1)}_{23}$ and $Y^{(1)}_{\tilde{2}3}$, further independent terms with the same fields exist, originating from the different $SU(2)_L$ singlets that can be constructed. Regarding $SU(3)_c$, except for the case when two fields are identical, there are two possibilities to contract the color indices tracing back to the two singlets in Eq.~\ref{eq:SU(3)_singlets}. We distinguish them by different couplings $Y$ and $Y^\prime$ for full generality. This yields the Lagrangian
\begin{align*}
    \begin{aligned}
        \mathcal{L}_{4 \Phi} 
        = &\phantom{+1} \frac{1}{2}\sum_{a} Y^{(1)}_{a} \left( \Phi_{a, c_1}^\dagger \Phi_{a, c_1} \right)\left(\Phi_{a, c_2}^\dagger \Phi_{a, c_2}\right) \\
        &+ \frac{1}{2} \sum_{a = 2, \tilde{2}, 3} Y^{(3)}_{a}  \left( \Phi_{a, c_1}^\dagger \Phi_{a, c_2} \right) \left(\Phi_{a, c_2}^\dagger \Phi_{a, c_1}\right)\\
        &+ \frac{1}{2} Y^{(5)}_{3} \left( \Phi_{3, c_1}^{I\dagger} \Phi_{3, c_1}^{J}  \Phi_{3, c_2}^{I\dagger} \Phi_{3, c_2}^{J}\right) \\
        	&+\sum_{a \neq b} \left( Y^{(1)}_{ab} \ \delta_{c_1 c_2} \delta_{c_3 c_4} + Y^{\prime(1)}_{ab} \ \delta_{c_1 c_4} \delta_{c_2 c_3} \right)  \left( \Phi_{a, c_1}^\dagger \Phi_{a, c_2}\right) \left( \Phi_{b, c_3}^\dagger \Phi_{b, c_4} \right) \\
         &+\left( Y^{(3)}_{2\tilde{2}} \ \delta_{c_1 c_2} \delta_{c_3 c_4} + Y^{\prime(3)}_{2\tilde{2}} \ \delta_{c_1 c_4} \delta_{c_2 c_3} \right)  \left( \Phi_{2, c_1}^\dagger \Phi_{\tilde{2}, c_2}\right) \left( \Phi_{\tilde{2}, c_3}^\dagger \Phi_{2, c_4} \right) \\
      \end{aligned}
\end{align*}
\begin{align}
	\begin{aligned}
        &+ \sum_{a = 2, \tilde{2}}\left( Y^{(3)}_{a3} \ \delta_{c_1 c_2} \delta_{c_3 c_4} + Y^{\prime(3)}_{a3} \ \delta_{c_1 c_4} \delta_{c_2 c_3} \right)\left(\Phi_{a, c_1}^\dagger \sigma^I \Phi_{a, c_2} \right)\left(\Phi_{3, c_3}^{J\dagger} i\epsilon^{IJK} \Phi_{3, c_4}^K \right) \\
        &+ \Bigg[ \phantom{+} \sum_{a =   2, \tilde{2} } \left(Y_{1aa3} \ \delta_{c_1 c_2} \delta_{c_3 c_4} + Y^\prime_{1aa3} \ \delta_{c_1 c_4} \delta_{c_2 c_3} \right) \Phi^\dagger_{1, c_1}  \left(\Phi_{a, c_3}^\dagger \big(\sigma \cdot \Phi_{3, c_4} \big) \Phi_{a, c_2} \right)  \\
        &\phantom{12 +\Bigg[}+ \left(Y_{\tilde{1} \tilde{2} 2 3}\  \delta_{c_1 c_2} \delta_{c_3 c_4} + Y^\prime_{\tilde{1} \tilde{2} 2 3} \ \delta_{c_1 c_4} \delta_{c_2 c_3} \right) \Phi_{\tilde{1}, c_1}^\dagger  \left(\Phi_{2, c_3}^\dagger \big(\sigma \cdot \Phi_{3, c_4} \big)\Phi_{\tilde{2}, c_2} \right) \\
         &\phantom{12 + \Bigg[}+ \left(Y_{1 \tilde{1} \tilde{2} 2} \ \delta_{c_1 c_2} \delta_{c_3 c_4} + Y^\prime_{1 \tilde{1} \tilde{2} 2}\  \delta_{c_1 c_4} \delta_{c_2 c_3} \right) \Phi^\dagger_{1, c_1} \Phi_{\tilde{1}, c_2} \big(\Phi_{\tilde{2}, c_3}^\dagger  \Phi_{2, c_4}\big) \\
         &\phantom{12 + \Bigg[}+ \frac{1}{2} Y_{1 3 1 3} \ \delta_{c_1 c_2} \delta_{c_3 c_4} \left(\Phi^\dagger_{1, c_1} \Phi^I_{3, c_2} \Phi^\dagger_{1, c_3} \Phi^I_{3, c_4} \right) \\
         &\phantom{12 + \Bigg[}+ Y_{1 3 3 3} \ \delta_{c_1 c_2} \delta_{c_3 c_4} \left(\Phi^\dagger_{1, c_1} \Phi^I_{3, c_2} \Phi^{J \dagger}_{3, c_3} \Phi^K_{3, c_4} i\epsilon^{IJK} \right) \text{~ + h.c.} \Bigg]\,,
        \label{eq:LQ_quartic_interactions}
    \end{aligned}
\end{align}
with $a, b \in \left \{1, \tilde{1}, 2, \tilde{2}, 3 \right \}$ unless stated otherwise. Note that we set $Y^{(1)}_{ab} = Y^{(1)}_{ba}$ and $Y^{(1)\prime}_{ab} = Y^{(1)\prime}_{ba}$ for $a \neq b$, since they correspond to the same terms. The Feynman rules are expressed using only one of them, i.e. $Y^{(1)}_{\tilde{1}3}$, but not $Y^{(1)}_{3\tilde{1}}$.

\clearpage
\section{Methods}
\label{sec:methods}

The determination of the Feynman rules to be presented in Chapters \ref{sec:SMInteractions} and \ref{sec:LQInteractions} is carried out using the Mathematica package FeynRules 2.3.36~\cite{Alloul:2013bka}. All model files and the Mathematica notebook needed to display and export the Feynman rules can be found online at GitLab. In the following, the structure of the code files will be discussed. For more information about FeynRules, the interested reader is referred to Ref.~\cite{Alloul:2013bka}.

\subsection{GitLab Project}
This GitLab project can be accessed under

\begin{center} \href{https://gitlab.com/lucschnell/SLQrules}{https://gitlab.com/lucschnell/SLQrules}. \end{center}
 \href{https://gitlab.com/lucschnell/SLQrules/-/blob/master/SLQrules.nb}{SLQrules.nb} contains a Mathematica notebook that is used to determine the Feynman rules and to export them to MadGraph or FeynArts. The FeynRules model files can be found in the folder \href{https://gitlab.com/lucschnell/SLQrules/-/tree/master/SLQrules}{SLQrules}. In order to use them, download the entire folder and add it to the FeynRules directory FeynRules/Models on your computer. We also provide prefabricated MOD and UFO model files that can be used in FeynArts \cite{Hahn:2000kx} and MadGraph \cite{Alwall:2014hca}. 


\subsubsection{Mathematica Notebook}
The notebook \href{https://gitlab.com/lucschnell/SLQrules/-/blob/master/SLQrules.nb}{SLQrules.nb} is used to display and export the Feynman rules. Having loaded the model files, the different parts of the Lagrangian can be accessed via the FeynRules variable names shown in Table~\ref{tab:Lagrangians}.
\begin{table}[ht!]
\renewcommand{\arraystretch}{1.1}
    \centering
    \begin{tabular}{c | c c c c c c}
         Lagrangian & Variable name\\ \hline
    	$\mathcal{L}^{\text{LQ}}$ & LQall \\
	$\mathcal{L}_{2\Phi}$ & LQ2Phi \\
	$\mathcal{L}_{kin}$& LQkin \\
	$\mathcal{L}_{f}$ &LQf \\
	$\mathcal{L}_{3\Phi}$ &LQ3Phi \\
	$\mathcal{L}_{4\Phi}$ &LQ4Phi \\
    \end{tabular}\,.
    \caption{Variable names of the Lagrangians defined in Sec.~\ref{sec:LQLagrangian}. }
\label{tab:Lagrangians}
\end{table} \\
\clearpage
For example, 
\begin{lstlisting}[label=lst:load]
FeynmanRules[LQall];
\end{lstlisting} 
is used to determine the Feynman rules for the entire Lagrangian $\mathcal{L}^{\text{LQ}}$. The Feynman rules can be exported as UFO files for MadGraph. This is done via the command
\begin{lstlisting}[label=lst:load]
FeynmanGauge = False;
WriteUFO[LSM + LQall, Output -> "SLQrules-UFO"];
\end{lstlisting} 
The boolean FeynmanGauge on line~1 is used to switch between Feynman and unitary gauge. The files are exported to SLQrules/SLQrules-UFO in your FeynRules directory. We also provide ready-to-use UFO files for $\mathcal{L}^{\text{LQ}}$, these can be found under \href{https://gitlab.com/lucschnell/SLQrules/-/tree/master/UFO/SLQrules-UFO}{SLQrules/UFO} in the GitLab project. Similarly, the export for FeynArts is carried out using 
\begin{lstlisting}[label=lst:load]
FeynmanGauge = False;
WriteFeynArtsOutput[LSM + LQall, Output -> "SLQrules-FA"];
\end{lstlisting}
We provide prefabricated FeynArts files under \href{https://gitlab.com/lucschnell/SLQrules/-/tree/master/FeynArts/SLQrules-FA}{SLQrules/FeynArts}. Additionally, the export to Sherpa, CalcHep or Whizard is possible. The corresponding commands can be found in \href{https://gitlab.com/lucschnell/SLQrules/-/blob/master/SLQrules.nb}{SLQrules.nb}, but we opt not to include prefabricated files for these applications. The package NLOCT 1.0~\cite{Degrande:2014vpa} even allows for the automatic determination of UV and $R_2$ counterterms, making the export of NLO models possible. We include corresponding commands in \href{https://gitlab.com/lucschnell/SLQrules/-/blob/master/SLQrules.nb}{SLQrules.nb}, but note that these determinations become highly time-consuming for complex models (i.e.~involving LQ mixing). In this work we only consider models at tree-level.

\subsubsection{Model Files}
\label{sec:modelFiles}
The FeynRules model files are located in the folder \href{https://gitlab.com/lucschnell/SLQrules/-/tree/master/SLQrules}{SLQrules}. They are based on existing model files for the SM~\cite{Alloul:2013bka} and the individual scalar LQ representations~\cite{Dorsner:2018ynv}. The scalar LQ fields defined in Ref.~\cite{Dorsner:2018ynv} were merged into the model file \href{https://gitlab.com/lucschnell/SLQrules/-/blob/master/SLQrules/SLQrules.fr}{SLQrules/SLQrules.fr} and their interactions with two SM quarks, as well as the LQ-LQ-Higgs(-Higgs), LQ-LQ-LQ(-Higgs) and LQ-LQ-LQ-LQ interactions were added. The file \href{https://gitlab.com/lucschnell/SLQrules/-/blob/master/SLQrules/SLQrules.fr}{SLQrules.fr} consists of three parts: 
\begin{enumerate}[(i)]
\item the parameter definitions,
\item the definition of the LQ fields and 
\item the LQ Lagrangian. 
\end{enumerate}
To exemplify (i), a generic parameter definition is shown in Listing \ref{lst:parameter}. The full list of all parameter definitions is given in the \href{https://gitlab.com/lucschnell/SLQrules/-/blob/master/SLQrules/README.md}{README.md} file of \href{https://gitlab.com/lucschnell/SLQrules/-/tree/master/SLQrules}{SLQrules}.

\begin{lstlisting}[label=lst:parameter,caption=Parameter definition (coupling $Y_{13}$)]
Y13 == {
        ParameterType    -> External,
	      ComplexParameter -> True,
    	  Indices          -> {},
	      BlockName        -> Y13,
        TeX              -> Subscript[Y, "13"],
    	  Value            -> 1.0,
	      InteractionOrder -> {QED, 2},
    	  Description      -> "S3-S1 scalar leptoquark mixing 
	                           coupling"
},
\end{lstlisting}
\vspace{5px}
We set the values of all couplings to unity per default. The QED interaction order of $Y_{13}$ is set to 2, since this interaction involves two Higgs fields. An example for (ii), the definition of a LQ field, is given in Listing~\ref{lst:field}. The original fields $\Phi_a$ before the  diagonalization of the mass terms are defined as unphysical fields, whereas the fields $\hat{\Phi}_a$ represent the physical fields in the FeynRules code. In this way, $\Phi_a$ can be used when defining LQ Lagrangians (see below) and the FeynRules code automatically replaces them by the corresponding mass eigenstates $\hat{\Phi}_a$ and the rotation matrices $W^q$
\begin{equation}
	\Phi^q_a = W_{n_bn_a}^{q*} \hat{\Phi}_b^{q}
\end{equation}
 Such a replacement is defined for $\Phi_1$ on the lines~25-30 of Listing~\ref{lst:field}. 
\vspace{5px}
\begin{lstlisting}[label=lst:field,caption=Leptoquark field definition (singlet field $\Phi_{1}$)]
(* physical fields *)
S[100] == {
    ClassName        -> S1m13hat,
    Mass             -> {m1m13hat, Internal},
    Width            -> {W1m13hat, Internal},
    SelfConjugate    -> False,
    PropagatorLabel  -> "S1m13hat",
    PropagatorType   -> ScalarDash,
    PropagatorArrow  -> None,
    QuantumNumbers   -> {Q -> -1/3},
    Indices          -> {Index[Colour]},
    ParticleName     -> "S1m13hat",
    AntiParticleName -> "S1m13hat~",
    FullName         -> "S1m13hat"
},

(* unphysical fields *)
S[101] == {
    ClassName        -> S1m13,
    Unphysical       -> True,
    SelfConjugate    -> False,
    QuantumNumbers   -> {Y -> -1/3},
    Indices          -> {Index[Colour]},
    Definitions      -> {S1m13[cc_] :>   HC[W13mat[1,1]] 
                                         S1m13hat[cc] 
                                       + HC[W13mat[2,1]] 
                                         R2tm13hat[cc] 
                                       + HC[W13mat[3,1]] 
                                         S3m13hat[cc]}
},
\end{lstlisting}
\vspace{5px}

Finally, to exemplify the definition of (iii), the Lagrangian for the LQ-LQ-LQ-Higgs interaction
\begin{equation}
Y_{\tilde{2}33} \ \left(\Phi_{\tilde{2}, c_1}^\intercal i \sigma_2 \sigma^I H\right) \left( \Phi^J_{3, c_2} i\epsilon^{IJK}\Phi^K_{3, c_3} \right) \text{~+ h.c.} \,,
\end{equation}
is given in Listing \ref{lst:lagrangian}. Single lowercase letters ($a,b,c$) are used to refer to the weak doublet indices, single capital letters ($D,E,F$) to refer to the weak triplet indices and double lowercase letters ($aa,bb,cc$) are color indices. As described above, so-called unphysical fields ($\Phi_{\tilde2}, \Phi_3$) are used in the definition of the Lagrangian, which are later replaced automatically by the physical fields ($\hat{\Phi}_{\tilde{2}}^{+2/3}, \hat{\Phi}_{\tilde{2}}^{-1/3}, \hat{\Phi}_{3}^{-1/3}, \hat{\Phi}_{3}^{+2/3}$ and  $\hat{\Phi}_{3}^{-4/3}$ in this case).
\vspace{5px}
\begin{lstlisting}[label=lst:lagrangian,caption=Lagrangian definition (LQ interaction corresponding to $Y_{\tilde{2}33}$).]
L2t33NonHC := Module[ {a,b,c,D,E,F,aa,bb,cc},
ExpandIndices[ Y2t33 R2t[a,aa] Eps[a,b] 
		2*Ta[D,b,c] Phi[c] S3[E,bb] I*Eps[D,E,F] S3[F,cc] 
		Eps[aa,bb,cc], 
		FlavorExpand->{SU2D, SU2W}]];
L2t33 := L2t33NonHC + HC[L2t33NonHC];
\end{lstlisting}
The assignment of FlavorExpand on line~5 makes sure that both $\Phi_{\tilde{2}}$ and $\Phi_3$ are expanded into their $SU(2)_L$ components.

\clearpage
\section{SM Feynman Rules}
\label{sec:SMInteractions}


\subsection{External Fields}
The Feynman rules for the external legs are listed in this subsection. See Ref.~\cite{Denner:1992vza} for details. $u^{(s)}$ and $v^{(s)}$ denote the spinors for particles and antiparticles, respectively, and $s$ is the spin. The polarization tensors for the gauge bosons are denoted by $\epsilon_\mu$. The circles denote the generic reminder of the Feynman diagram. 
\begin{center}
\begin{table}[!h]
    \centering
\begin{tabular}{>{\centering\arraybackslash}m{3cm} m{9.5cm}}
    \multicolumn{2}{c}{External Fermions} \\
    \hline
        Diagram & Feynman Rule \\
        \hline
        \vspace{10 px}
        \includegraphics[scale=0.025]{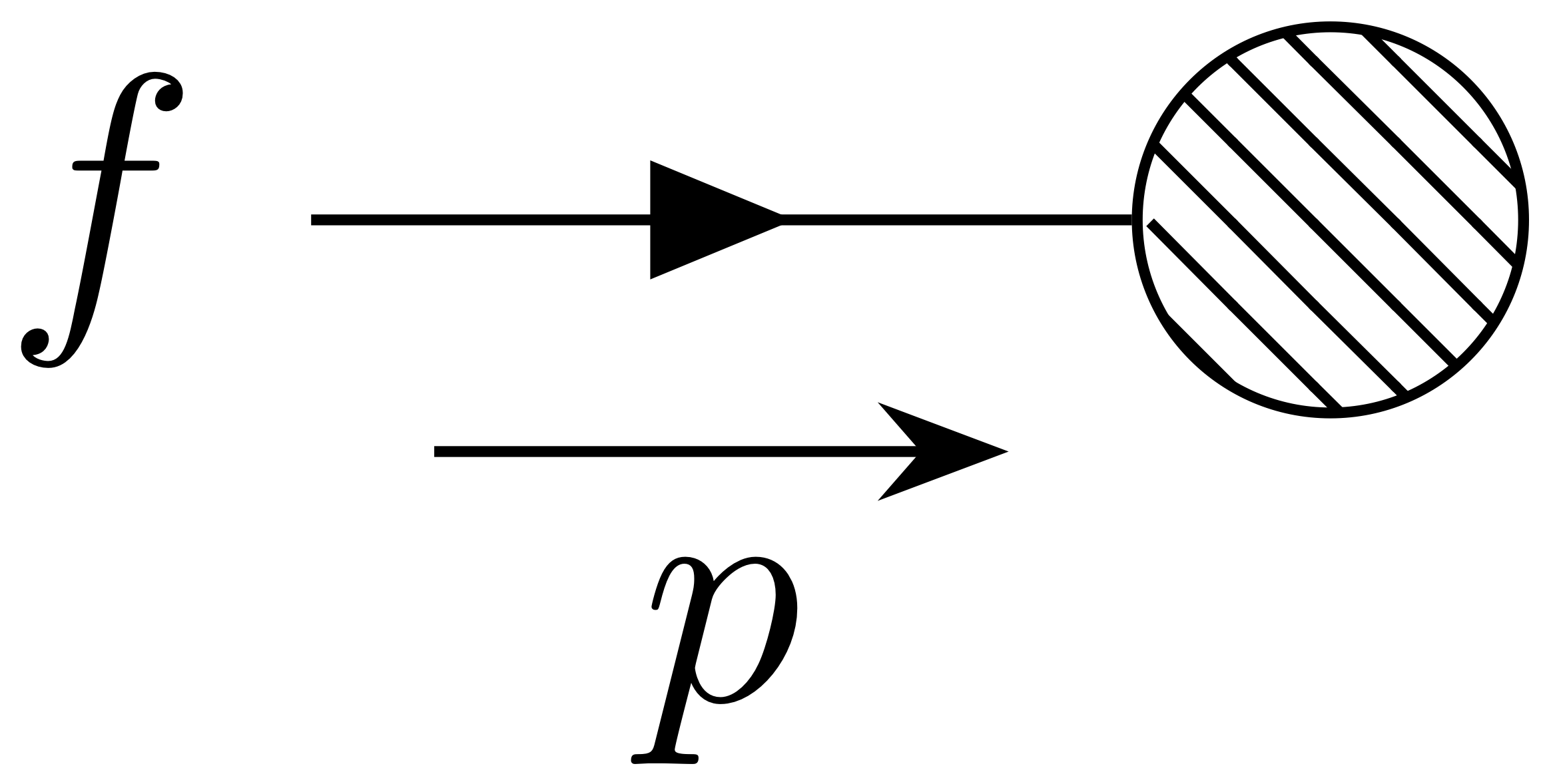} &
        \vspace{5px}
        $u^{(s)}(\vec{p})$ \newline \\ 
        \vspace{10 px}
        \includegraphics[scale=0.025]{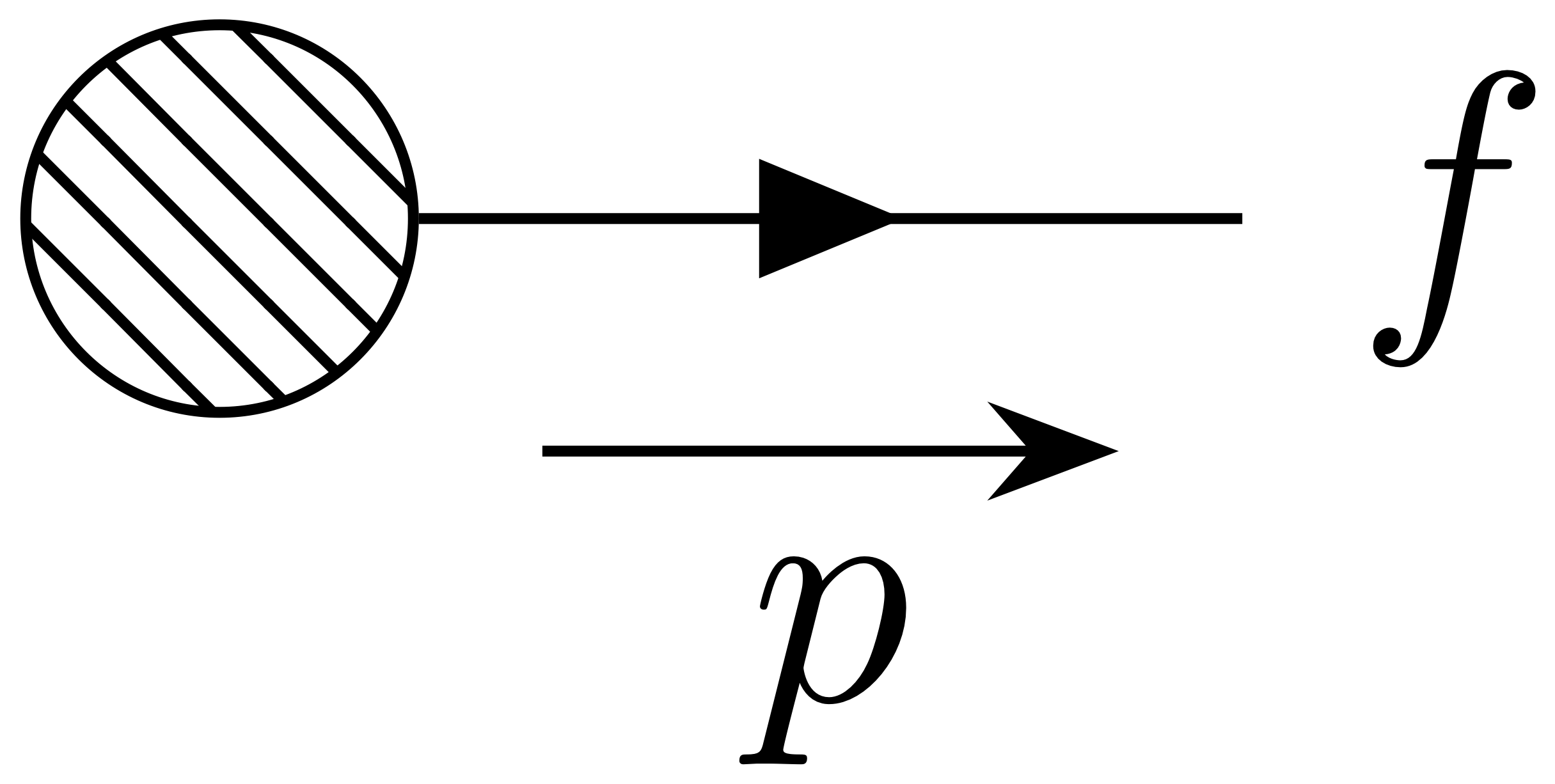} &
        \vspace{5px}
        $\bar{u}^{(s)}(\vec{p})$ \newline \\
        \vspace{10 px}
        \includegraphics[scale=0.025]{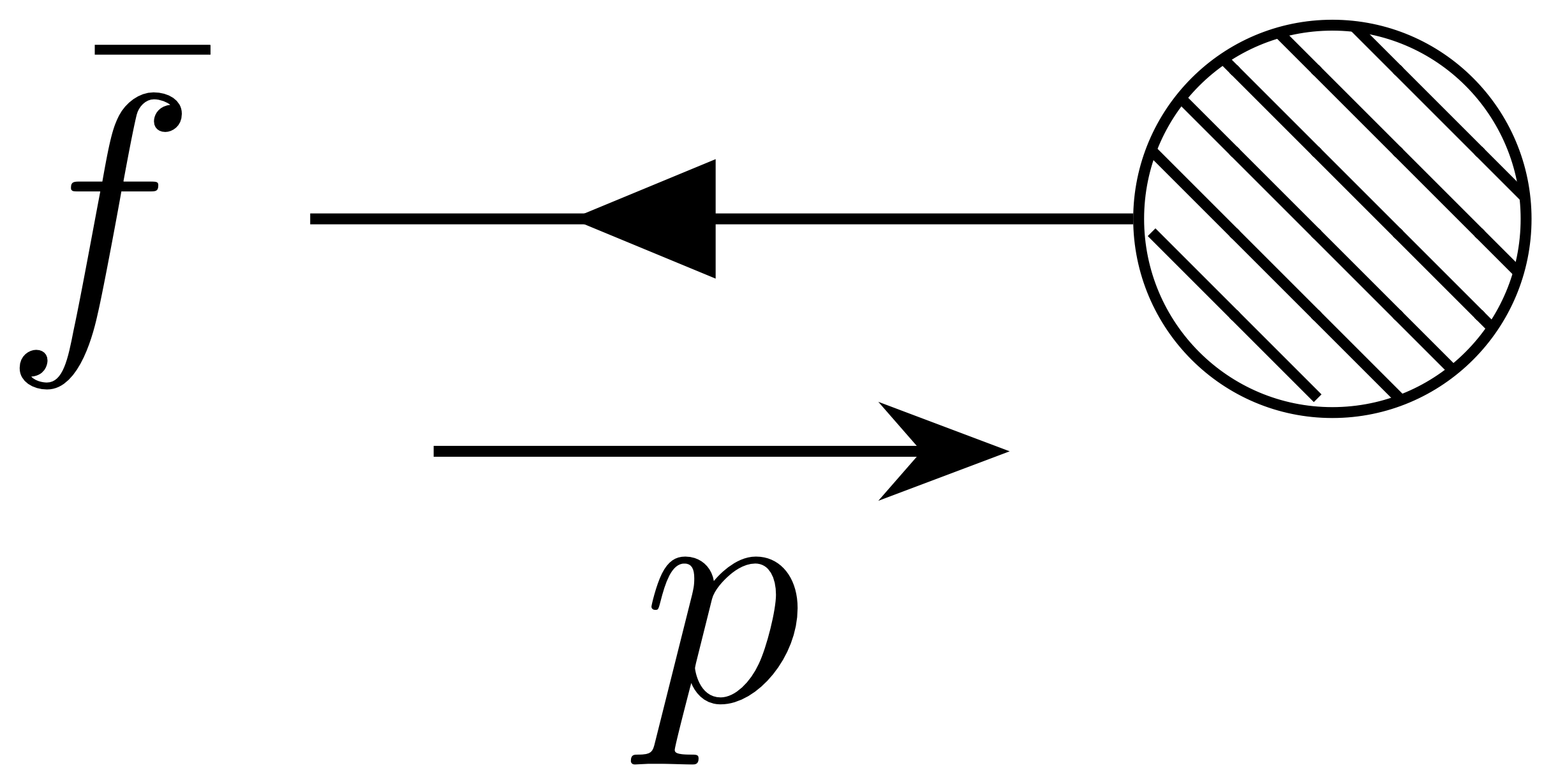} &
        \vspace{5px}
        $\bar{v}^{(s)}(\vec{p})$ \newline \\
        \vspace{10 px}
        \includegraphics[scale=0.025]{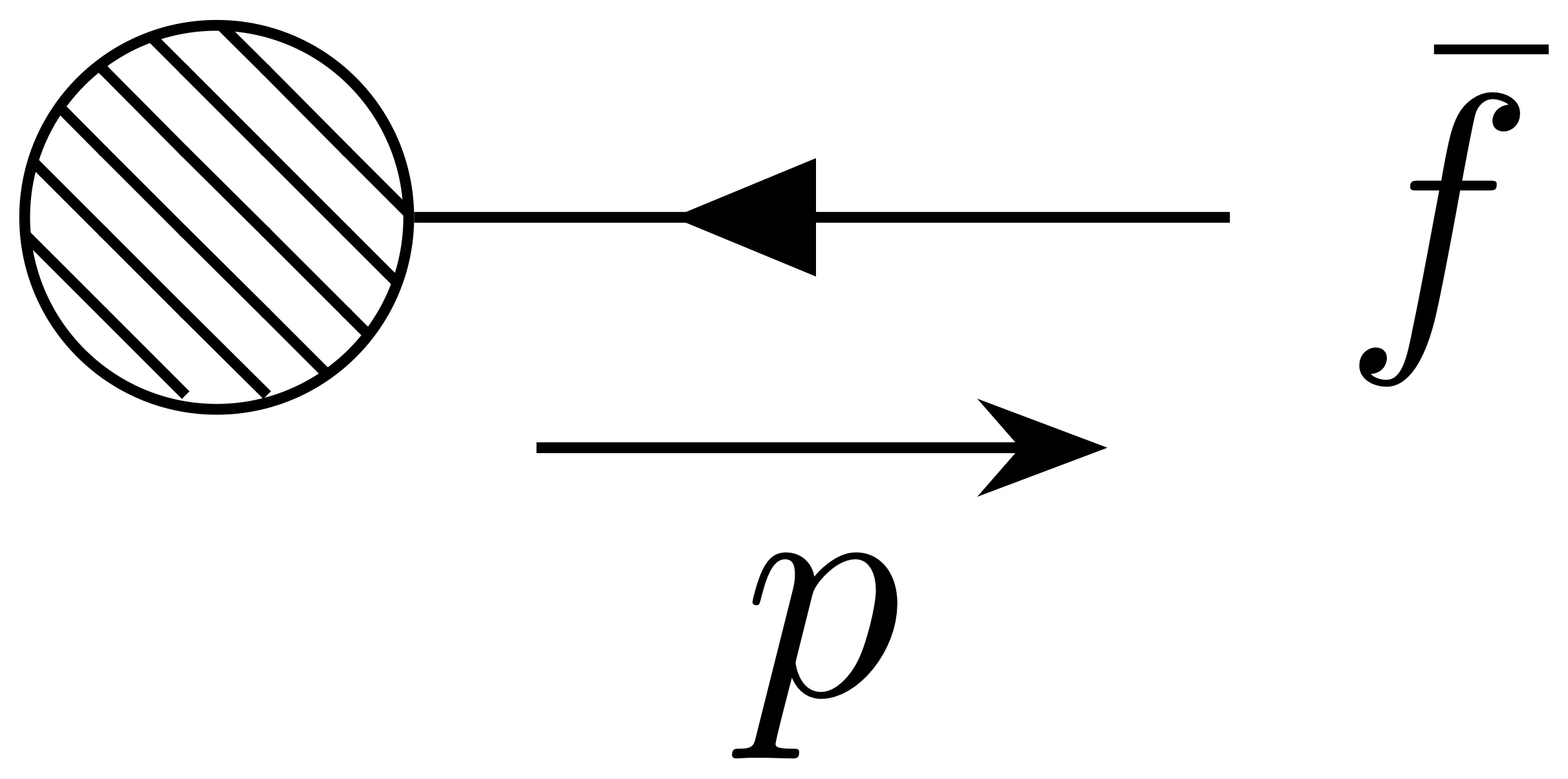} &
        \vspace{5px}
        $v^{(s)}(\vec{p})$ \newline \\
        \hline
    \end{tabular}
\end{table}
\end{center}

\begin{center}
\begin{table}[!h]
    \centering\begin{tabular}{ >{\centering\arraybackslash}m{3cm} m{9.5cm}} 
        \multicolumn{2}{c}{External Gauge Bosons} \\
        \hline
        Diagram & Feynman Rule \\
        \hline
        \vspace{10 px}
        \includegraphics[scale=0.025]{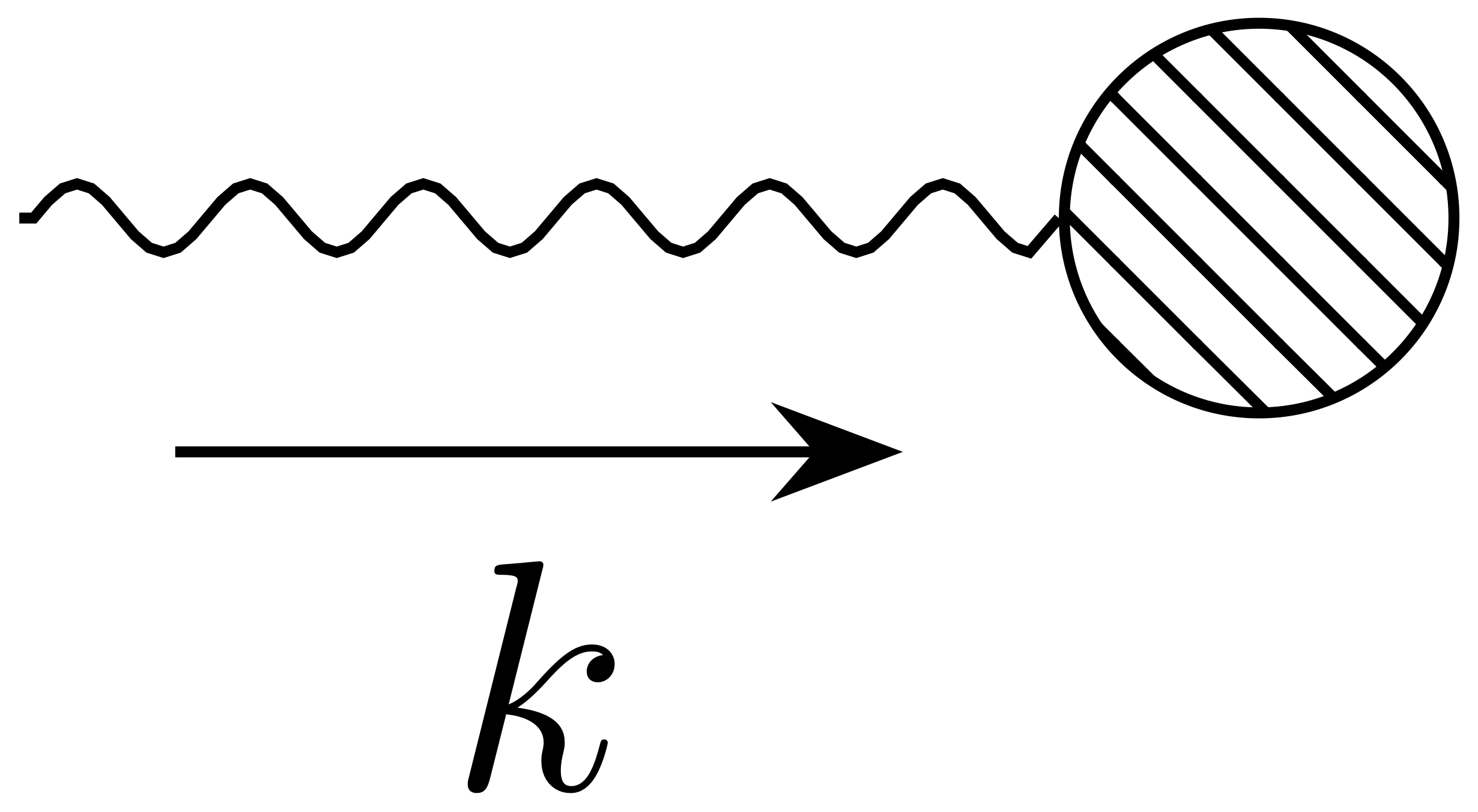} \newline 
        \includegraphics[scale=0.025]{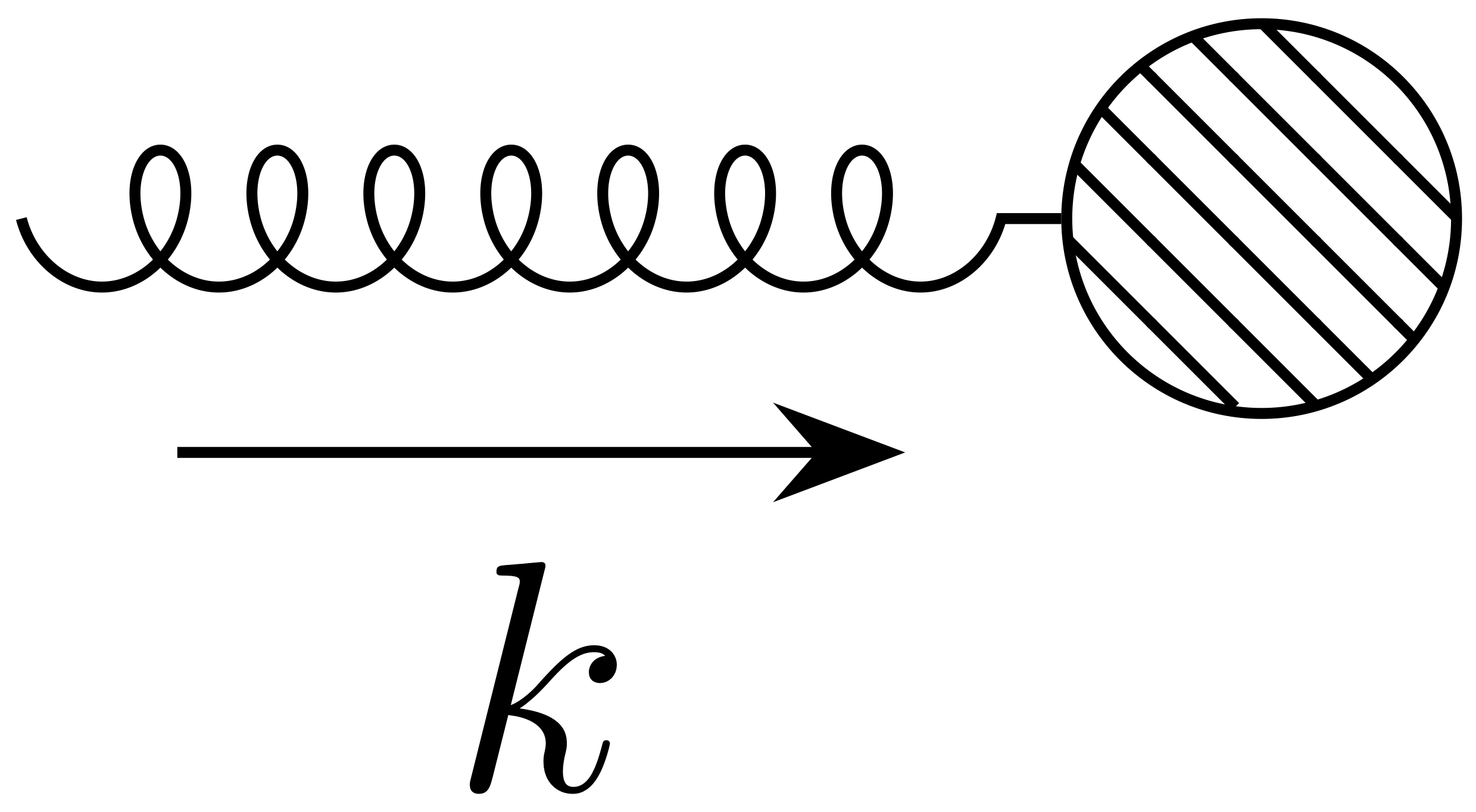} ~~~ &
        \vspace{5px}
        $ \epsilon_\mu(\vec{k})$ \newline \\
        \vspace{10 px}
        \includegraphics[scale=0.025]{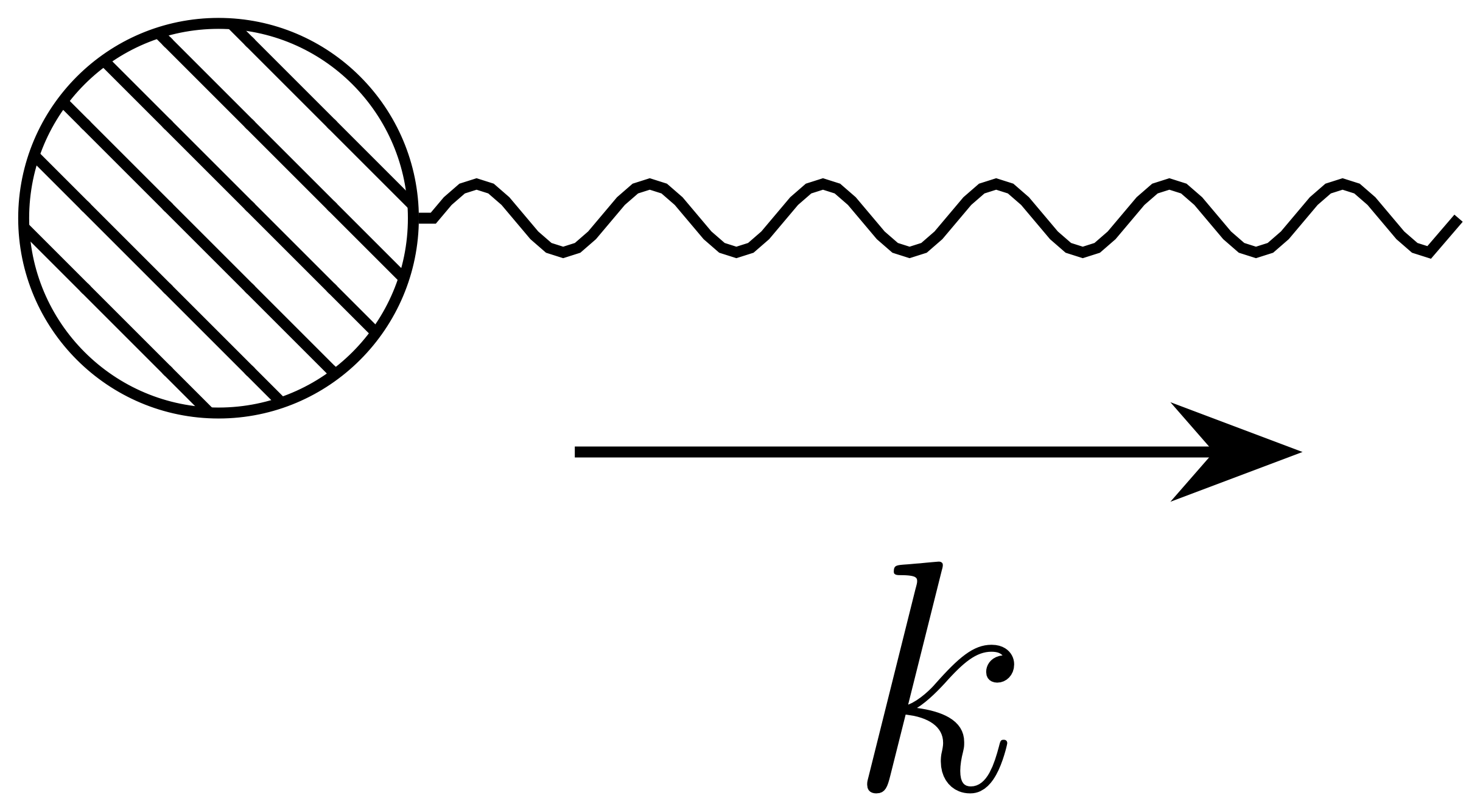} \newline
        \includegraphics[scale=0.025]{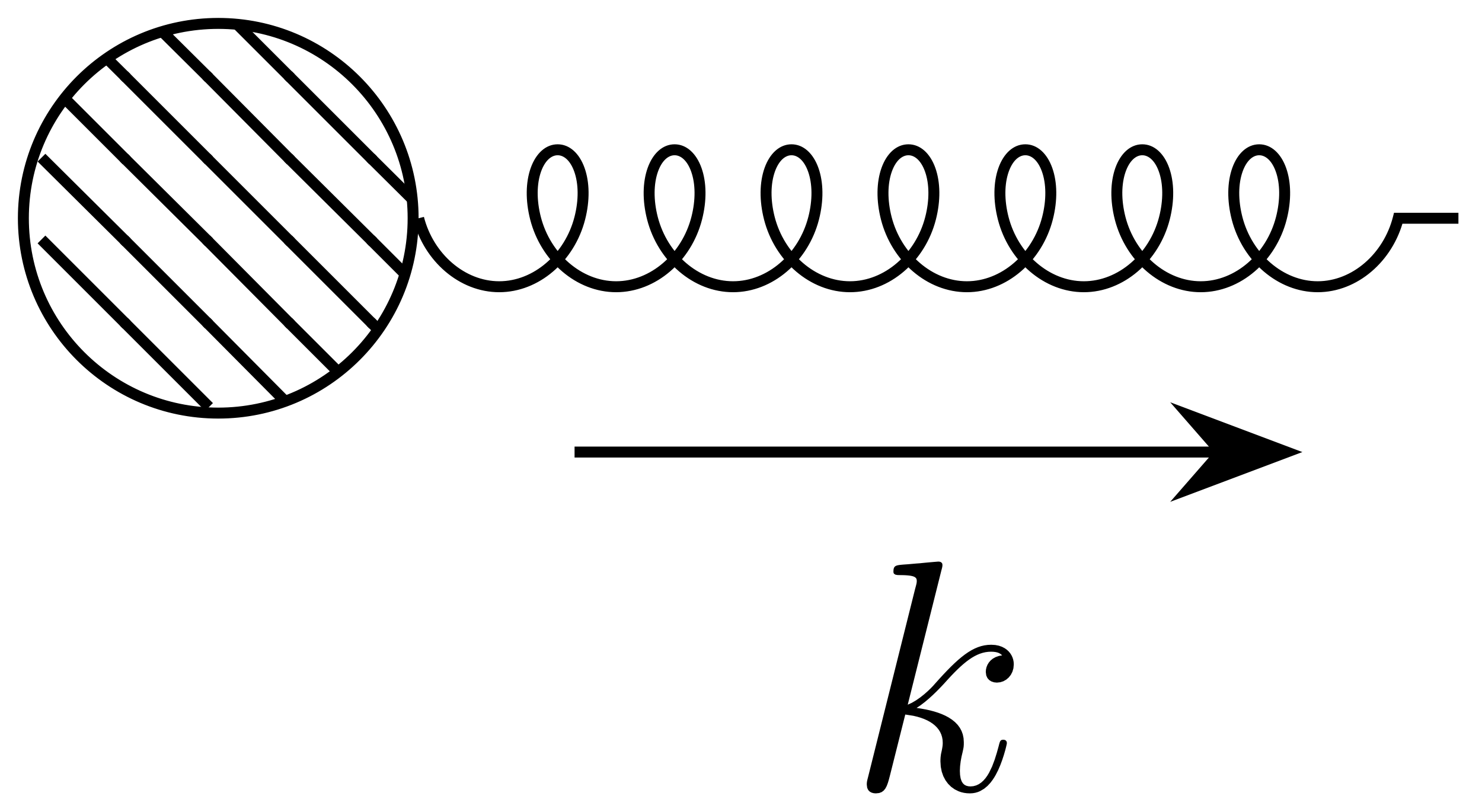} ~~~ &
        \vspace{5px}
        $\epsilon_\mu^*(\vec{k})$ \newline \\
        \hline
    \end{tabular}
\end{table}
\end{center}

\begin{center}
\begin{table}[!h]
    \centering\begin{tabular}{ >{\centering\arraybackslash}m{3cm} m{9.5cm}} 
        \multicolumn{2}{c}{External Higgs and Goldstone Fields} \\
    \hline
        Diagram & Feynman Rule \\
        \hline
        \vspace{10 px}
        \includegraphics[scale=0.025]{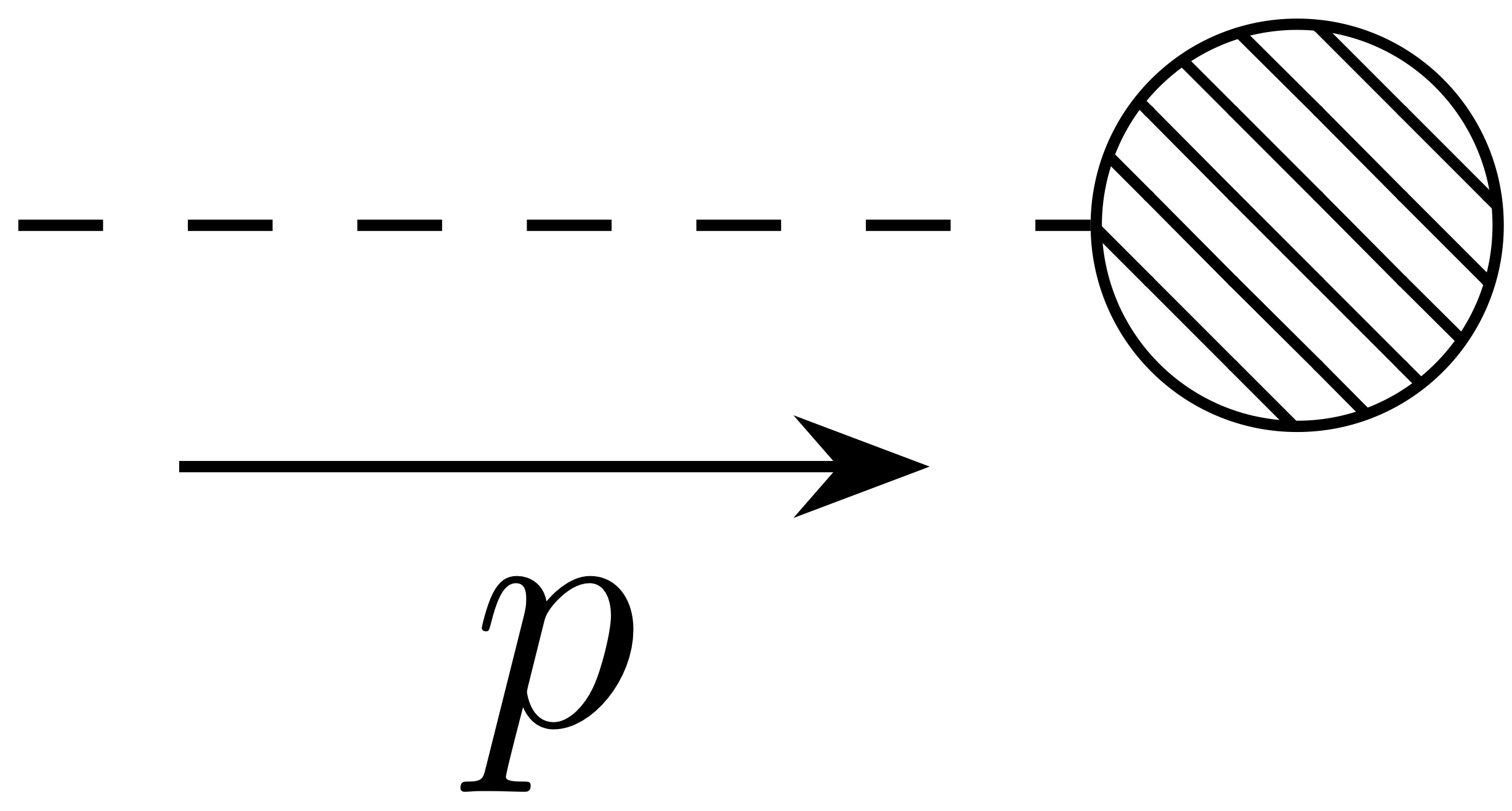}&
        \vspace{10 px}
        1 \newline \\
        \vspace{10 px}
        \includegraphics[scale=0.025]{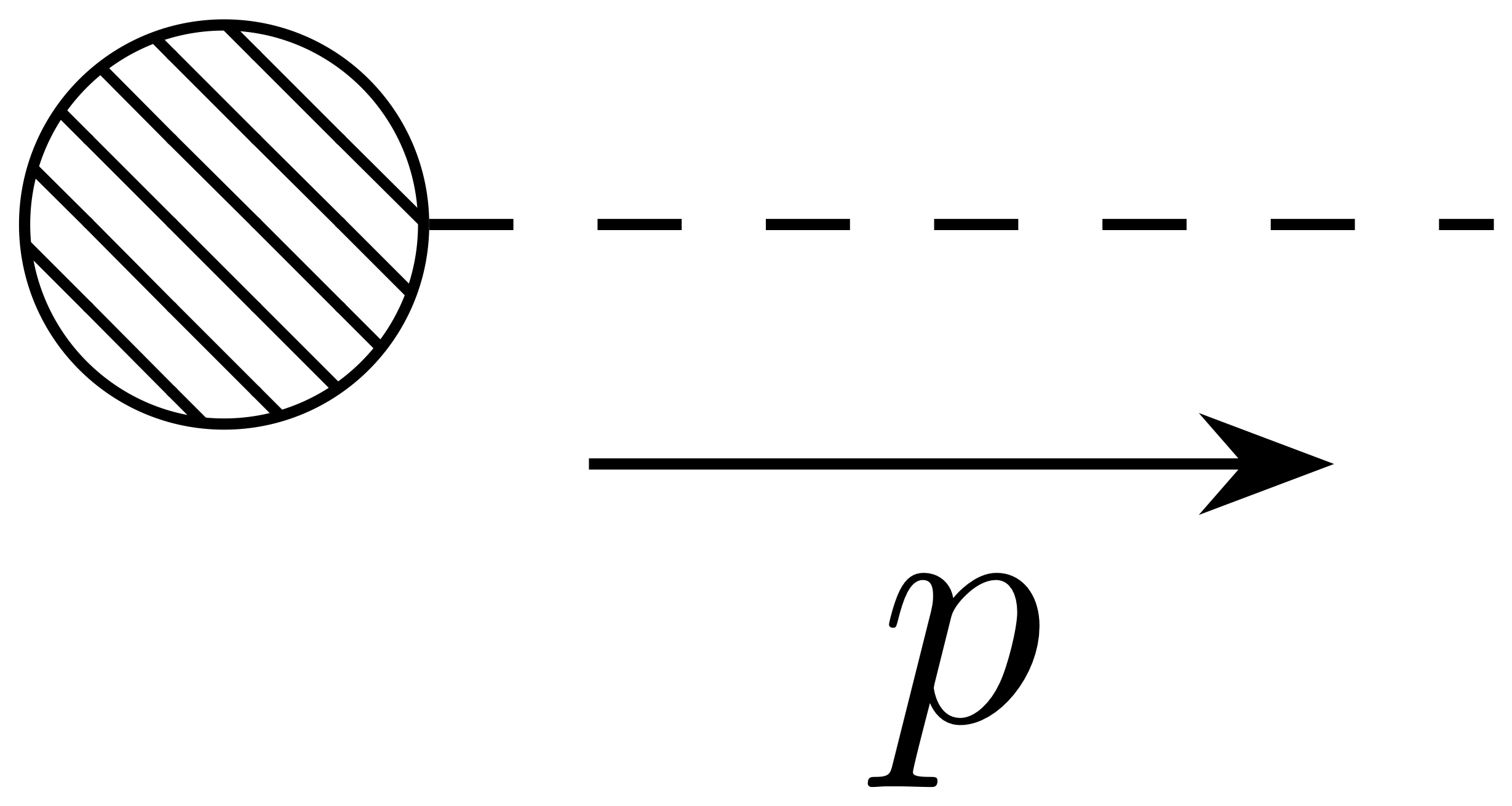}&
        \vspace{10 px}
        1 \newline \\
        \hline
        \hline
    \end{tabular}
\end{table}
\end{center}

\clearpage
\subsection{Propagators}
The propagators for the SM particles are given below. We work in Feynman gauge. The fermion masses $m_{\ell_i}, m_{u_i}, m_{d_i}$ are defined in Eq.~(\ref{eq:fermionMasses}), the gauge boson masses $m_{W}, m_{Z}$ in Eq.~(\ref{eq:bosonMasses}).

\begin{center}
\begin{table}[!h]
    \centering\begin{tabular}{>{\centering\arraybackslash}m{3.5cm} m{9cm}} 
        \multicolumn{2}{c}{Fermion Propagators} \\
    \hline
        Diagram & Feynman Rule \\
        \hline
        \vspace{10 px}
        \includegraphics[scale=0.03]{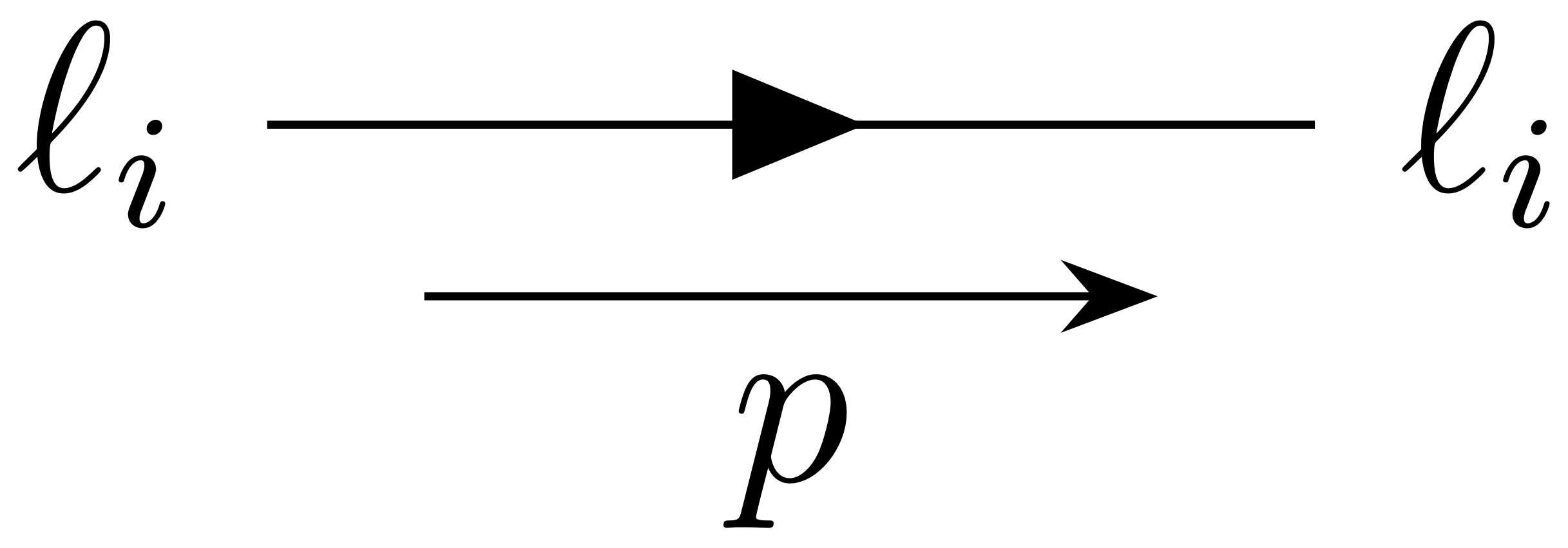}&
        \vspace{10 px}
        $\dfrac{i \left(\slashed{p} + m_{\ell_i} \right)}{p^2 - m^2_{\ell_i} +i\epsilon}$ \newline \\
        \vspace{10 px}
        \includegraphics[scale=0.03]{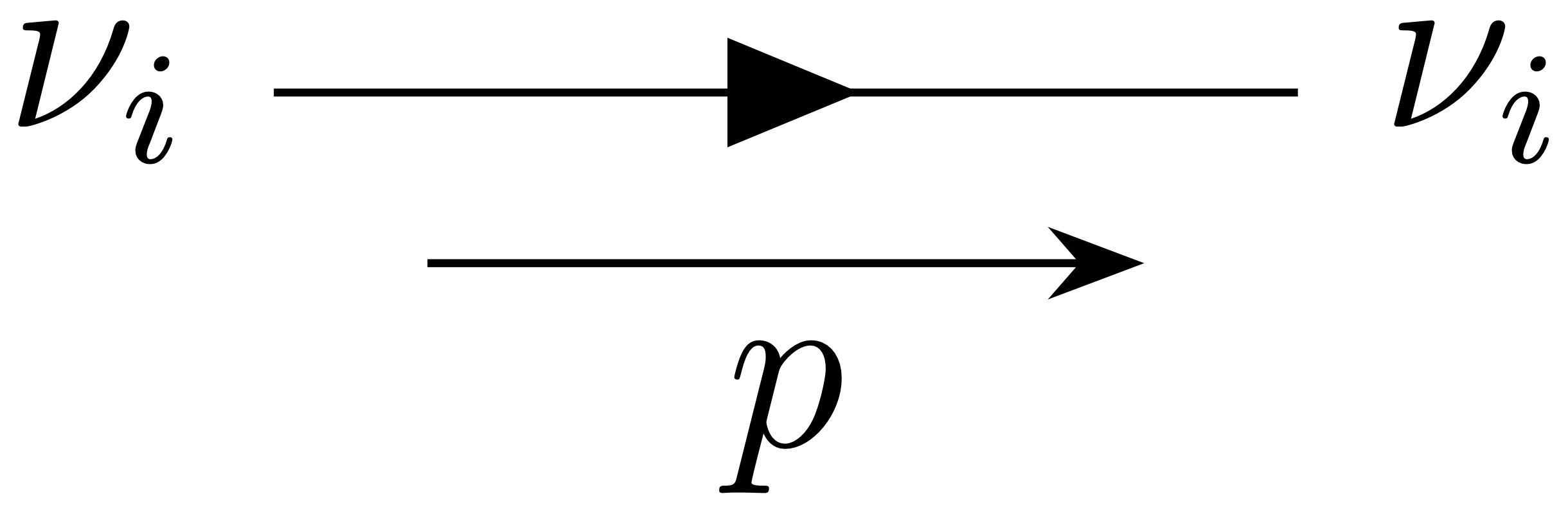}&
        \vspace{10 px}
        $\dfrac{i \slashed{p}}{p^2+i\epsilon}$ \newline \\
        \vspace{10 px}
        \includegraphics[scale=0.03]{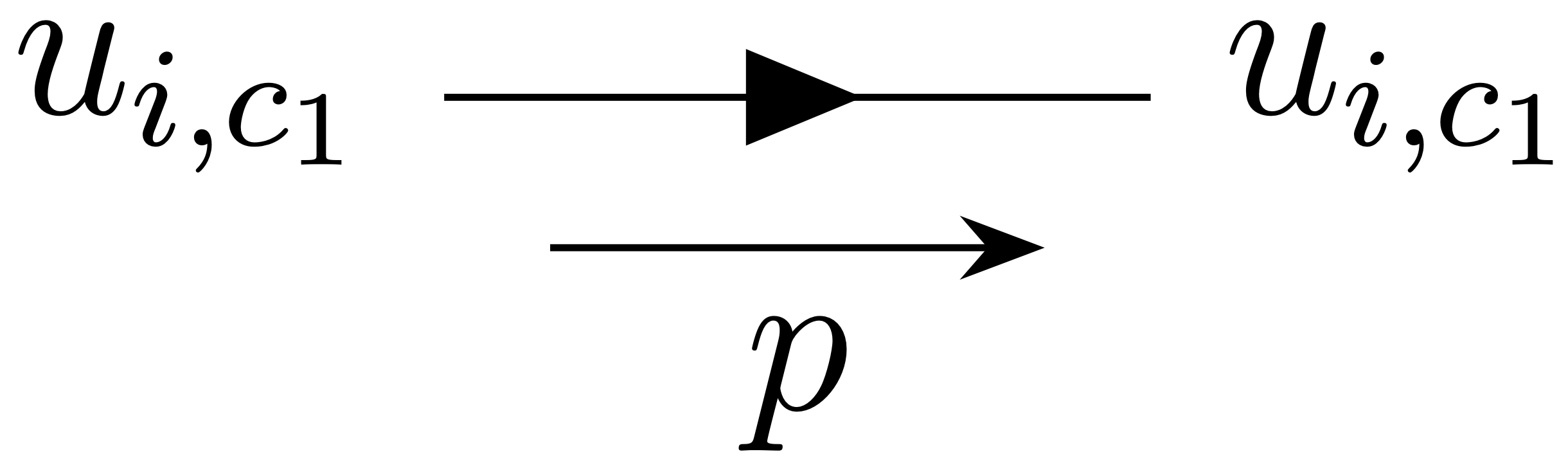}&
        \vspace{10 px}
        $\dfrac{i \left(\slashed{p} + m_{u_i} \right)}{p^2 - m^2_{u_i} +i\epsilon}$ \newline \\
        \vspace{10 px}
        \includegraphics[scale=0.03]{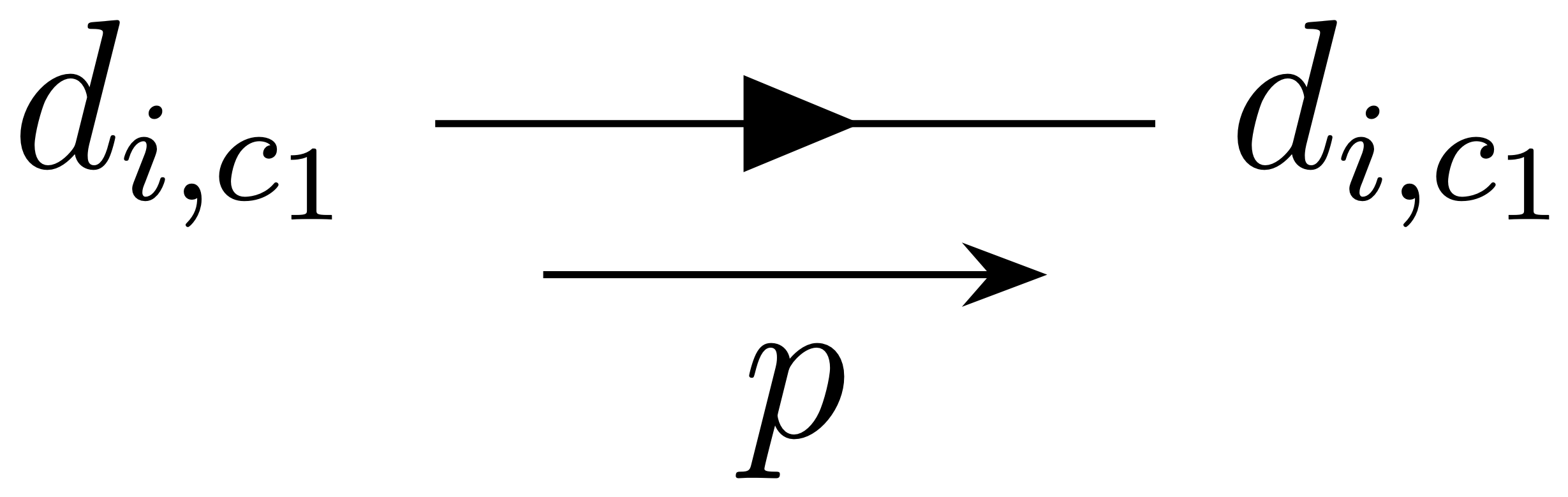}&
        \vspace{10 px}
        $\dfrac{i \left(\slashed{p} + m_{d_i} \right)}{p^2 - m^2_{d_i} +i\epsilon}$ \newline \\
        \hline
    \end{tabular}
\end{table}
\end{center}

\begin{center}
\begin{table}[!h]
    \centering\begin{tabular}{>{\centering\arraybackslash}m{3.5cm} m{9cm}} 
    \multicolumn{2}{c}{Gauge Boson Propagators} \\
    \hline
        Diagram & Feynman Rule \\
        \hline
        \vspace{10 px}
        \includegraphics[scale=0.03]{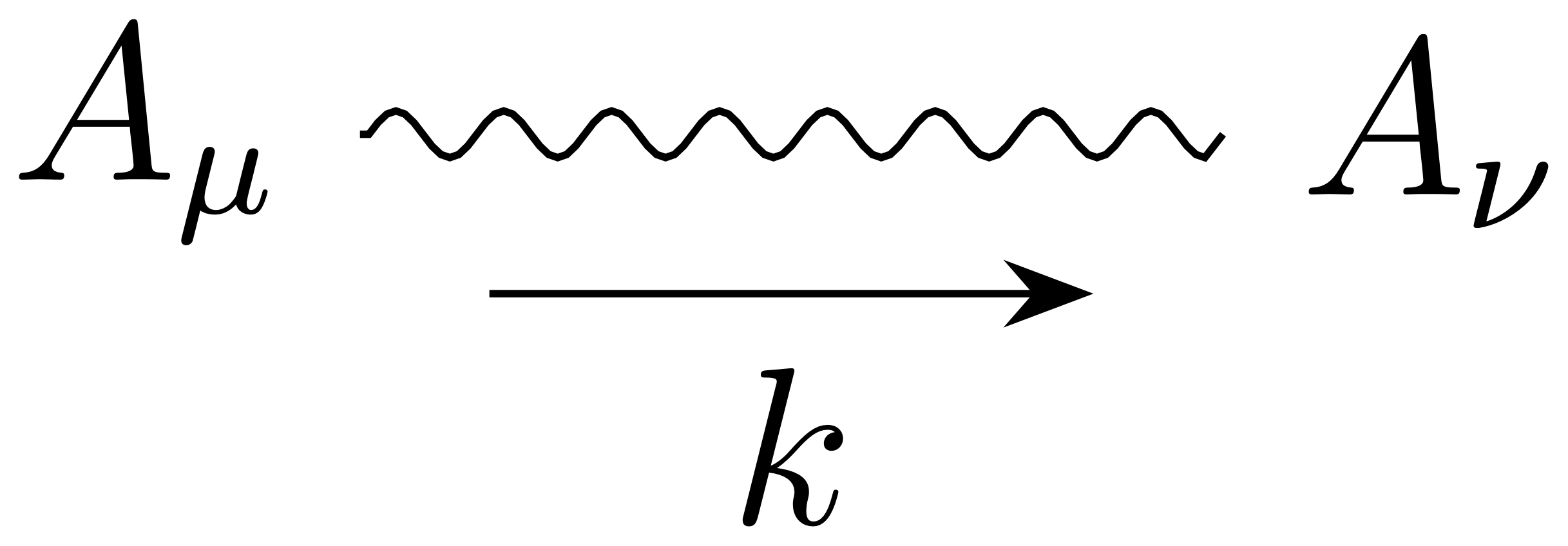}&
        \vspace{10 px}
        $\dfrac{-i g^{\mu \nu} }{k^2 + i\epsilon}$ \newline \\
        \vspace{10 px}
        \includegraphics[scale=0.03]{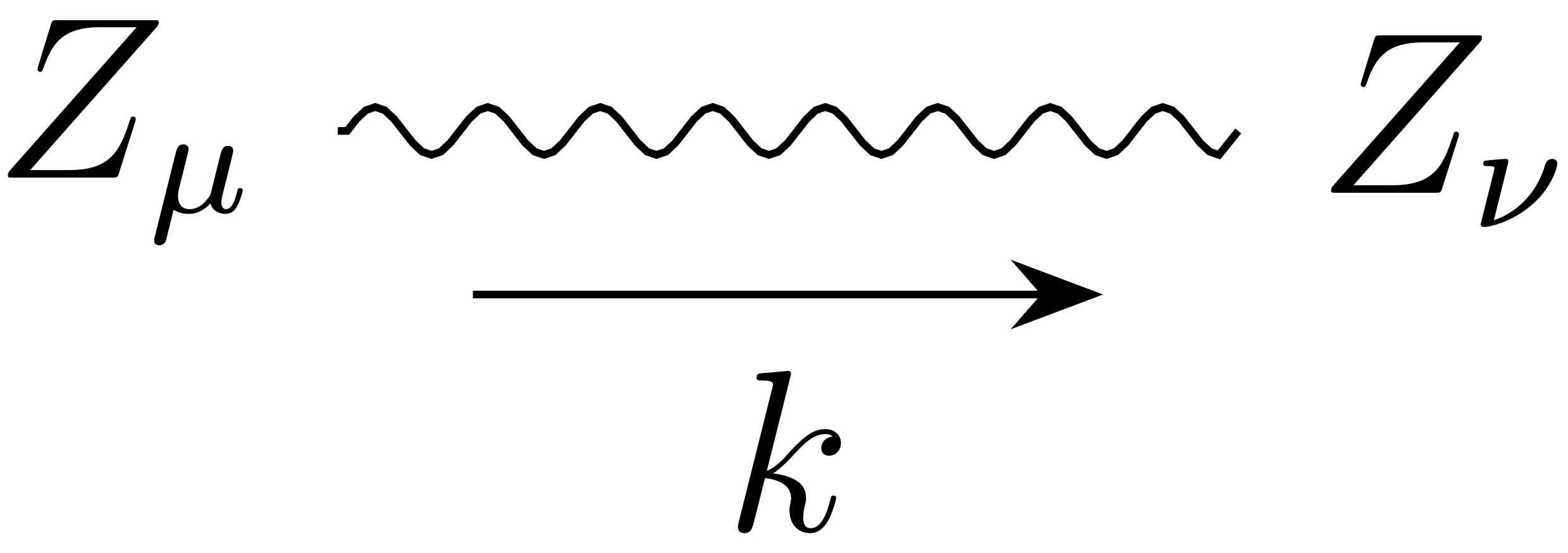}&
        \vspace{10 px}
        $\dfrac{-i g^{\mu \nu}}{k^2 - m_Z^2 + i\epsilon}$ \newline \\
        \vspace{10 px}
        \includegraphics[scale=0.03]{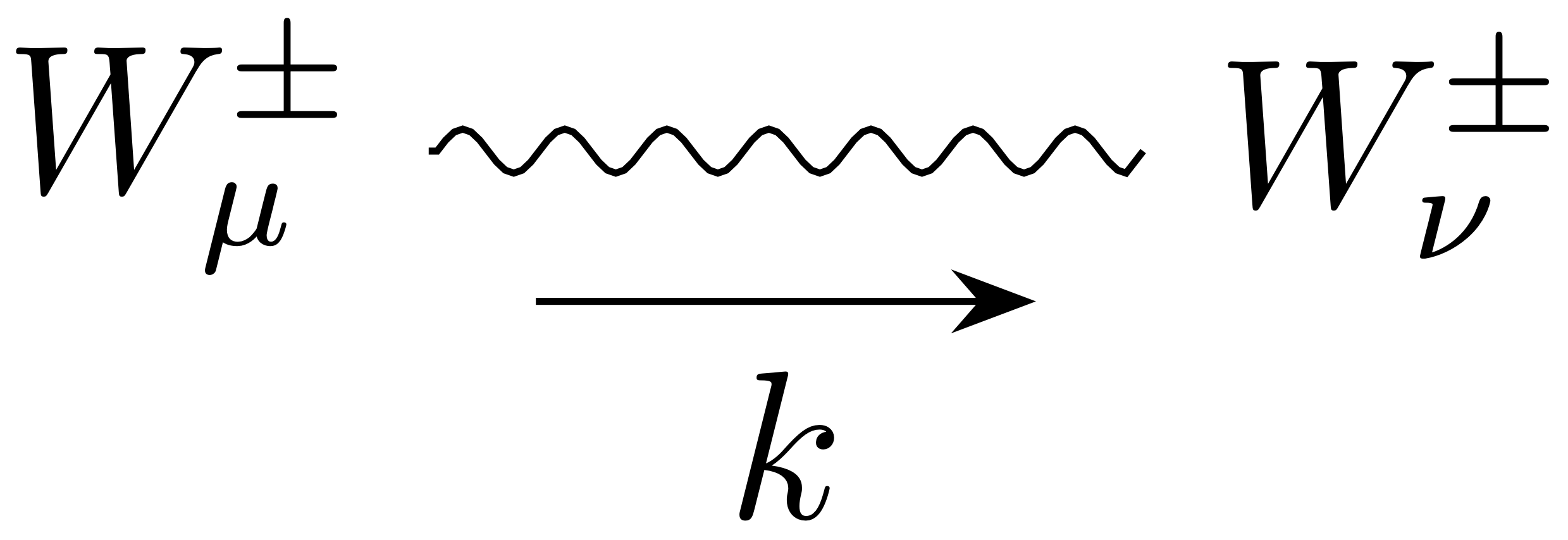}&
        \vspace{10 px}
        $\dfrac{-i g^{\mu \nu} }{k^2 - m_W^2 + i\epsilon}$ \newline \\
        \vspace{10 px}
        \includegraphics[scale=0.03]{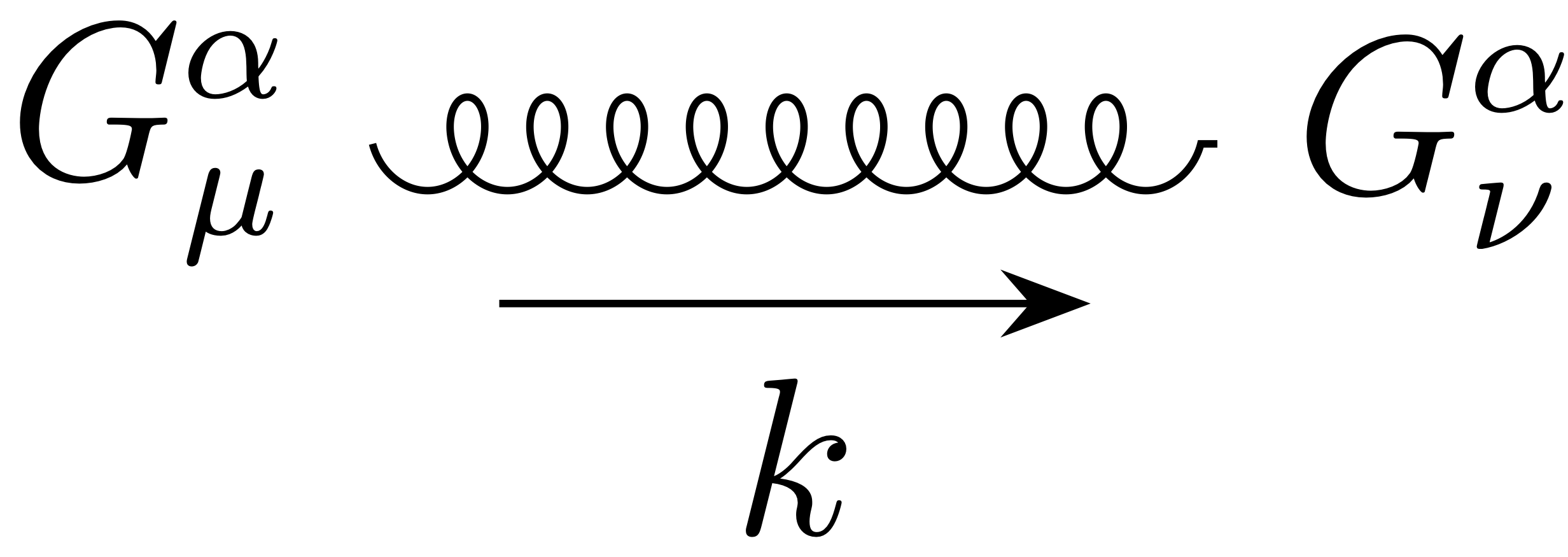}&
        \vspace{10 px}
                $\dfrac{-i g^{\mu \nu} }{k^2 + i\epsilon}$ \newline \\
        \hline
    \end{tabular}
\end{table}
\end{center}

\begin{center}
\begin{table}[!h]
    \centering\begin{tabular}{ >{\centering\arraybackslash}m{3.5cm} m{9cm}} 
    \multicolumn{2}{c}{Higgs and Goldstone Boson Propagators} \\
    \hline
        Diagram & Feynman Rule \\
        \hline
        \vspace{10 px}
        \includegraphics[scale=0.03]{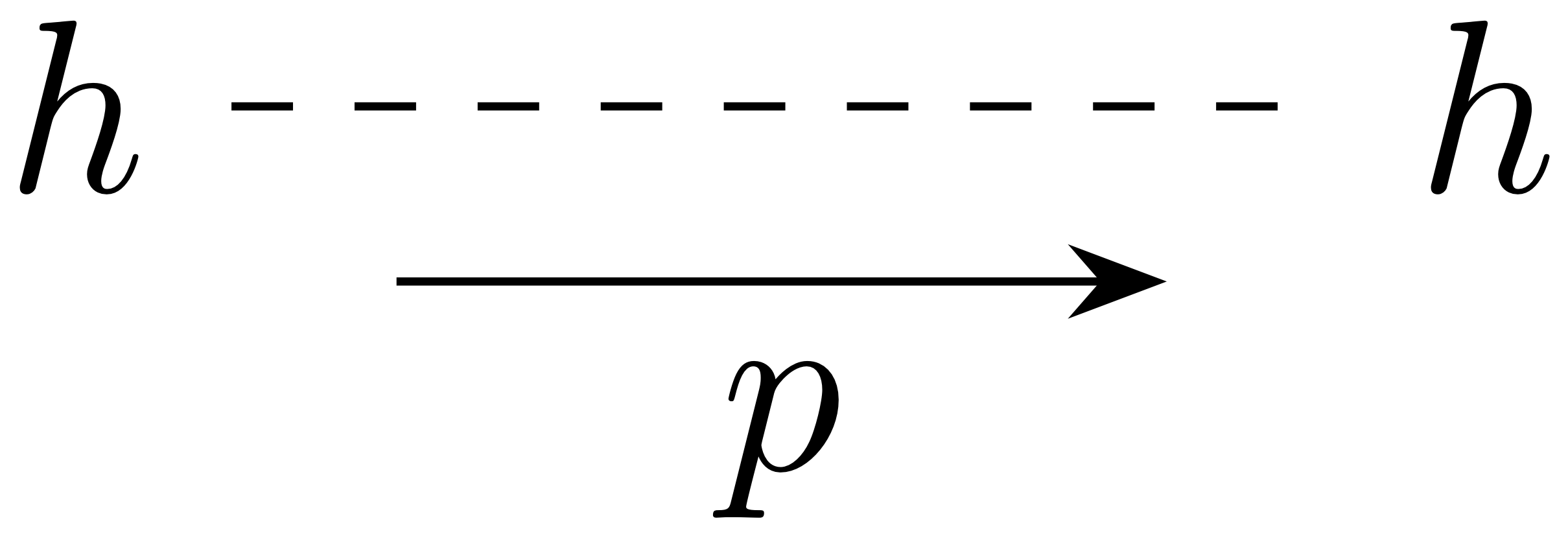}&
        \vspace{10 px}
        $\dfrac{i}{p^2 - m_h^2 + i \epsilon}$ \newline \\
        \vspace{10 px}
        \includegraphics[scale=0.03]{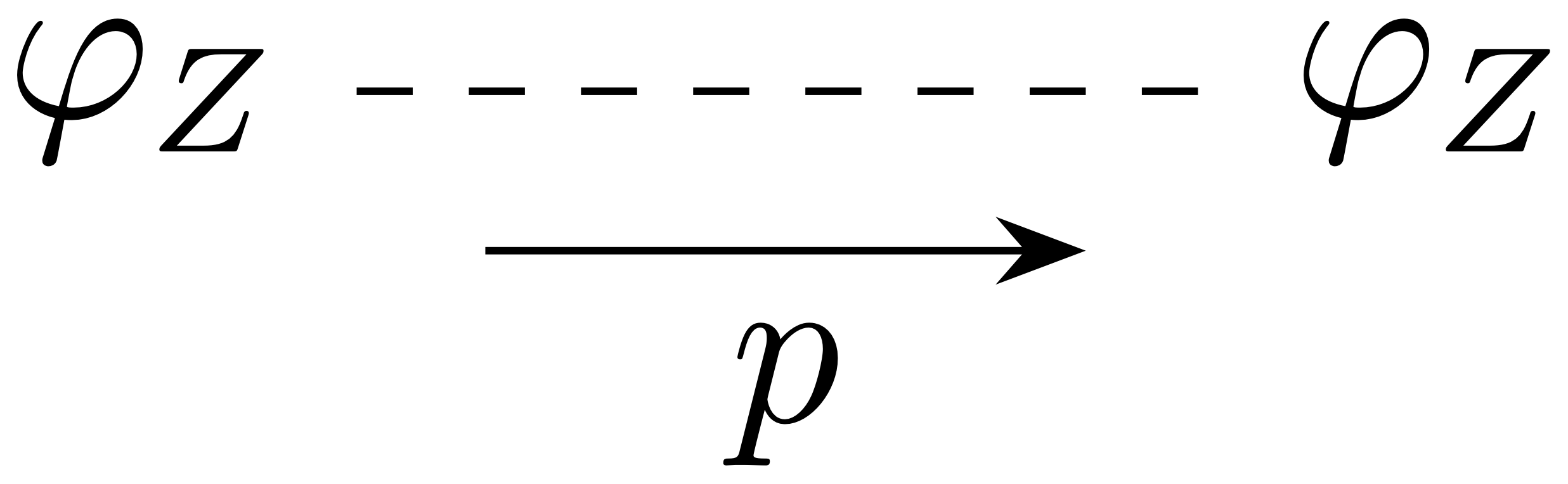}&
        \vspace{10 px}
        $\dfrac{i}{p^2 - m_Z^2 +i \epsilon}$ \newline \\
        \vspace{10 px}
        \includegraphics[scale=0.03]{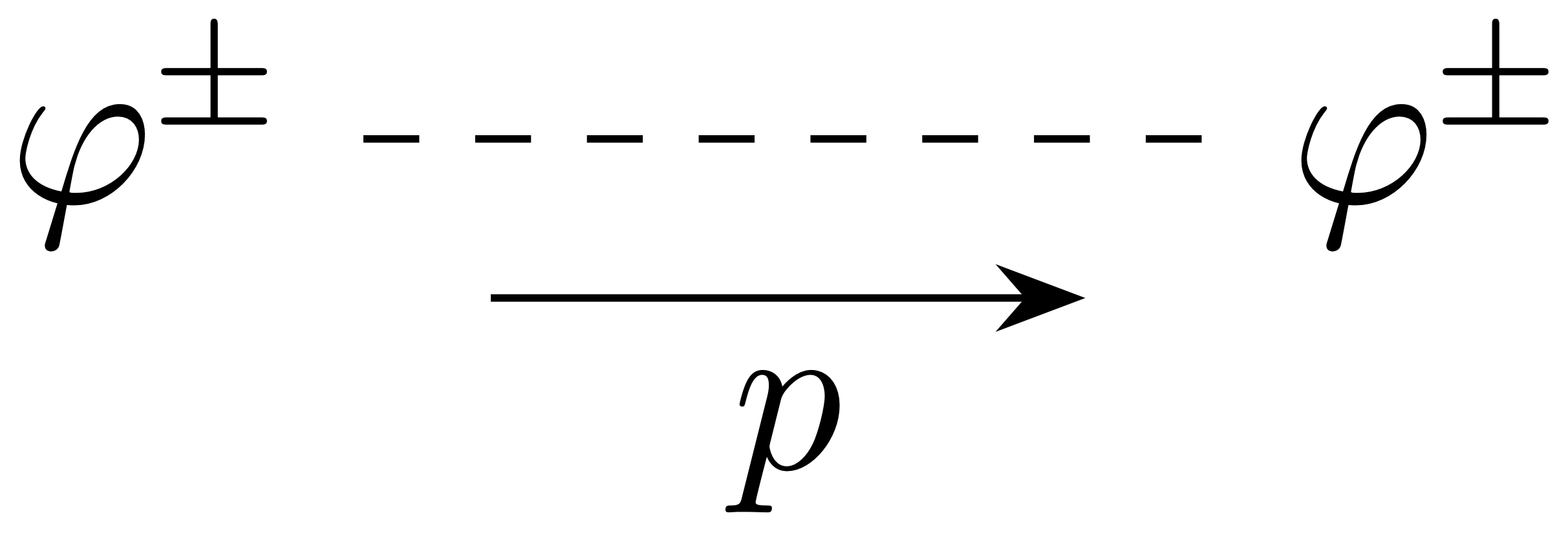}&
        \vspace{10 px}
        $\dfrac{i}{p^2 - m_W^2 +i \epsilon}$ \newline \\
        \hline
    \end{tabular}
\end{table}
\end{center}

\begin{center}
\begin{table}[!h]
    \centering\begin{tabular}{>{\centering\arraybackslash}m{3.5cm} m{9cm}} 
    \multicolumn{2}{c}{Ghost Propagators} \\
    \hline
        Diagram & Feynman Rule \\
        \hline
        \vspace{10 px}
        \includegraphics[scale=0.03]{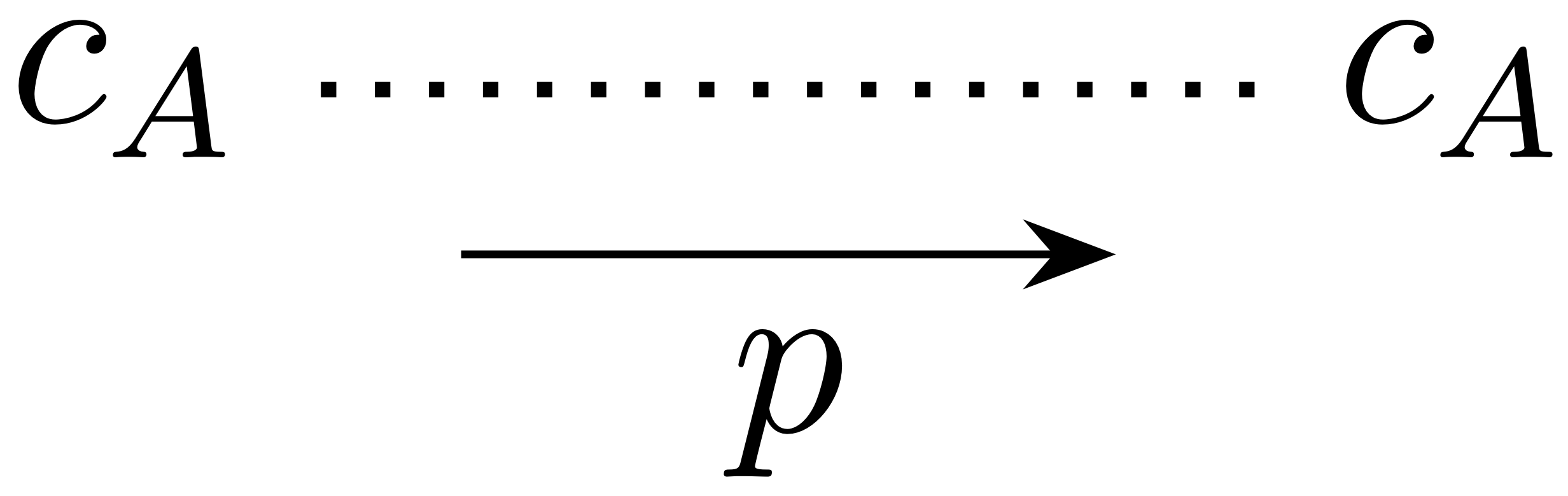}&
        \vspace{10 px}
        $\dfrac{i}{k^2 + i \epsilon}$\newline \\
        \vspace{10 px}
        \includegraphics[scale=0.03]{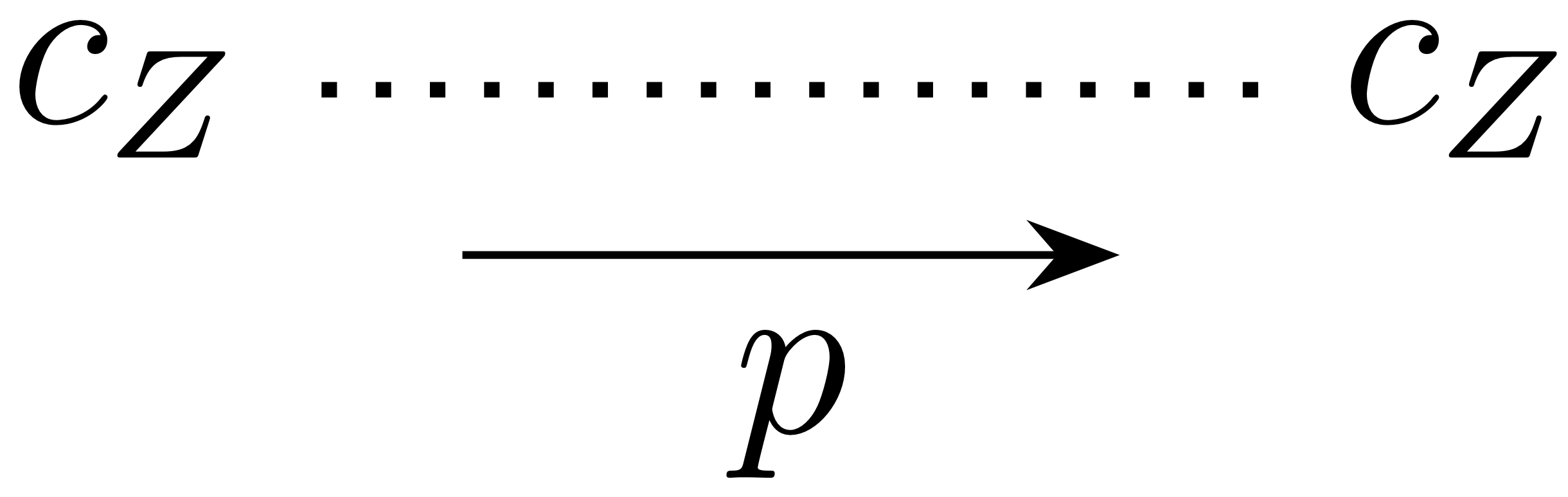}&
        \vspace{10 px}
        $\dfrac{i}{k^2 - m_Z^2 + i \epsilon}$ \newline \\
        \vspace{10 px}
        \includegraphics[scale=0.03]{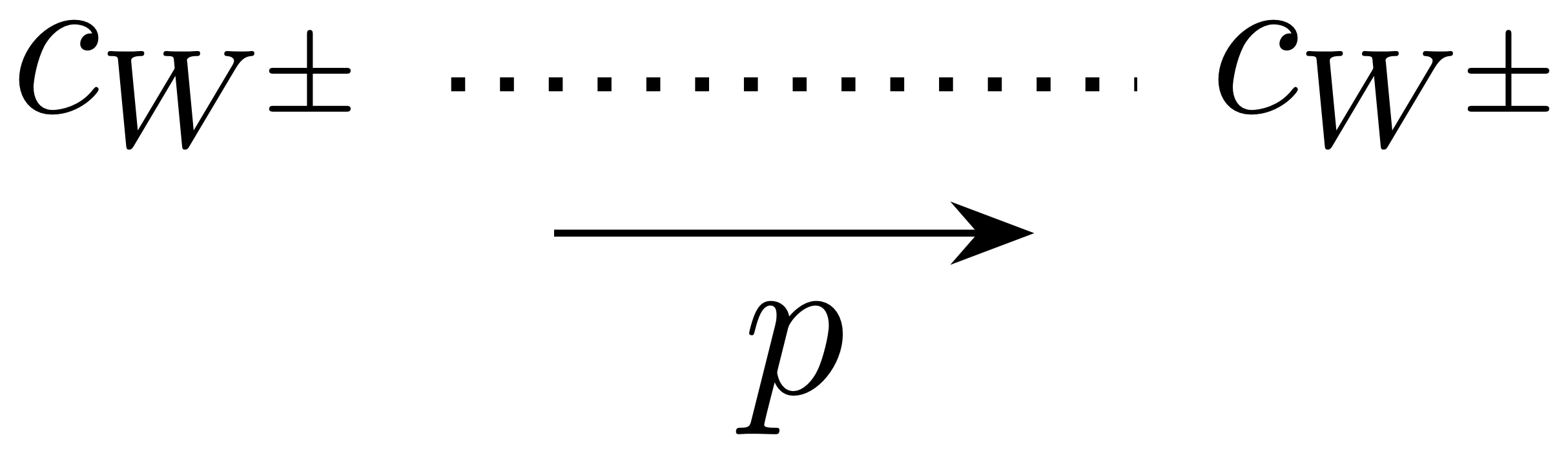}&
        \vspace{10 px}
        $\dfrac{i}{k^2 - m_W^2 + i \epsilon}$ \newline \\
        \vspace{10 px}
        \includegraphics[scale=0.03]{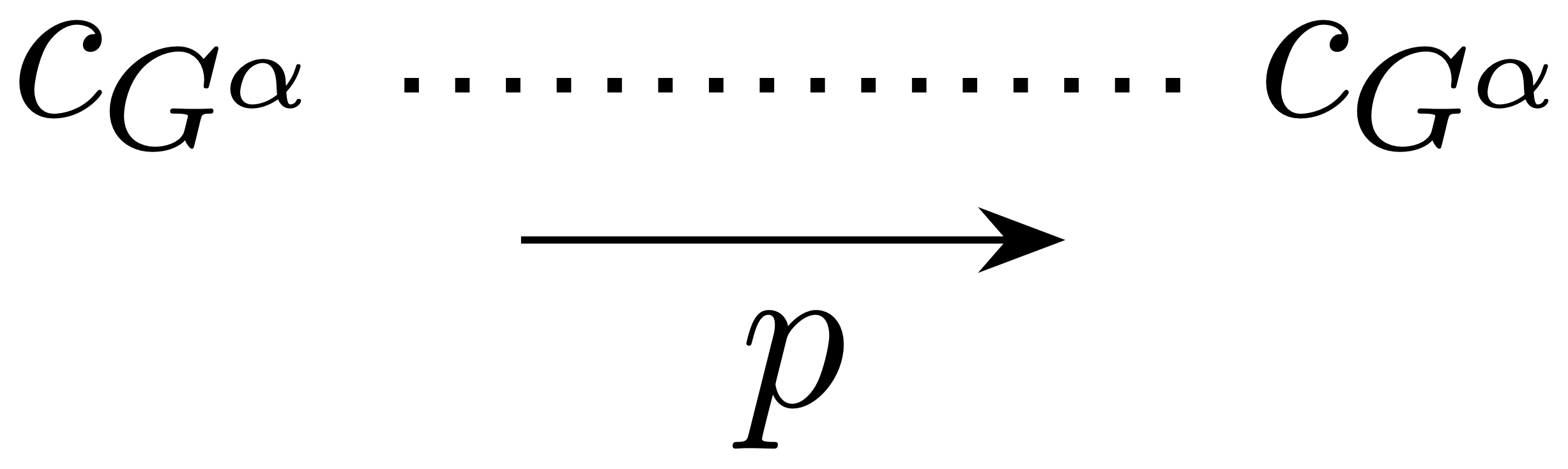}&
        \vspace{10 px}
        $\dfrac{i}{k^2 + i \epsilon}$ \newline \\
        \hline
        \hline
    \end{tabular}
\end{table}
\end{center} 
\pagebreak

\clearpage

\subsection{SM Fermions}
The Feynman rules for the interactions of SM fermions with gauge bosons are given below. All fields are defined to be incoming. The bar labels an antifermion. We omit the color indices whenever they just involve trivial contractions.

\begin{center}
\begin{table}[!h]
    \centering\begin{tabular}{>{\centering\arraybackslash}m{4cm} m{8.5cm}} 
    \multicolumn{2}{c}{Interactions with Photons} \\
    \hline
        Diagram & Feynman Rule \\
        \hline
        \vspace{5 px}
        \includegraphics[height=2.25cm]{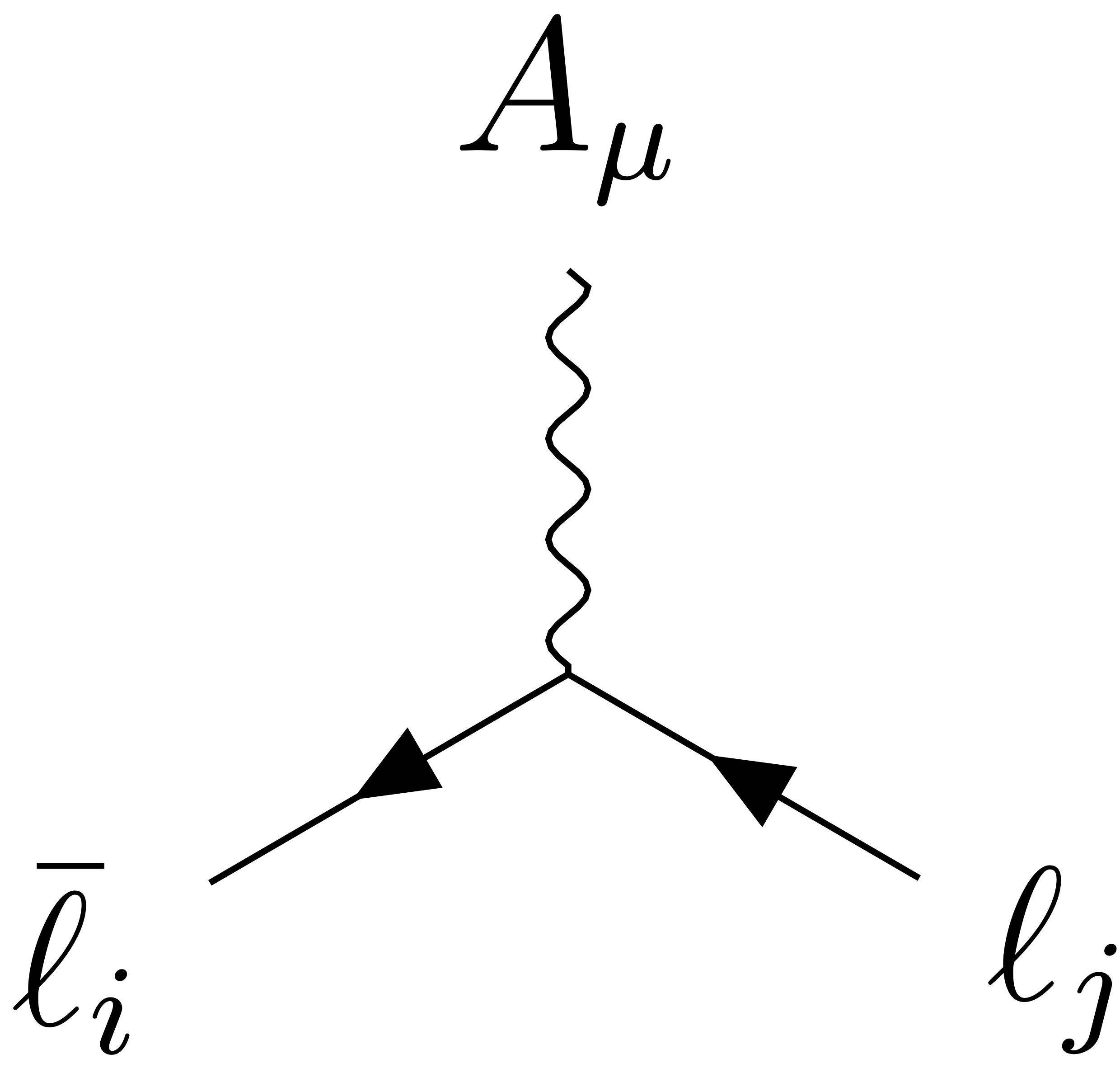}&
        $- \ i e \delta_{ij} \gamma^{\mu}$ \\
        \vspace{5 px}
        \includegraphics[height=2.25cm]{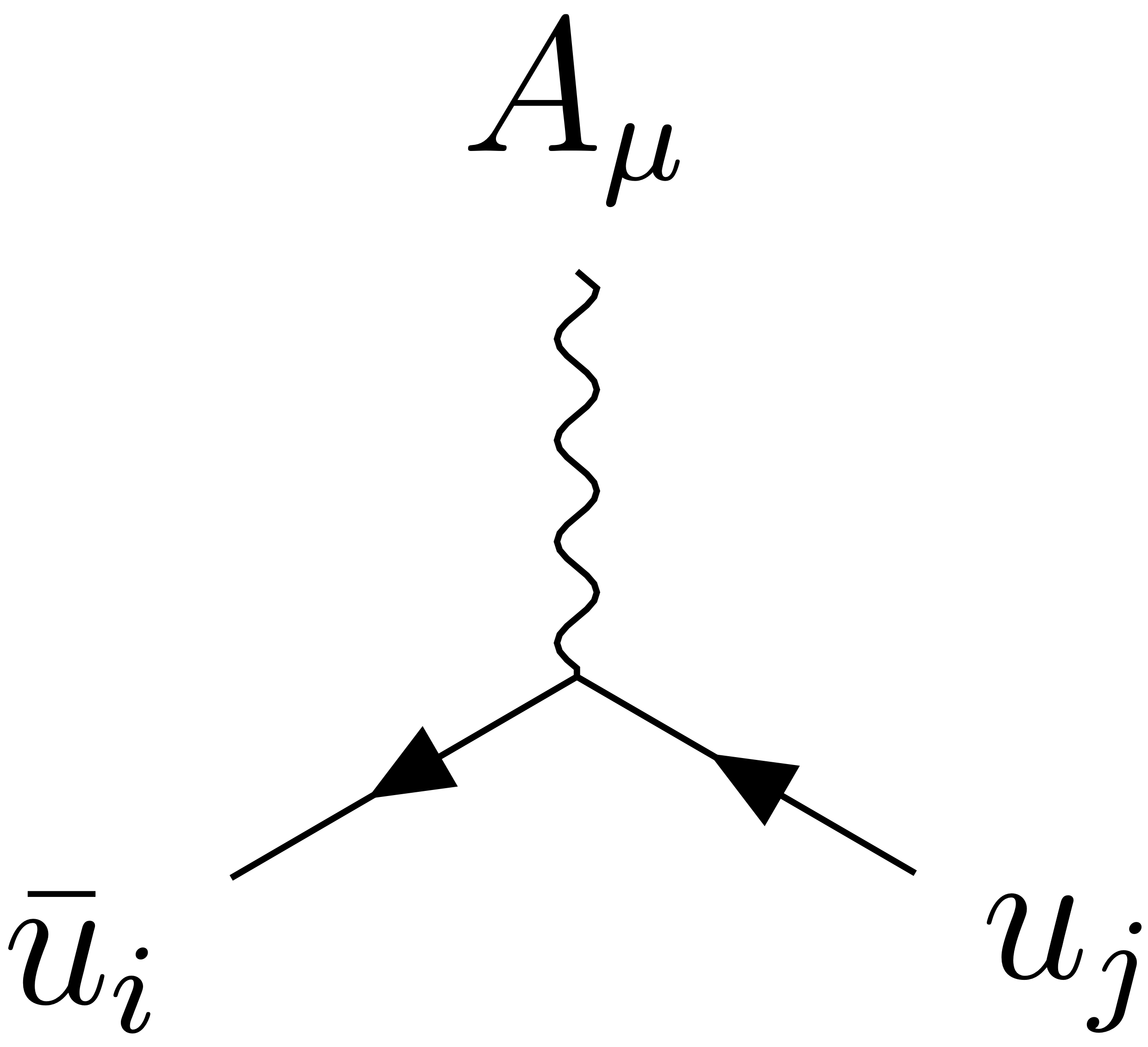}&
        $+ \ \dfrac{2i}{3} e \delta_{ij} \gamma^{\mu}$\\
        \vspace{5 px}
        \includegraphics[height=2.25cm]{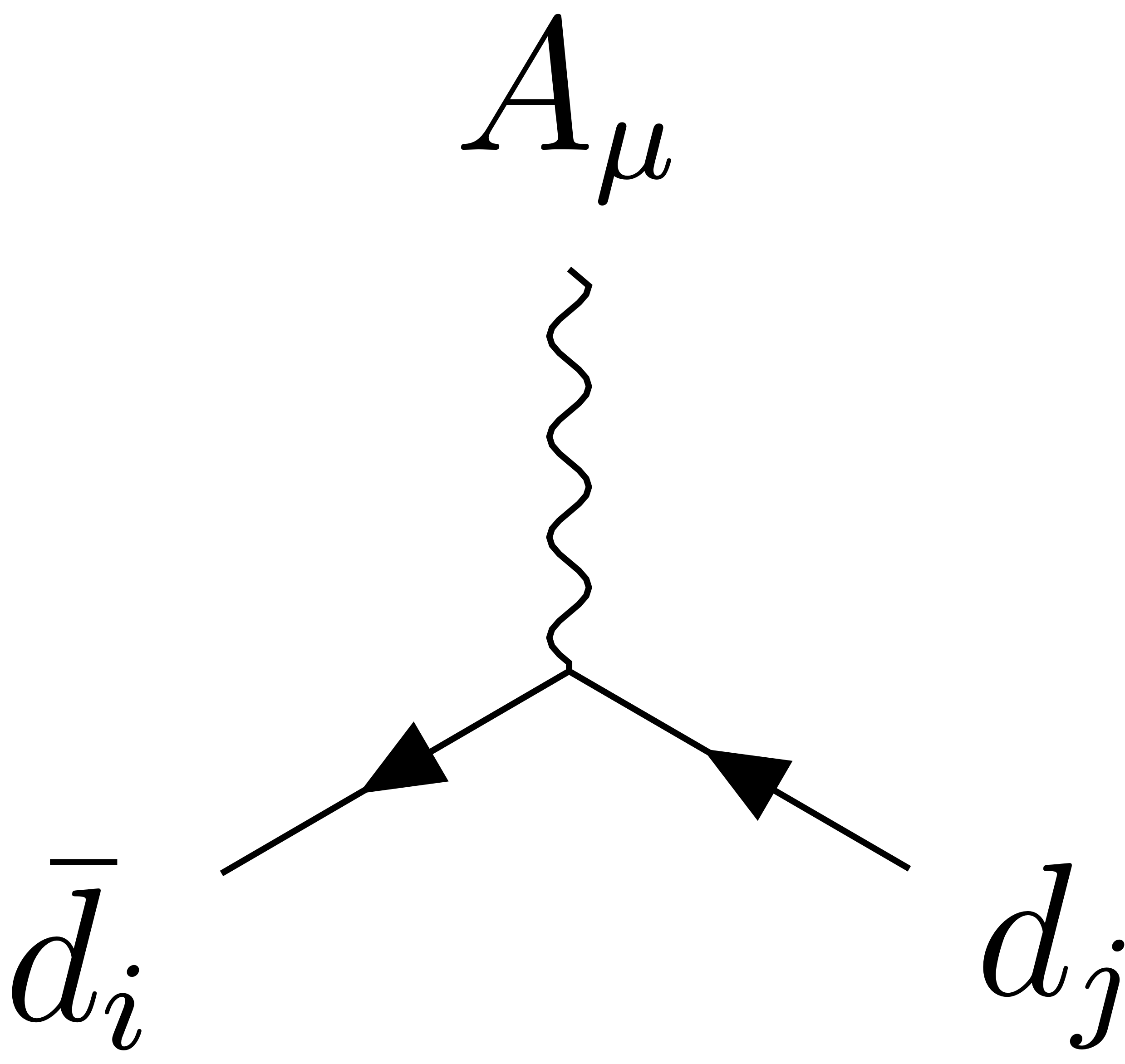}&
        $- \ \dfrac{i}{3} e \delta_{ij} \gamma^{\mu}$\\
        \hline
    \end{tabular}
\end{table}
\end{center}

\begin{center}
\begin{table}[!h]
    \centering\begin{tabular}{>{\centering\arraybackslash}m{4cm} m{8.5cm}} 
        \multicolumn{2}{c}{Interactions with $Z$ Bosons} \\
    \hline
        Diagram & Feynman Rule \\
        \hline
        \vspace{5 px}
        \includegraphics[height=2.25cm]{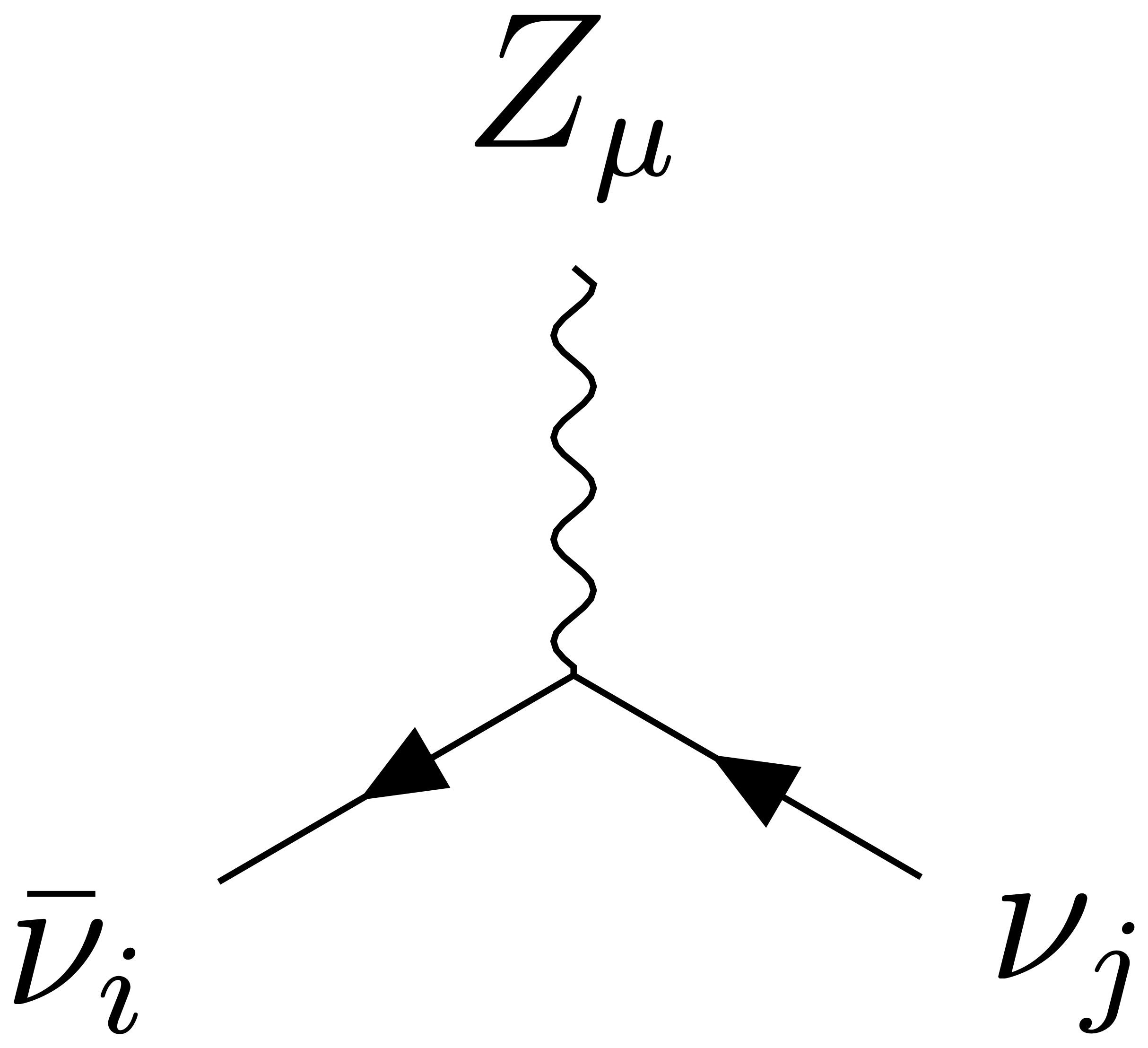}&
        $+ \ \dfrac{i}{2} \dfrac{e}{s_w c_w} \delta_{ij} \gamma^{\mu} \ \text{P}_{\text{L}}$\\
        \vspace{5 px}
        \includegraphics[height=2.25cm]{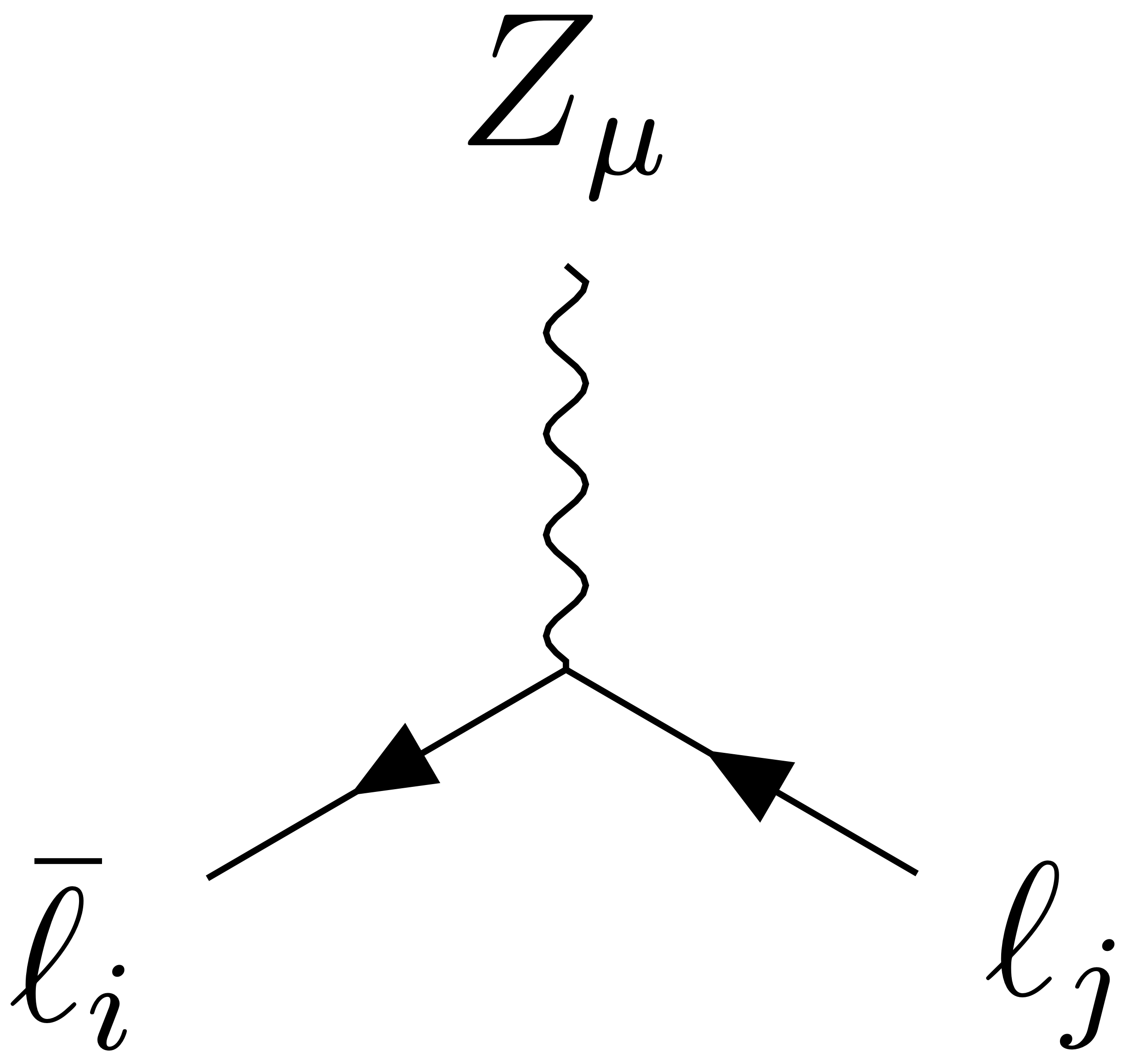}&
        $+ \ i \dfrac{e}{s_w c_w} \delta_{ij} \gamma^{\mu} \left(-\dfrac{1}{2} + s_w^2 \right) \ \text{P}_{\text{L}} \vspace{5px} \newline
        + i e \dfrac{s_w}{c_w} \delta_{ij} \gamma^{\mu} \  \text{P}_{\text{R}}$\\
        \vspace{5 px}
        \includegraphics[height=2.25cm]{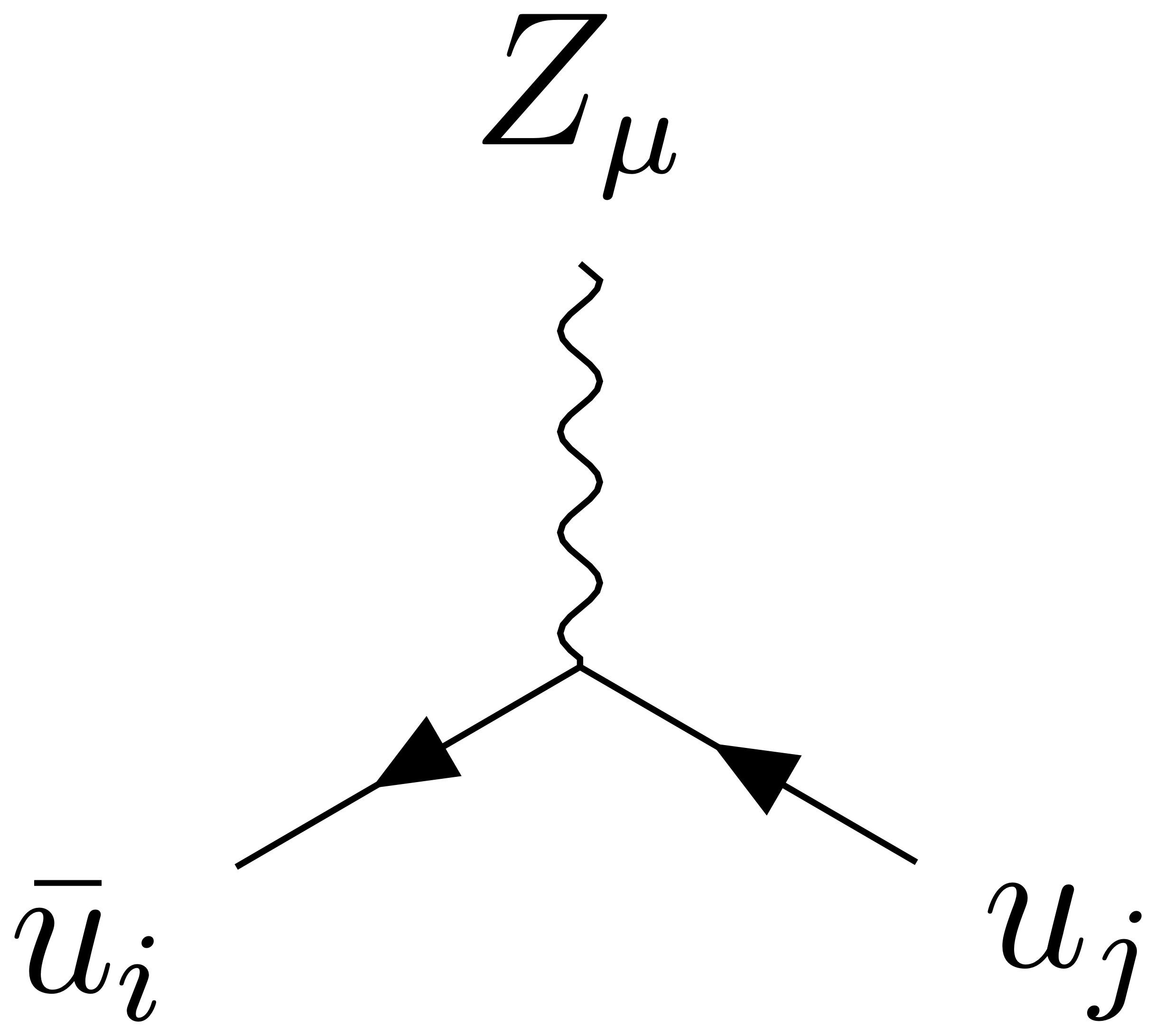}&
        $+ \ i \dfrac{e}{s_w c_w} \delta_{ij} \gamma^{\mu} \left(\dfrac{1}{2} - \dfrac{2}{3}s_w^2 \right) \ \text{P}_{\text{L}} \vspace{5px} \newline
        - \dfrac{2i}{3} e \dfrac{s_w}{c_w} \delta_{ij} \gamma^{\mu} \  \text{P}_{\text{R}}$\\
        \vspace{5 px}
        \includegraphics[height=2.25cm]{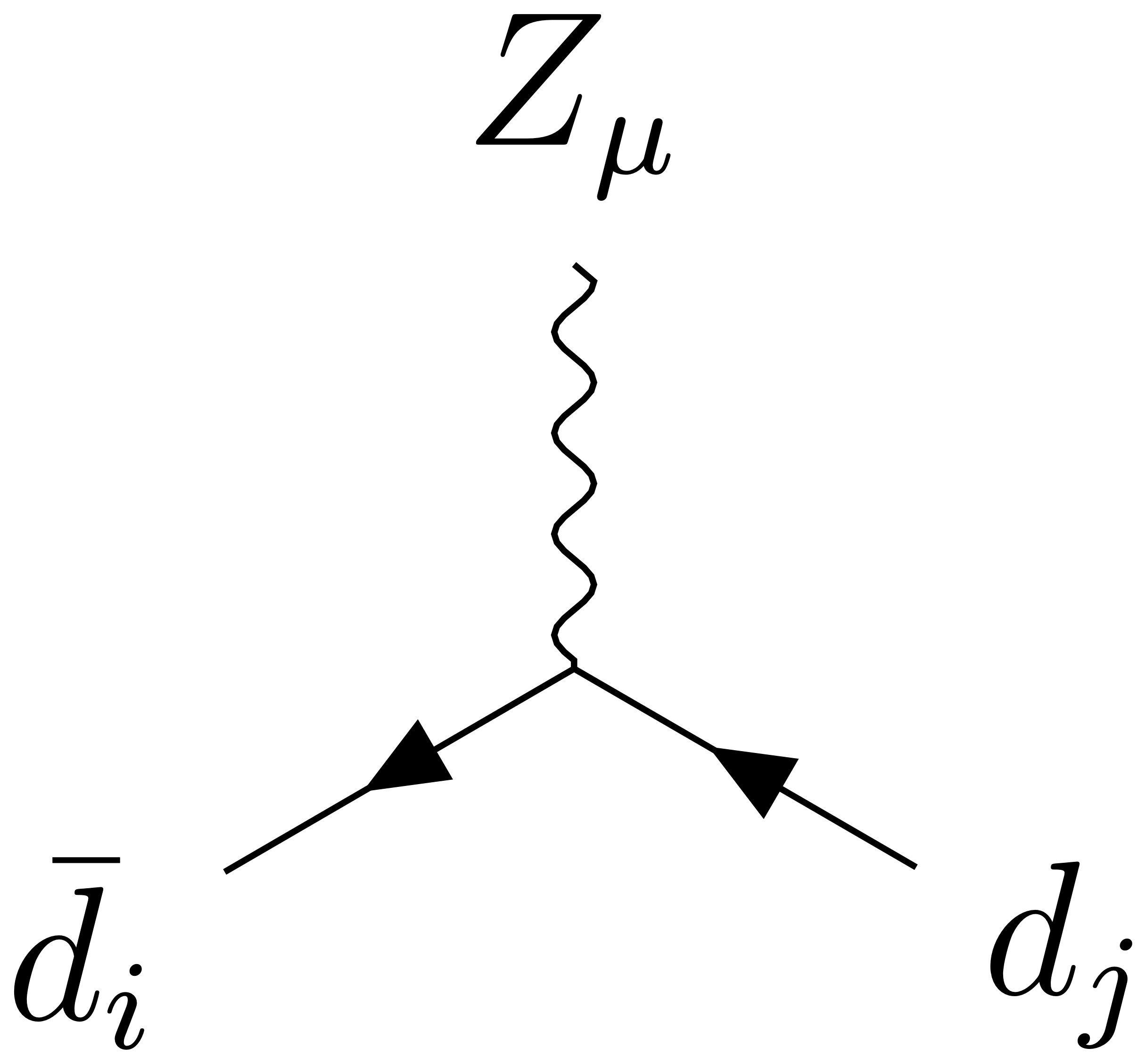}&
        $+ \ i \dfrac{e}{s_w c_w} \delta_{ij} \gamma^{\mu} \left(-\dfrac{1}{2} + \dfrac{1}{3}s_w^2 \right) \ \text{P}_{\text{L}}  \vspace{5px} \newline
        + \dfrac{i}{3} e\dfrac{s_w}{c_w} \delta_{ij} \gamma^{\mu} \ \text{P}_{\text{R}}$\\
        \hline
    \end{tabular}
\end{table}
\end{center}

\begin{center}
\begin{table}[!h]
    \centering\begin{tabular}{>{\centering\arraybackslash}m{4cm} m{8.5cm}} 
        \multicolumn{2}{c}{Interactions with $W$ Bosons} \\
    \hline
        Diagram & Feynman Rule \\
        \hline
        \vspace{5 px}
        \includegraphics[height=2.25cm]{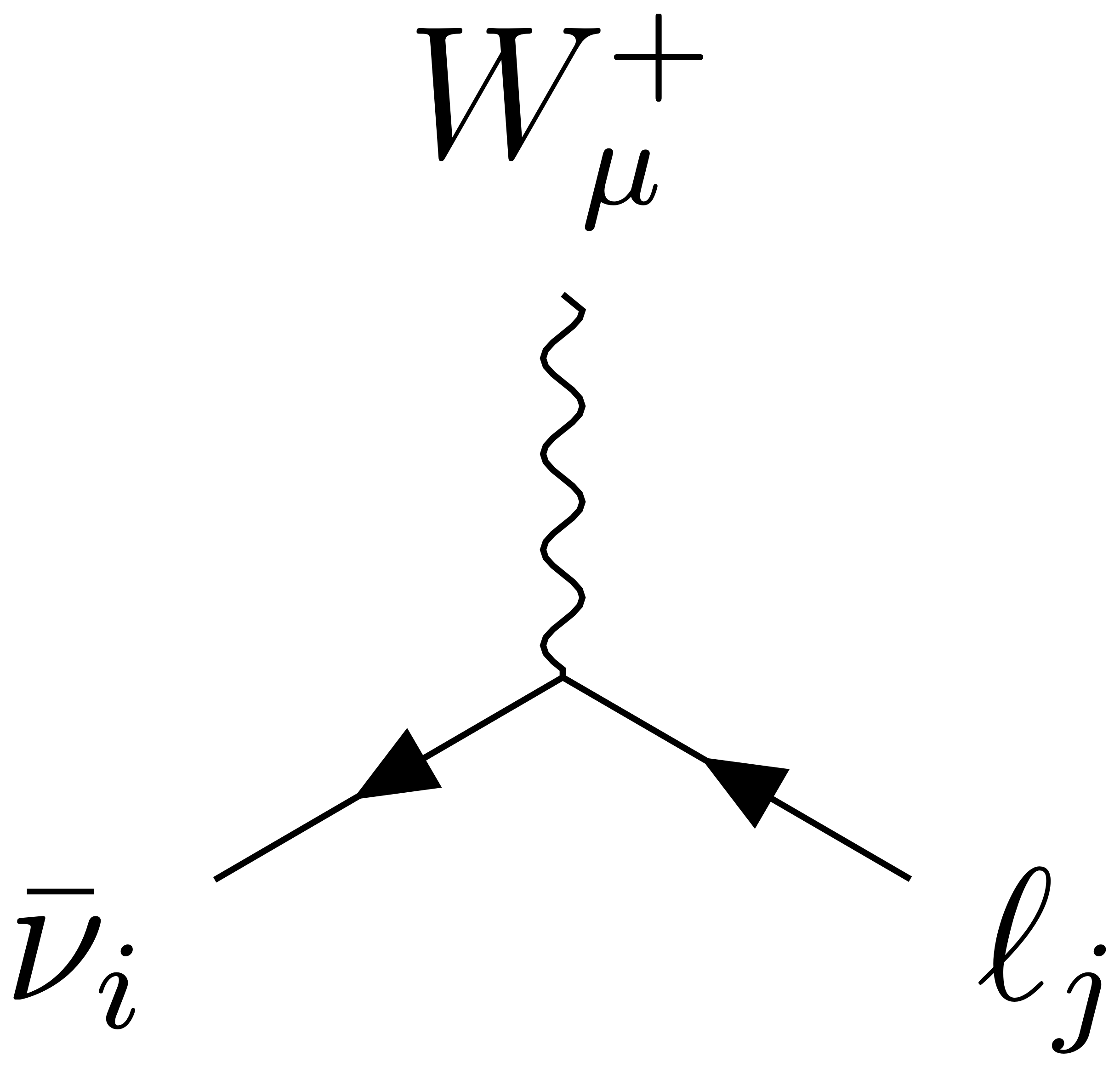}&
        $+ \ \dfrac{i}{\sqrt{2}} \dfrac{e}{s_w} \delta_{ij} \gamma^{\mu} \ \text{P}_{\text{L}}$\\
        \vspace{5 px}
        \includegraphics[height=2.25cm]{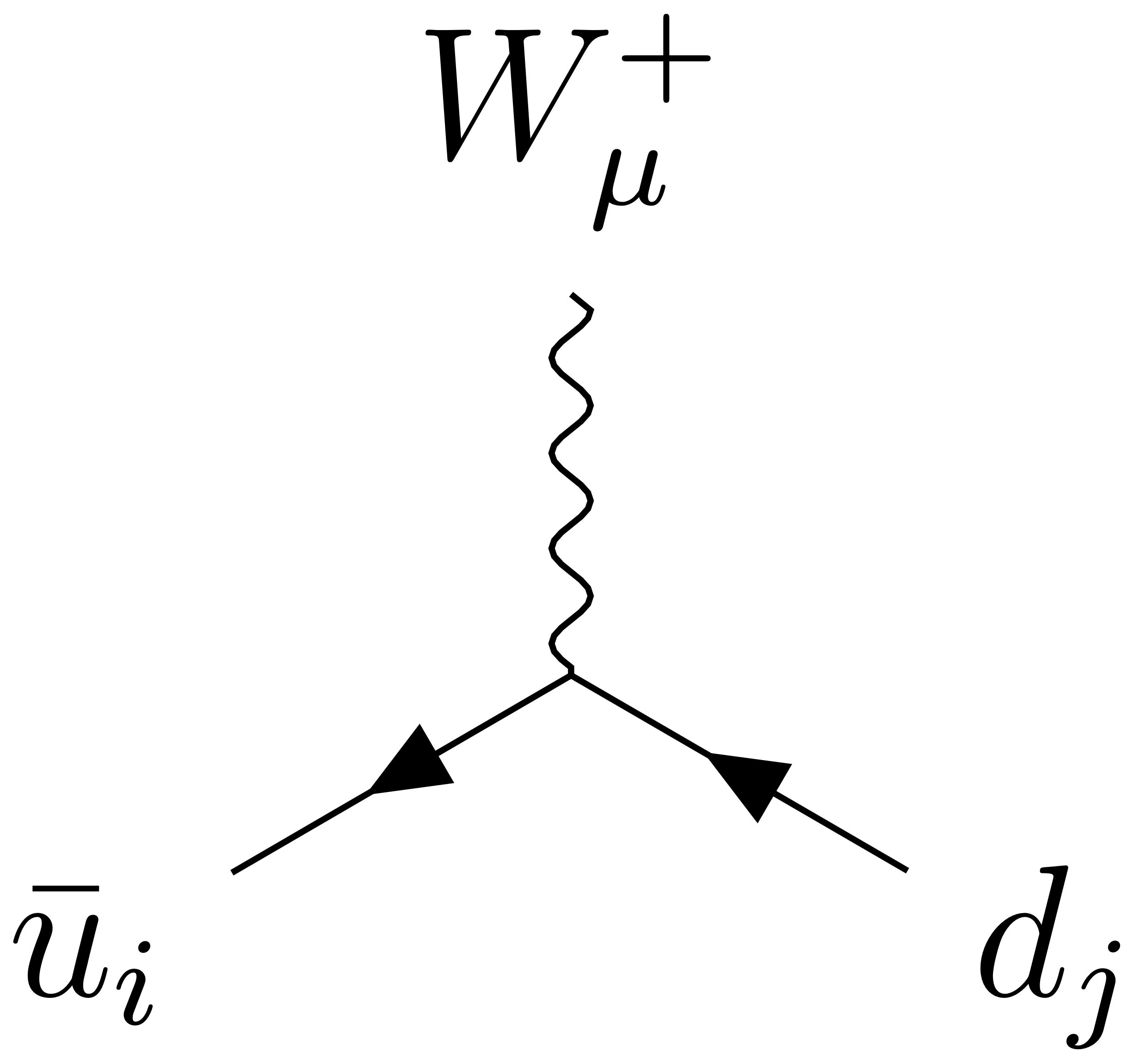}&
        $+ \ \dfrac{i}{\sqrt{2}} \dfrac{e}{s_w} V_{ij} \gamma^{\mu} \   \text{P}_{\text{L}}$ \\
        \hline
    \end{tabular}
\end{table}
\end{center}

\begin{center}
\begin{table}[!h]
    \centering\begin{tabular}{>{\centering\arraybackslash}m{4cm} m{8.5cm}} 
        \multicolumn{2}{c}{Interactions with Gluons} \\
    \hline
        Diagram & Feynman Rule \\
        \hline
        \vspace{5 px}
        \includegraphics[height=2.25cm]{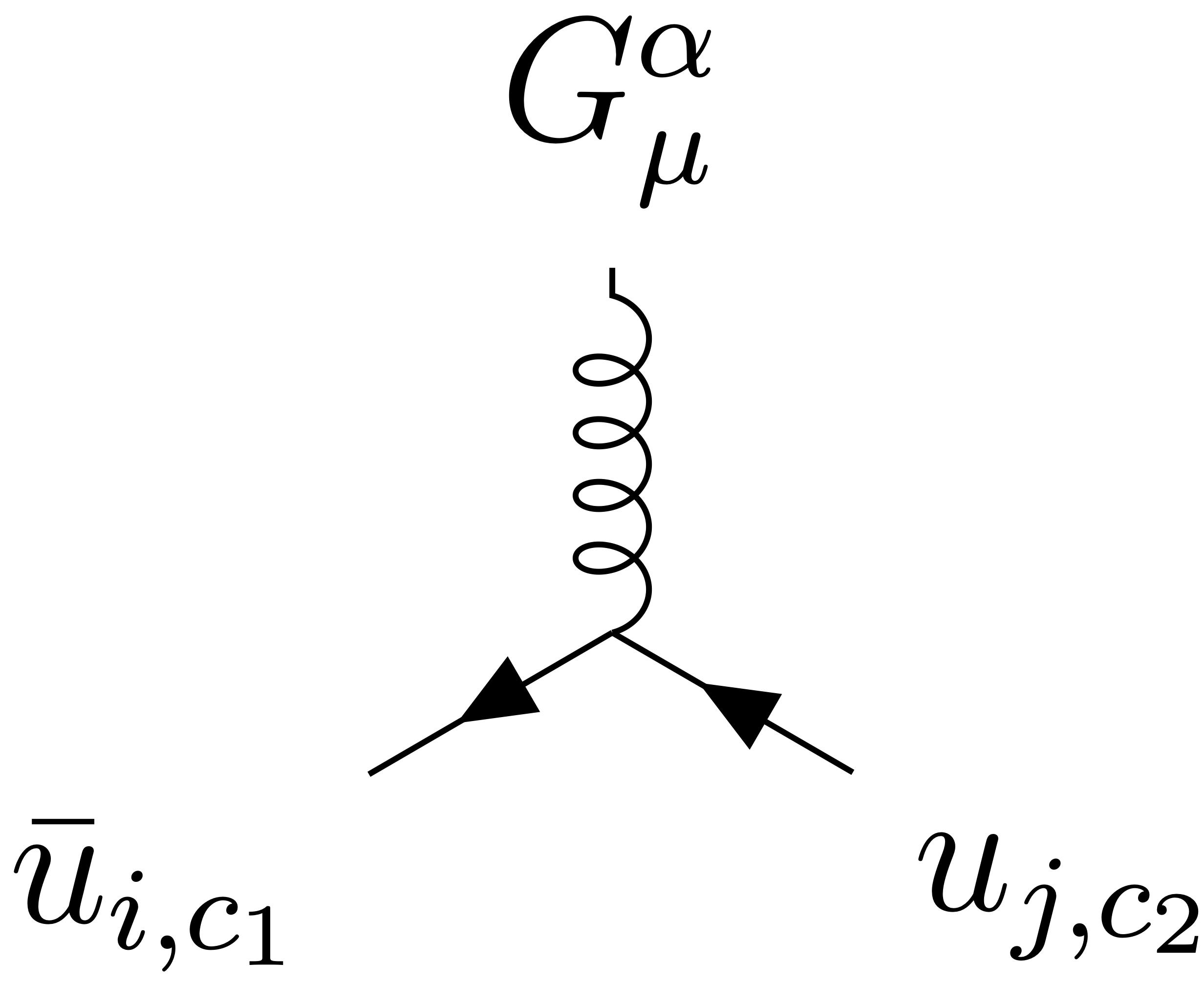}&
        $+ \ i g_s \delta_{ij}  T^{\alpha}_{c_1 c_2} \gamma^{\mu}$\\
        \vspace{5 px}
        \includegraphics[height=2.25cm]{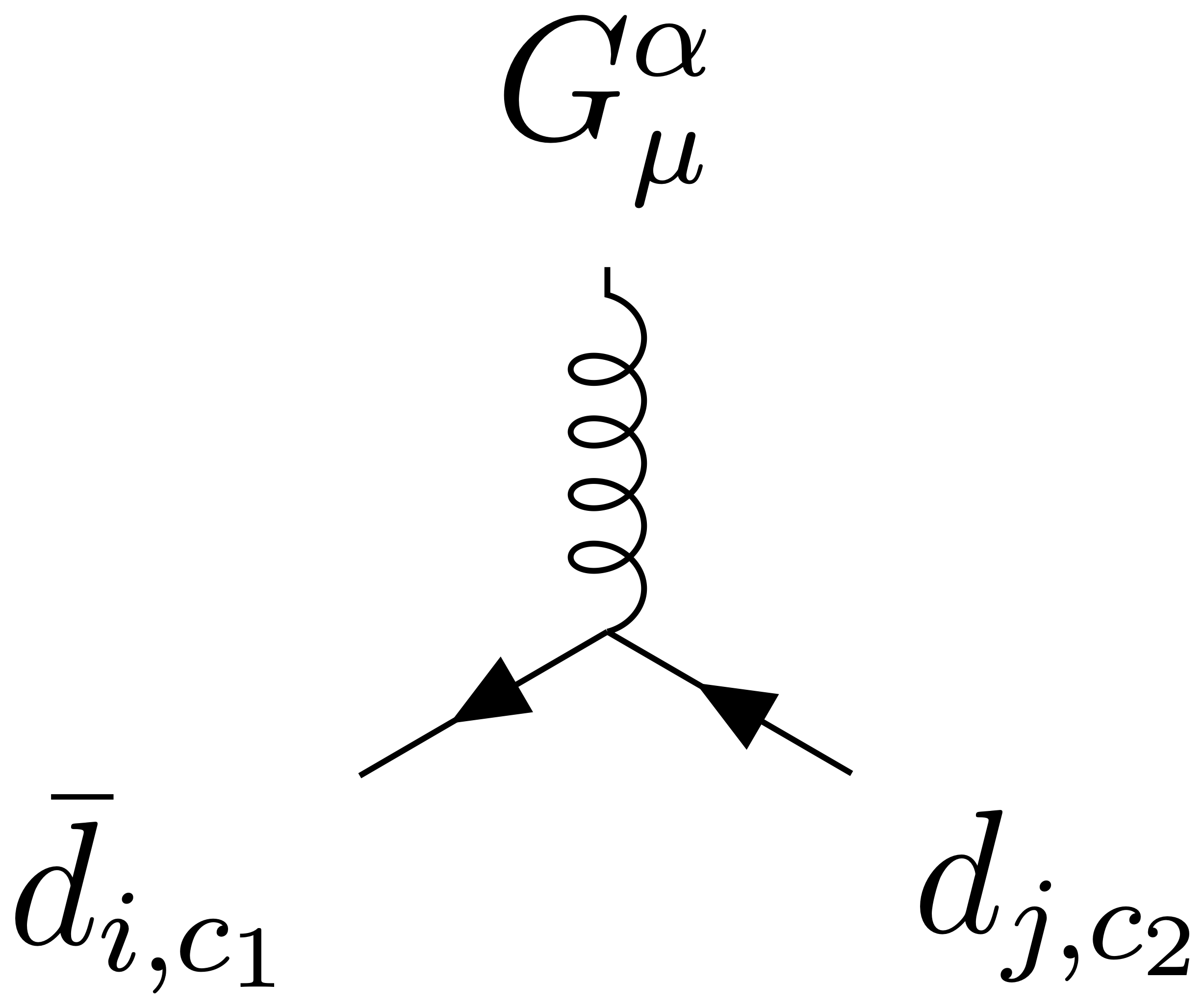}&
        $+ \ i g_s \delta_{ij}  T^{\alpha}_{c_1 c_2} \gamma^{\mu}$\\
        \hline
        \hline
    \end{tabular}
\end{table}
\end{center}

\clearpage
\subsection{SM Gauge Bosons}
The Feynman rules for SM gauge boson self-interactions as well as the SM gauge boson interactions with ghost fields are given below.

\begin{center}
\begin{table}[!h]
    \centering\begin{tabular}{>{\centering\arraybackslash}m{4cm} m{8.5cm}} 
            \multicolumn{2}{c}{Triple Self-Interactions} \\
    \hline
        Diagram & Feynman Rule \\
        \hline
        \vspace{5 px}
        \includegraphics[height=2.25cm]{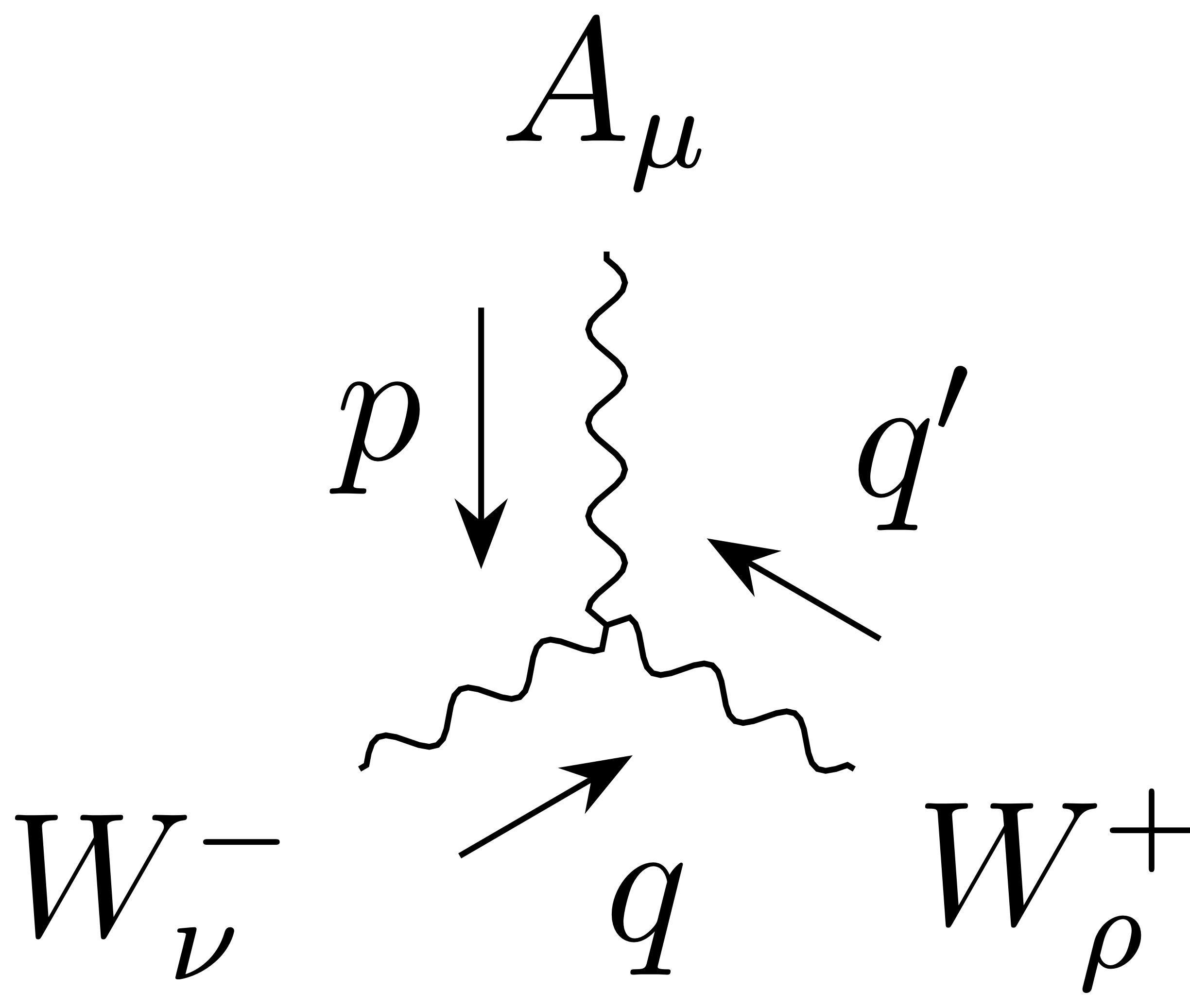}&
        $+ \ i e \eta^{\mu\rho} \left(q^\prime - p\right)^\nu \vspace {5px} 
        + i e \eta^{\mu\nu} \left(p - q\right)^{\rho}  \vspace {5px} \newline
        + i e \eta^{\nu \rho} \left(q - q^\prime\right)^{\mu}
        $\\
        \vspace{5 px}
        \includegraphics[height=2.25cm]{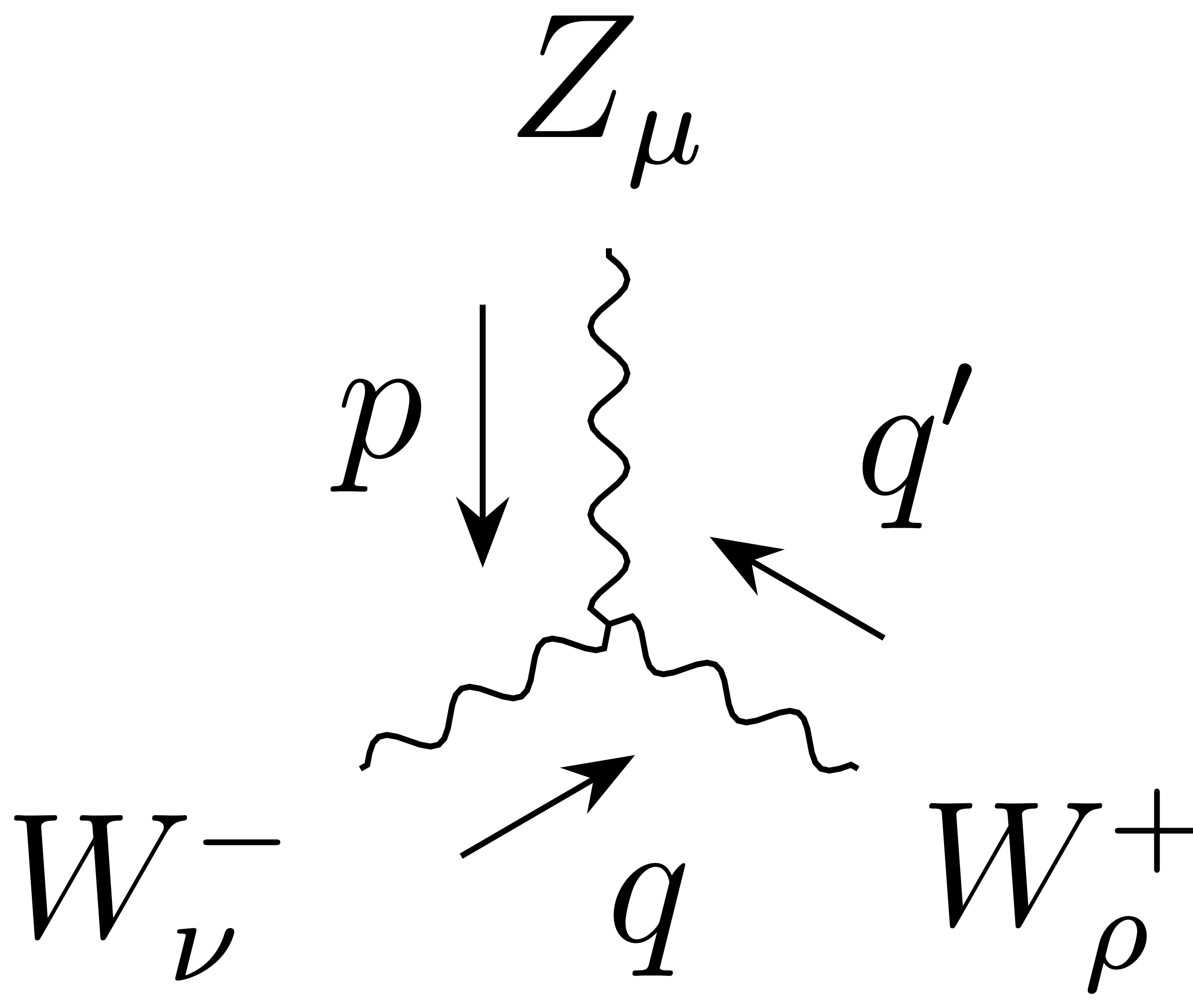}&
        $+ \ i e \dfrac{c_w}{s_w} \eta^{\mu\rho} \left(q^\prime - p\right)^\nu  \vspace {5px} \newline
        + i e\dfrac{c_w}{s_w}  \eta^{\mu\nu} \left(p - q\right)^{\rho}  \vspace {5px} \newline
        + i e\dfrac{c_w}{s_w}  \eta^{\nu \rho} \left(q - q^\prime\right)^{\mu}
        $ \vspace{5px} \\
        \vspace{5 px}
        \includegraphics[height=2.25cm]{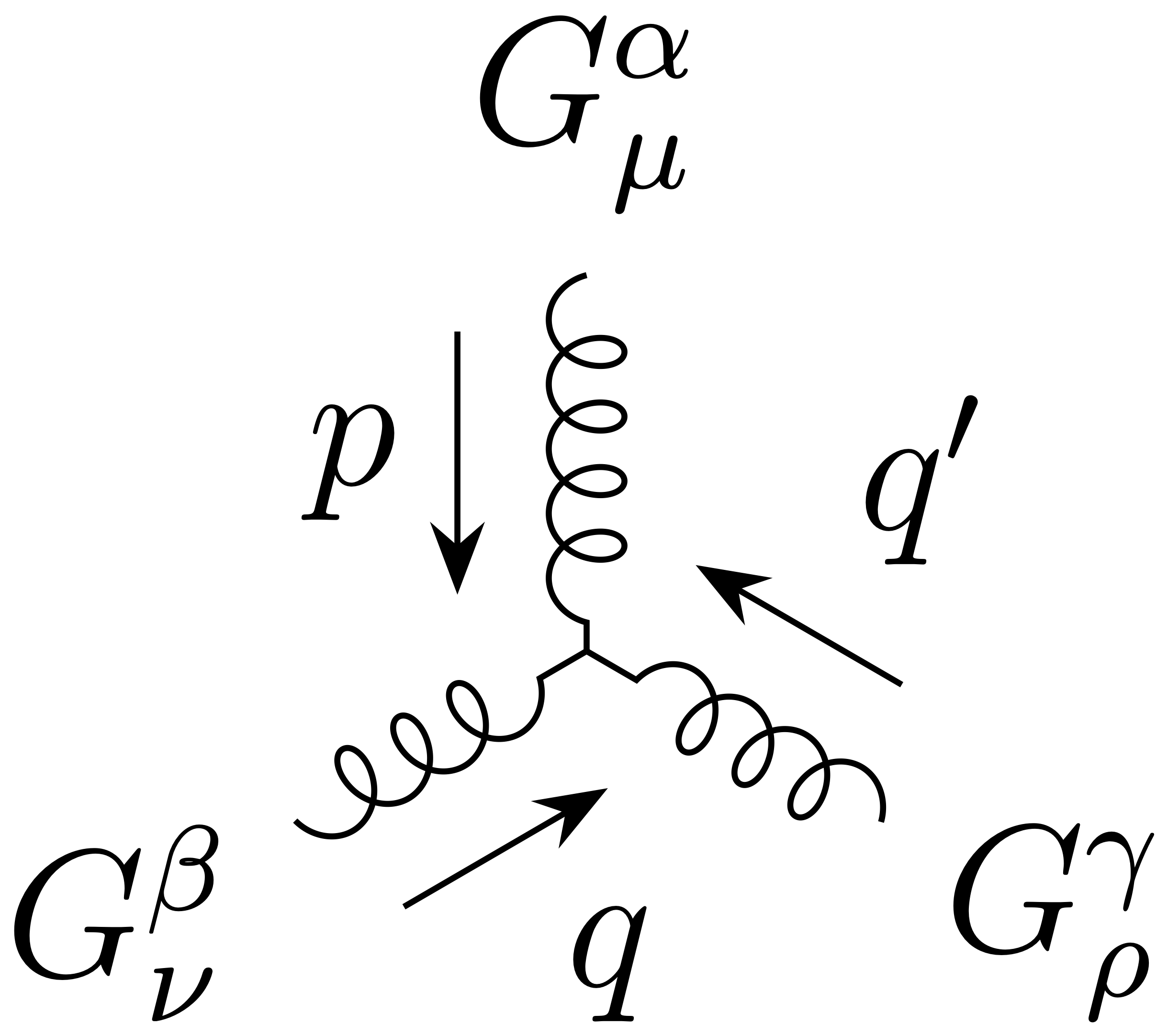}&
        $+ \ g_s f^{\alpha \beta \gamma} \eta^{\mu\nu} \left(p - q \right)^\rho \vspace{5px} \newline
        + g_s f^{\alpha \beta \gamma} \eta^{\mu\rho} \left(q^\prime - p \right)^\nu \vspace{5px} \newline
        + g_s f^{\alpha \beta \gamma} \eta^{\nu\rho} \left(q - q^\prime \right)^\mu $\\
        \hline
    \end{tabular}
\end{table}
\end{center}

\begin{center}
\begin{table}[!h]
   \centering\begin{tabular}{>{\centering\arraybackslash}m{5cm} m{7.5cm}} 
            \multicolumn{2}{c}{Quartic Self-Interactions} \\
    \hline
        Diagram & Feynman Rule \\
        \hline
        \vspace{5 px}
        \includegraphics[height=3.5cm]{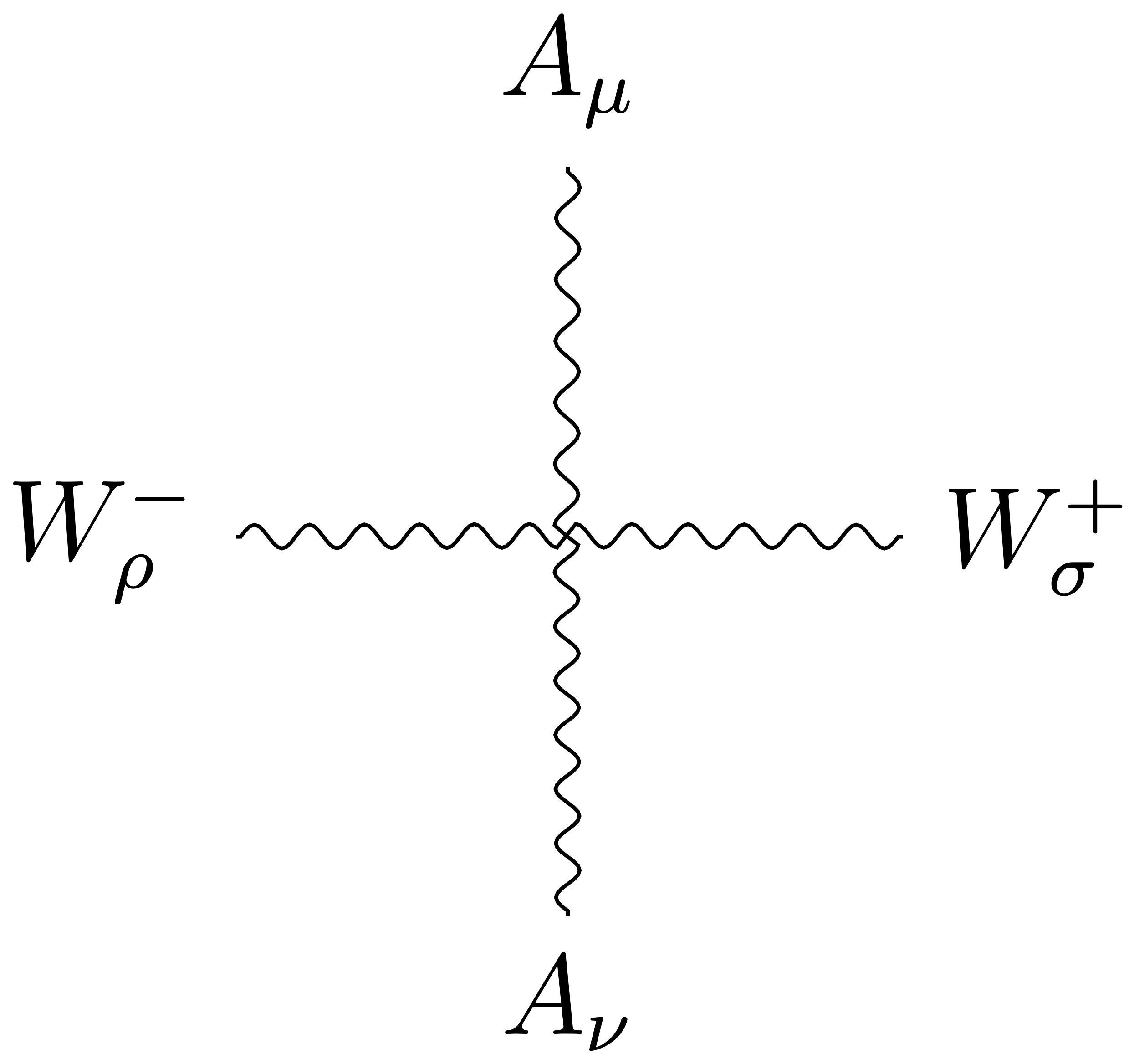}&
        $+ \ i e^2 \eta^{\mu\rho} \eta^{\nu\sigma} \vspace{5px} \newline
        + i e^2 \eta^{\mu \sigma} \eta^{\nu \rho} \vspace{5px} \newline
        - 2 i e^2 \eta^{\mu \nu} \eta^{\sigma \rho}  $\\
        \vspace{5 px}
        \includegraphics[height=3.5cm]{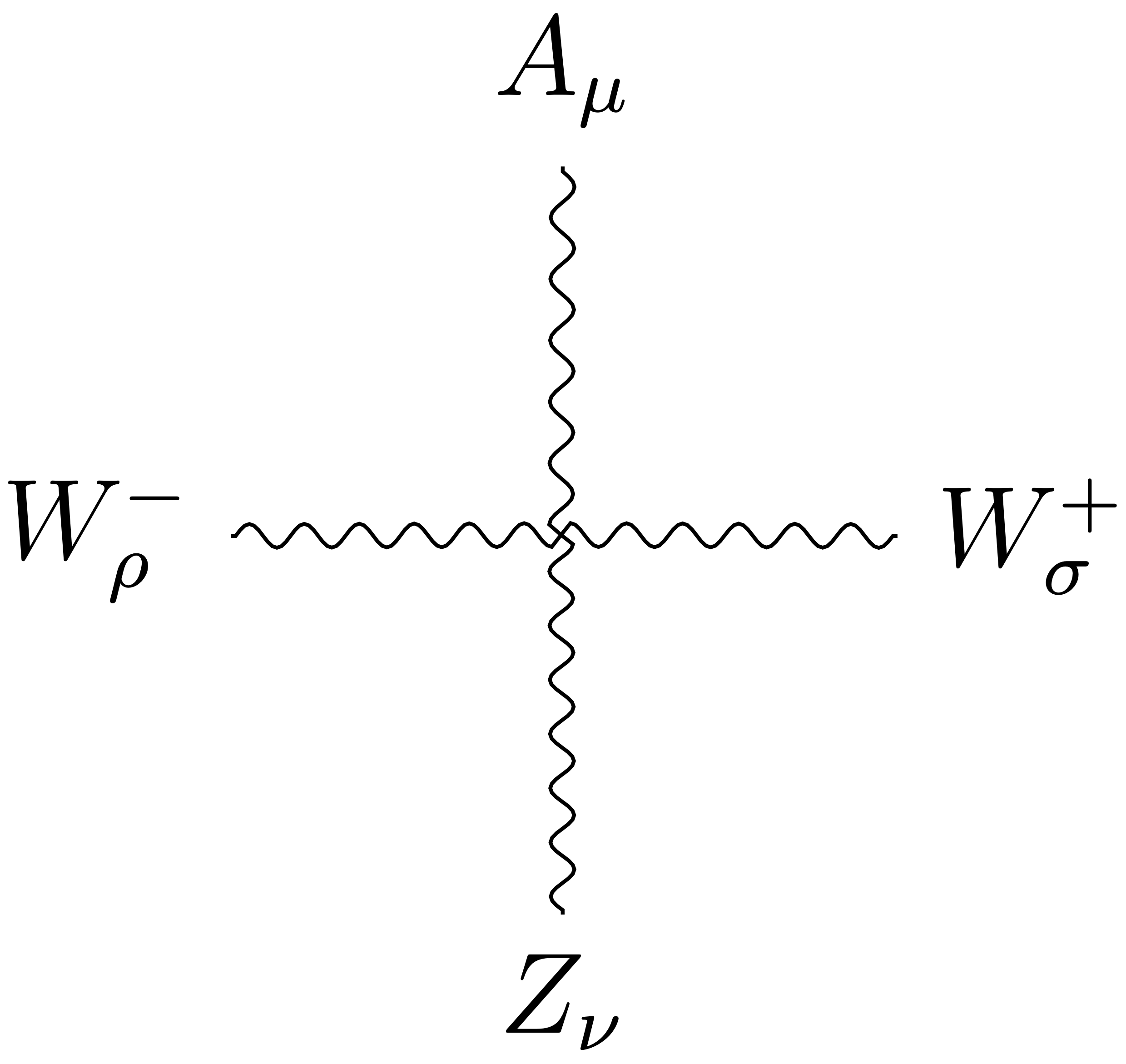}&
        $+ \ i e^2 \dfrac{c_w}{s_w} \eta^{\mu\rho} \eta^{\nu\sigma} \vspace{5px} \newline
        + i e^2 \dfrac{c_w}{s_w} \eta^{\mu \sigma} \eta^{\nu \rho} \vspace{5px} \newline
        - 2 i e^2 \dfrac{c_w}{s_w} \eta^{\mu \nu} \eta^{\rho \sigma}$\\
        \vspace{5 px}
        \includegraphics[height=3.5cm]{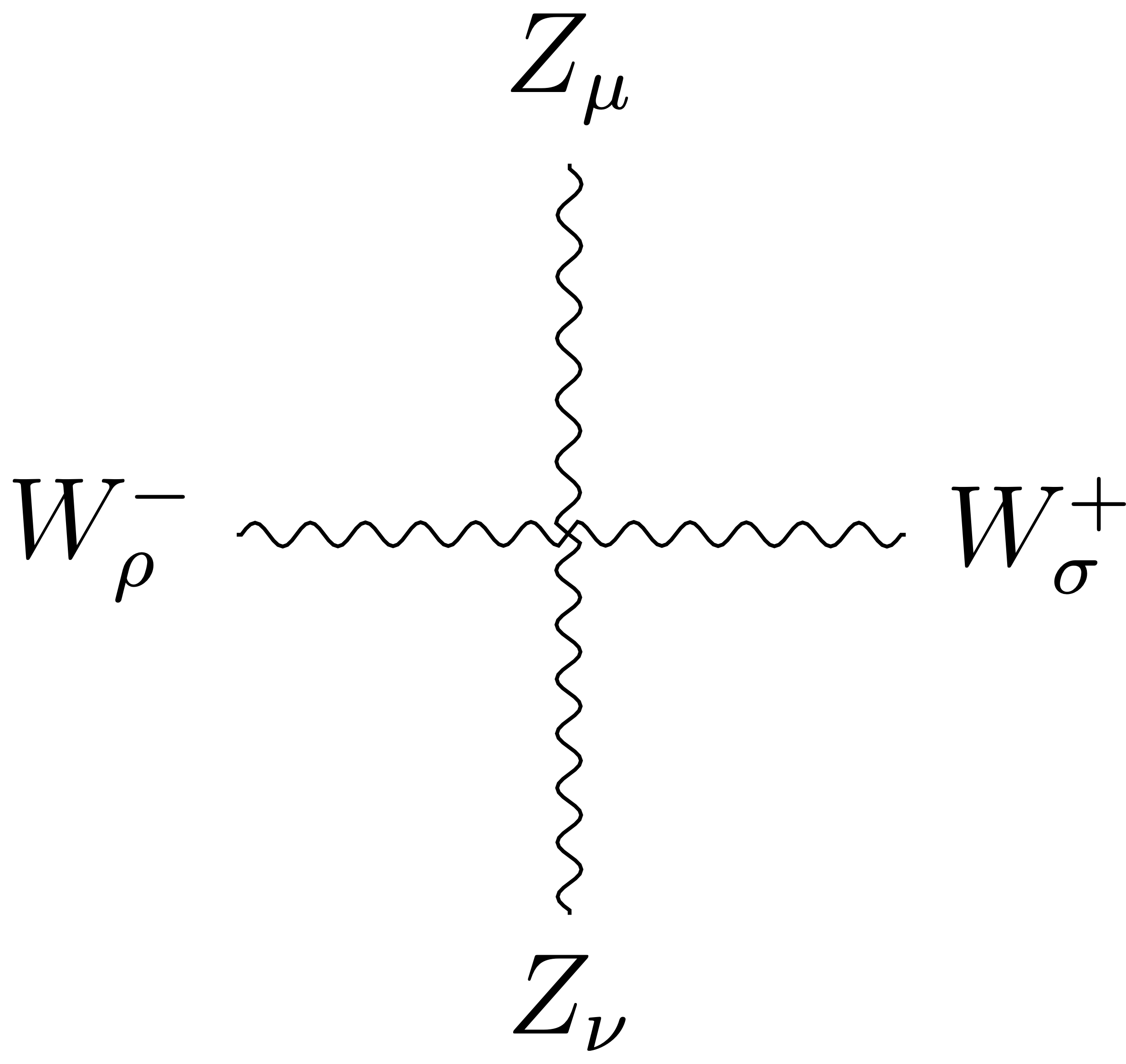}&
        $+ \ i e^2 \left(\dfrac{c_w}{s_w}\right)^2 \eta^{\mu\rho} \eta^{\nu\sigma} \vspace{5px} \newline
        + i e^2 \left(\dfrac{c_w}{s_w}\right)^2 \eta^{\mu \sigma} \eta^{\nu \rho} \vspace{5px} \newline
        - 2 i e^2 \left(\dfrac{c_w}{s_w}\right)^2 \eta^{\mu \nu} \eta^{\rho \sigma}$\\
        \vspace{5 px}
        \includegraphics[height=3.5cm]{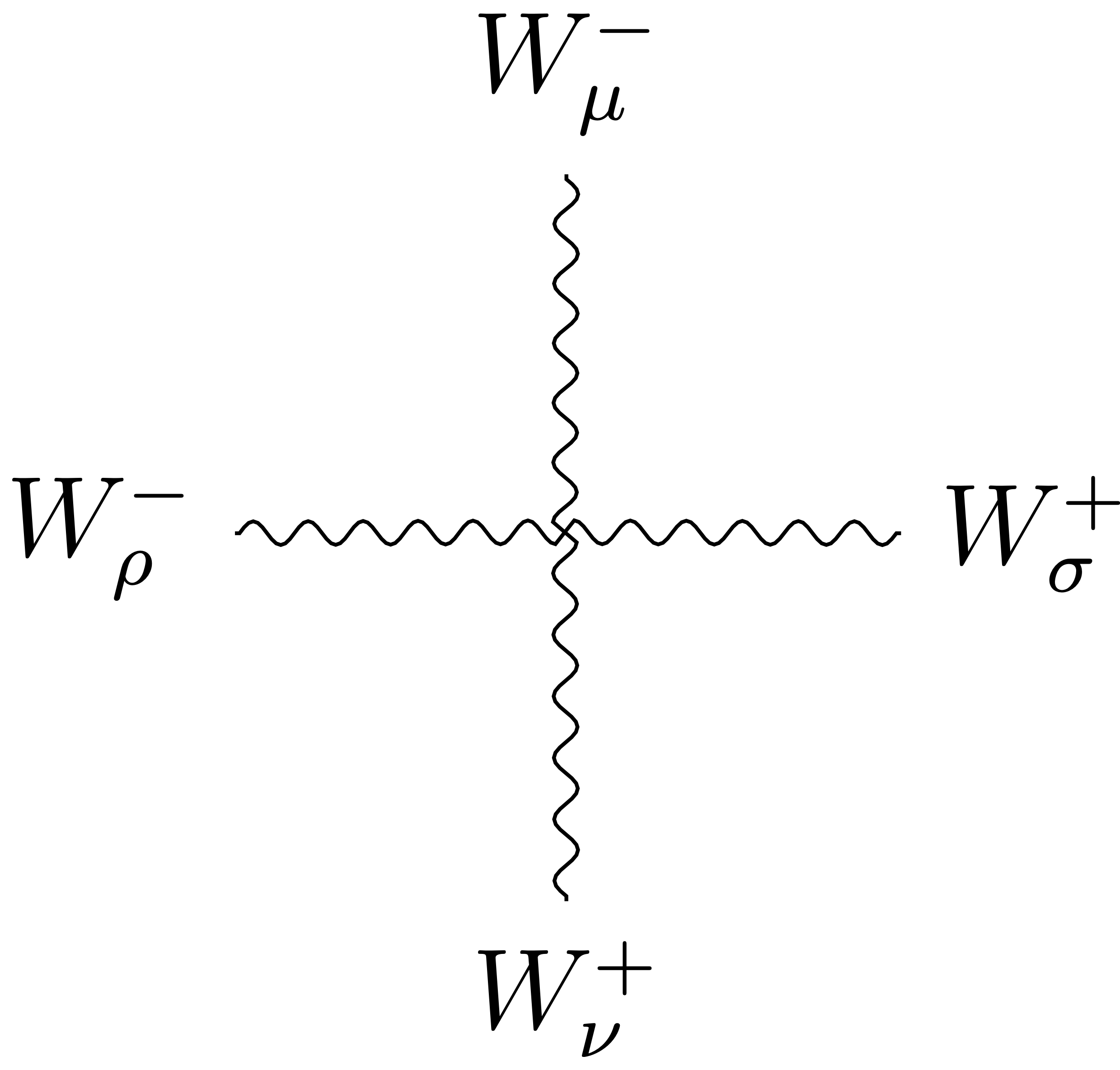}&
        $- \ i \left(\dfrac{e}{s_w}\right)^2 \eta^{\mu\nu} \eta^{\rho\sigma} \vspace{5px} \newline
        - i \left(\dfrac{e}{s_w}\right)^2 \eta^{\mu \sigma} \eta^{\nu \rho} \vspace{5px} \newline
        + 2 i \left(\dfrac{e}{s_w}\right)^2 \eta^{\mu \rho} \eta^{\nu \sigma}$ \vspace{5px}\\
        \hline
    \end{tabular}
\end{table}
\end{center}

\begin{center}
\begin{table}[!h]
   \centering\begin{tabular}{>{\centering\arraybackslash}m{5cm} m{7.5cm}} 
            \multicolumn{2}{c}{Quartic Self-Interactions} \\
    \hline
        Diagram & Feynman Rule \\
        \hline
        \vspace{5 px}
        \includegraphics[height=3.5cm]{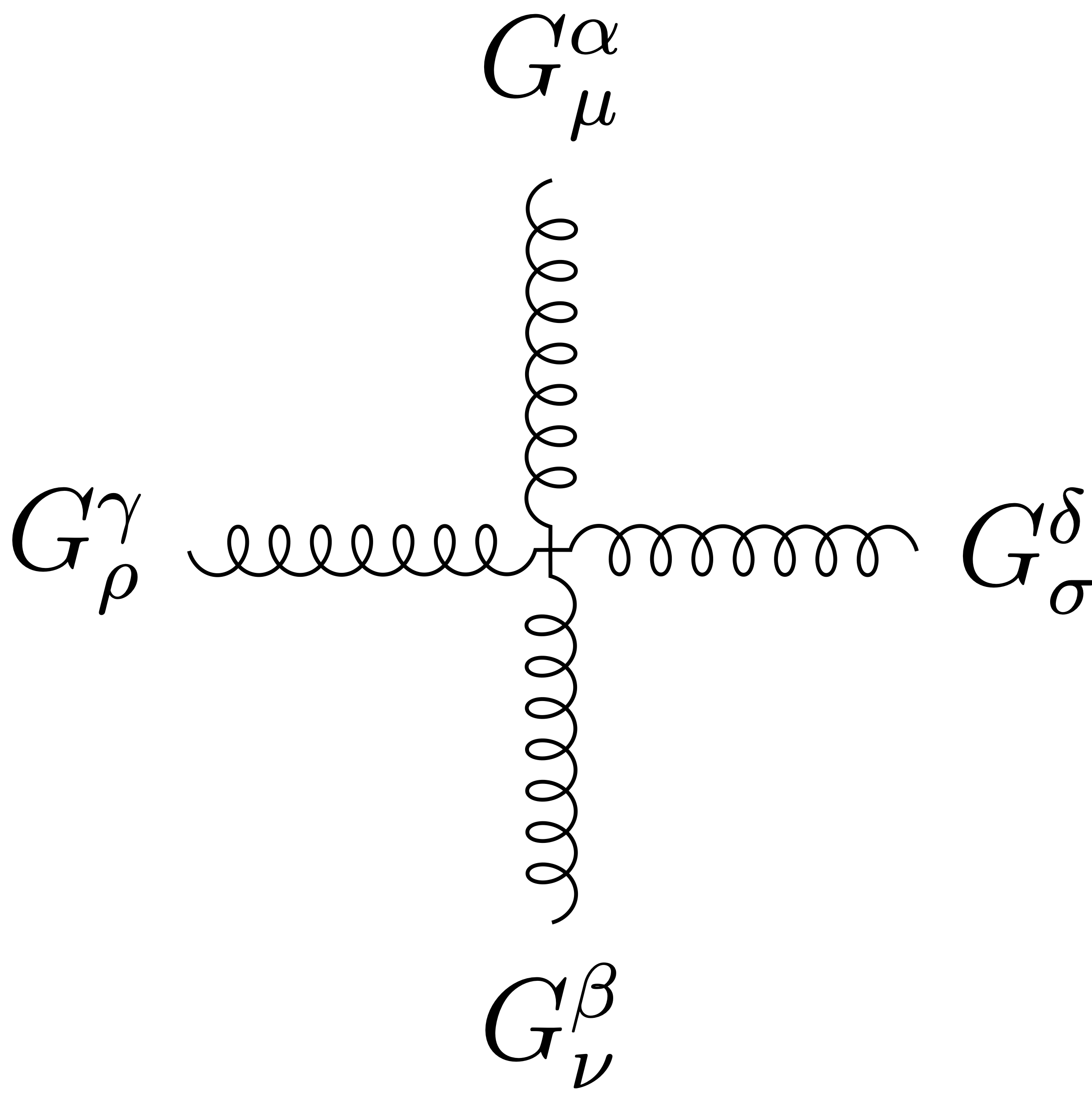}&
        $+ \ i g_s^2 \ \eta^{\mu\sigma} \eta^{\nu\rho} \left(f^{\alpha \gamma \epsilon}f^{\beta \delta \epsilon} + f^{\alpha \beta \epsilon}f^{\gamma \delta \epsilon}\right)\vspace{5px} \newline
        + \ i g_s^2 \ \eta^{\mu\rho} \eta^{\nu\sigma} \left(f^{\alpha \delta \epsilon}f^{\beta \gamma \epsilon} - f^{\alpha \beta \epsilon}f^{\gamma \delta \epsilon}\right)\vspace{5px} \newline
        - \ i g_s^2 \ \eta^{\mu\nu} \eta^{\rho\sigma} \left(f^{\alpha \delta \epsilon}f^{\beta \gamma \epsilon} + f^{\alpha \gamma \epsilon}f^{\beta \delta \epsilon}\right)$\\
        \hline
    \end{tabular}
\end{table}
\end{center}

\begin{center}
\begin{table}[!h]
    \centering\begin{tabular}{>{\centering\arraybackslash}m{4cm} m{8.5cm}} 
            \multicolumn{2}{c}{Triple Interactions with Ghosts} \\
    \hline
        Diagram & Feynman Rule \\
        \hline
        \vspace{5 px}
        \includegraphics[height=2.25cm]{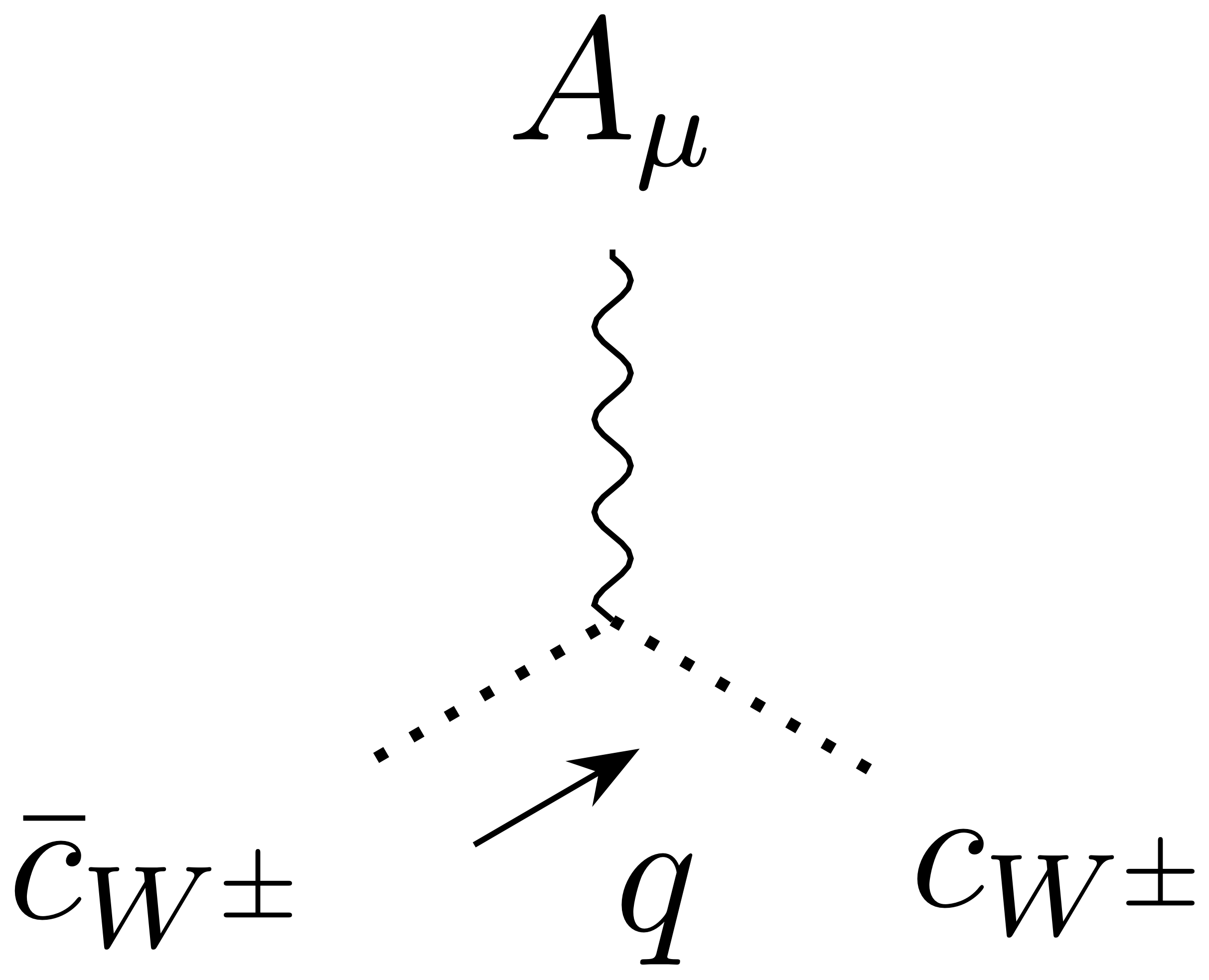}&
        $ \mp \ ie q^\mu$\\
        \vspace{5 px}
        \includegraphics[height=2.25cm]{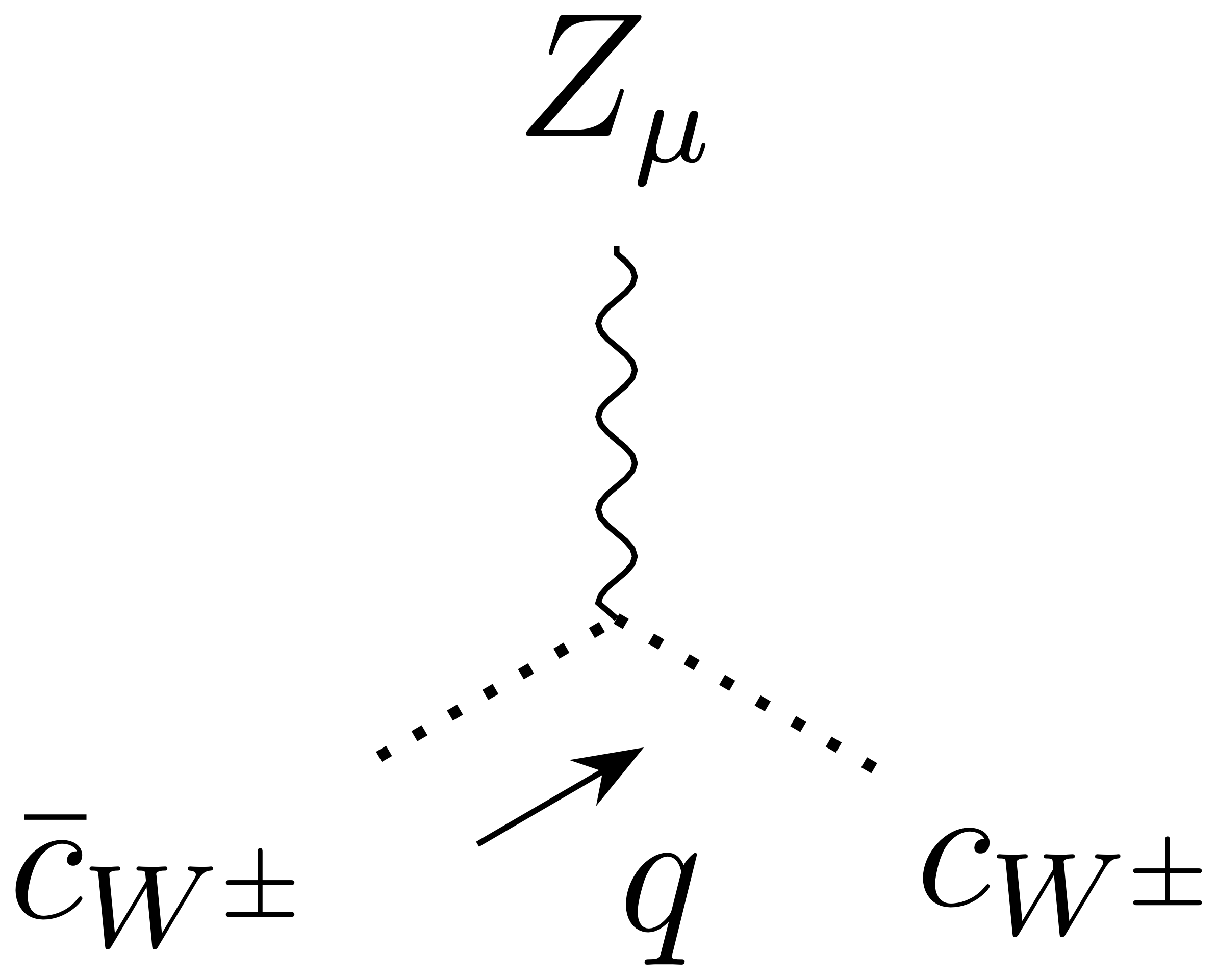}&
        $\mp \ i e\dfrac{c_w}{s_w} q^\mu$\\
        \vspace{5 px}
        \includegraphics[height=2.25cm]{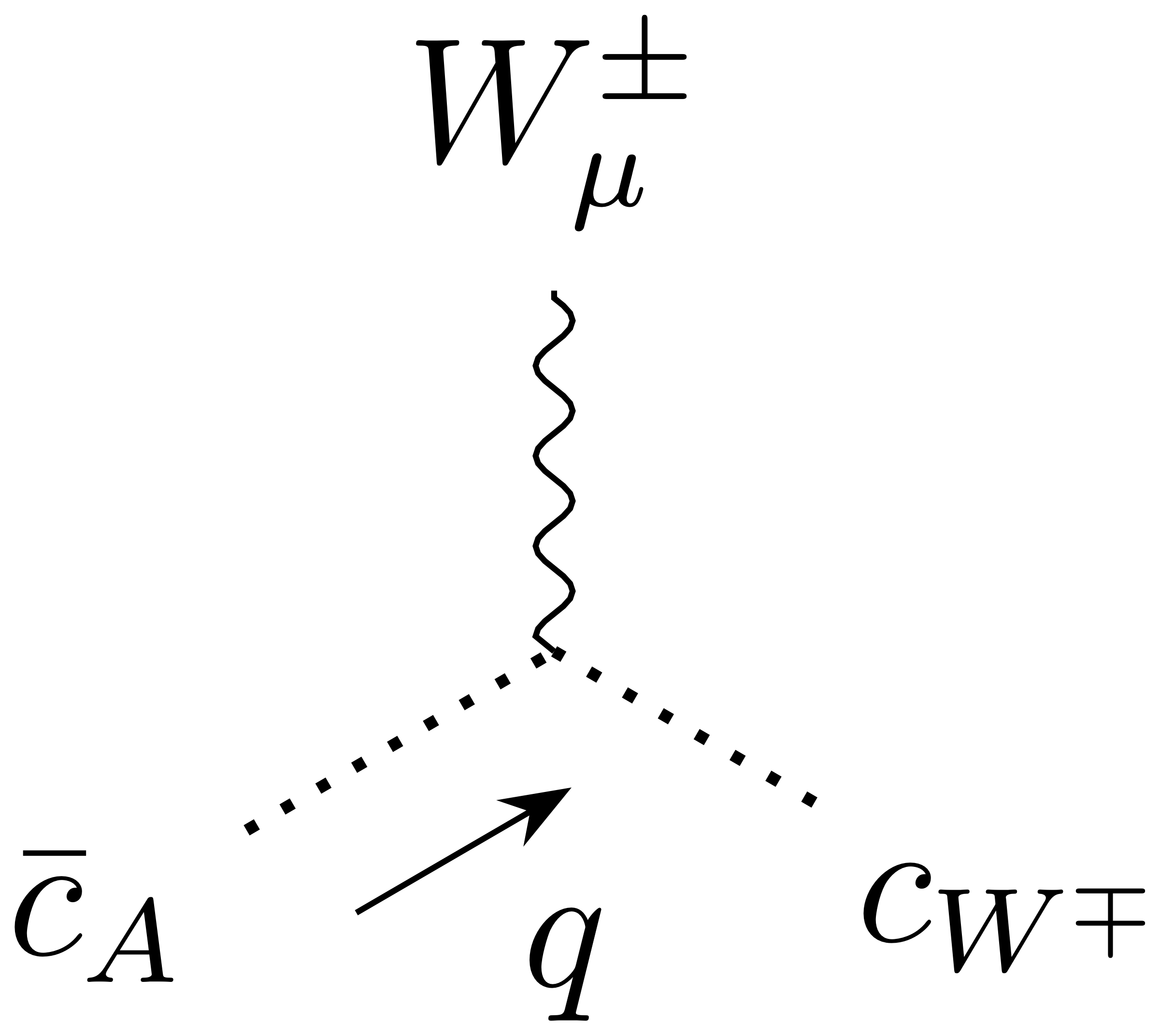}&
        $ \mp \ ie q^\mu$\\
        \hline
    \end{tabular}
\end{table}
\end{center}

\begin{center}
\begin{table}[!h]
    \centering\begin{tabular}{>{\centering\arraybackslash}m{4cm} m{8.5cm}} 
    \multicolumn{2}{c}{Triple Interactions with Ghosts} \\
    \hline
        Diagram & Feynman Rule \\
        \hline
        \vspace{5 px}
        \includegraphics[height=2.25cm]{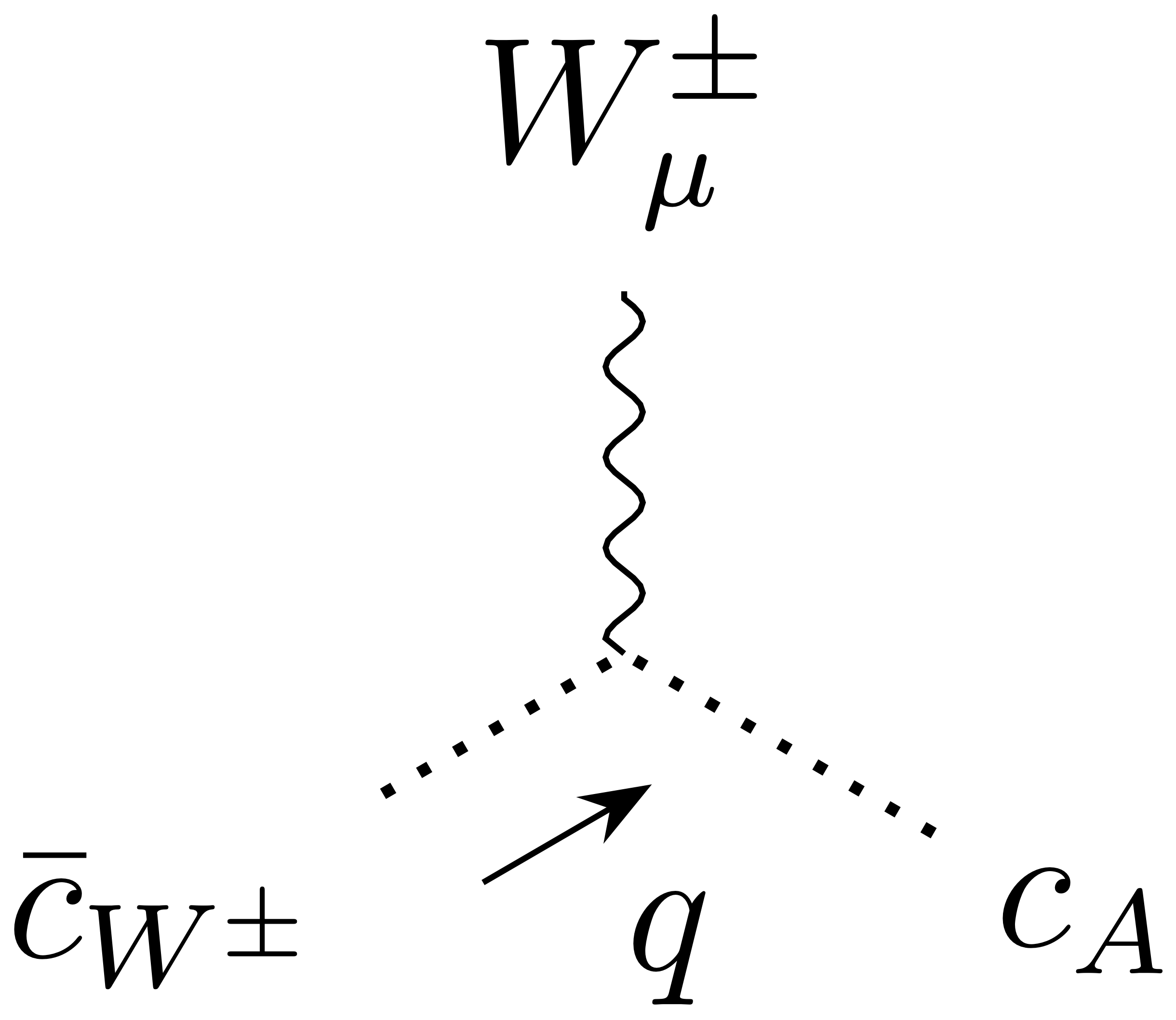}&
        $ \pm \ ie q^\mu$\\
        \vspace{5 px}
        \includegraphics[height=2.25cm]{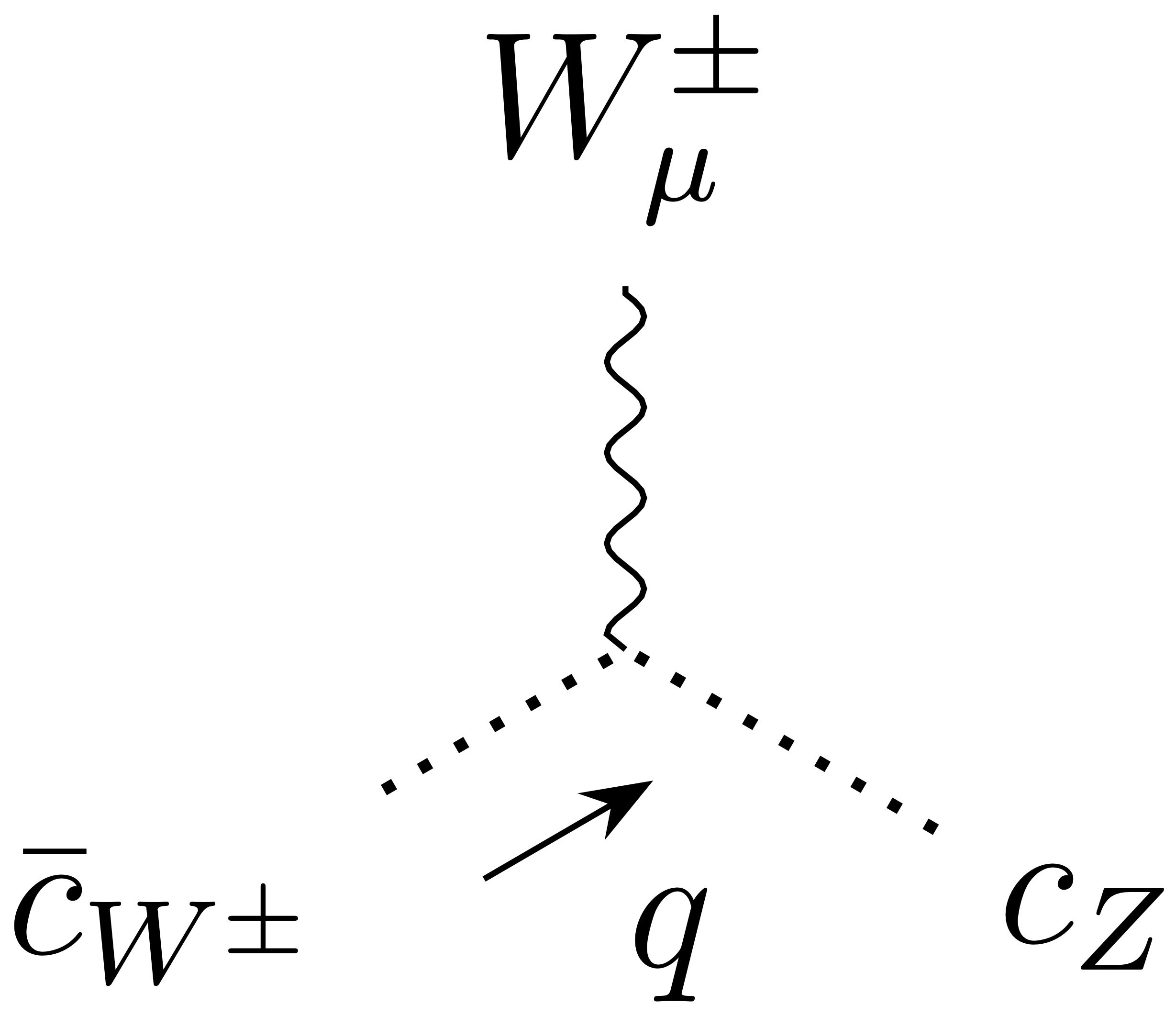}&
        $ \pm \ ie \dfrac{c_w}{s_w} q^\mu$\\
        \vspace{5 px}
        \includegraphics[height=2.25cm]{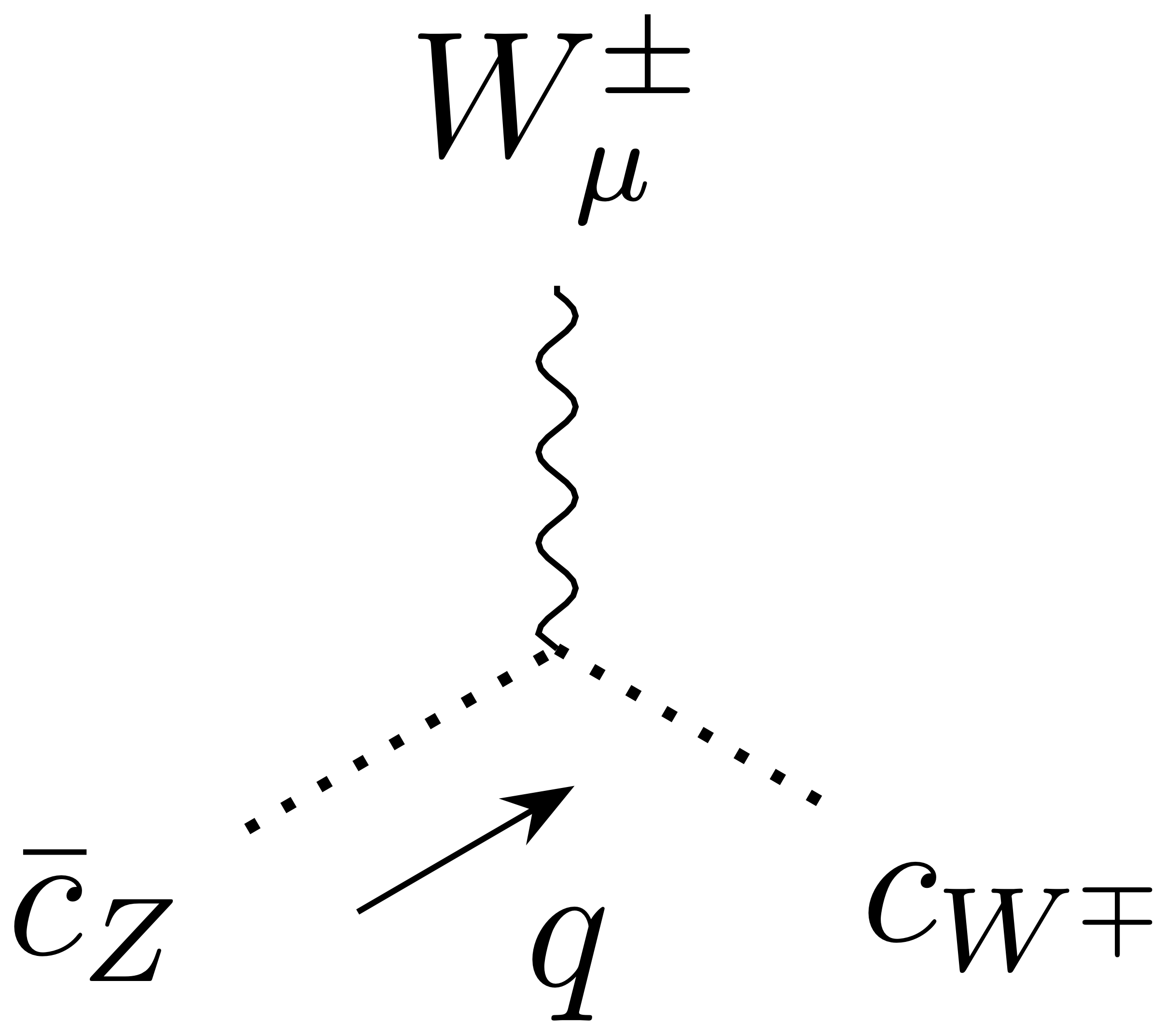}&
        $ \pm \ i e\dfrac{c_w}{s_w} q^\mu$\\
        \vspace{5 px}
        \includegraphics[height=2.25cm]{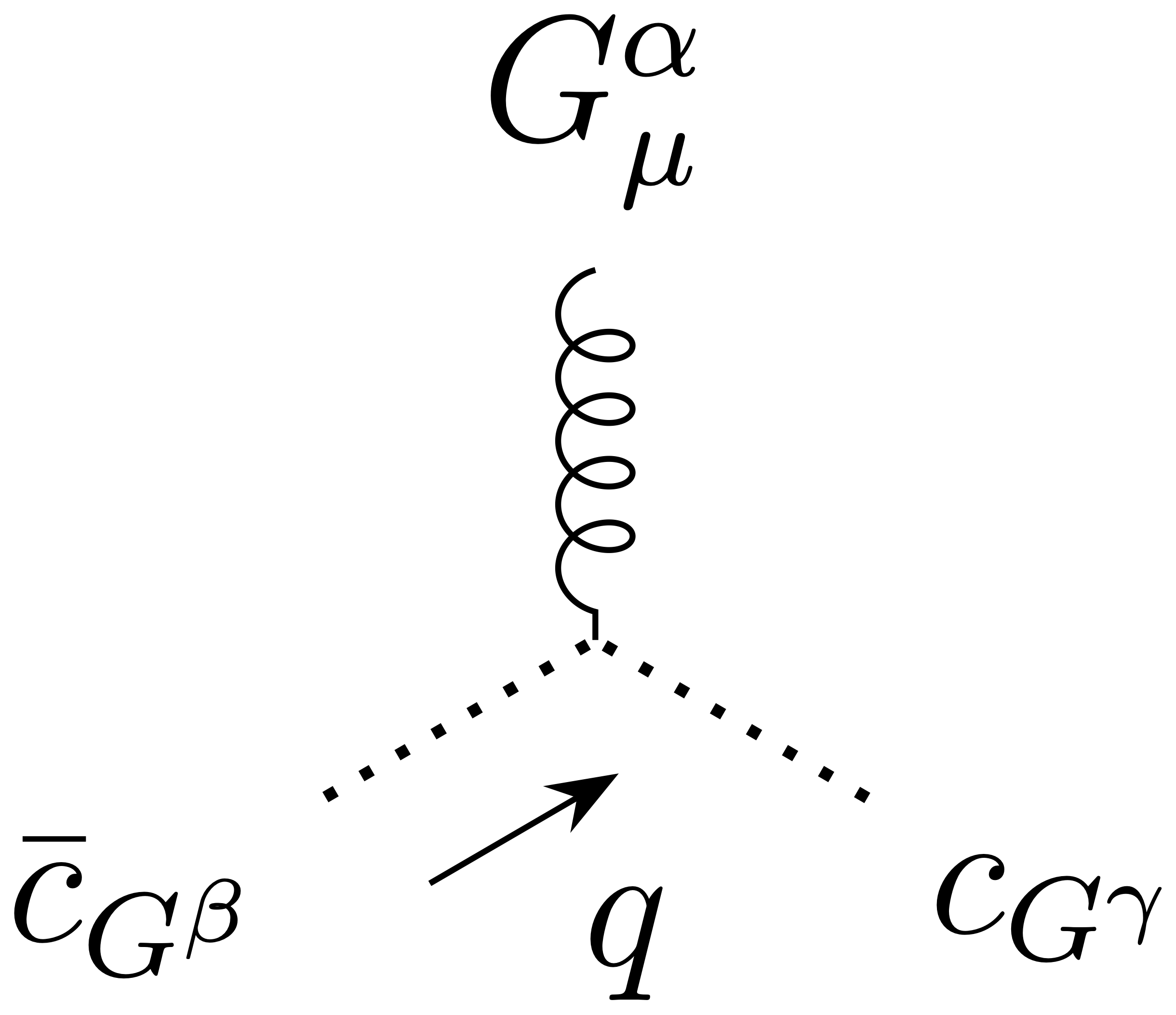}&
        $ - \ g_s f^{\alpha \beta \gamma} q^\mu$\\
        \hline
        \hline
    \end{tabular}
\end{table}
\end{center}

\clearpage
\subsection{Higgs Interactions}
The Feynman rules for the Higgs field self-interactions as well as the Higgs field interactions with the SM gauge bosons, fermions and ghost fields are given below.
\begin{center}
\begin{table}[!ht]
    \centering\begin{tabular}{>{\centering\arraybackslash}m{4cm} m{8.5cm}} 
        \multicolumn{2}{c}{Triple Self-Interaction} \\
    \hline
        Diagram & Feynman Rule \\
        \hline
        \vspace{5 px}
        \includegraphics[height=2.25cm]{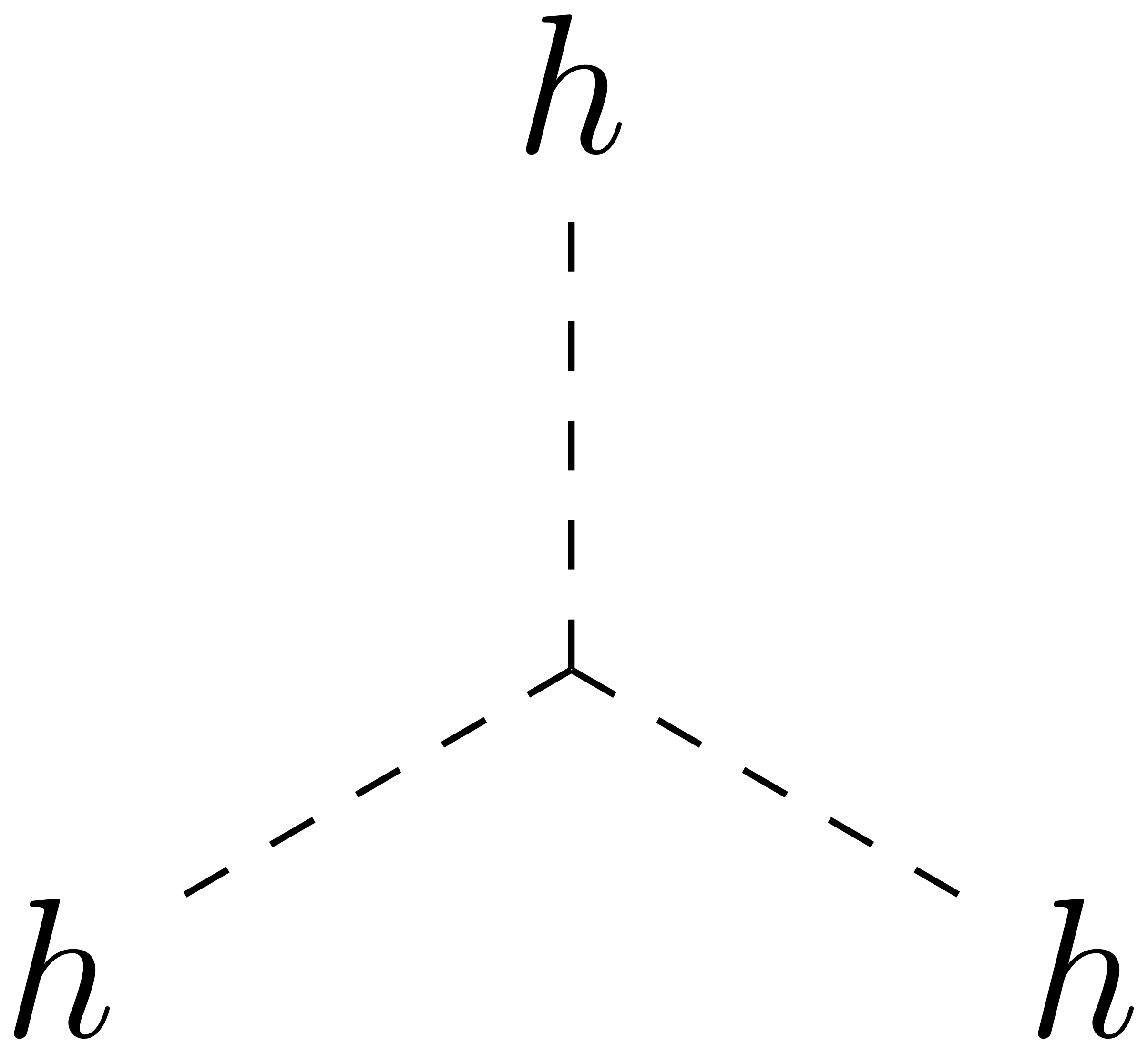}&
        $ -3i \dfrac{m_h^2}{v}$\\
        \hline
    \end{tabular}
\end{table}
\end{center}

\begin{center}
\begin{table}[!ht]
    \centering\begin{tabular}{>{\centering\arraybackslash}m{5cm} m{7.5cm}} 
    \multicolumn{2}{c}{Quartic Self-Interaction} \\
    \hline
        Diagram & Feynman Rule \\
        \hline
        \vspace{5 px}
        \includegraphics[height=3.5cm]{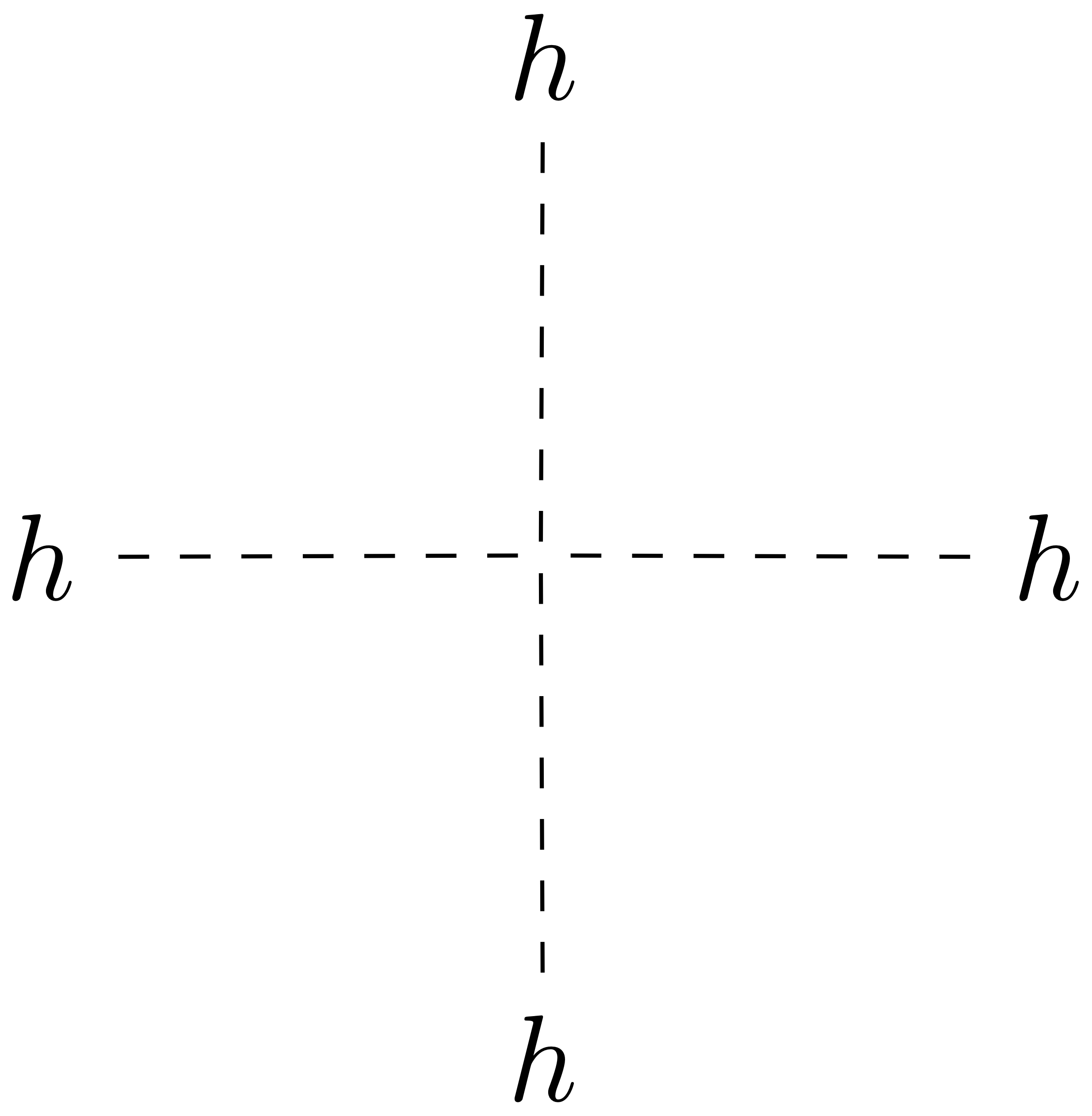}&
        $-3i \dfrac{m_h^2}{v^2}$\\
        \hline
    \end{tabular}
\end{table}
\end{center}

\begin{center}
\begin{table}[!ht]
    \centering\begin{tabular}{>{\centering\arraybackslash}m{4cm} m{8.5cm}} 
    \multicolumn{2}{c}{Interactions with SM Fermions} \\
    \hline
        Diagram & Feynman Rule \\
        \hline
        \vspace{5 px}
        \includegraphics[height=2.25cm]{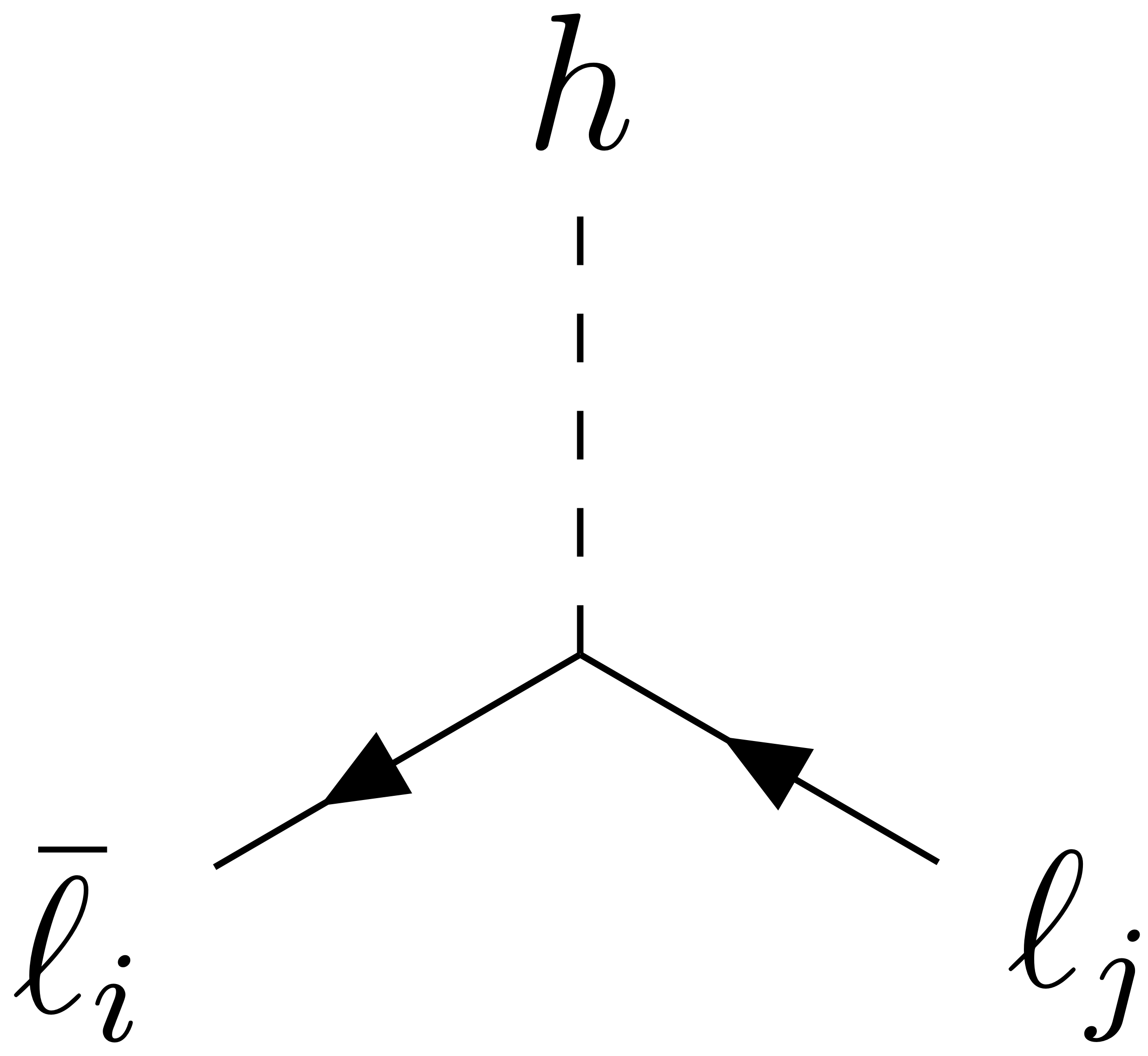}&
        $- \ i \dfrac{m_{\ell_i}}{v} \delta_{ij}$\\
        \vspace{5 px}
        \includegraphics[height=2.25cm]{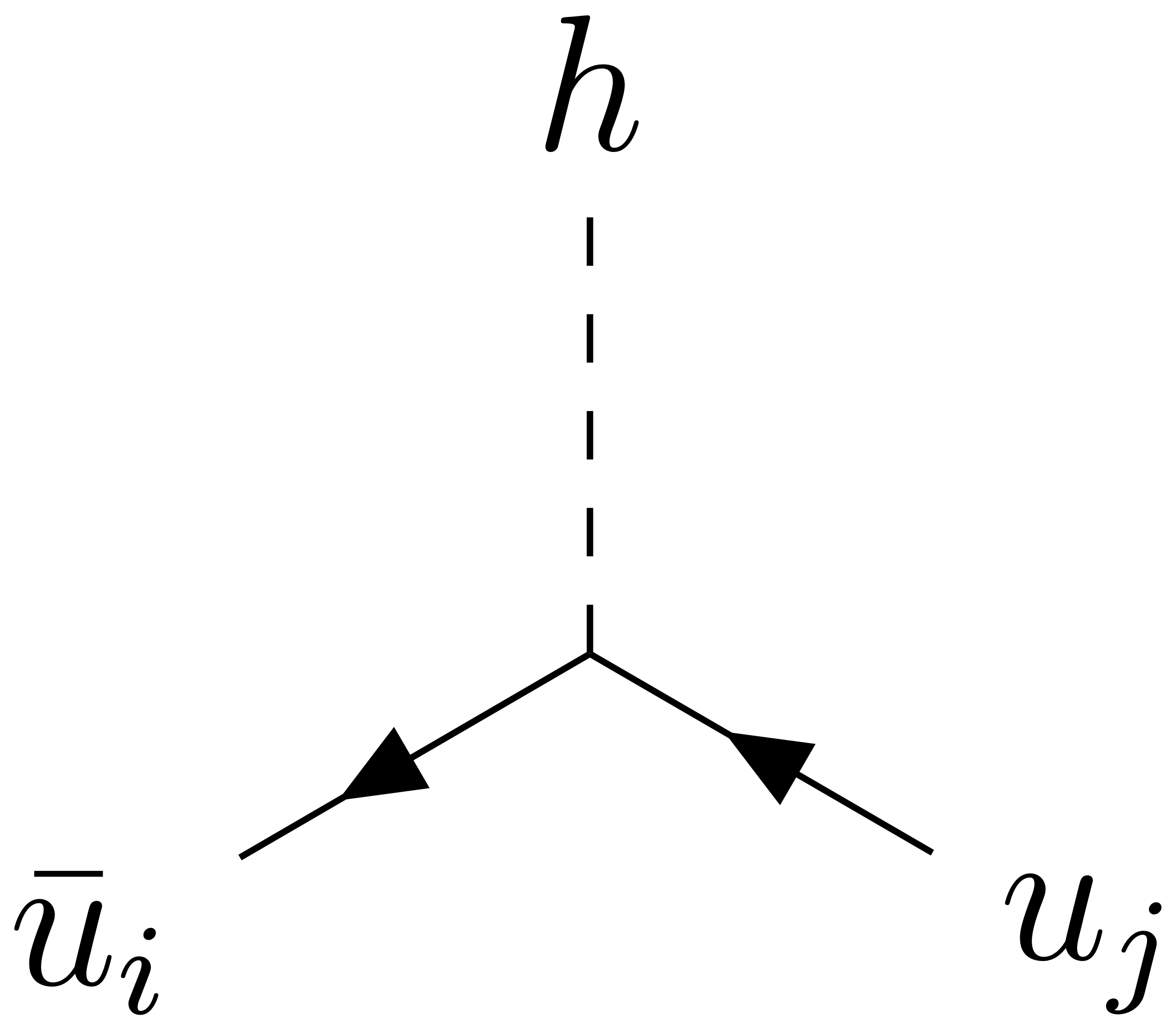}&
        $- \ i \dfrac{m_{u_i}}{v} \delta_{ij}$\\
        \vspace{5 px}
        \includegraphics[height=2.25cm]{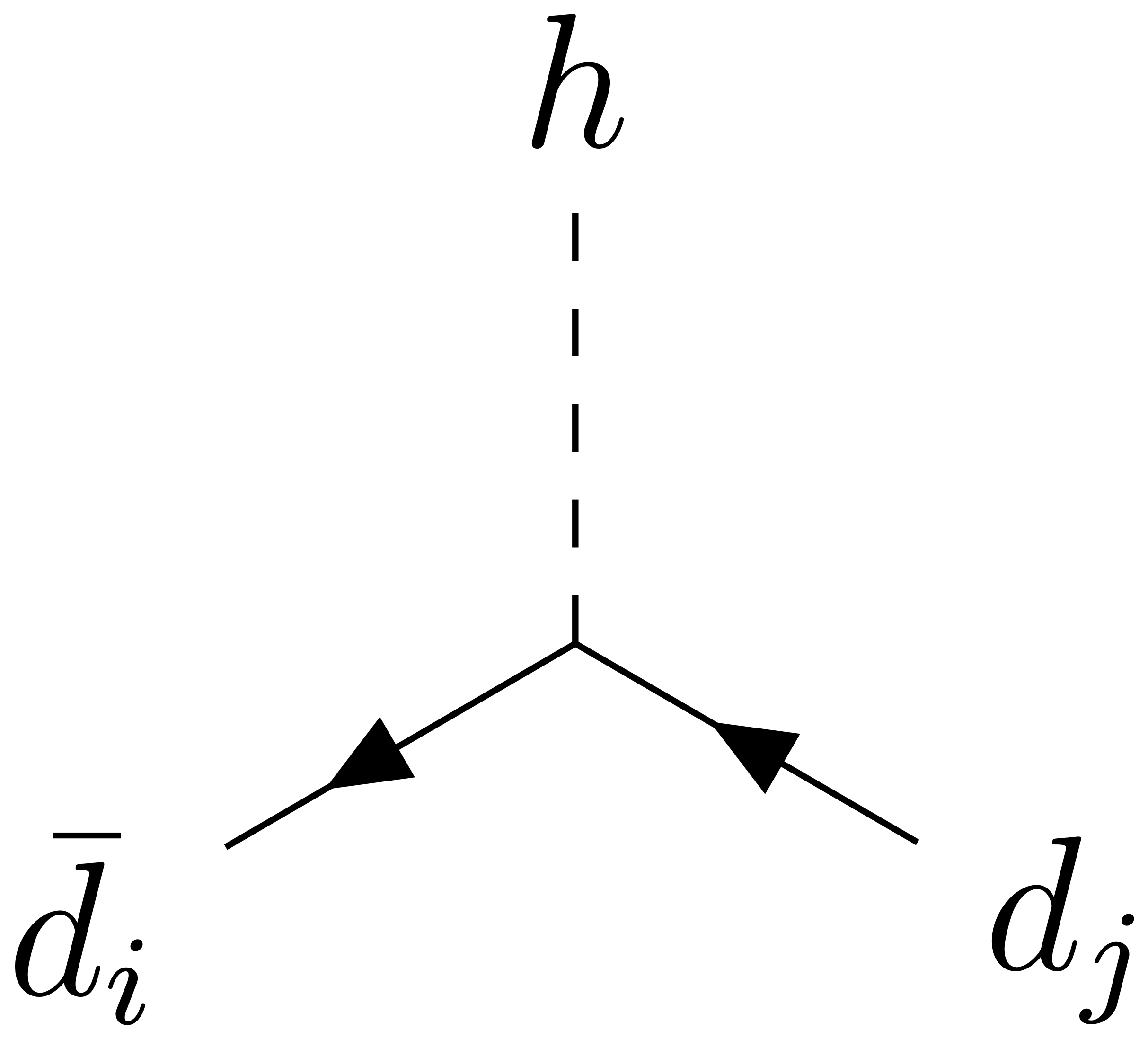}&
        $- \ i \dfrac{m_{d_i}}{v} \delta_{ij}$\\
        \hline
    \end{tabular}
\end{table}
\end{center}

\begin{center}
\begin{table}[!ht]
    \centering\begin{tabular}{>{\centering\arraybackslash}m{4cm} m{8.5cm}} 
    \multicolumn{2}{c}{Triple Interactions with SM Gauge Bosons} \\
    \hline
        Diagram & Feynman Rule \\
        \hline
        \vspace{5 px}
        \includegraphics[height=2.25cm]{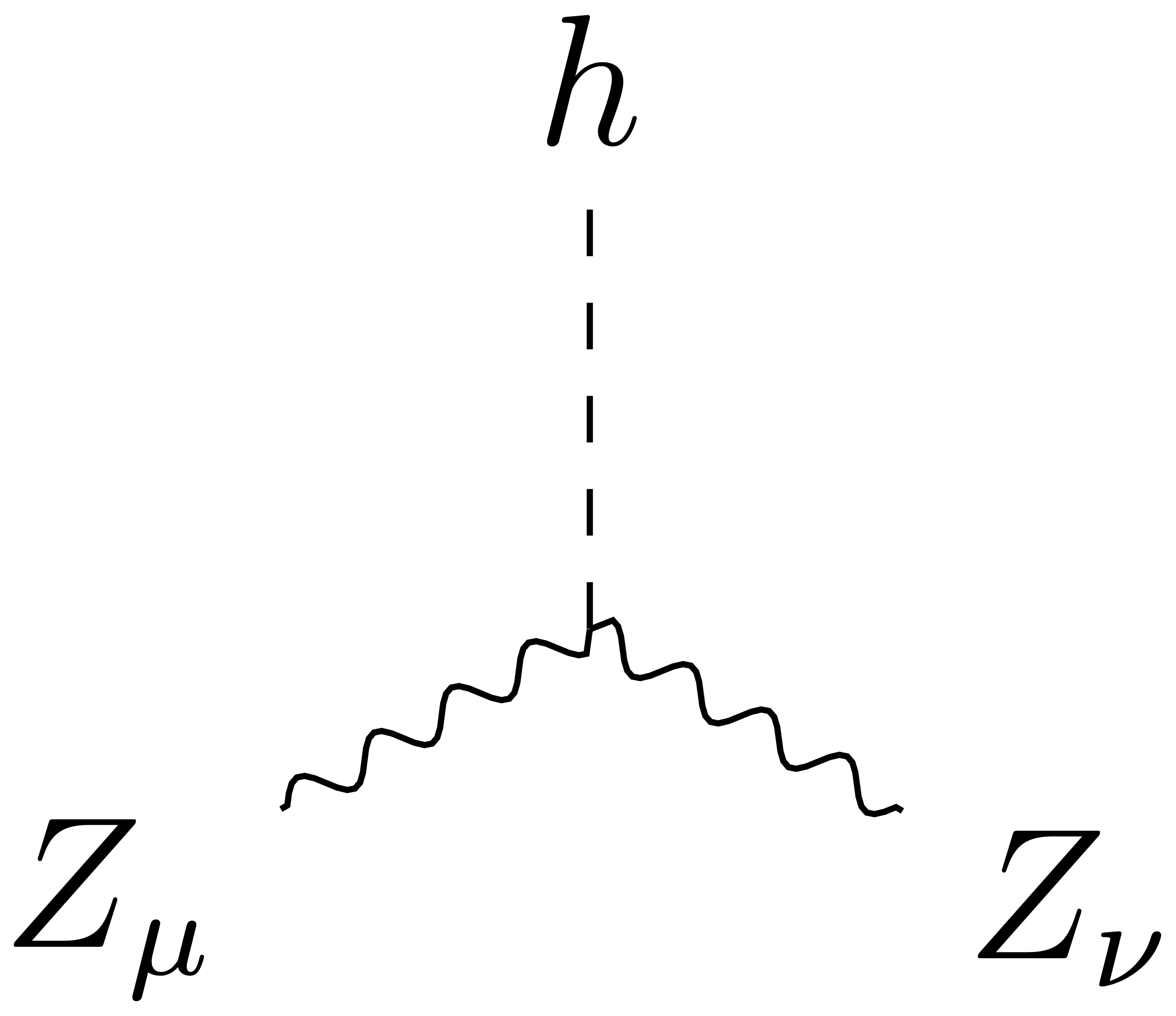}&
        $ + \ \dfrac{i}{2} \left(\dfrac{e}{c_w s_w}\right)^2 v \ \eta^{\mu\nu}$\\
        \vspace{5 px}
        \includegraphics[height=2.25cm]{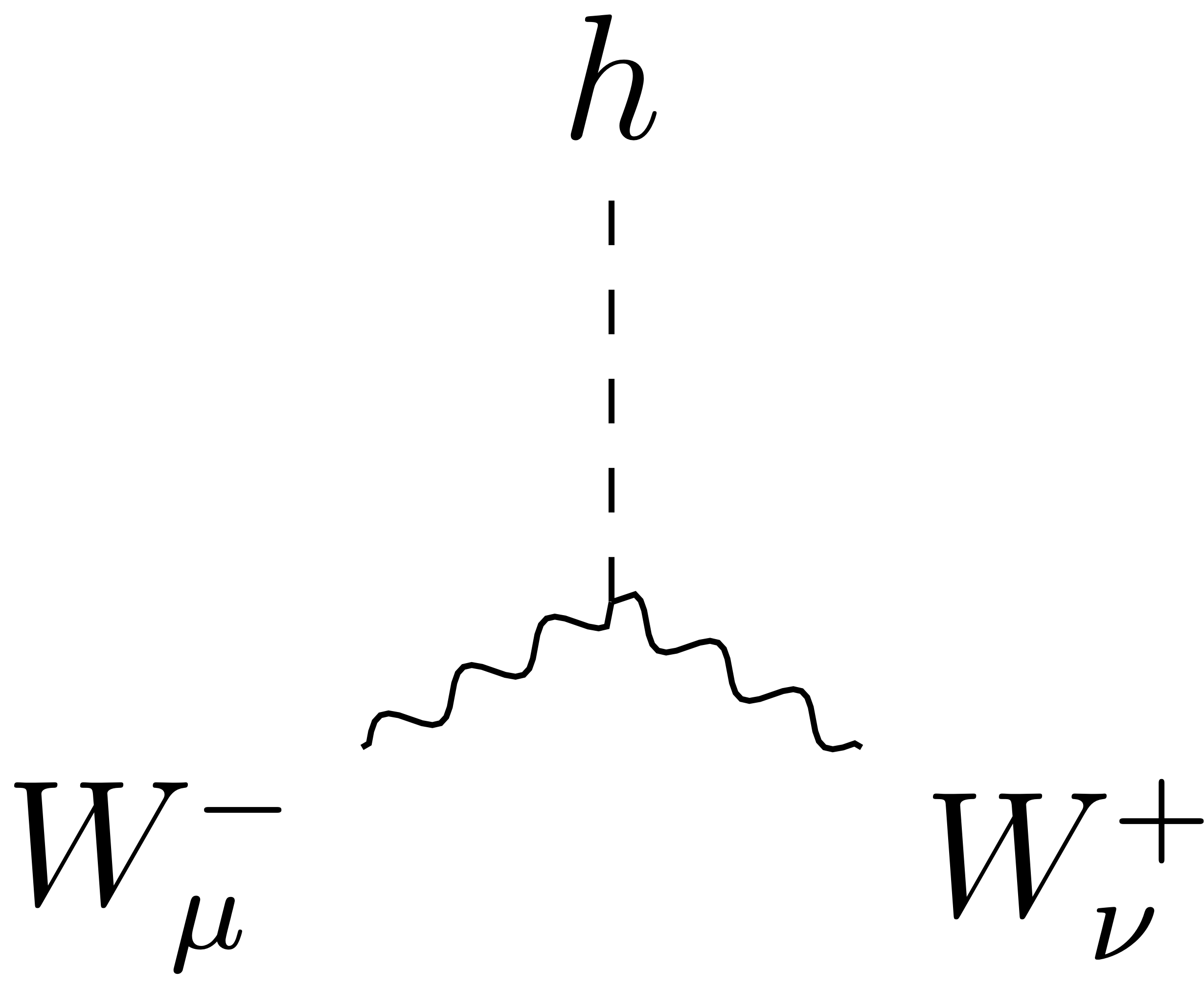}&
        $ + \ \dfrac{i}{2} \left(\dfrac{e}{s_w}\right)^2 v \ \eta^{\mu\nu}$\\
        \hline
    \end{tabular}
\end{table}
\end{center}

\begin{center}
\begin{table}[!ht]
    \centering\begin{tabular}{>{\centering\arraybackslash}m{5cm} m{7.5cm}} 
    \multicolumn{2}{c}{Quartic Interactions with SM Gauge Bosons} \\
    \hline
        Diagram & Feynman Rule \\
        \hline
        \vspace{5 px}
        \includegraphics[height=3.5cm]{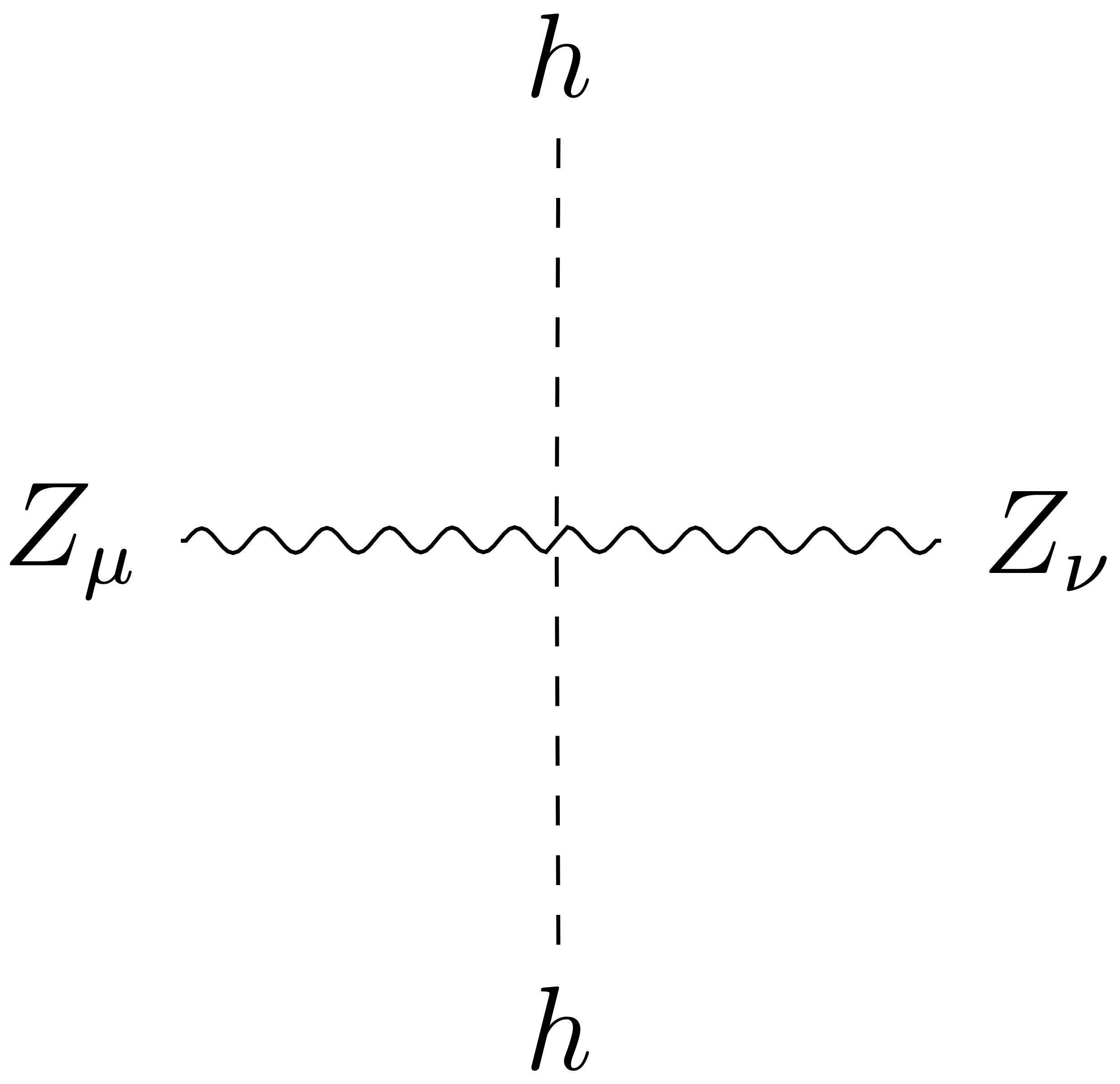}&
        $+ \ \dfrac{i}{2} \left(\dfrac{e}{c_w s_w}\right)^2 \eta^{\mu \nu}$\\
        \vspace{5 px}
        \includegraphics[height=3.5cm]{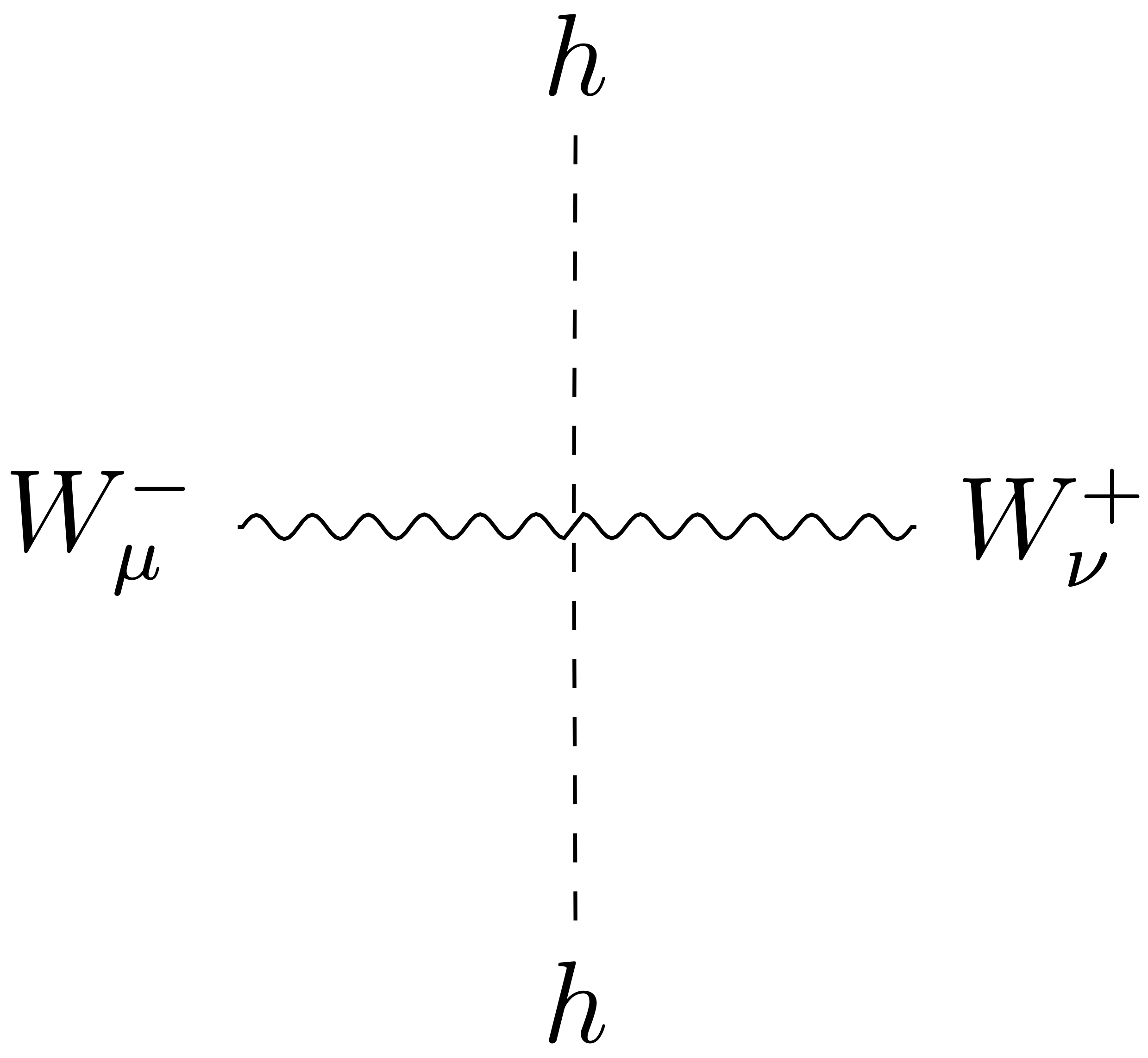}&
        $+ \ \dfrac{i}{2} \left(\dfrac{e}{s_w}\right)^2 \eta^{\mu \nu}$\\
        \hline
    \end{tabular}
\end{table}
\end{center}

\begin{center}
\begin{table}[!ht]
    \centering\begin{tabular}{>{\centering\arraybackslash}m{4cm} m{8.5cm}} 
    \multicolumn{2}{c}{Interactions with Ghosts} \\
    \hline
        Diagram & Feynman Rule \\
        \hline
        \vspace{5 px}
        \includegraphics[height=2.25cm]{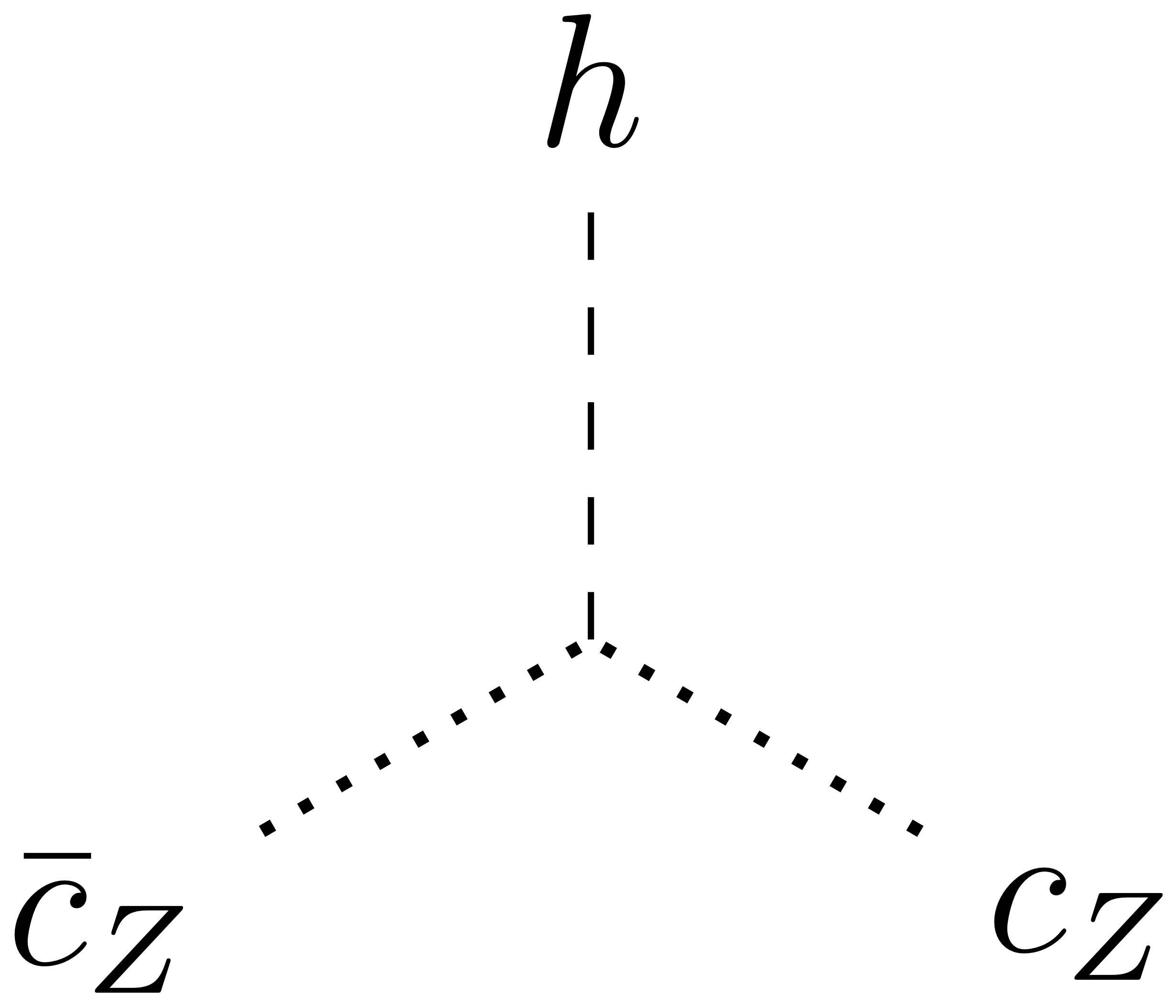}&
        $- \ \dfrac{i}{4} \left(\dfrac{e}{c_w s_w}\right)^2 v $\\
        \vspace{5 px}
        \includegraphics[height=2.25cm]{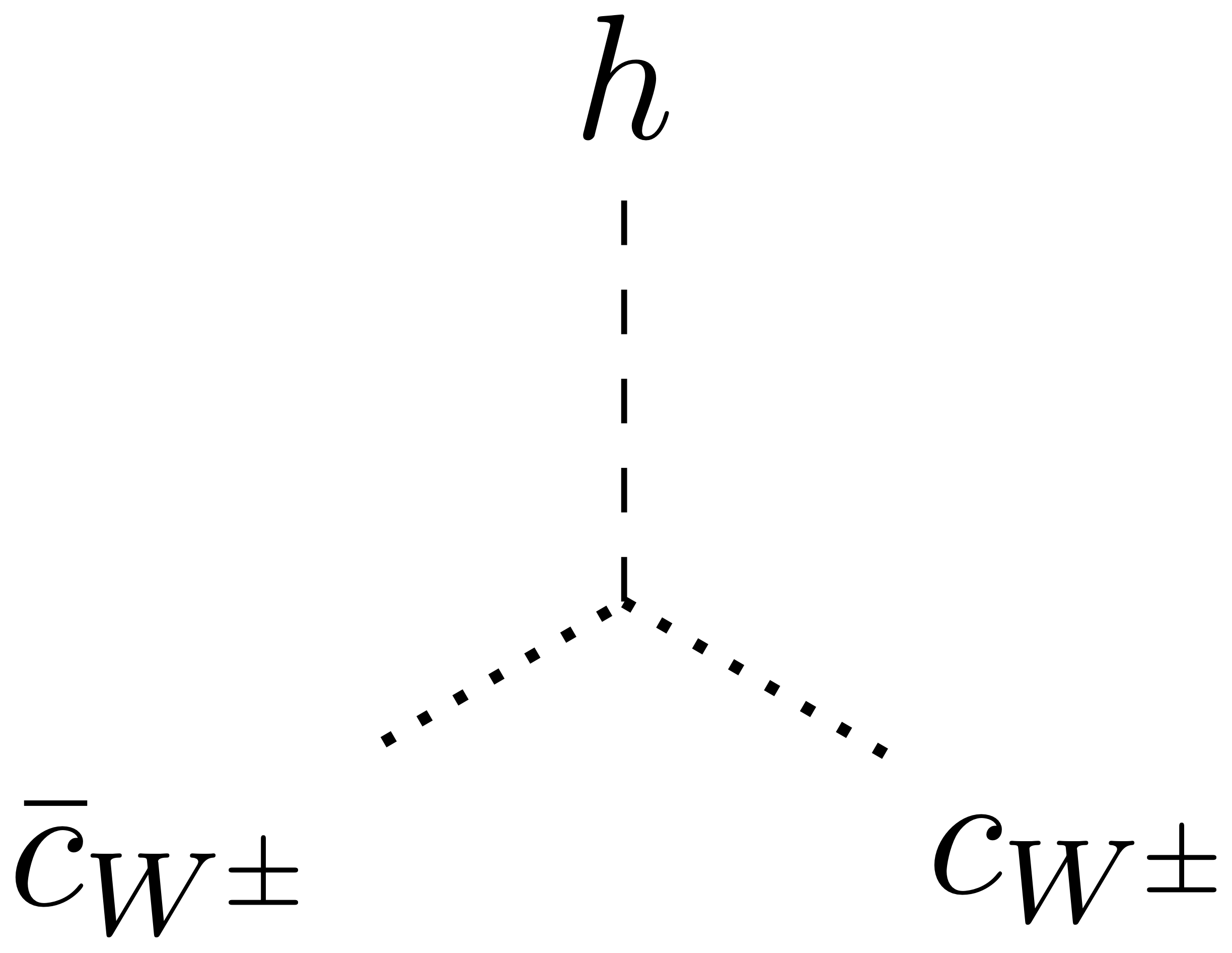}&
        $- \ \dfrac{i}{4} \left(\dfrac{e}{s_w}\right)^2 v $\\
        \hline
        \hline
    \end{tabular}
\end{table}
\end{center}

\clearpage
\subsection{Goldstone Interactions}
The diagrams for the Goldstone boson self-interactions as well as the Goldstone boson interactions with the Higgs field, SM gauge bosons, fermions and ghost fields are given below.

\begin{center}
\begin{table}[!ht]
    \centering\begin{tabular}{>{\centering\arraybackslash}m{4cm} m{8.5cm}} 
    \multicolumn{2}{c}{Triple Higgs Interactions} \\
    \hline
        Diagram & Feynman Rule \\
        \hline
        \vspace{5 px}
        \includegraphics[height=2.25cm]{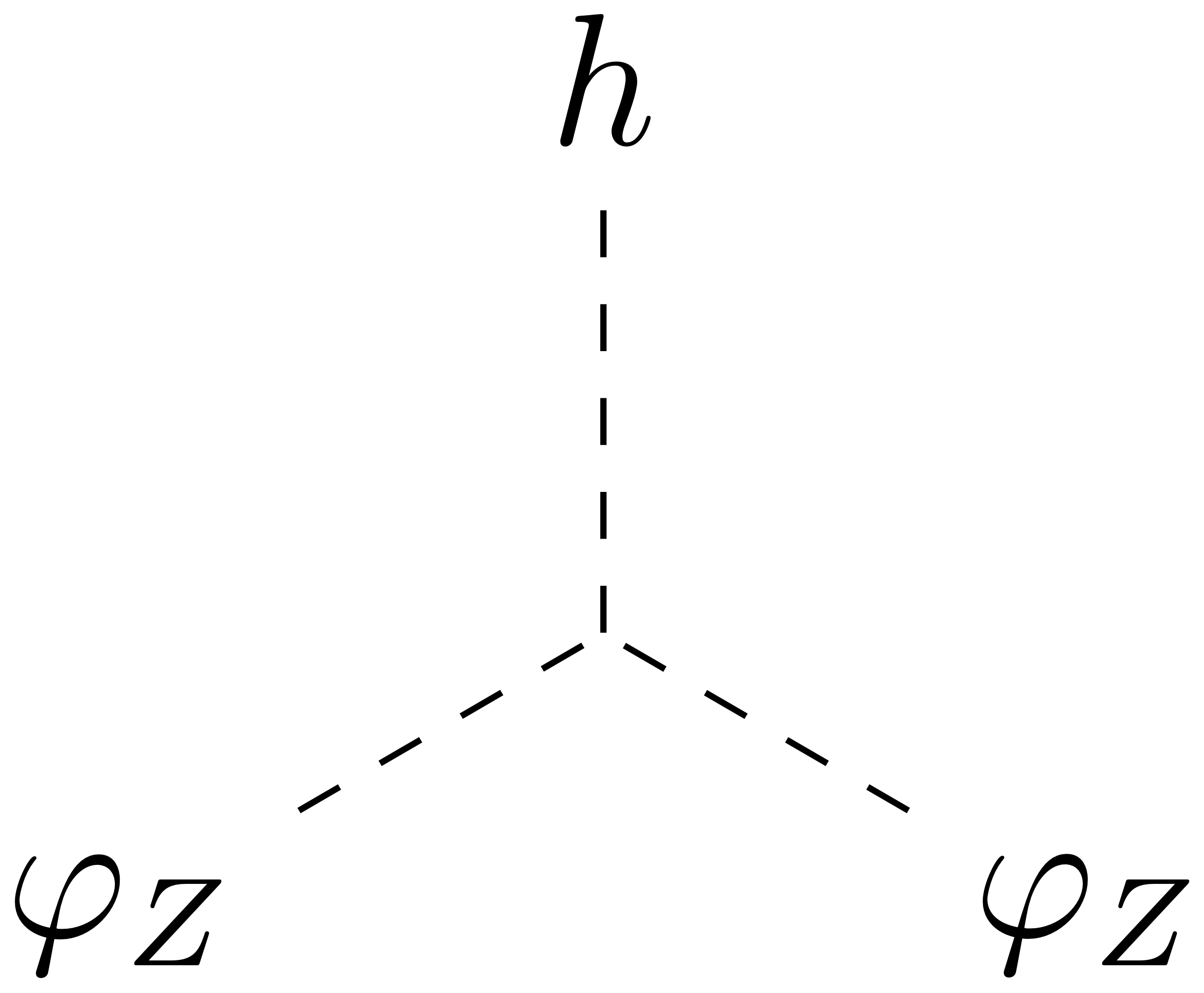}&
        $ - \ i \dfrac{m_h^2}{v}$\\
        \vspace{5 px}
        \includegraphics[height=2.25cm]{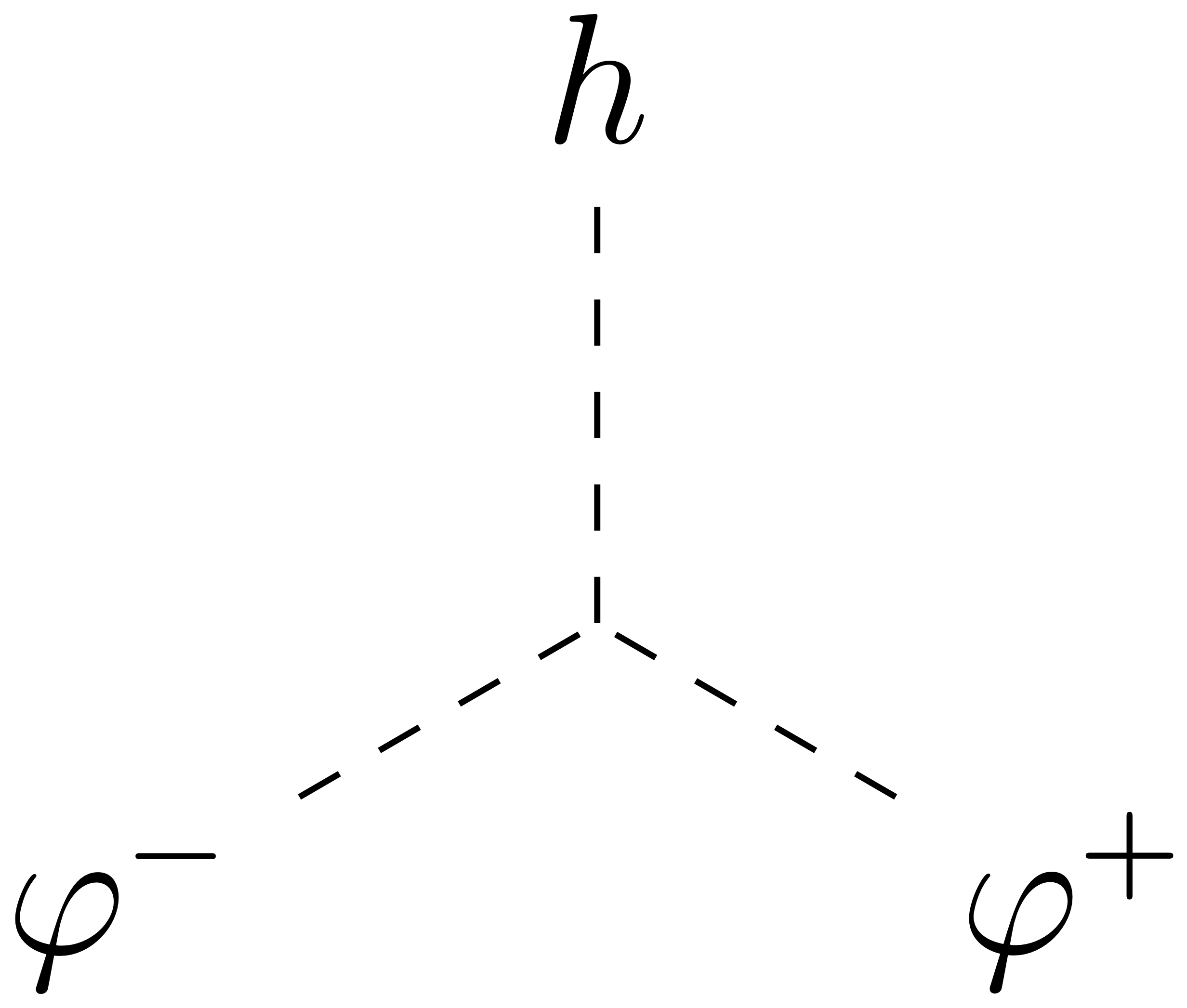}&
        $ - \ i \dfrac{m_h^2}{v}$\\
        \hline
    \end{tabular}
\end{table}
\end{center}

\begin{center}
\begin{table}[!ht]
    \centering\begin{tabular}{>{\centering\arraybackslash}m{5cm} m{7.5cm}} 
        \multicolumn{2}{c}{Quartic Higgs Interactions} \\
    \hline
        Diagram & Feynman Rule \\
        \hline
        \vspace{5 px}
        \includegraphics[height=3.5cm]{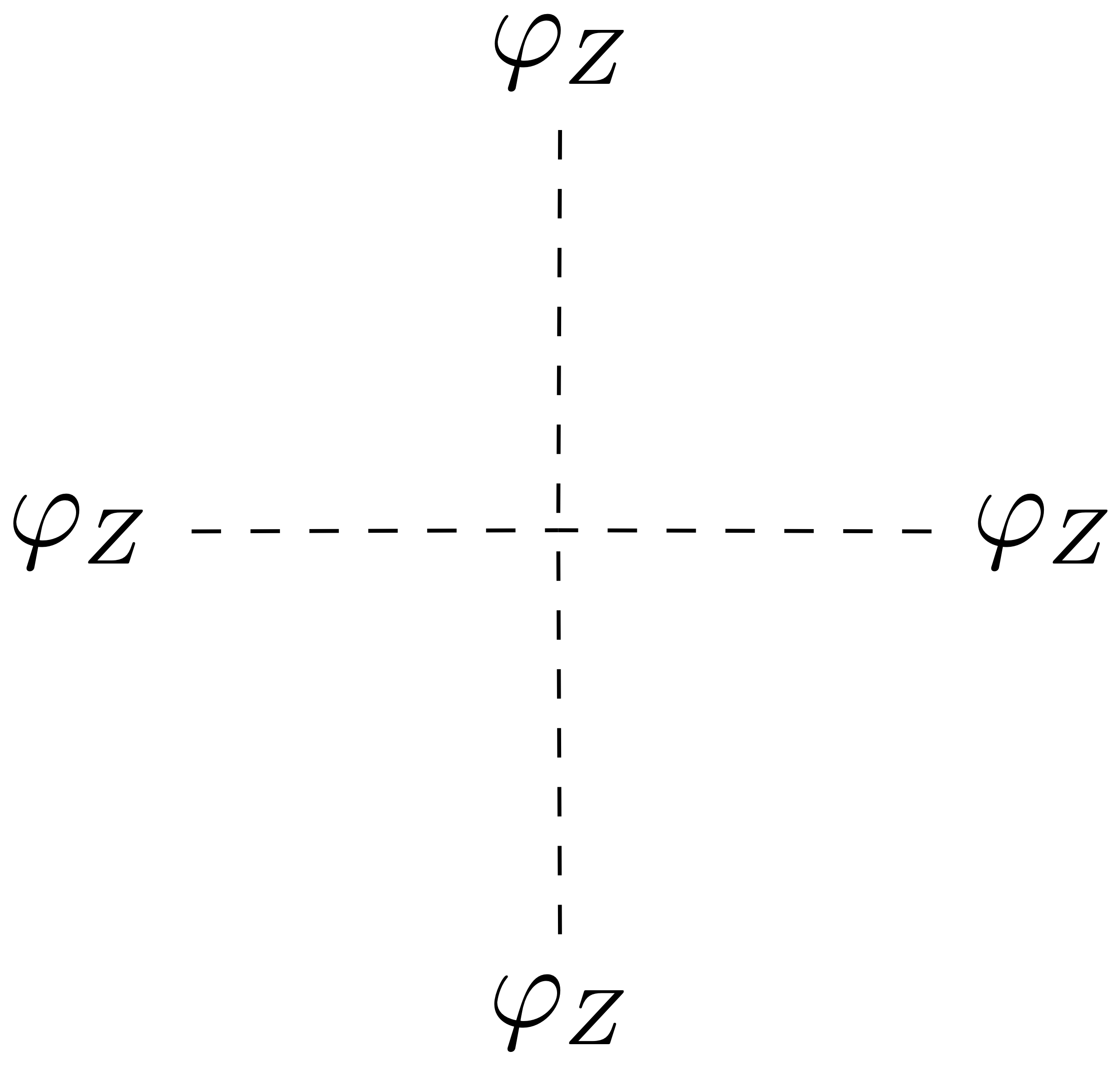}&
        $- \ 3i \dfrac{m_h^2}{v^2}$\\
        \vspace{5 px}
        \includegraphics[height=3.5cm]{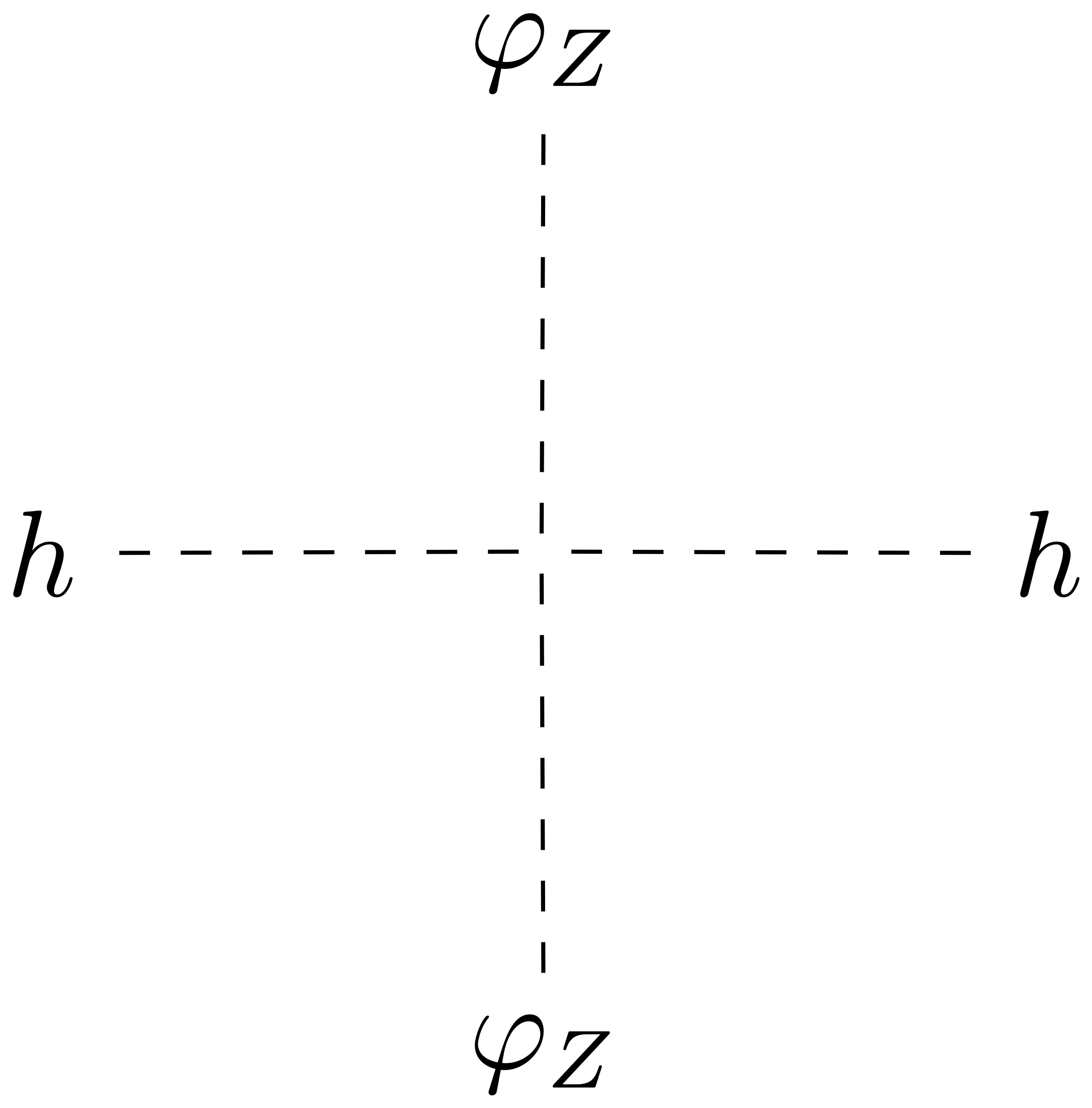}&
        $- \ i \dfrac{m_h^2}{v^2}$\\
        \vspace{5 px}
        \includegraphics[height=3.5cm]{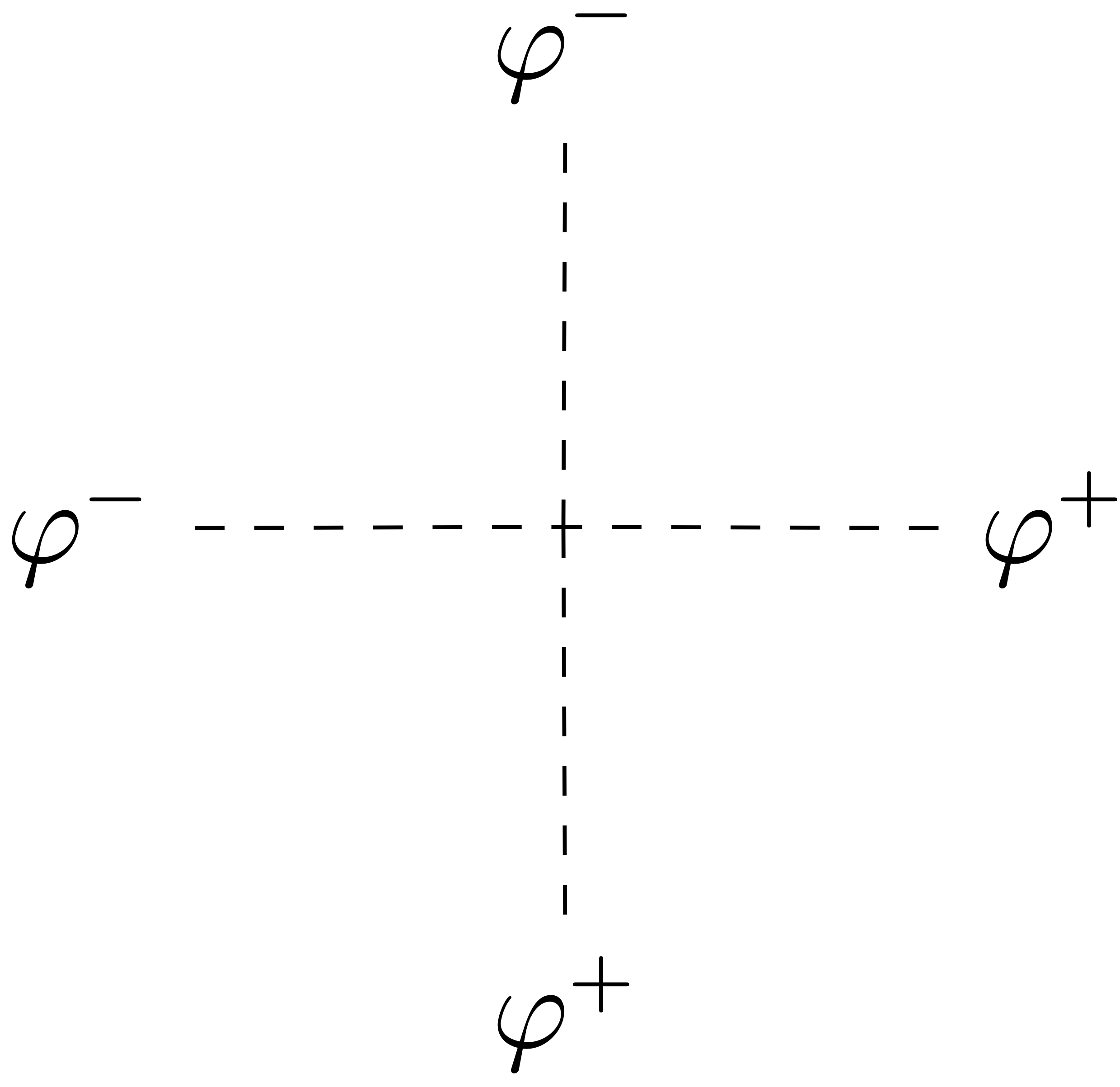}&
        $- \ 2i \dfrac{m_h^2}{v^2}$\\
        \vspace{5 px}
        \includegraphics[height=3.5cm]{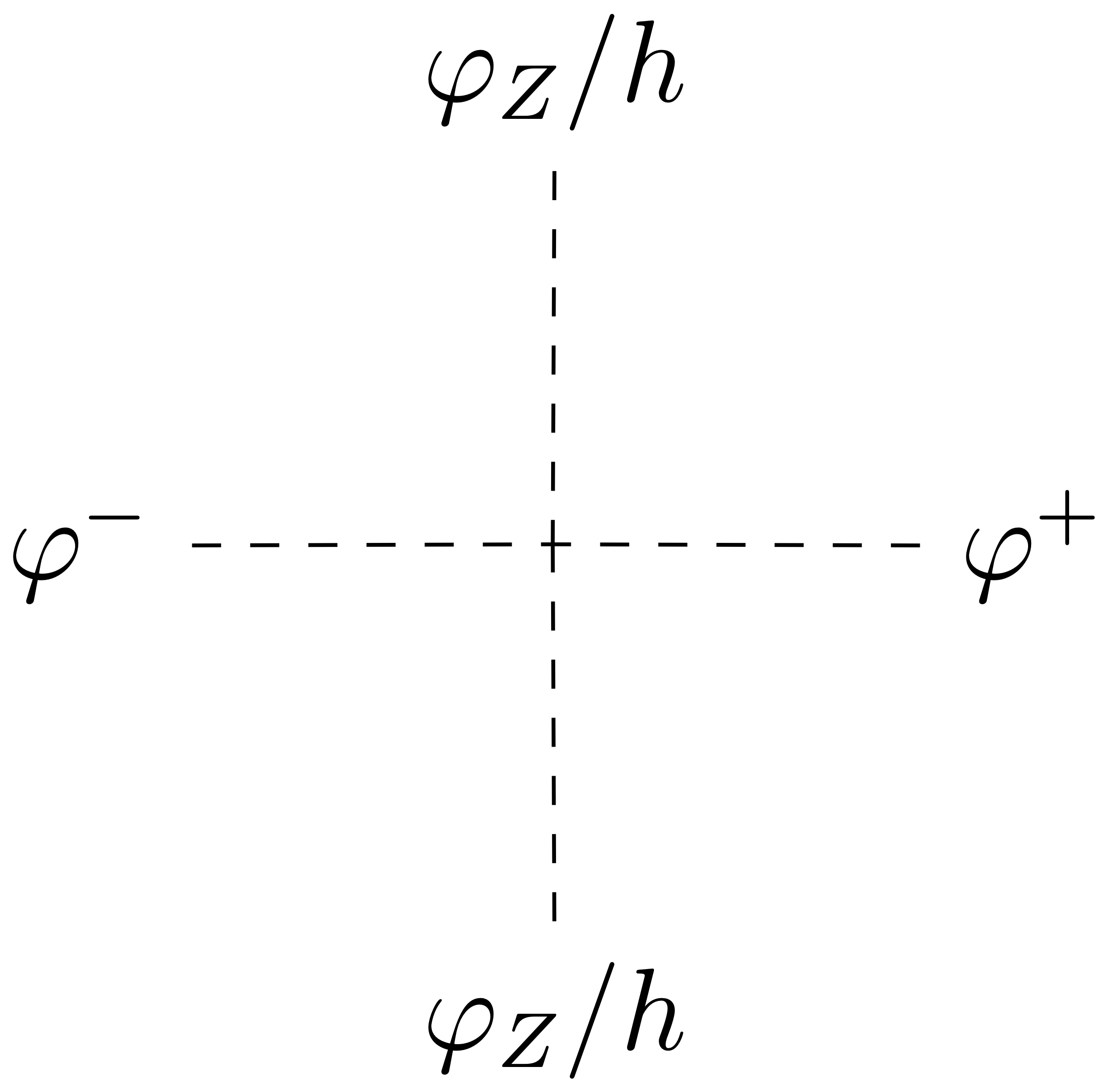}&
        $- \ i \dfrac{m_h^2}{v^2}$\\
        \hline
    \end{tabular}
\end{table}
\end{center}

\begin{center}
\begin{table}[!ht]
    \centering\begin{tabular}{>{\centering\arraybackslash}m{4cm} m{8.5cm}} 
        \multicolumn{2}{c}{Interactions with SM Fermions} \\
    \hline
        Diagram & Feynman Rule \\
        \hline
        \vspace{5 px}
        \includegraphics[height=2.25cm]{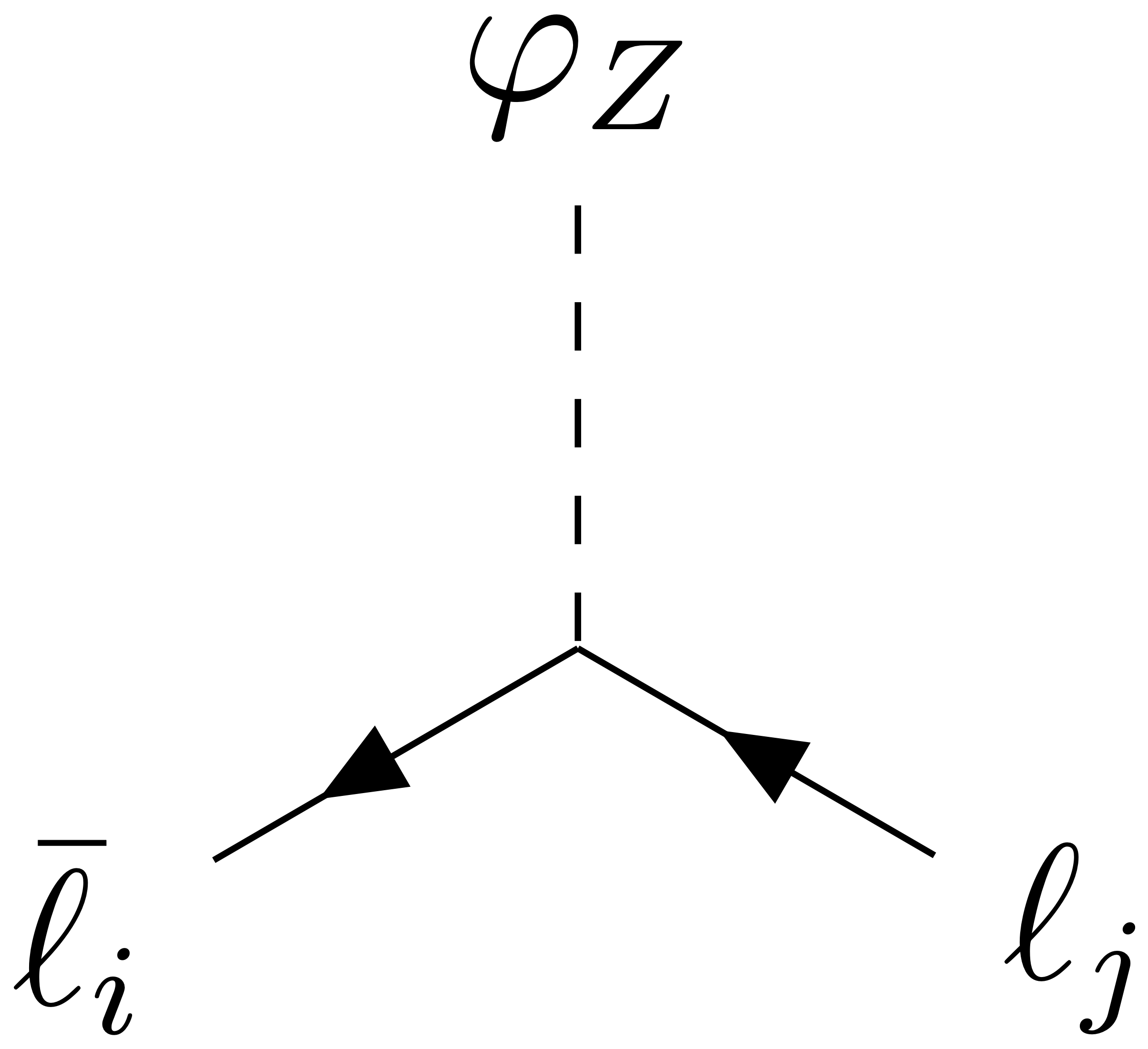}&
        $+ \dfrac{m_{\ell_i}}{v} \gamma^5 \delta_{ij}$\\
        \vspace{5 px}
        \includegraphics[height=2.25cm]{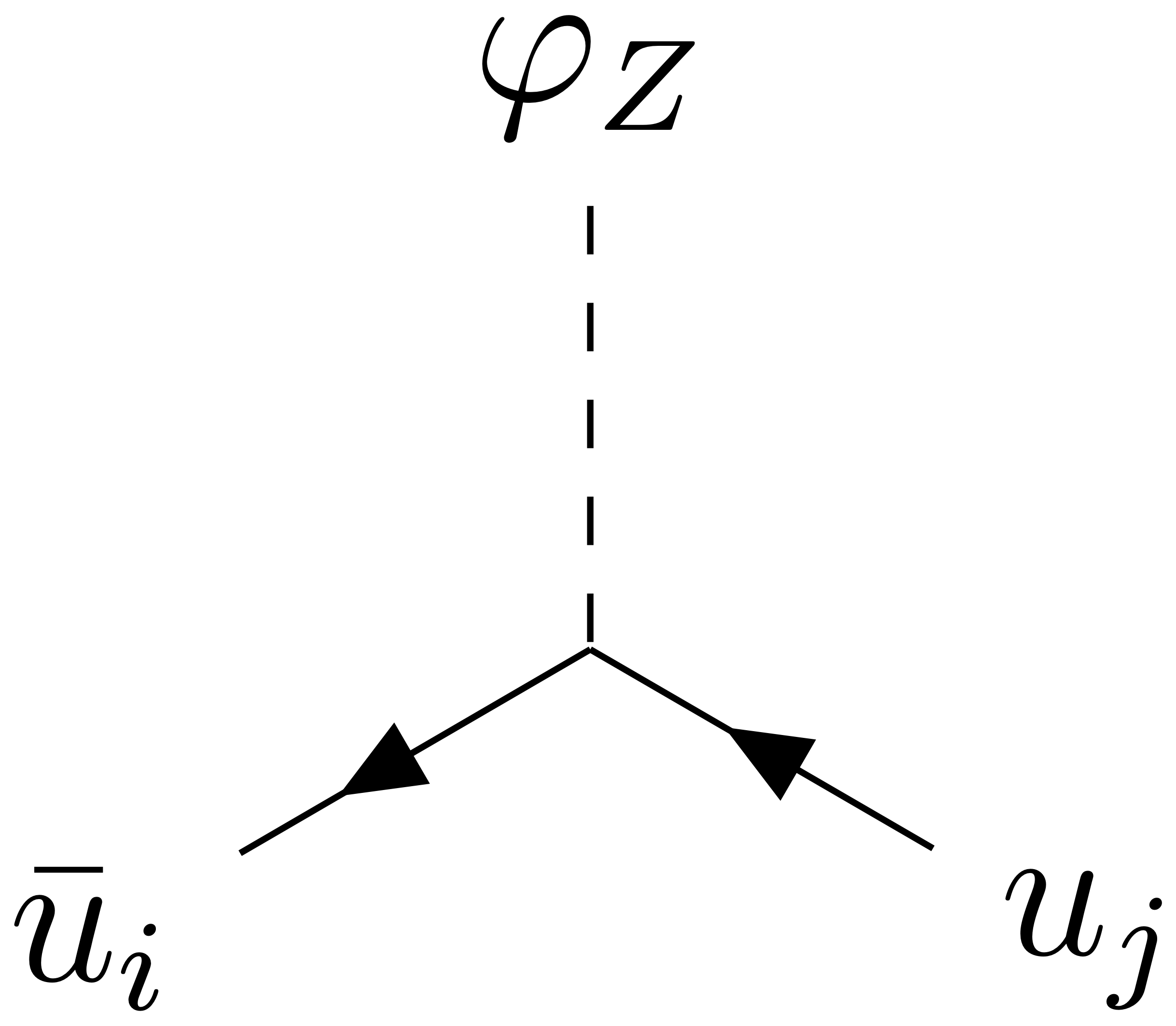}&
        $- \dfrac{m_{u_i}}{v} \gamma^5 \delta_{ij}
        $\\
        \vspace{5 px}
        \includegraphics[height=2.25cm]{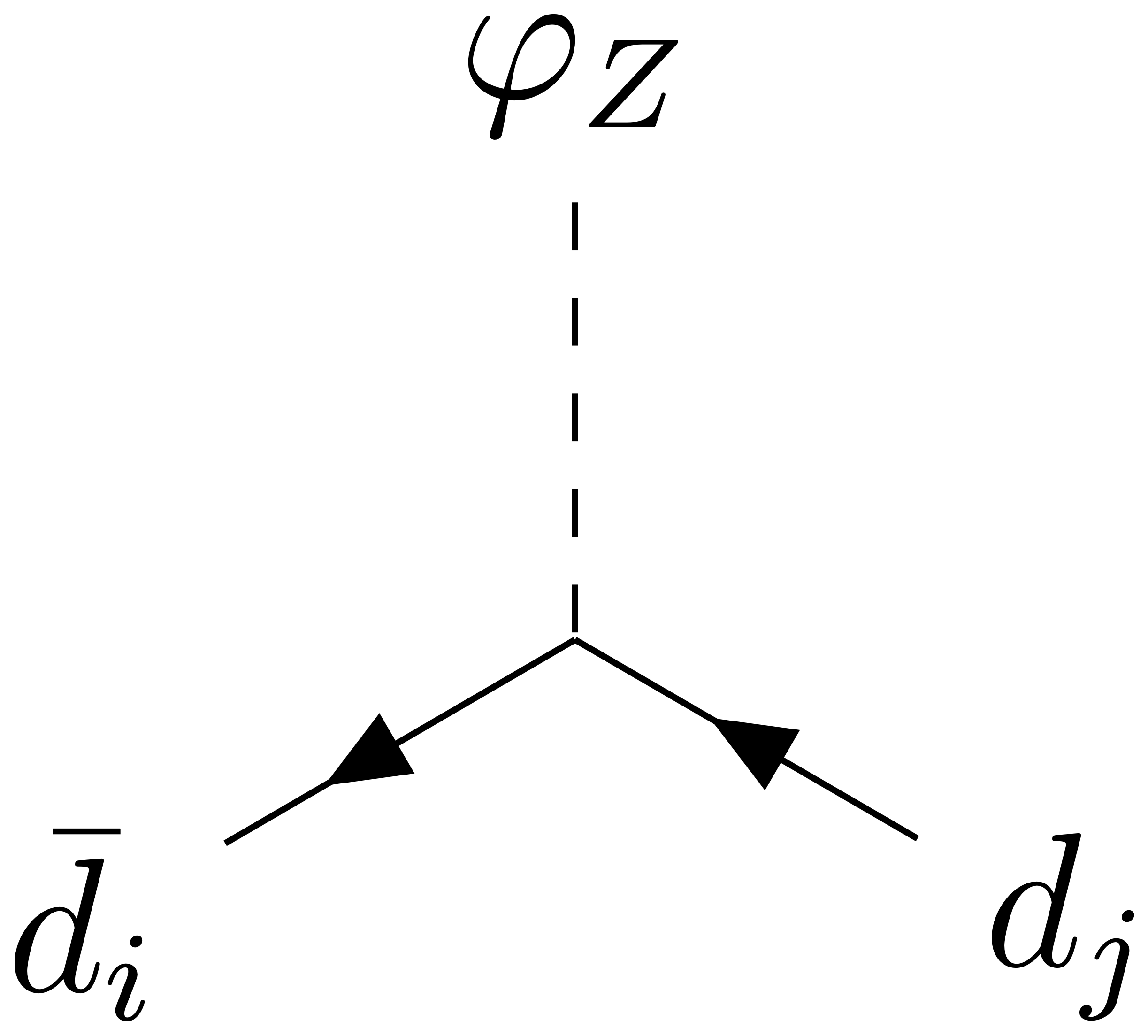}&
        $+ \dfrac{m_{d_i}}{v} \gamma^5 \delta_{ij}
        $\\
        \hline
    \end{tabular}
\end{table}
\end{center}

\begin{center}
\begin{table}[!ht]
    \centering\begin{tabular}{>{\centering\arraybackslash}m{4cm} m{8.5cm}} 
    \multicolumn{2}{c}{Interactions with SM Fermions} \\
    \hline
        Diagram & Feynman Rule \\
        \hline
        \vspace{5 px}
        \includegraphics[height=2.25cm]{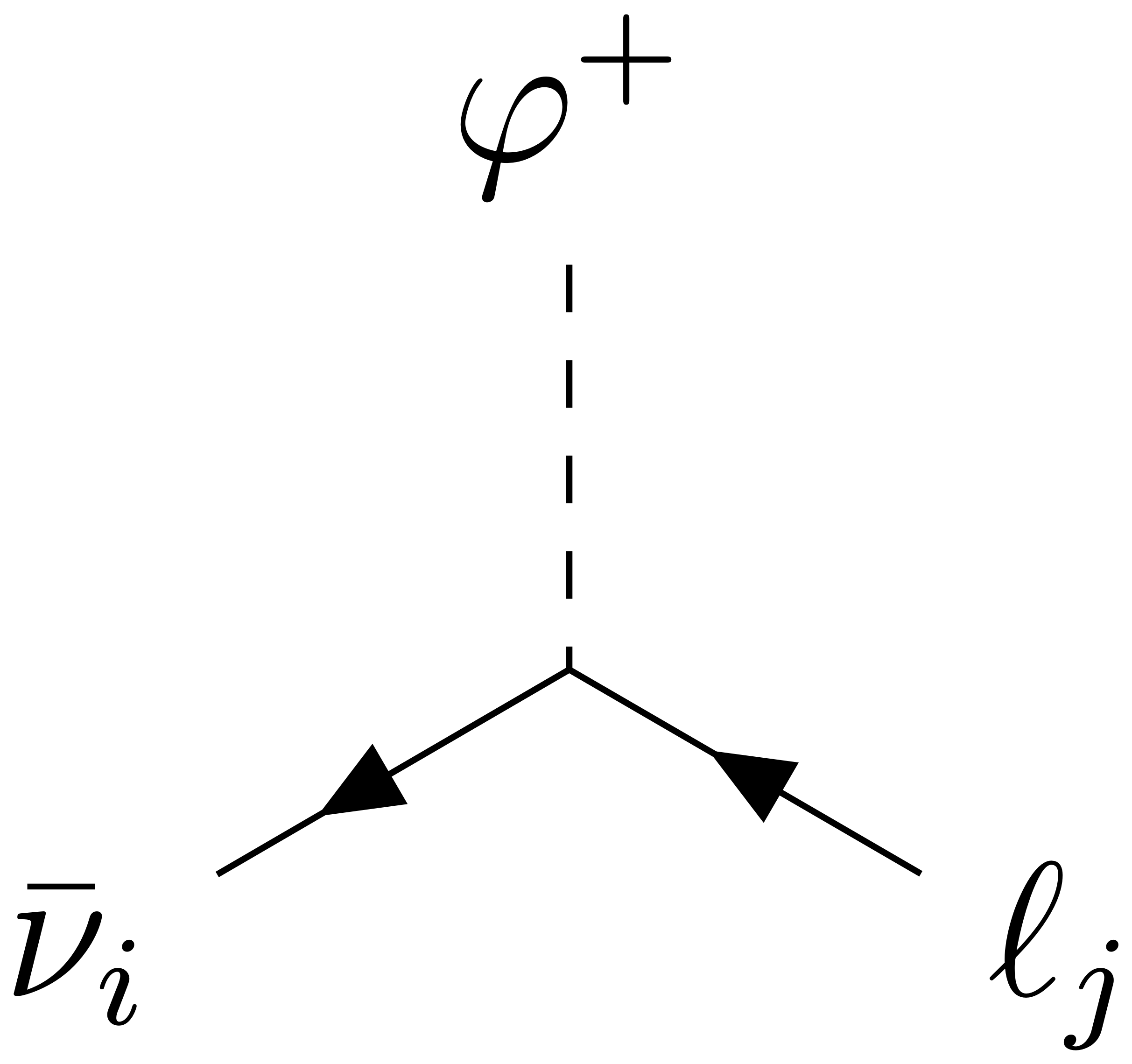}&
        $- \ \dfrac{\sqrt{2}}{v} m_{\ell_i} \delta_{ij}\  \text{P}_\text{R}$\vspace{5px}\\
        \vspace{10 px}
        \includegraphics[height=2.25cm]{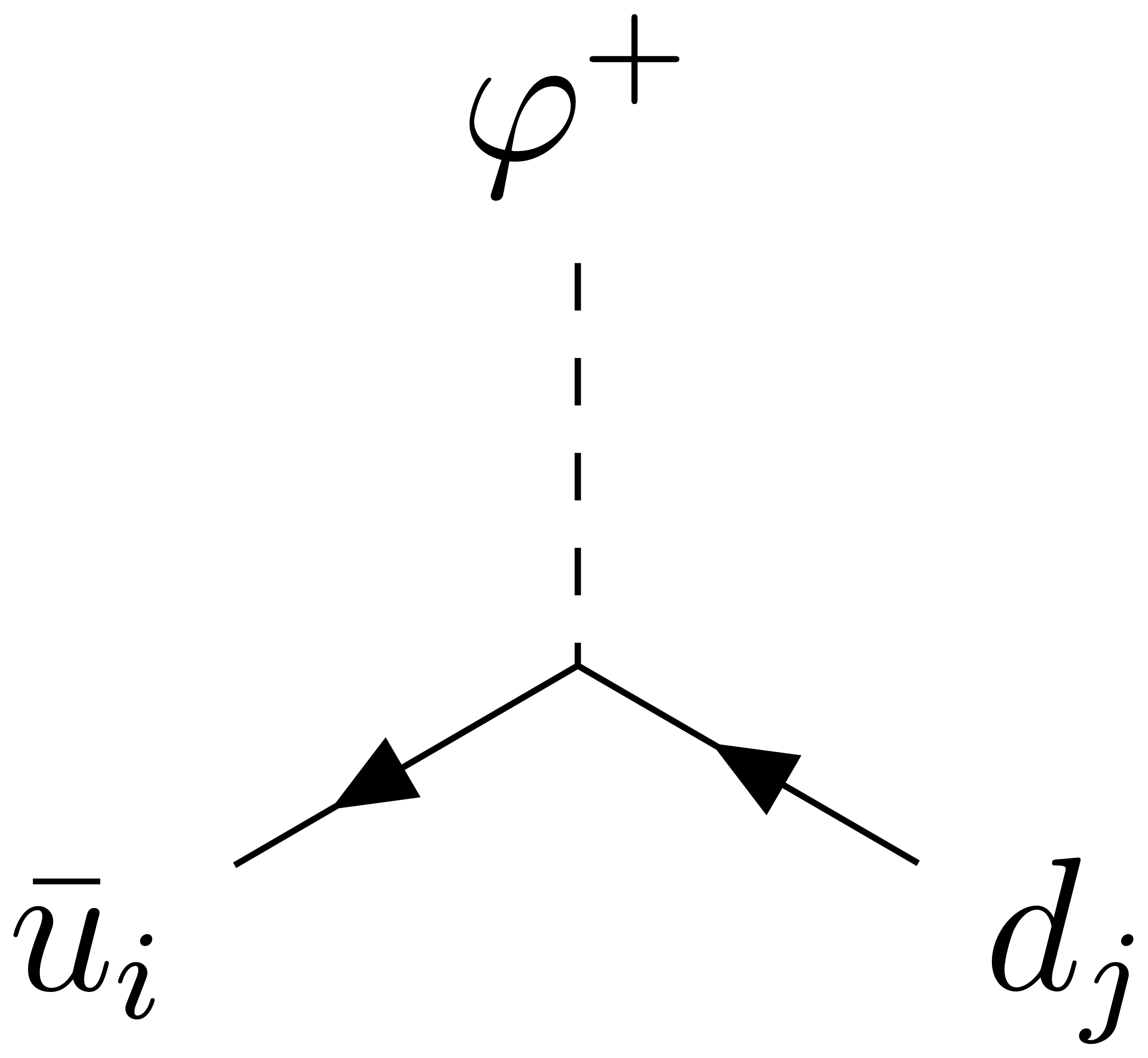}&
        \vspace{10 px}
        $+ \ \dfrac{\sqrt{2}}{v} V_{ij}  \left(m_{u_i}\text{P}_\text{L} - m_{d_i} \text{P}_\text{R} \right)$ \\
        \hline
    \end{tabular}
\end{table}
\end{center}

%

\begin{center}
\begin{table}[!ht]
    \centering\begin{tabular}{>{\centering\arraybackslash}m{4cm} m{8.5cm}} 
    \multicolumn{2}{c}{Triple Interactions with SM Gauge Bosons} \\
    \hline
        Diagram & Feynman Rule \\
        \hline
        \vspace{5 px}
        \includegraphics[height=2.25cm]{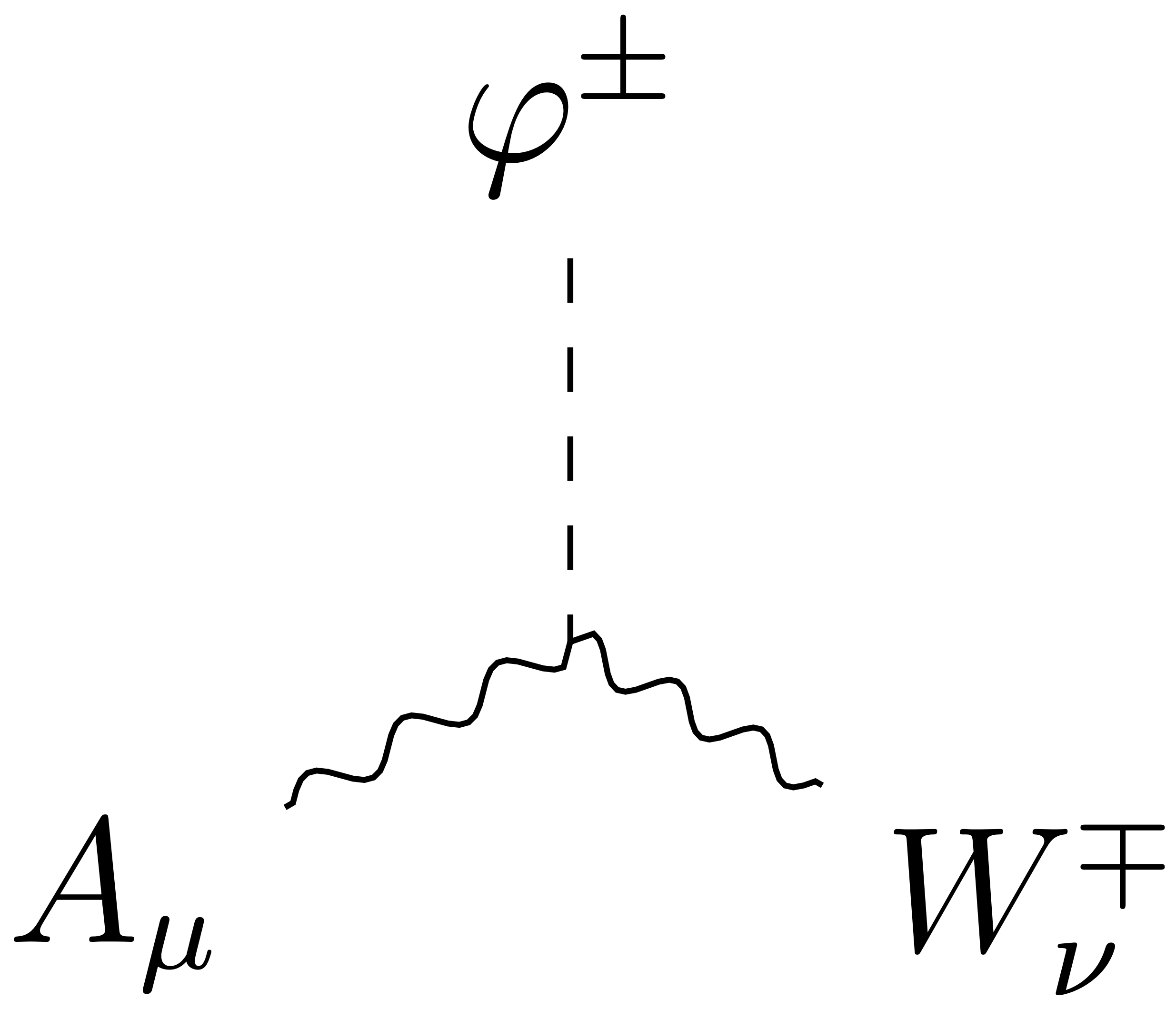}&
        $ \pm \ \dfrac{1}{2} \dfrac{e^2}{s_w} v \ \eta^{\mu\nu}$\\
        \vspace{5 px}
        \includegraphics[height=2.25cm]{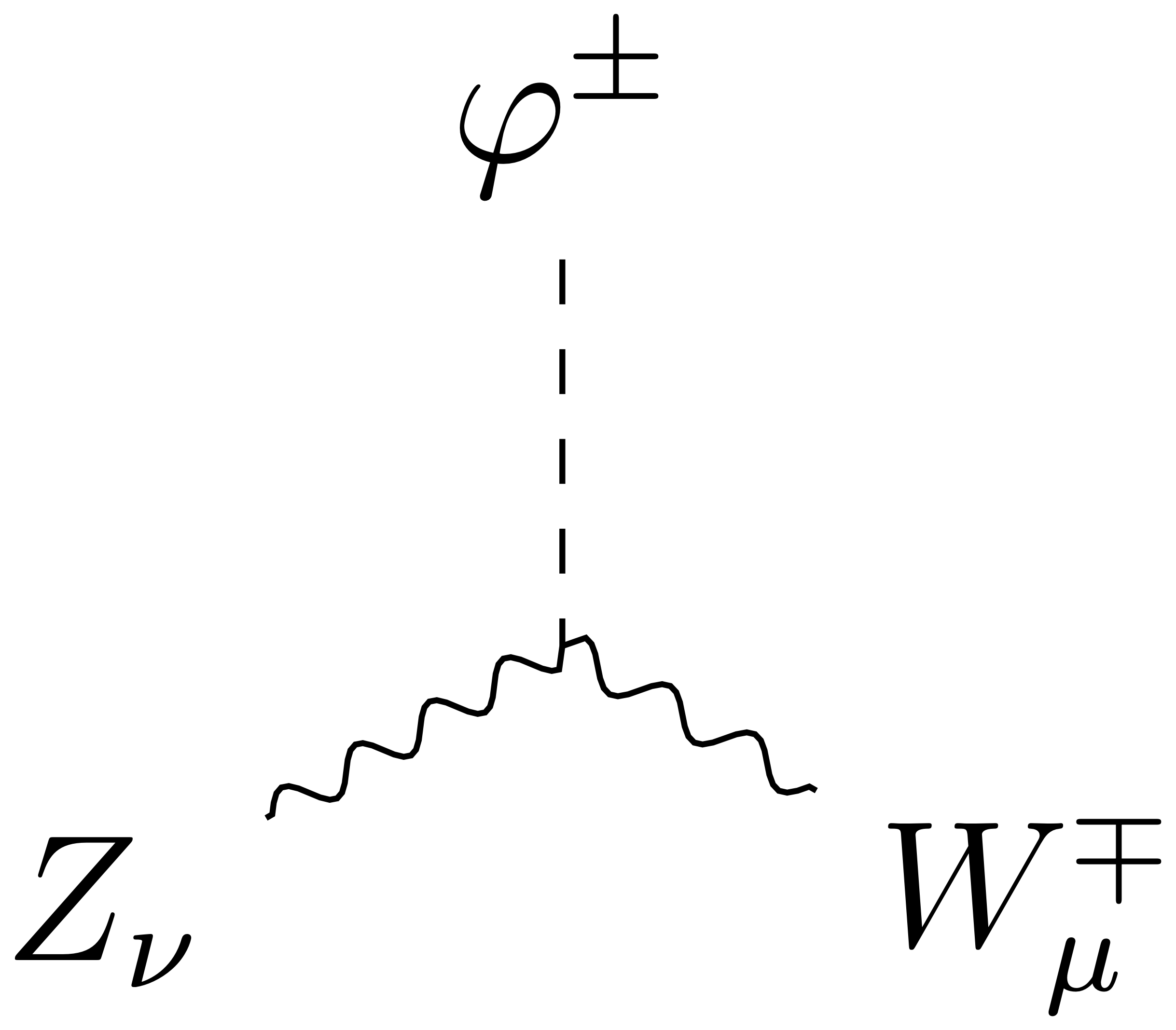}&
        $ \mp \ \dfrac{1}{2} \dfrac{e^2}{c_w} v \ \eta^{\mu\nu}$\\
        \vspace{5 px}
        \includegraphics[height=2.25cm]{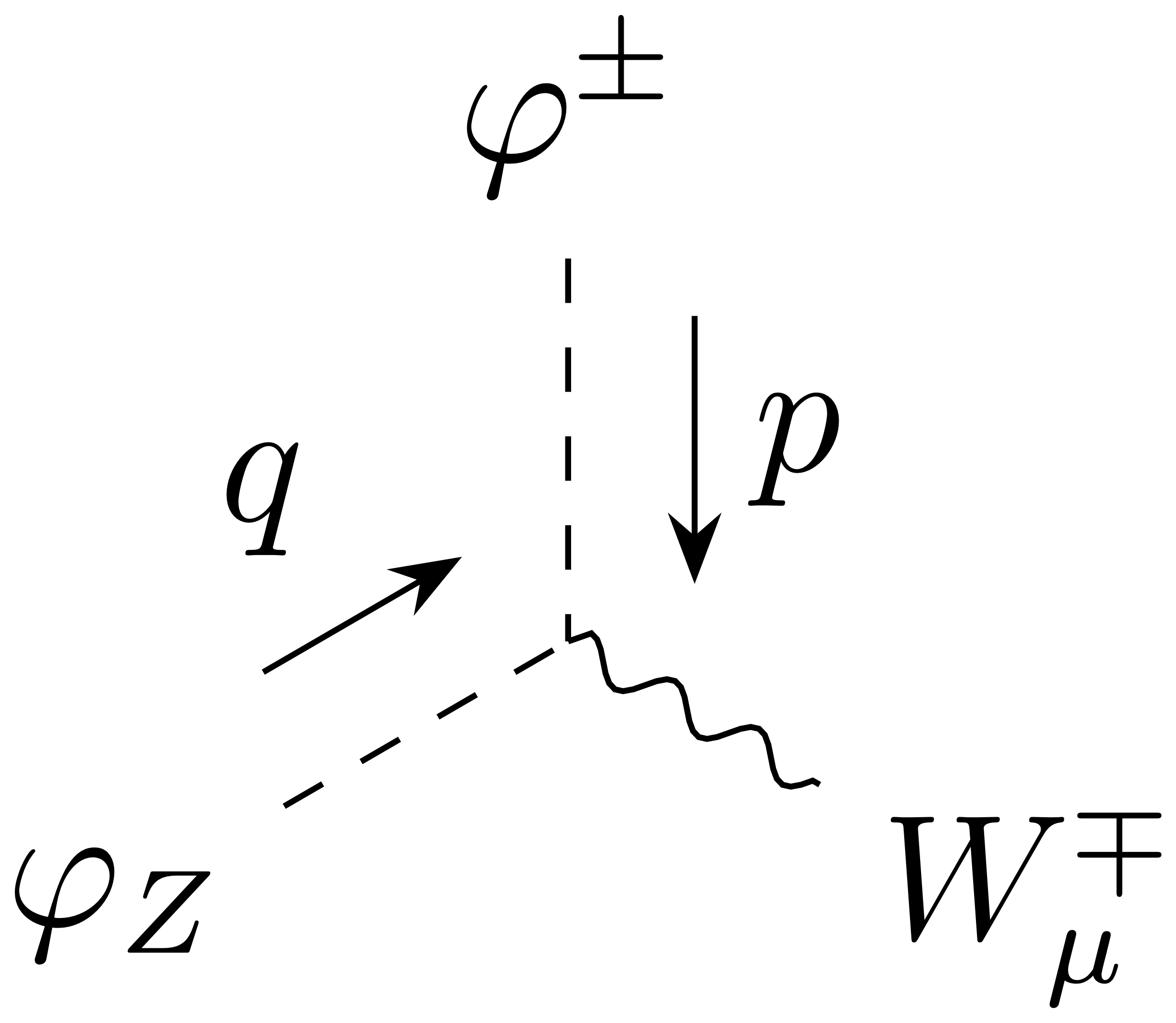}&
        $\mp \ \dfrac{i}{2} \dfrac{e}{s_w} \left( p^\mu - q^\mu \right)$\\
        \vspace{5 px}
        \includegraphics[height=2.25cm]{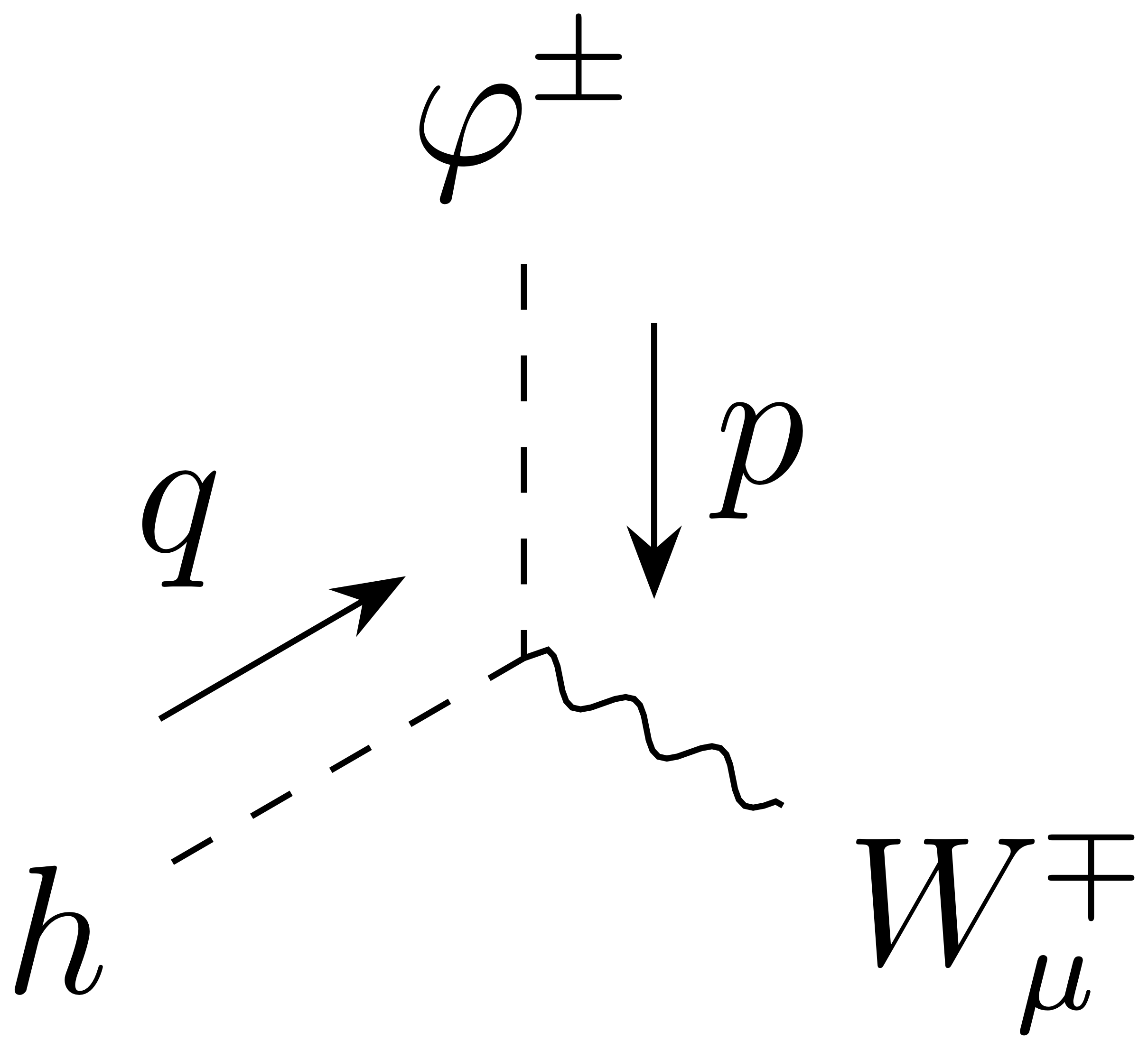}&
        $ + \ \dfrac{1}{2} \dfrac{e}{s_w} \left( p^\mu - q^\mu\right)$\\
        \vspace{5 px}
        \includegraphics[height=2.25cm]{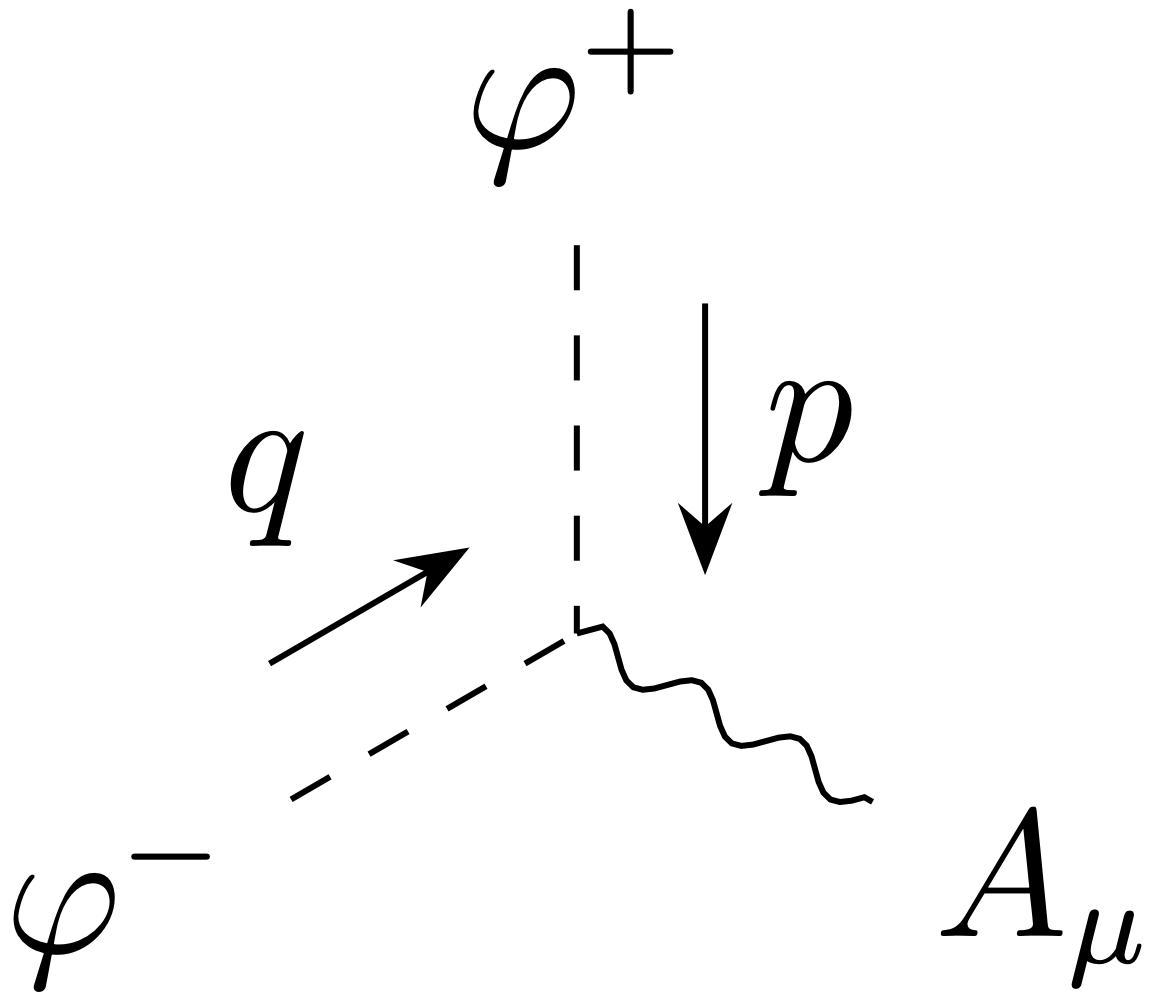}&
        $ + i e \left( p^\mu - q^\mu\right)$\\
        \hline
    \end{tabular}
\end{table}
\end{center}
\begin{center}
\begin{table}[!ht]
    \centering\begin{tabular}{>{\centering\arraybackslash}m{4cm} m{8.5cm}} 
    \multicolumn{2}{c}{Triple Interactions with SM Gauge Bosons} \\
    \hline
        Diagram & Feynman Rule \\
        \hline
        \vspace{5 px}
        \includegraphics[height=2.25cm]{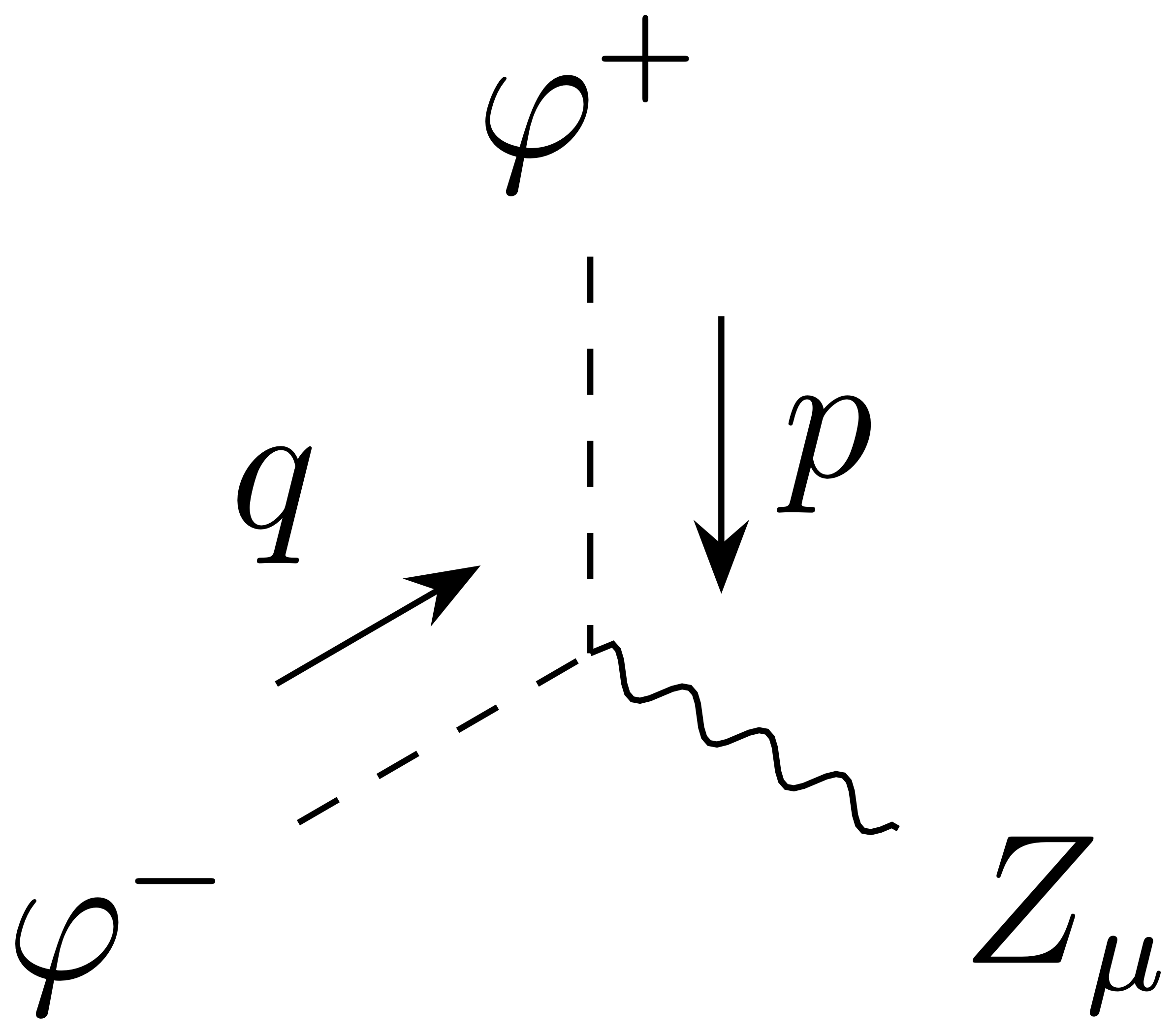}&
        $ + \dfrac{i}{2} \dfrac{e}{c_w s_w} \left( c_w^2 - s_w^2 \right) \left( p^\mu - q^\mu\right)$\\
        \vspace{5 px}
        \includegraphics[height=2.25cm]{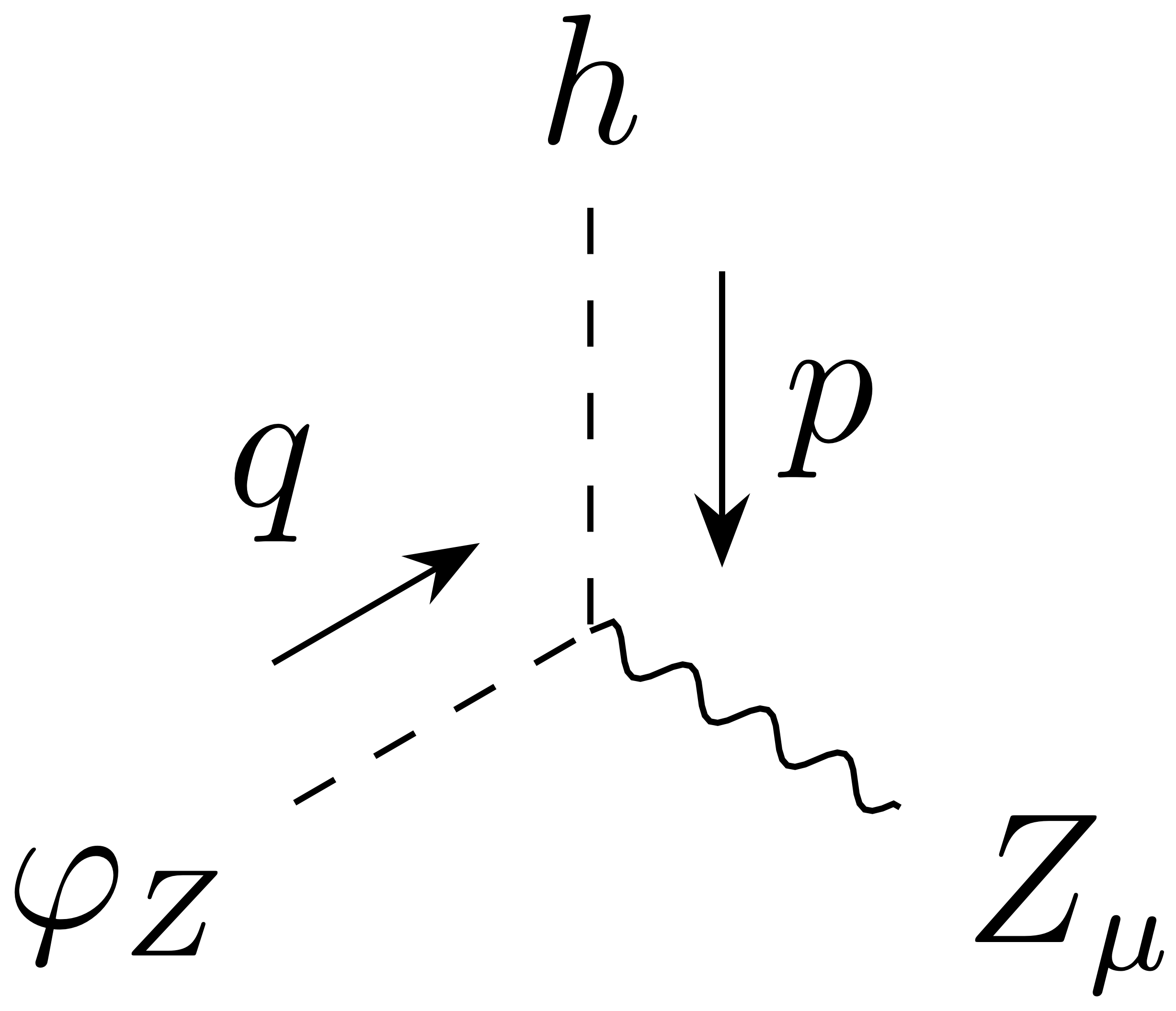}&
        $ - \dfrac{1}{2} \dfrac{e}{c_w s_w} \left(p^\mu - q^\mu \right)$\\
        \hline
    \end{tabular}
\end{table}
\end{center}

\begin{center}
\begin{table}[!ht]
    \centering\begin{tabular}{>{\centering\arraybackslash}m{5cm} m{7.5cm}} 
    \multicolumn{2}{c}{Quartic Interactions with SM Gauge Bosons} \\
    \hline
        Diagram & Feynman Rule \\
        \hline
        \vspace{5 px}
        \includegraphics[height=3.5cm]{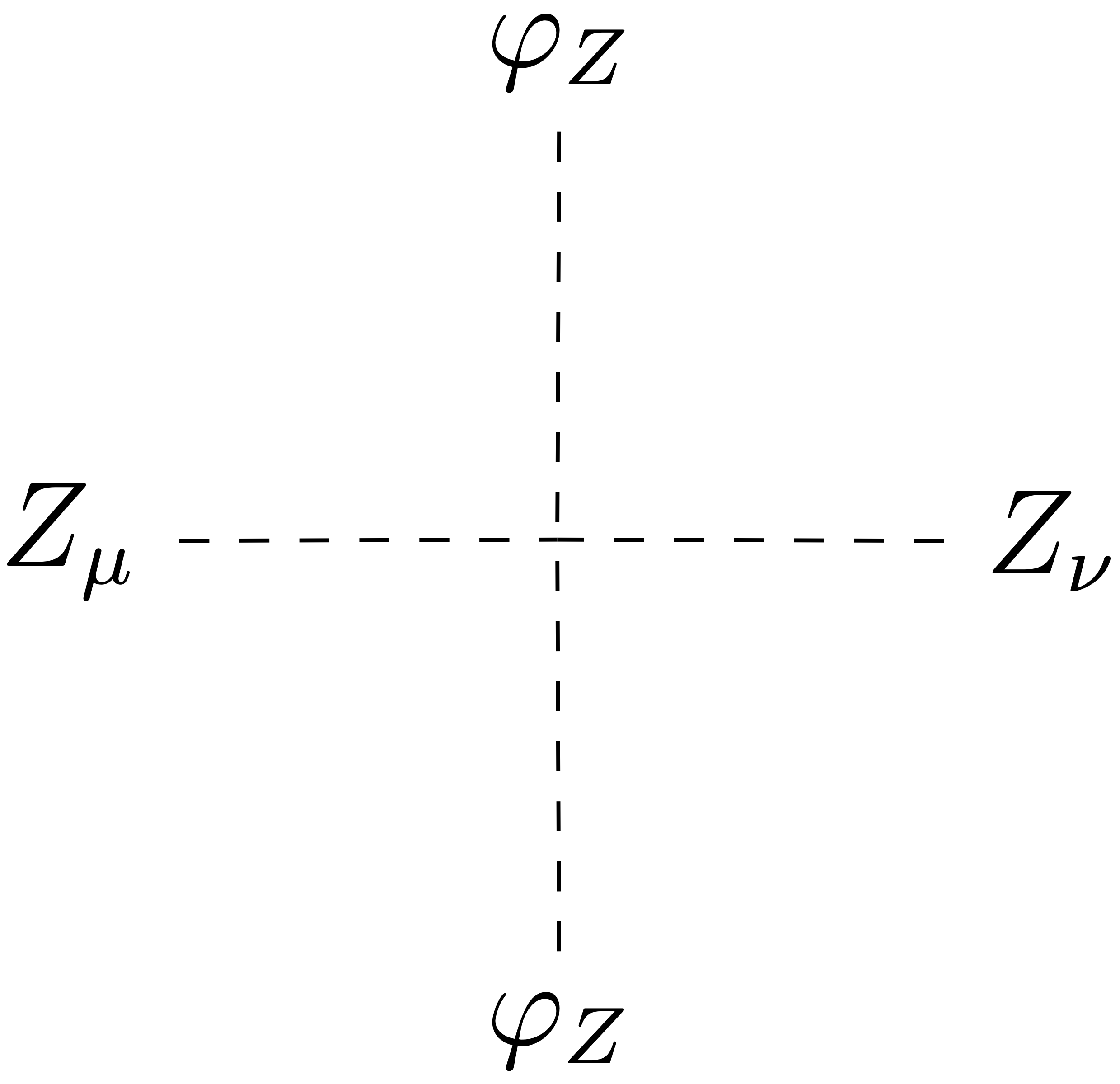}&
        $+ \ \dfrac{i}{2} \left(\dfrac{e}{c_w s_w}\right)^2 \eta^{\mu \nu}$\\
        \vspace{5 px}
        \includegraphics[height=3.5cm]{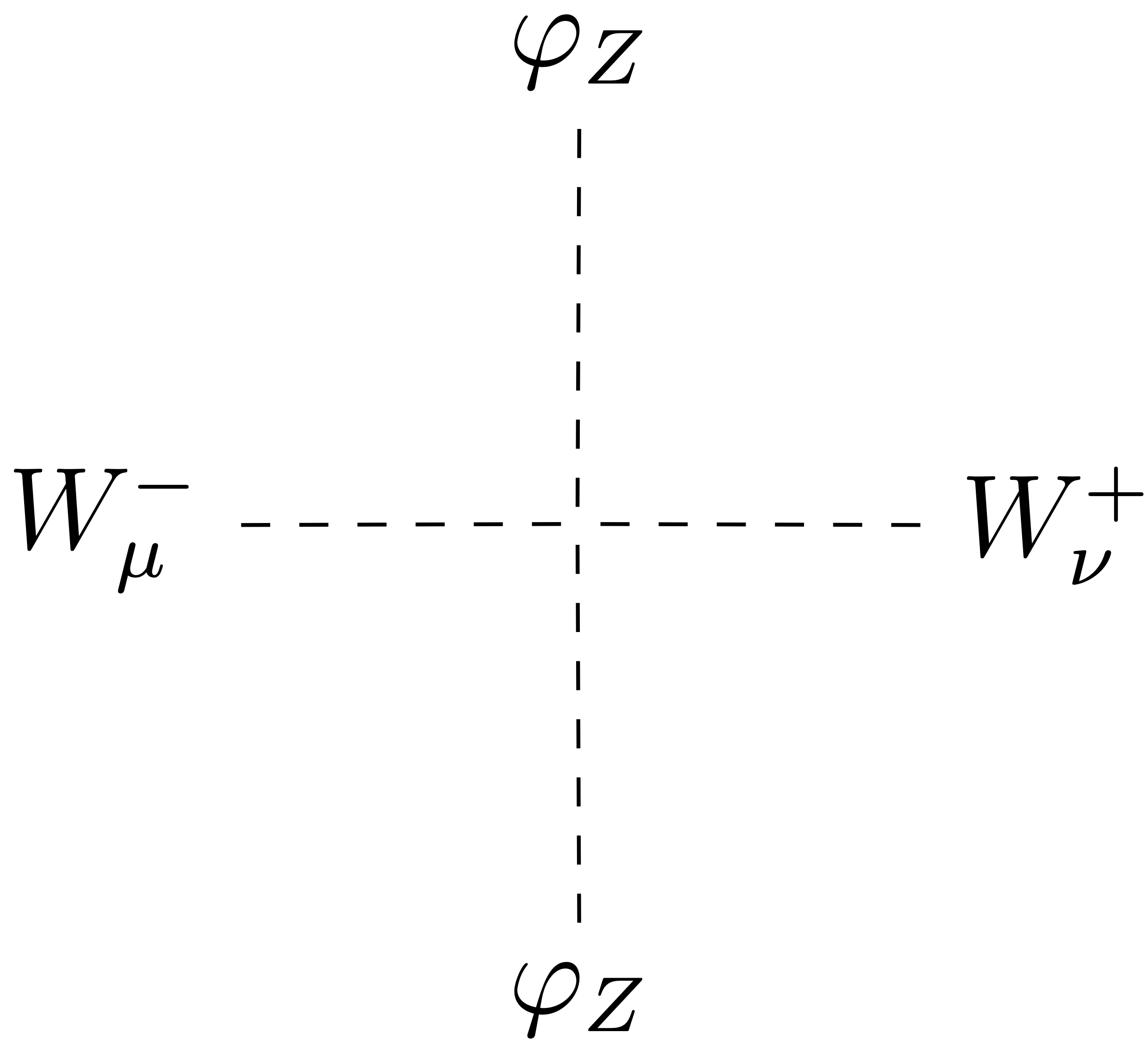}&
        $+ \ \dfrac{i}{2} \left(\dfrac{e}{s_w}\right)^2 \eta^{\mu \nu}$\\
        \hline
    \end{tabular}
\end{table}
\end{center}

\begin{center}
\begin{table}[!ht]
    \centering\begin{tabular}{>{\centering\arraybackslash}m{5cm} m{7.5cm}} 
    \multicolumn{2}{c}{Quartic Interactions with SM Gauge Bosons} \\
    \hline
         Diagram & Feynman Rule \\
        \hline
        \vspace{5 px}
        \includegraphics[height=3.5cm]{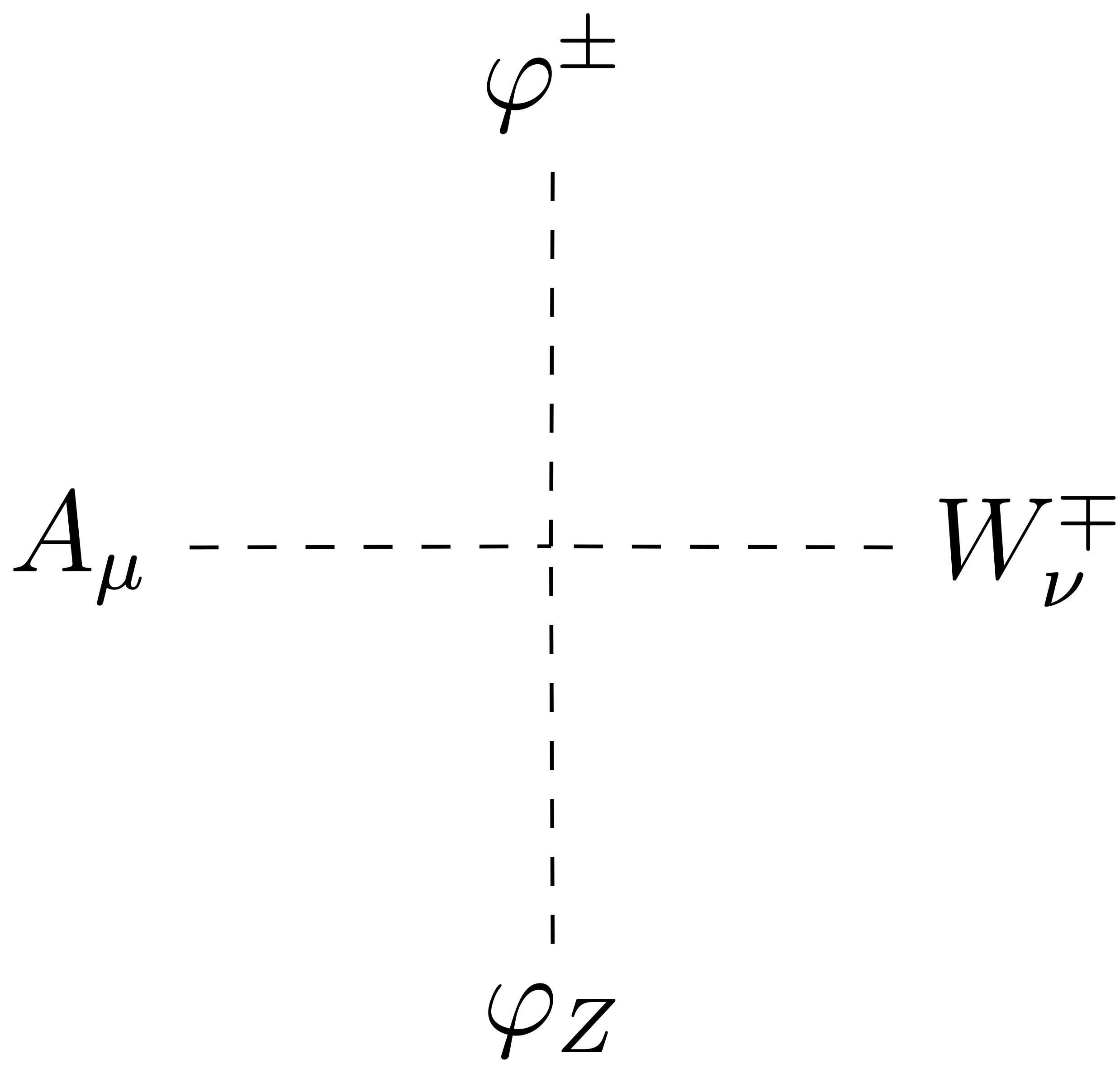}&
        $- \ \dfrac{i}{2} \dfrac{e^2}{s_w} \eta^{\mu \nu}$\\
        \vspace{5 px}
        \includegraphics[height=3.5cm]{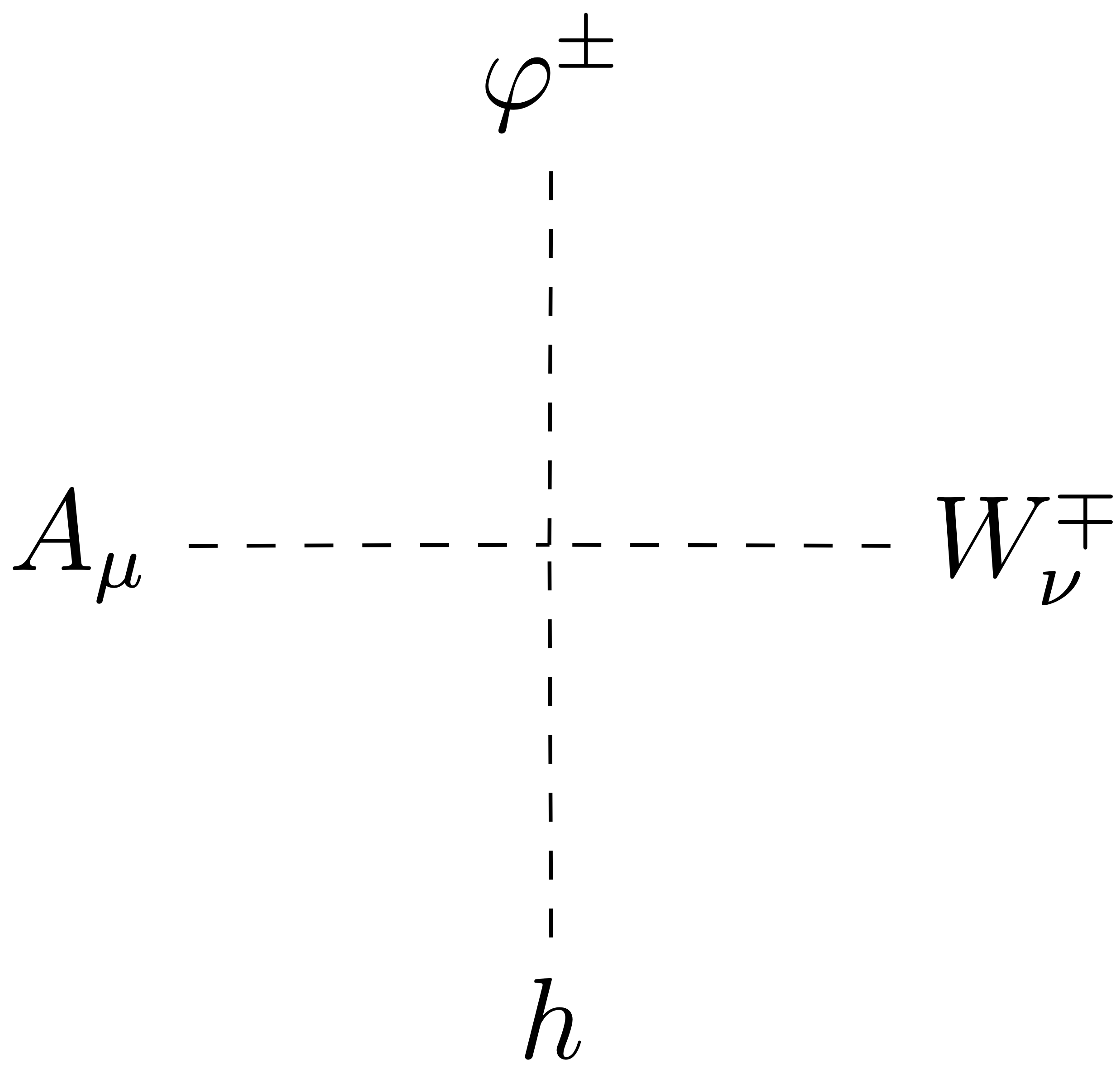}&
        $\pm \ \dfrac{1}{2} \dfrac{e^2}{s_w} \eta^{\mu \nu}$\\
        \vspace{5 px}
        \includegraphics[height=3.5cm]{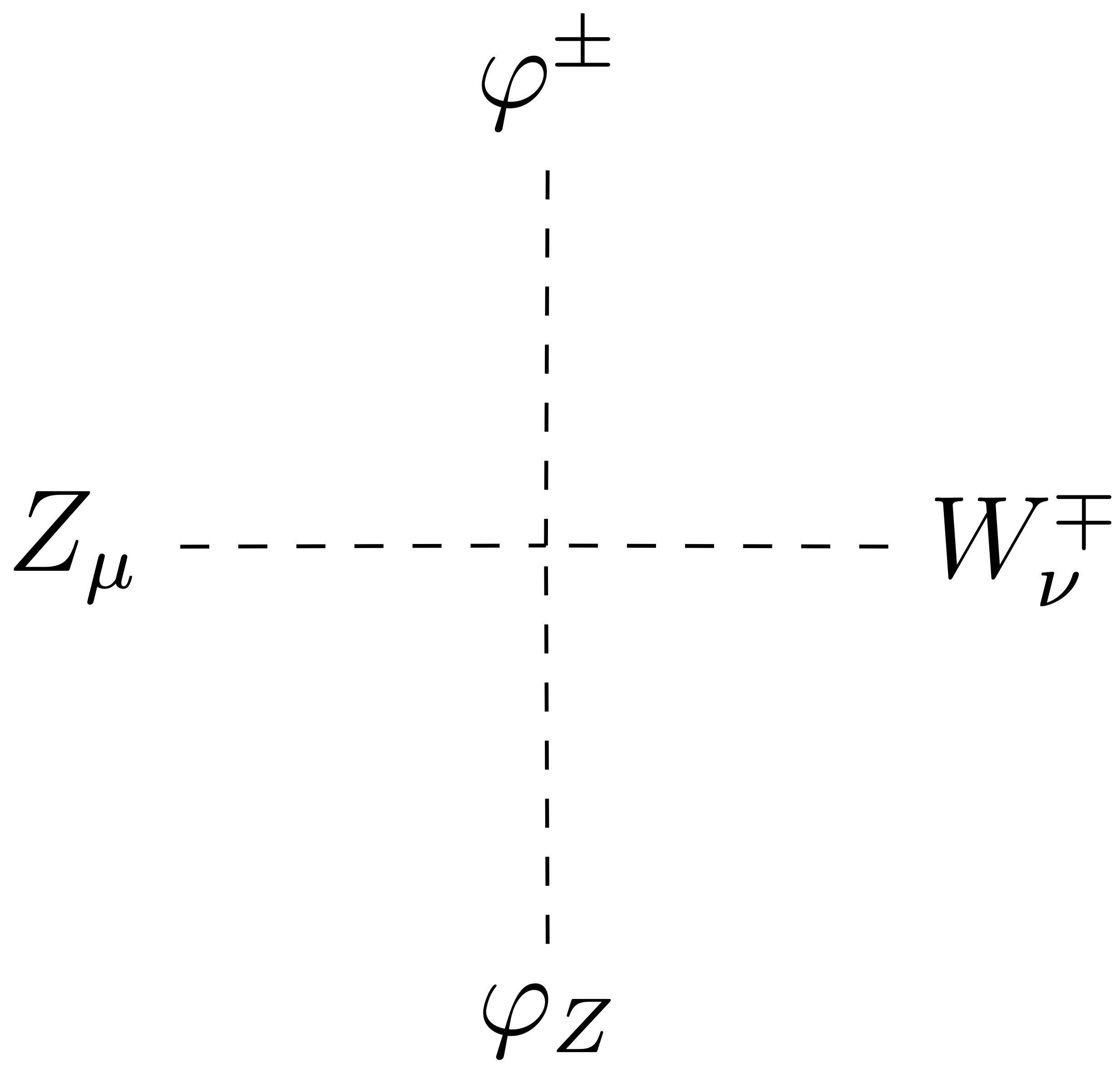}&
        $+ \ \dfrac{i}{2} \dfrac{e^2}{c_w} \eta^{\mu \nu}$\\
        \vspace{5 px}
        \includegraphics[height=3.5cm]{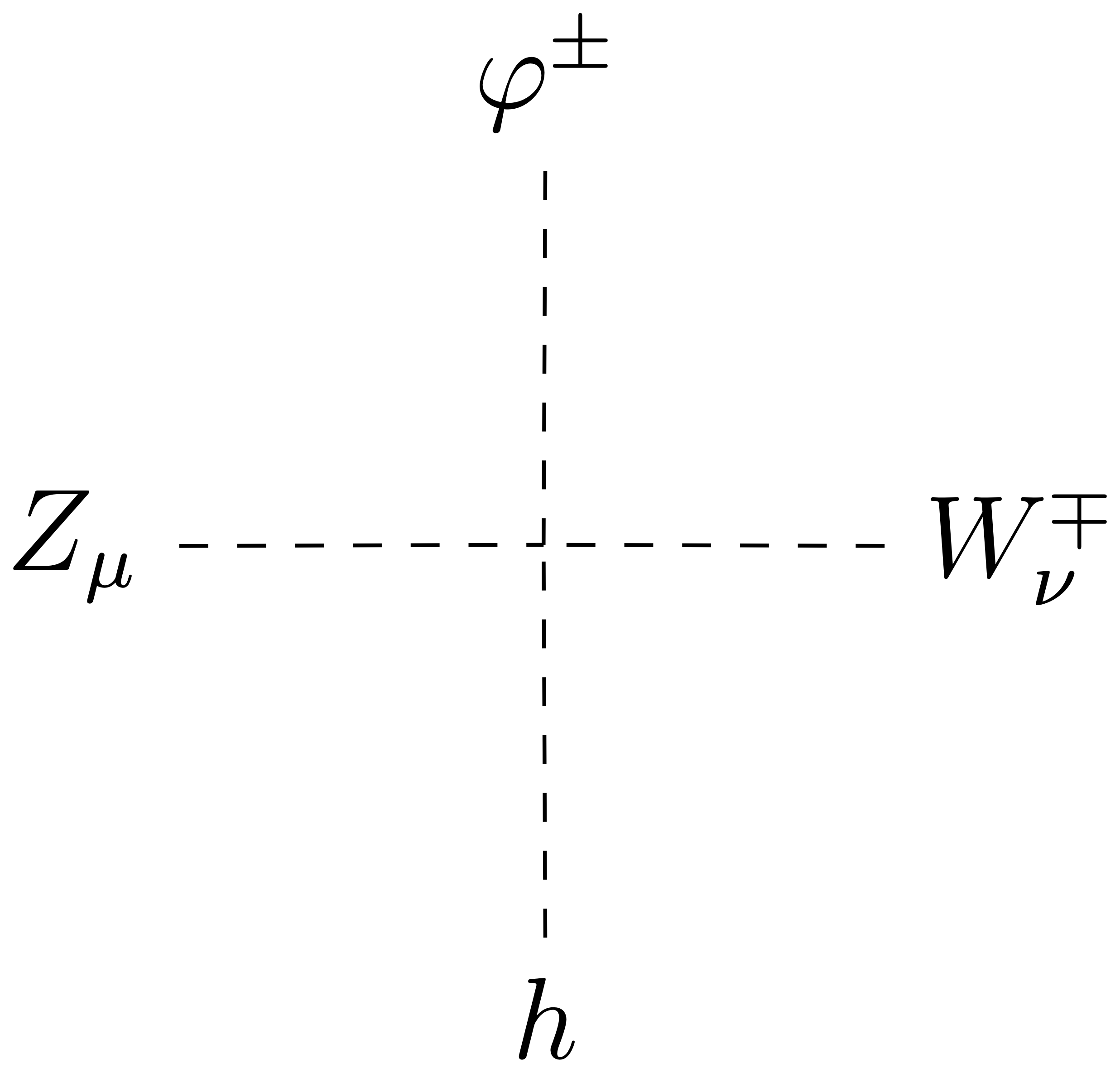}&
        $\mp \ \dfrac{1}{2} \dfrac{e^2}{c_w} \eta^{\mu \nu}$\\
        \hline
    \end{tabular}
\end{table}
\end{center}

\begin{center}
\begin{table}[!ht]
    \centering\begin{tabular}{ >{\centering\arraybackslash}m{5cm} m{7cm}} 
    \multicolumn{2}{c}{Quartic Interactions with SM Gauge Bosons} \\
    \hline
        Diagram & Feynman Rule \\
        \hline
        \vspace{5 px}
        \includegraphics[height=3.5cm]{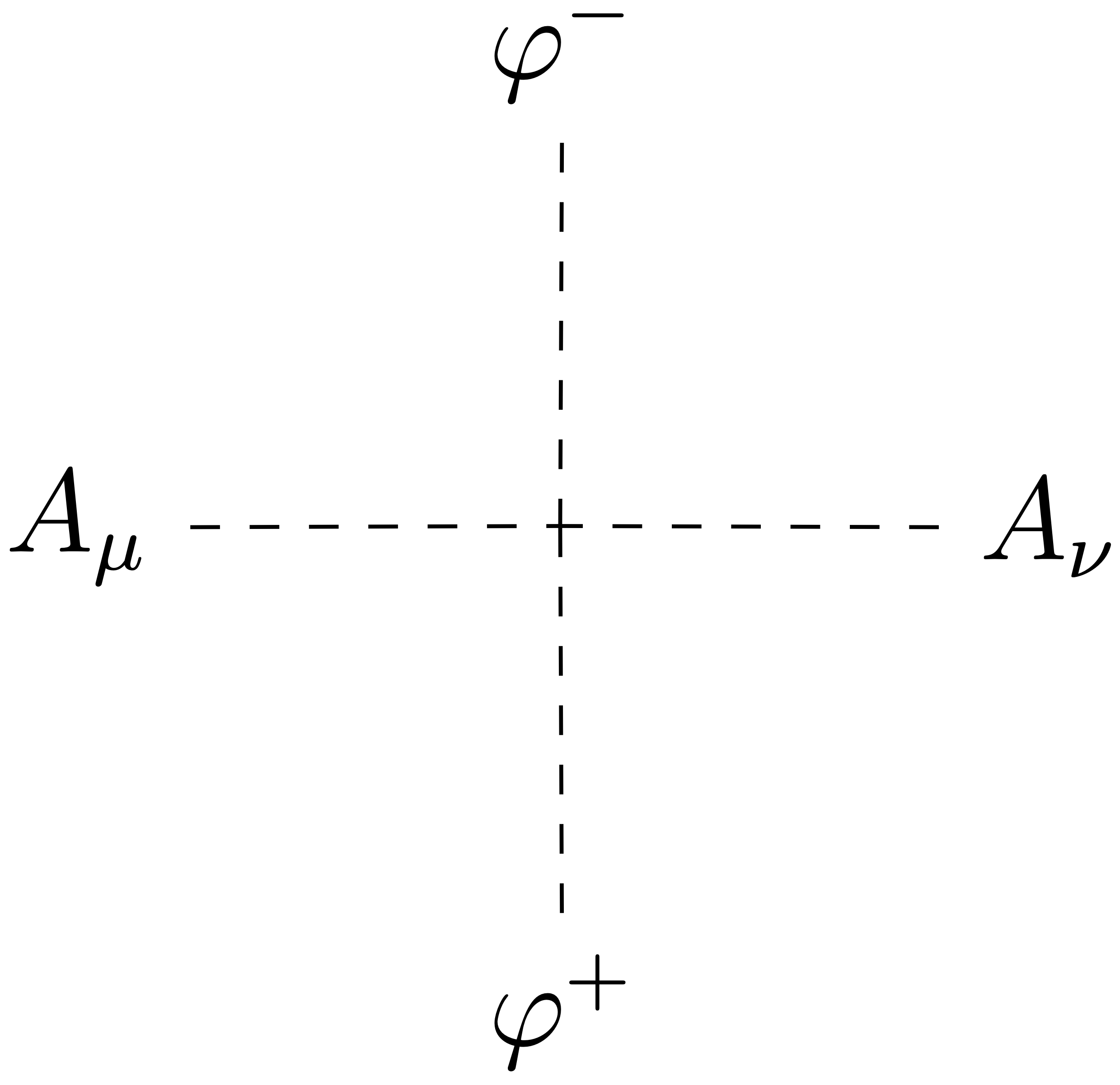}&
        $+ \ 2i e^2 \eta^{\mu \nu}$\\
        \vspace{5 px}
        \includegraphics[height=3.5cm]{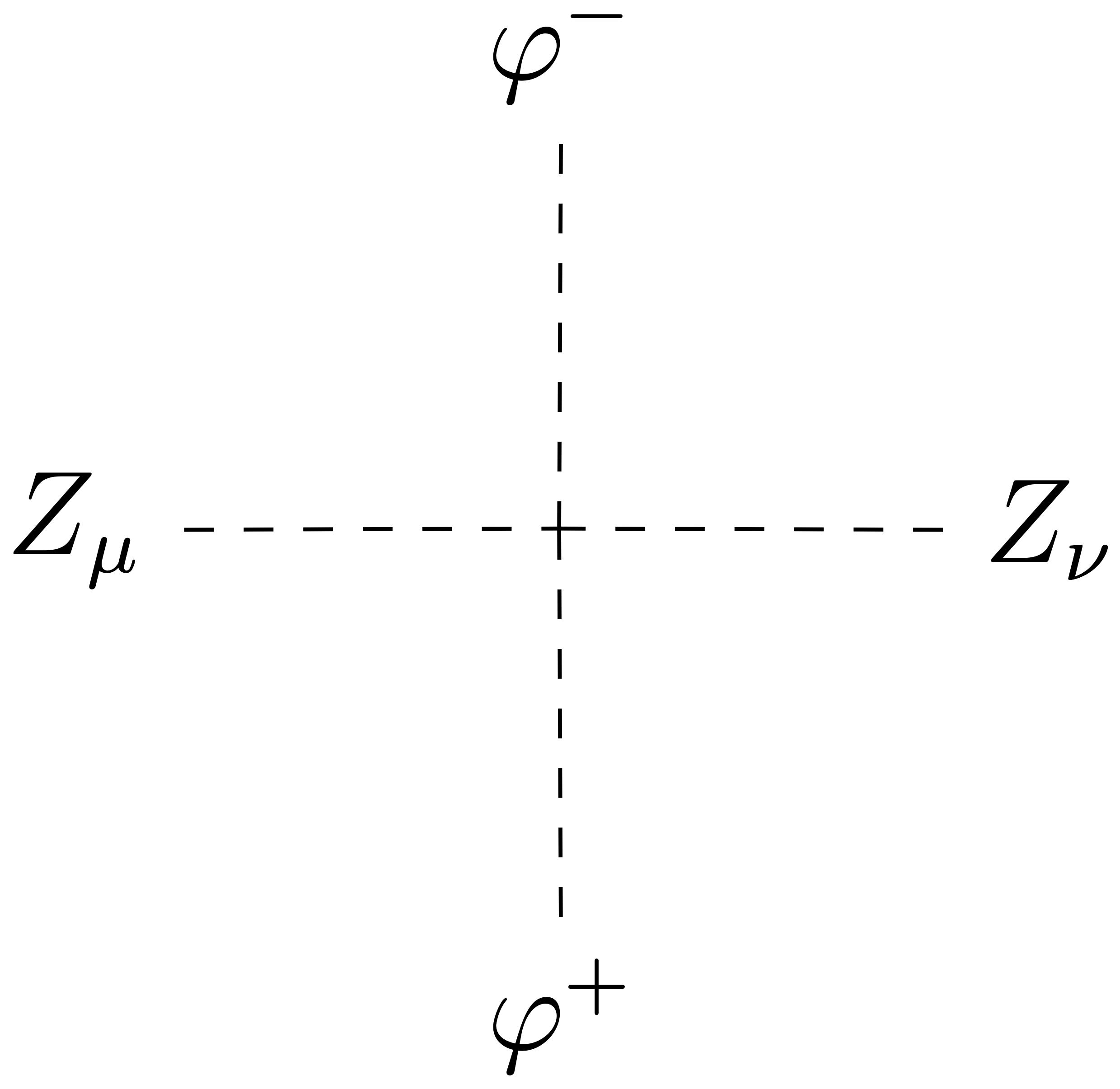}&
        $+ \ \dfrac{i}{2} \left(\dfrac{e}{s_w c_w}\right)^2 \eta^{\mu \nu} \left( c_w^2 - s_w^2 \right)^2$\\
        \vspace{5 px}
        \includegraphics[height=3.5cm]{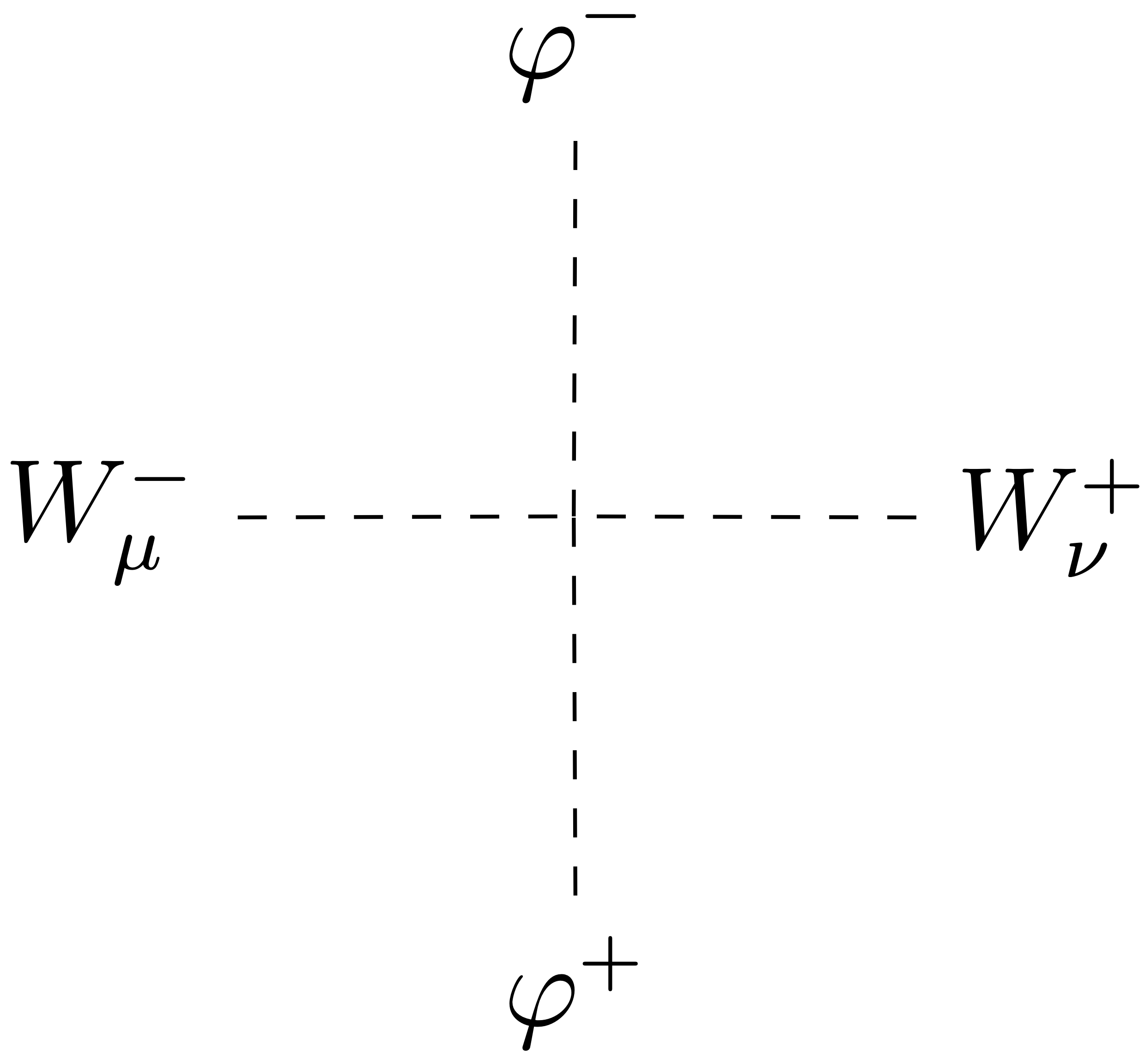}&
        $+ \ \dfrac{i}{2} \left(\dfrac{e}{s_w}\right)^2 \eta^{\mu \nu}$\\
        \vspace{5 px}
        \includegraphics[height=3.5cm]{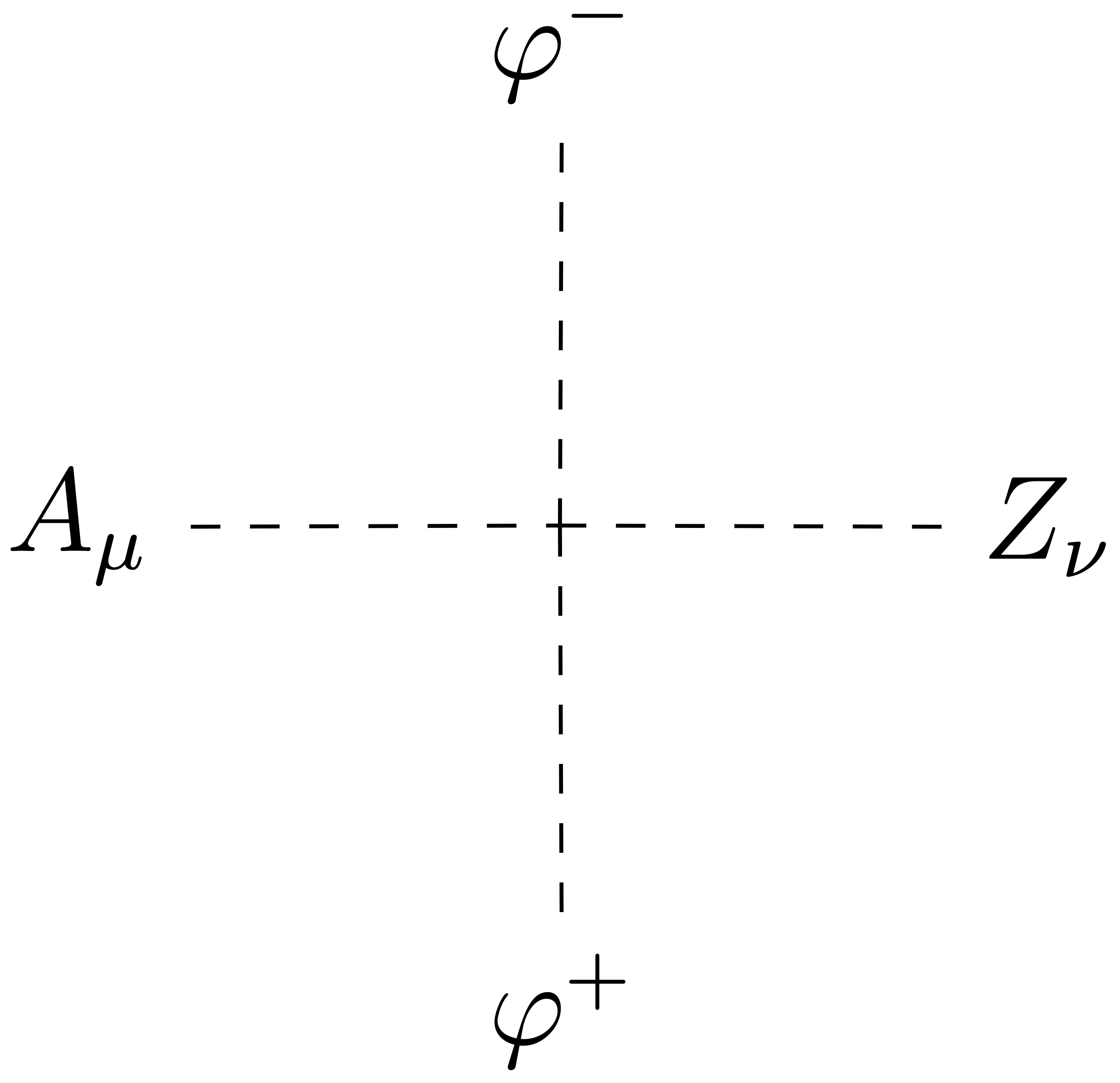}&
        $+ \ i \dfrac{e^2}{s_w c_w} \eta^{\mu \nu} \left(c_w^2 - s_w^2 \right)$\\
        \hline
    \end{tabular}
\end{table}
\end{center}

\begin{center}
\begin{table}[!ht]
    \centering\begin{tabular}{>{\centering\arraybackslash}m{4cm} m{8.5cm}} 
    \multicolumn{2}{c}{Triple Interactions with Ghosts} \\
    \hline
         Diagram & Feynman Rule \\
        \hline
        \vspace{5 px}
        \includegraphics[height=2.25cm]{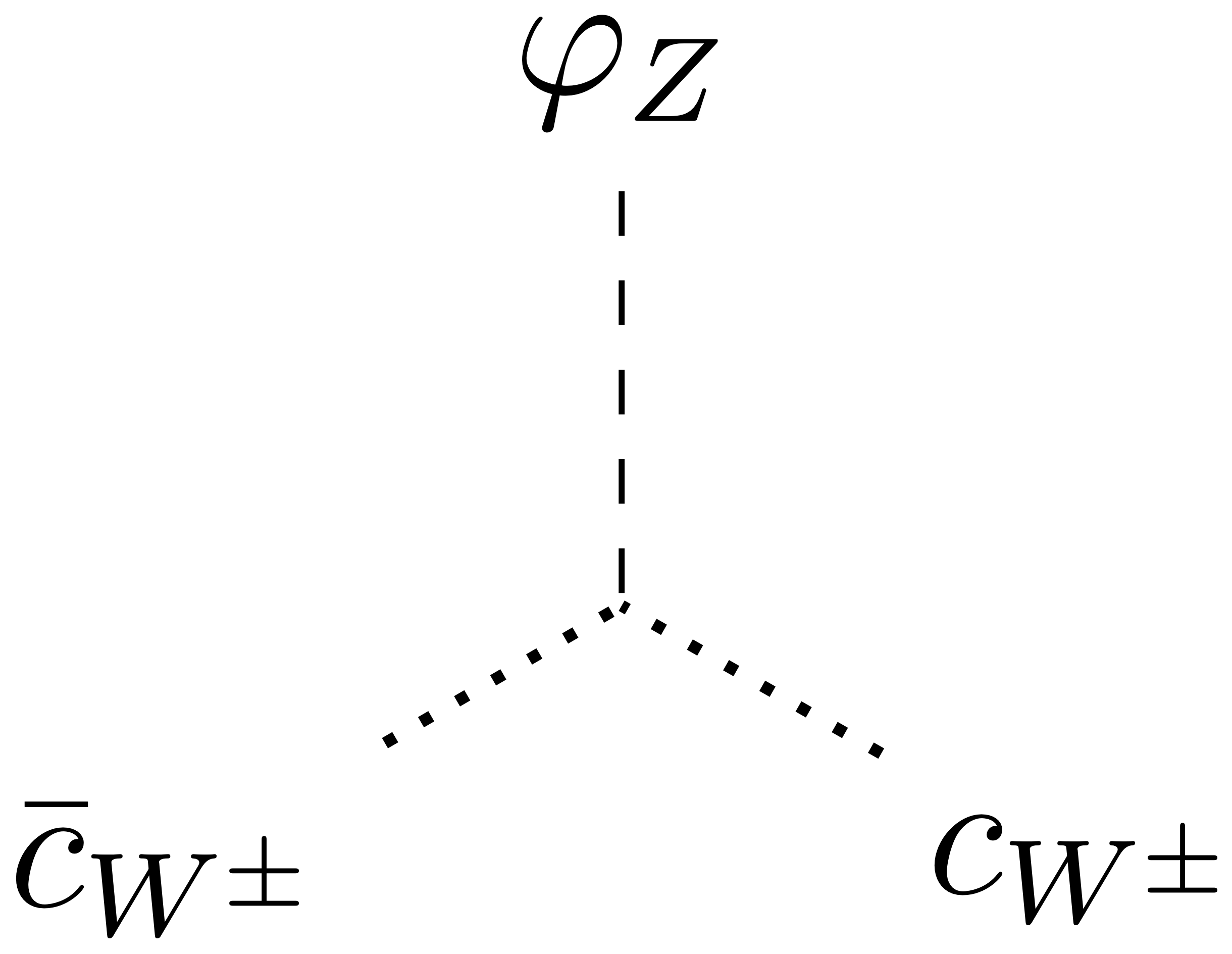}&
        $\pm\ \dfrac{1}{4} \left(\dfrac{e}{s_w}\right)^2 v $\\
        \vspace{5 px}
        \includegraphics[height=2.25cm]{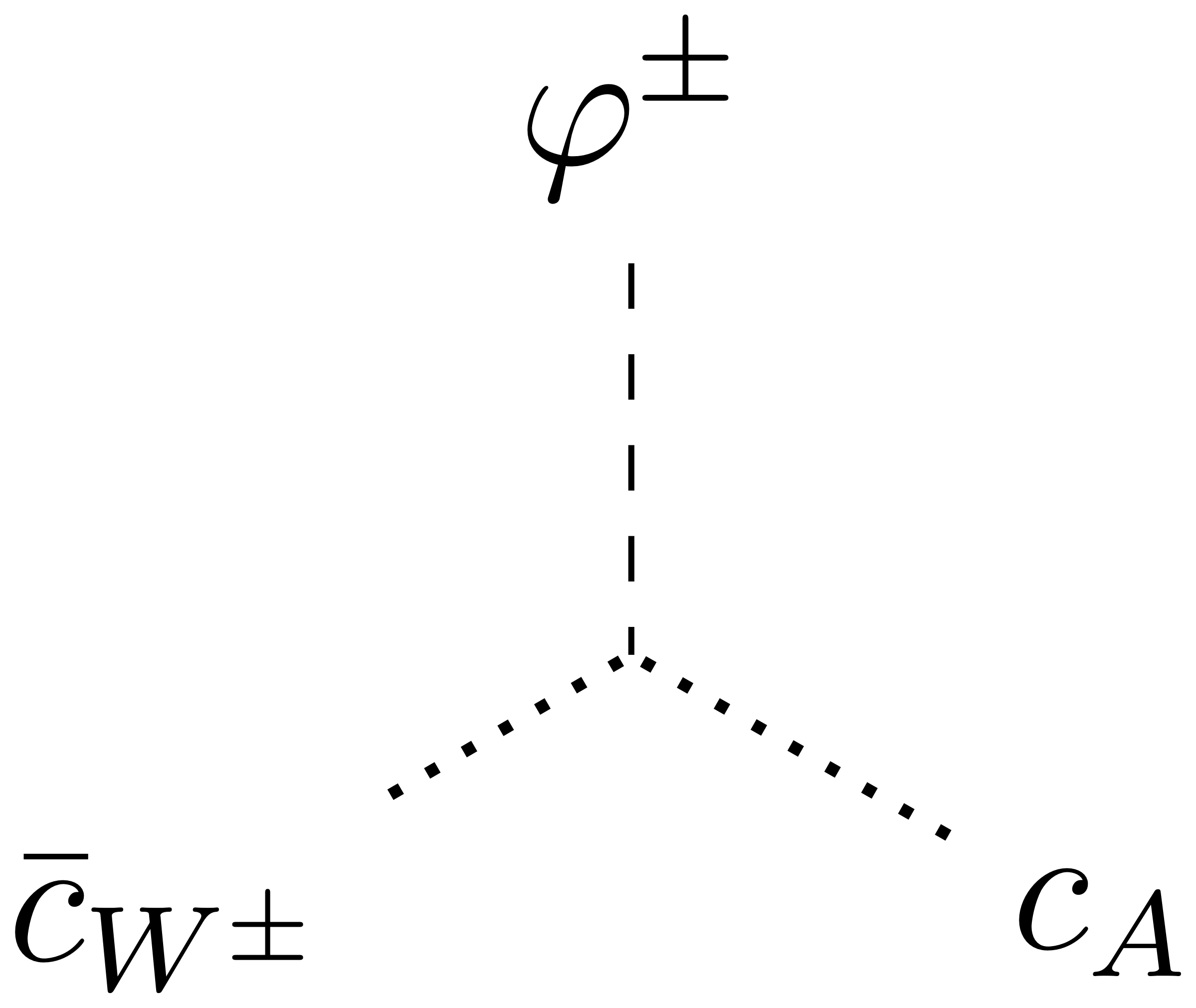}&
        $\mp \ \dfrac{1}{2} \dfrac{e^2}{s_w} v $\\
        \vspace{5 px}
        \includegraphics[height=2.25cm]{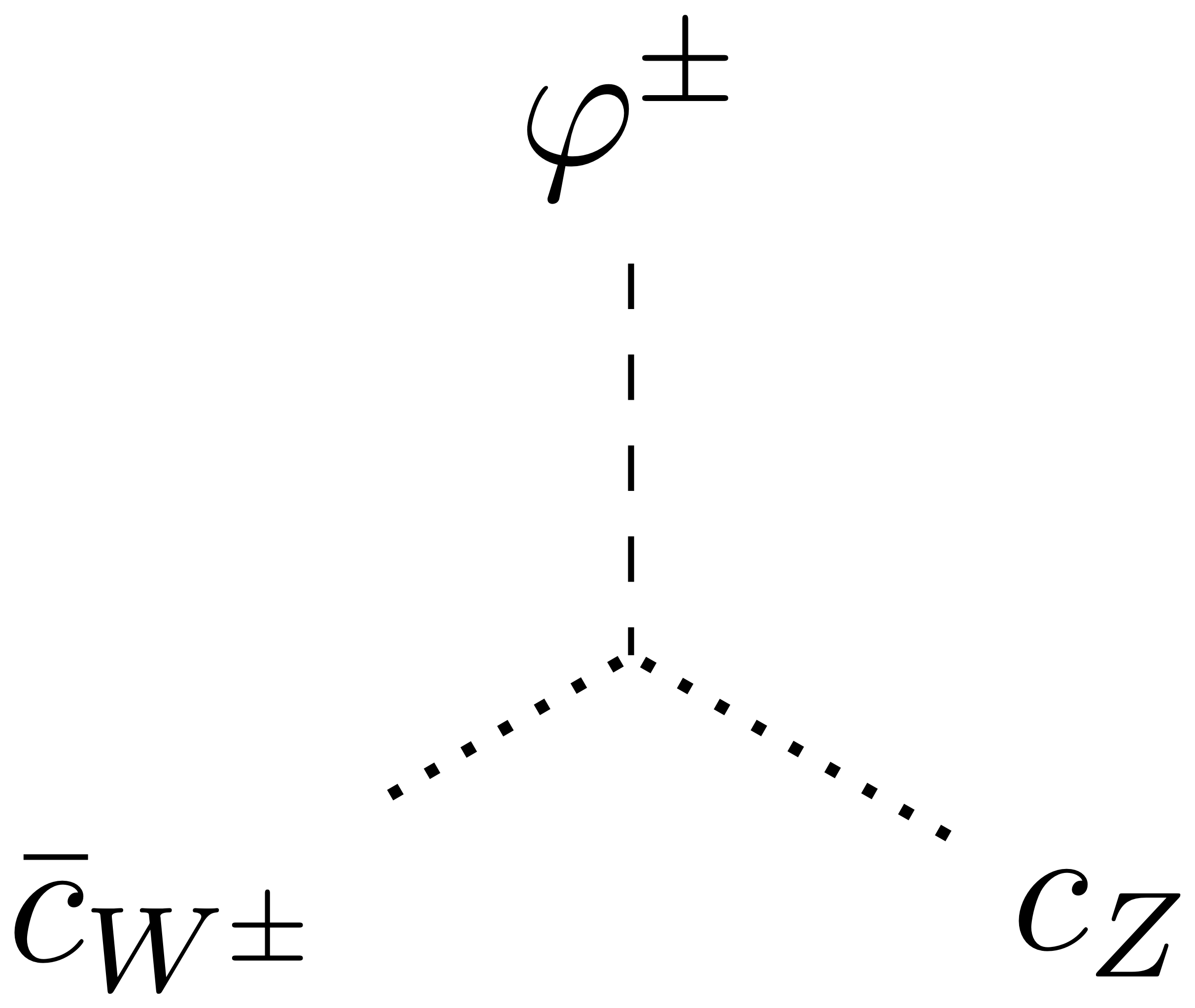}&
        $\mp \ \dfrac{1}{4} \dfrac{e^2}{c_w s_w^2} v \left(c_w^2 - s_w^2 \right)$\\
        \vspace{5 px}
        \includegraphics[height=2.25cm]{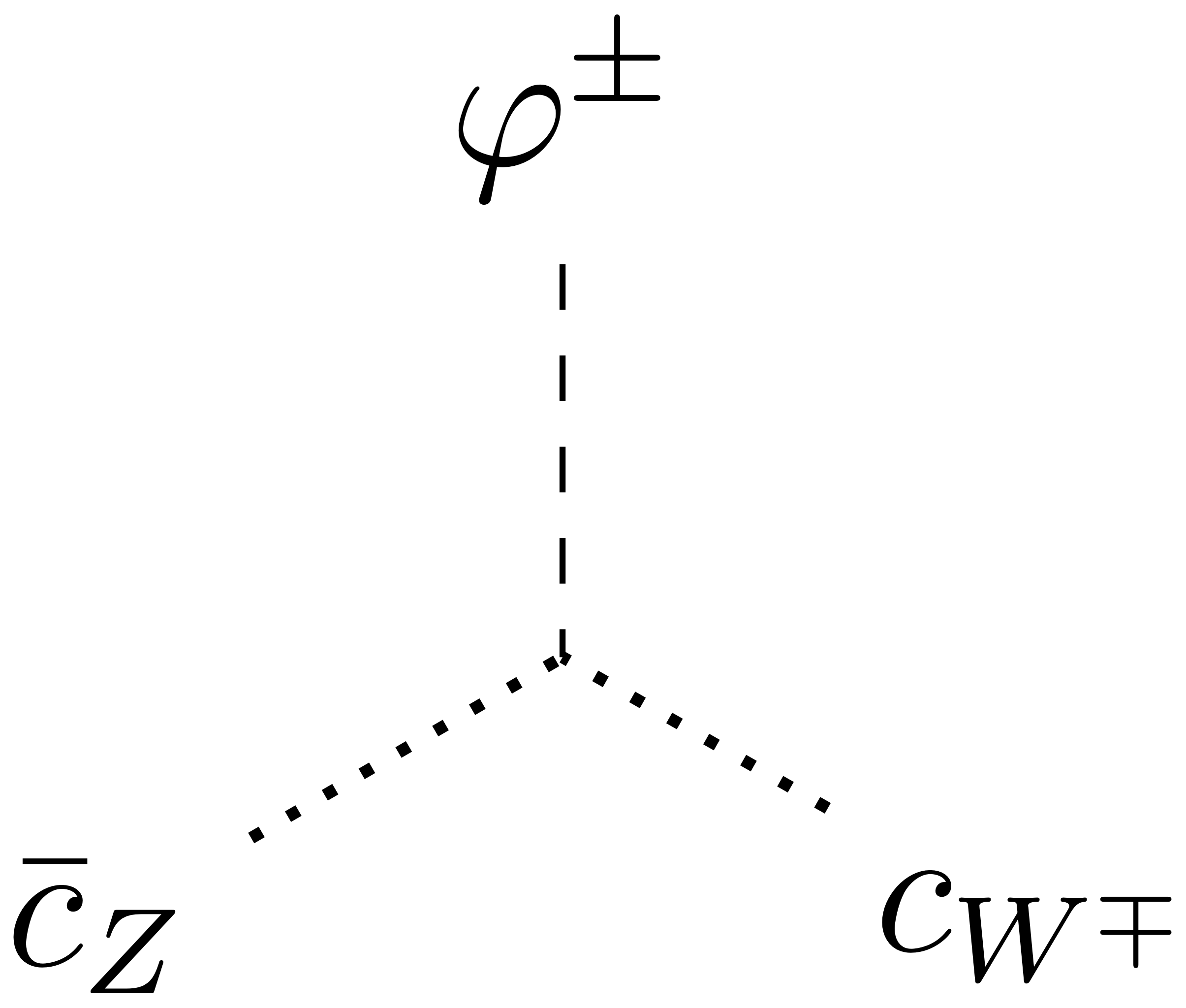}&
        $\pm\ \dfrac{1}{4} \dfrac{e^2}{c_w s_w^2} v $\\
        \hline
        \hline
    \end{tabular}
\end{table}
\end{center}

\clearpage
\section{Scalar LQ Feynman Rules}
\label{sec:LQInteractions}
In this section we state the Feynman rules for all LQ interactions. We write $\Phi^q_a$ with $q \in \{-\frac{4}{3}, -\frac{1}{3}, +\frac{2}{3},+\frac{5}{3} \}$ and $a \in \{1, \tilde{1}, 2, \tilde{2}, 3\}$. For $q = +\frac{5}{3}$, $\hat{\Phi}^{+5/3}$ is just a single field, therefore the subscript $a$ is omitted. The LQ electric charge eigenstate vectors are given in Eq.~(\ref{eq:ChargeVectors}).

\subsection{External Fields}
In this subsection, the Feynman rules for the external legs are listed. The circles denote the generic reminder of the Feynman diagram. 
\begin{center}
\begin{table}[!ht]
    \centering\begin{tabular}{ >{\centering\arraybackslash}m{4.0cm} m{8.5cm}} 
    \multicolumn{2}{c}{External LQs} \\
    \hline
        Diagram & Feynman Rule \\
        \hline
        \vspace{10px}
        \includegraphics[scale=0.025]{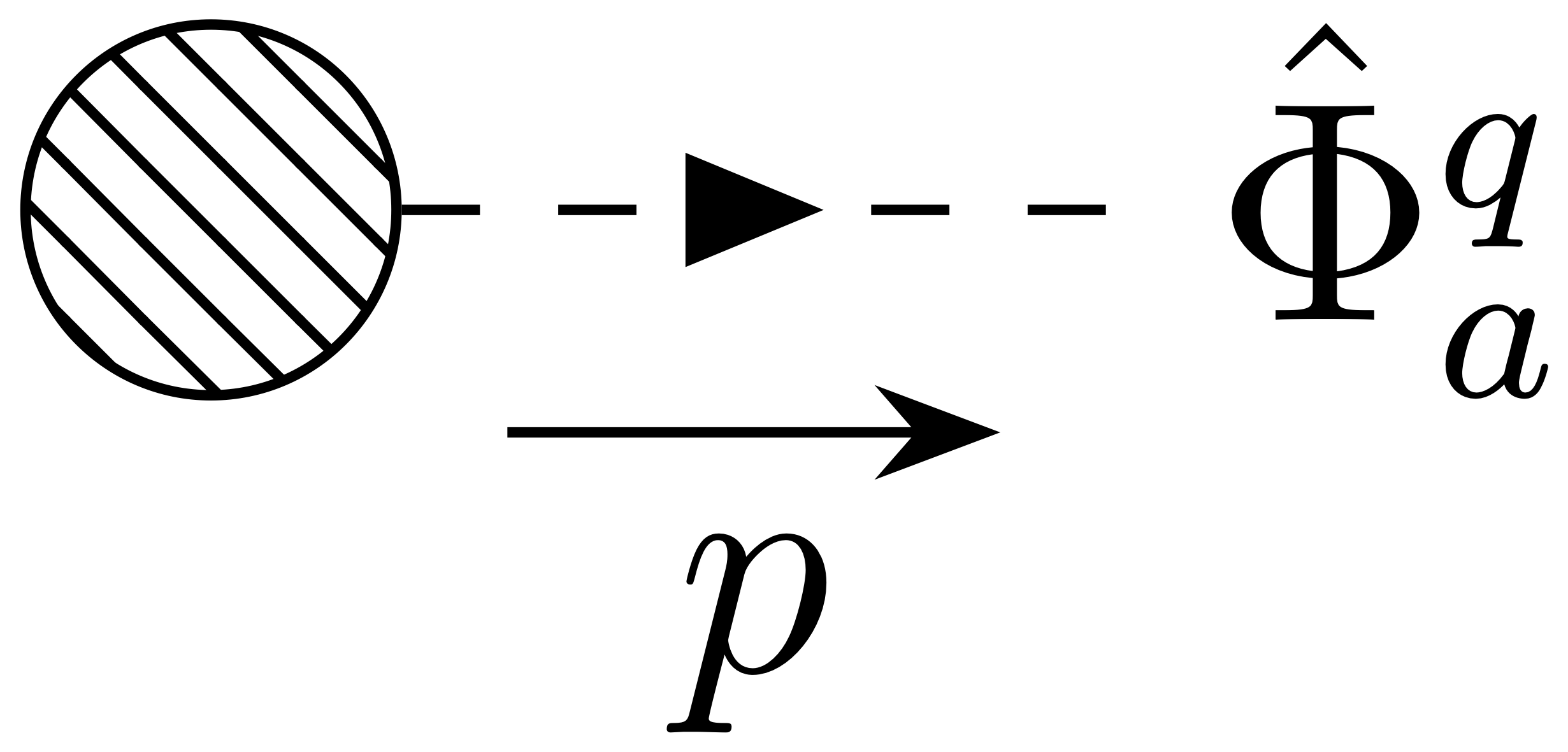} \newline
        \includegraphics[scale=0.025]{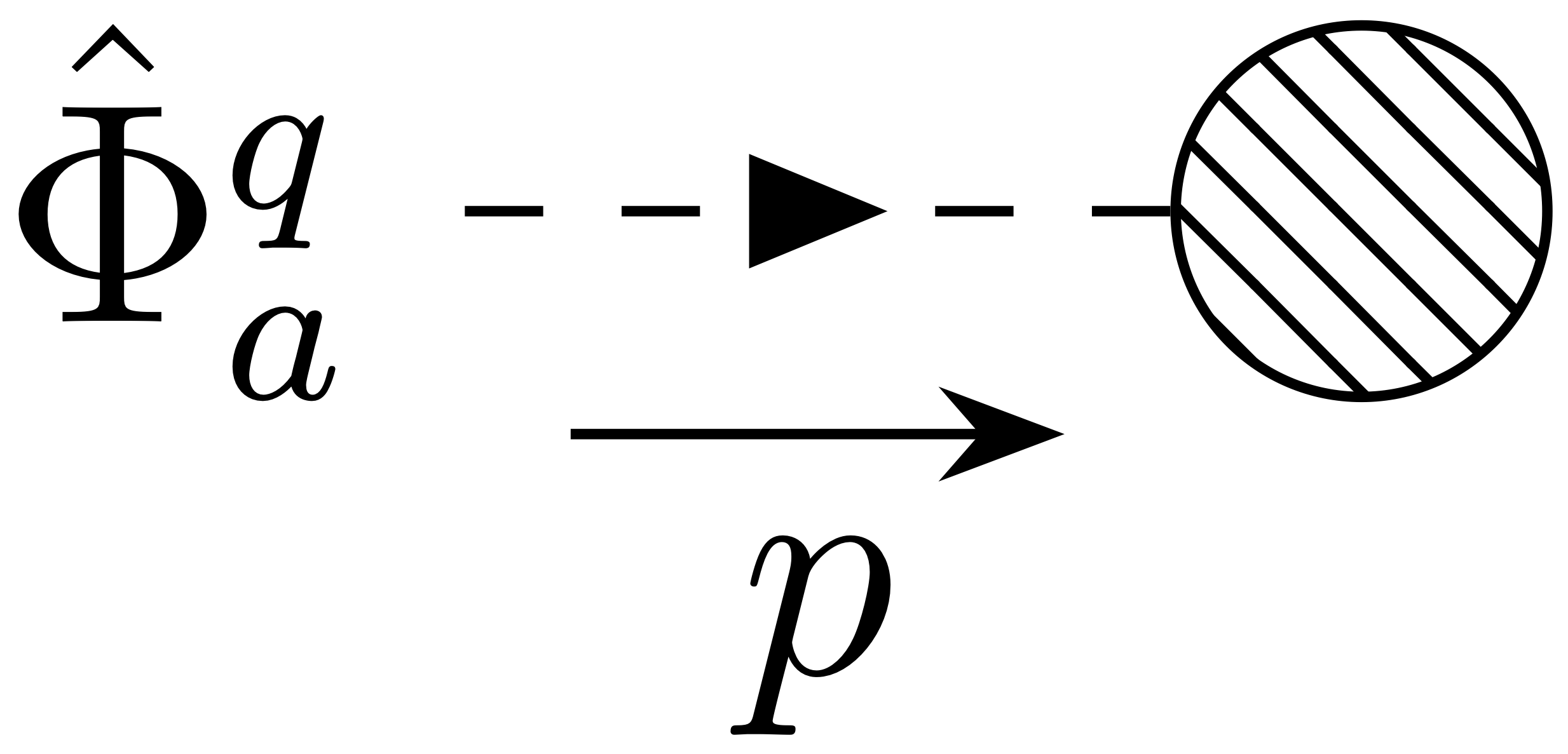} ~~~&
        \vspace{10px}
        1 \newline \\
        \hline
        \hline
    \end{tabular}
\end{table}
\end{center}

\subsection{Propagators}
The propagator for the LQ fields is listed below. The diagonalized mass matrices $\hat{\mathbbm{M}}^{q}_{ab}$ are given in Eq.~(\ref{eq:LQmassMatrices}). 
\begin{center}
\begin{table}[!ht]
    \centering\begin{tabular}{>{\centering\arraybackslash}m{4.0cm} m{8.5cm}} 
    \multicolumn{2}{c}{LQ Propagator} \\
    \hline
        Diagram & Feynman Rule \\
        \hline
        \vspace{10px}
        \includegraphics[scale=0.035]{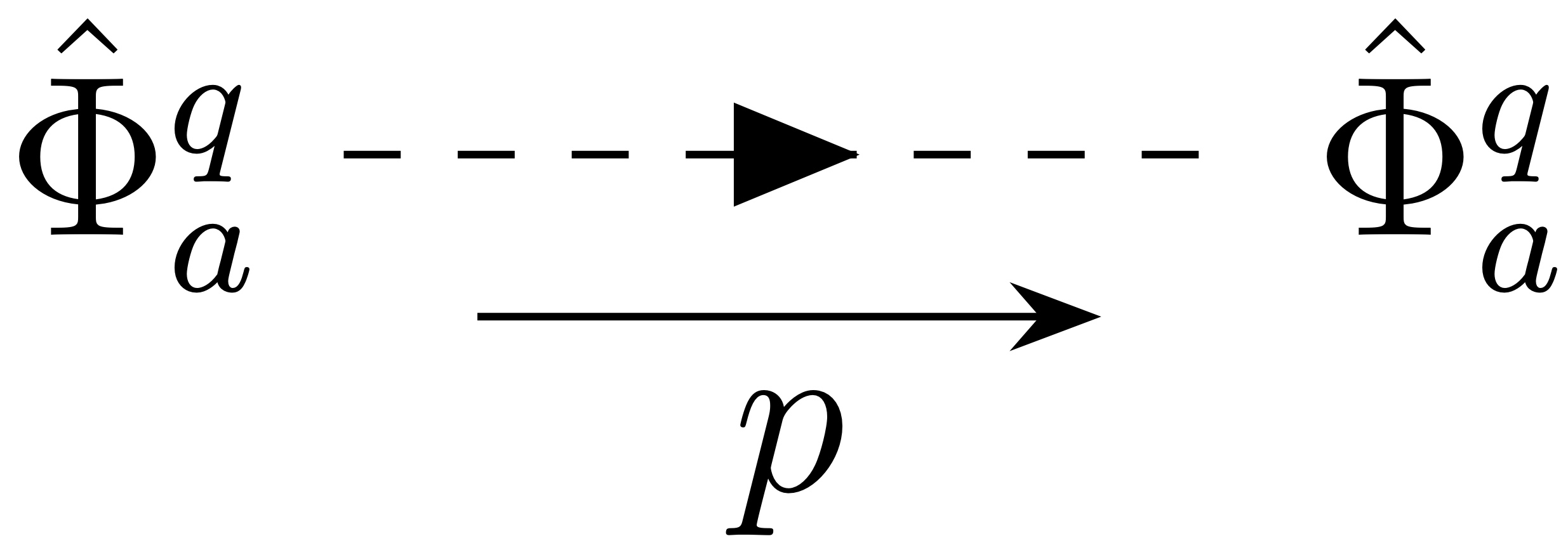}&
        \vspace{10px}
        $\dfrac{i}{p^2 - \left(\hat{\mathbbm{M}}^{q}_{aa}\right)^2 + i \epsilon }$ \newline \\
        \hline
        \hline
    \end{tabular}
\end{table}
\end{center}

\clearpage
\subsection{Interactions with SM Fermions}
\label{sec:LQfermions}
The Feynman rules for LQ interactions with the SM fermions are given below. All fields are defined to be incoming. We omit the color indices whenever they just involve trivial contractions. We express some of the rules using $n_a^q$, which is defined in Eq.~(\ref{eq:nqi}). Whenever $q$ is clear from the context, we write $n_a^q$ as $n_a$ to simplify the notation. The bar labels an antifermion. Note that while there is an equivalence between a fermion field $f$ and $\bar{f}^c$, stating Feynman rules for $\bar{f}^c$ makes sense in our treatment of charge-conjugate fermions. A detailed explanation is given in~\ref{sec:appendix}. 

\subsubsection{Charge 1/3}

\begin{center}
\begin{table}[!ht]
    \centering\begin{tabular}{>{\centering\arraybackslash}m{4cm} m{8.5cm}} 
    \multicolumn{2}{c}{Charge 1/3} \\
    \hline
        Diagram & Feynman Rule \\
        \hline
        \vspace{5px}
        \includegraphics[height=2.25cm]{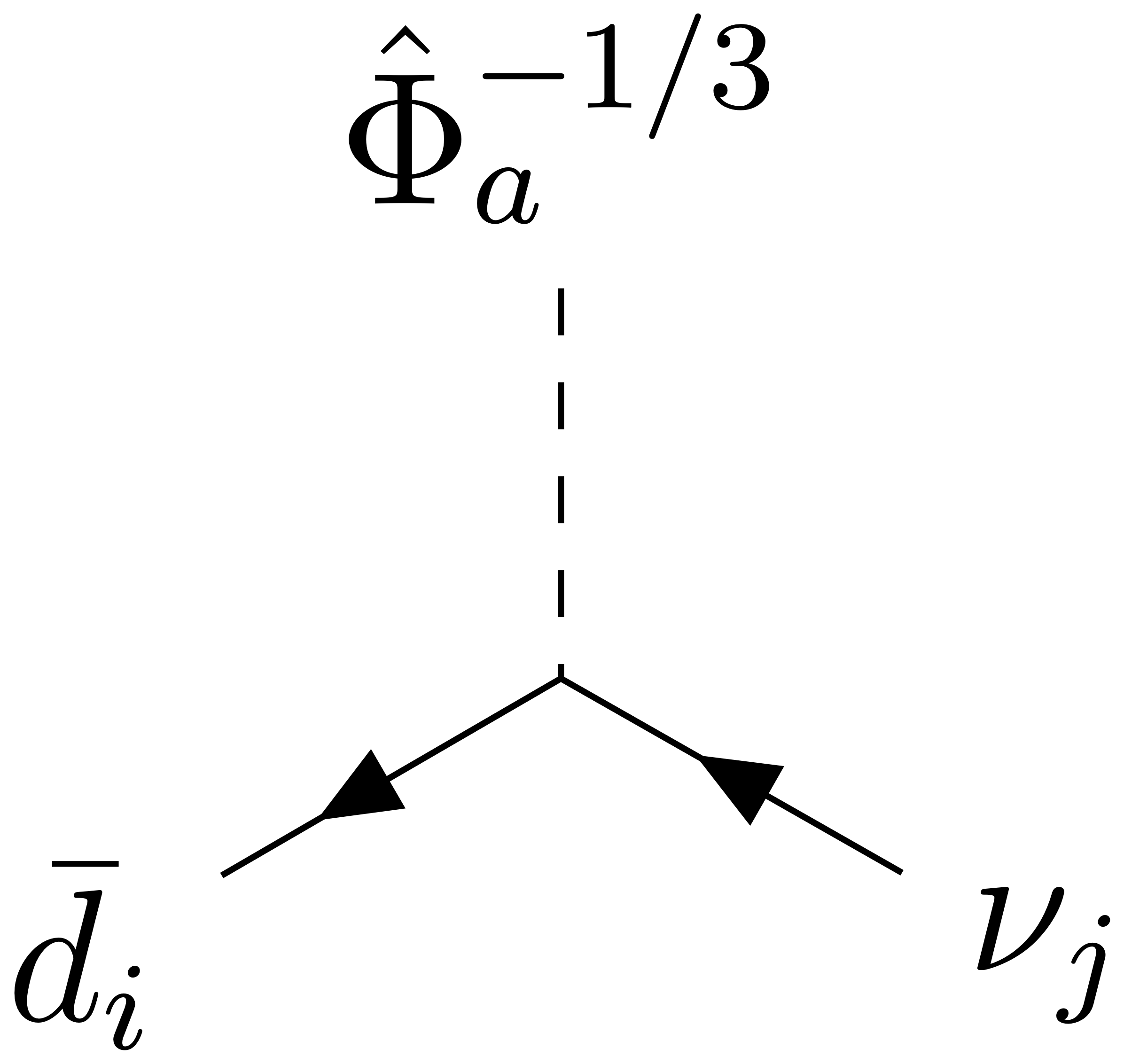}&
        $- \ i Y_{\tilde 2, ij}^{\text{RL}} \ W^{-1/3*}_{n_a2} \ \text{P}_{\text{L}} $  \newline \\
        \vspace{5px}
        \includegraphics[height=2.25cm]{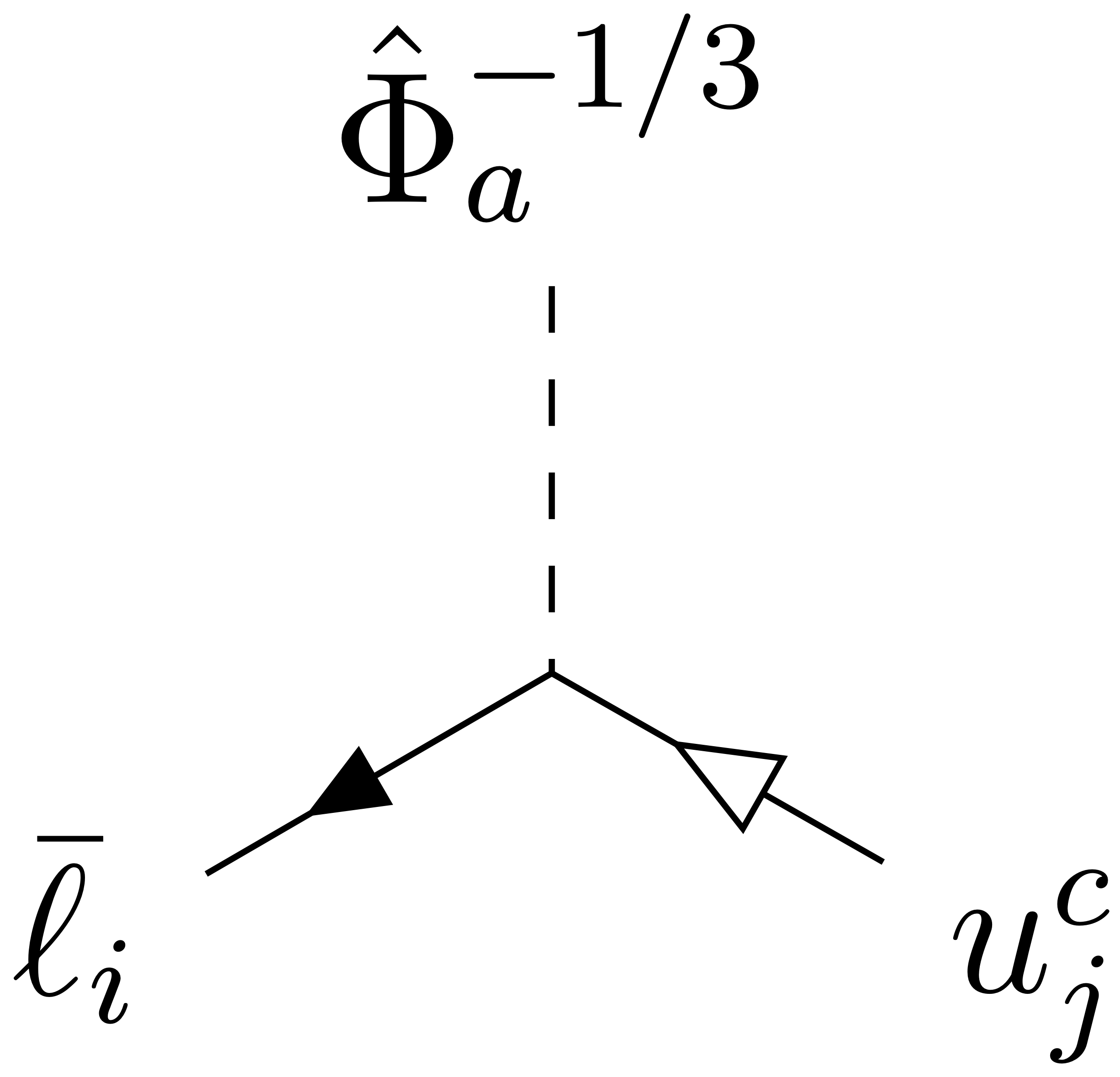}&
        $\begin{aligned}
        &+ \ i   Y_{1, ji}^{\text{LL}*}{} \ W^{-1/3*}_{n_a1} \vspace{5px}\ \text{P}_{\text{R}} \\
        &- \ i  Y_{3, ji}^{\text{LL}*} \ W^{-1/3*}_{n_a3} \vspace{5px} \  \text{P}_{\text{R}} \\
        &+ \  i Y_{1, ji}^{\text{RR}*} \ W^{-1/3*}_{n_a1} \  \text{P}_{\text{L}} \\
        \end{aligned}$  \newline \\
        \vspace{5px}
        \includegraphics[height=2.25cm]{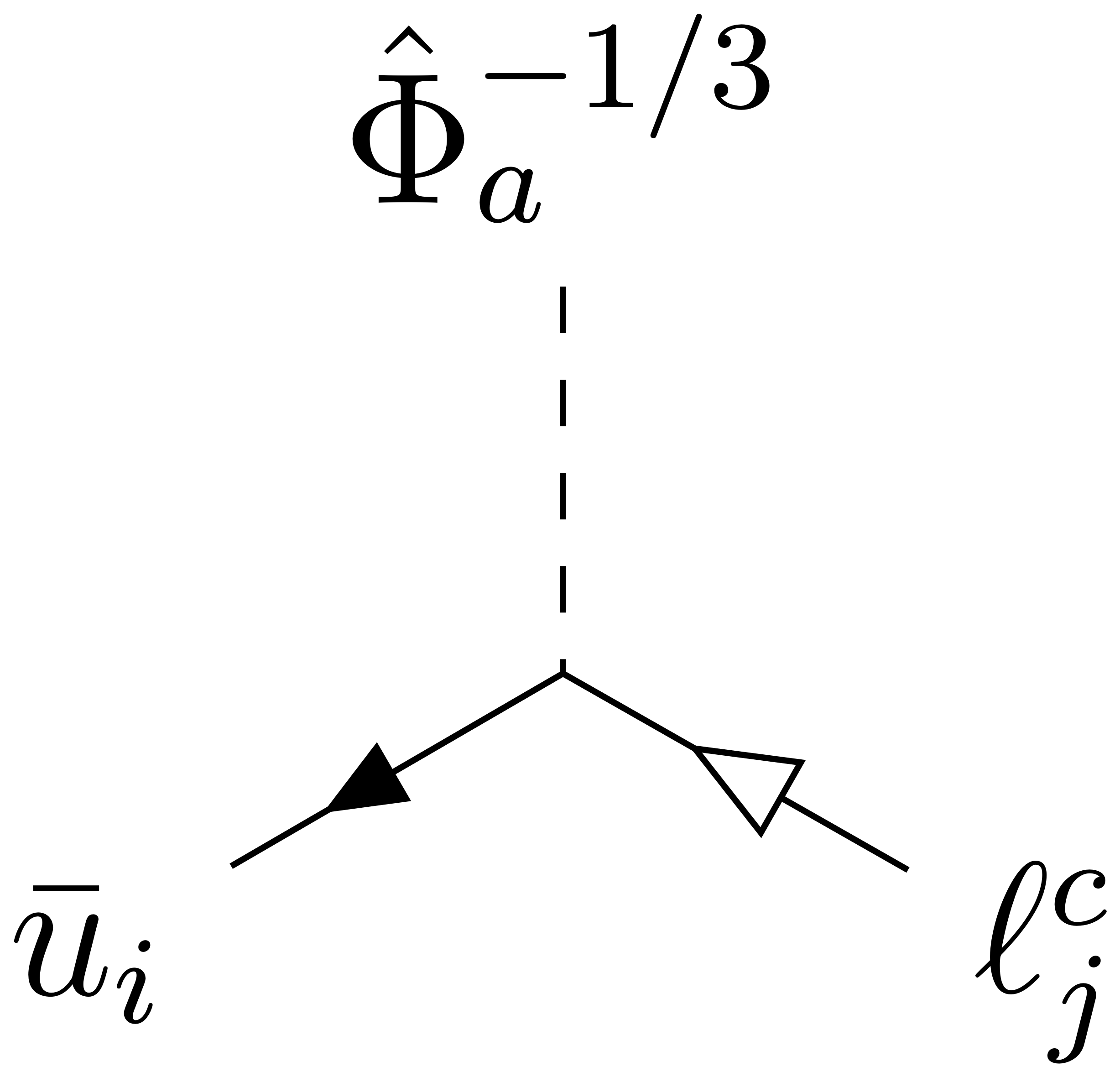}&
        $\begin{aligned}
        &+ \ i   Y_{1, ij}^{\text{LL}*}{} \ W^{-1/3*}_{n_a1} \vspace{5px}\ \text{P}_{\text{R}} \\
        &- \ i  Y_{3, ij}^{\text{LL}*} \ W^{-1/3*}_{n_a3} \vspace{5px} \  \text{P}_{\text{R}} \\
        &+ \ i Y_{1, ij}^{\text{RR}*} \ W^{-1/3*}_{n_a1} \  \text{P}_{\text{L}} \\
        \end{aligned}$  \newline \\
        \vspace{5px}
        \includegraphics[height=2.25cm]{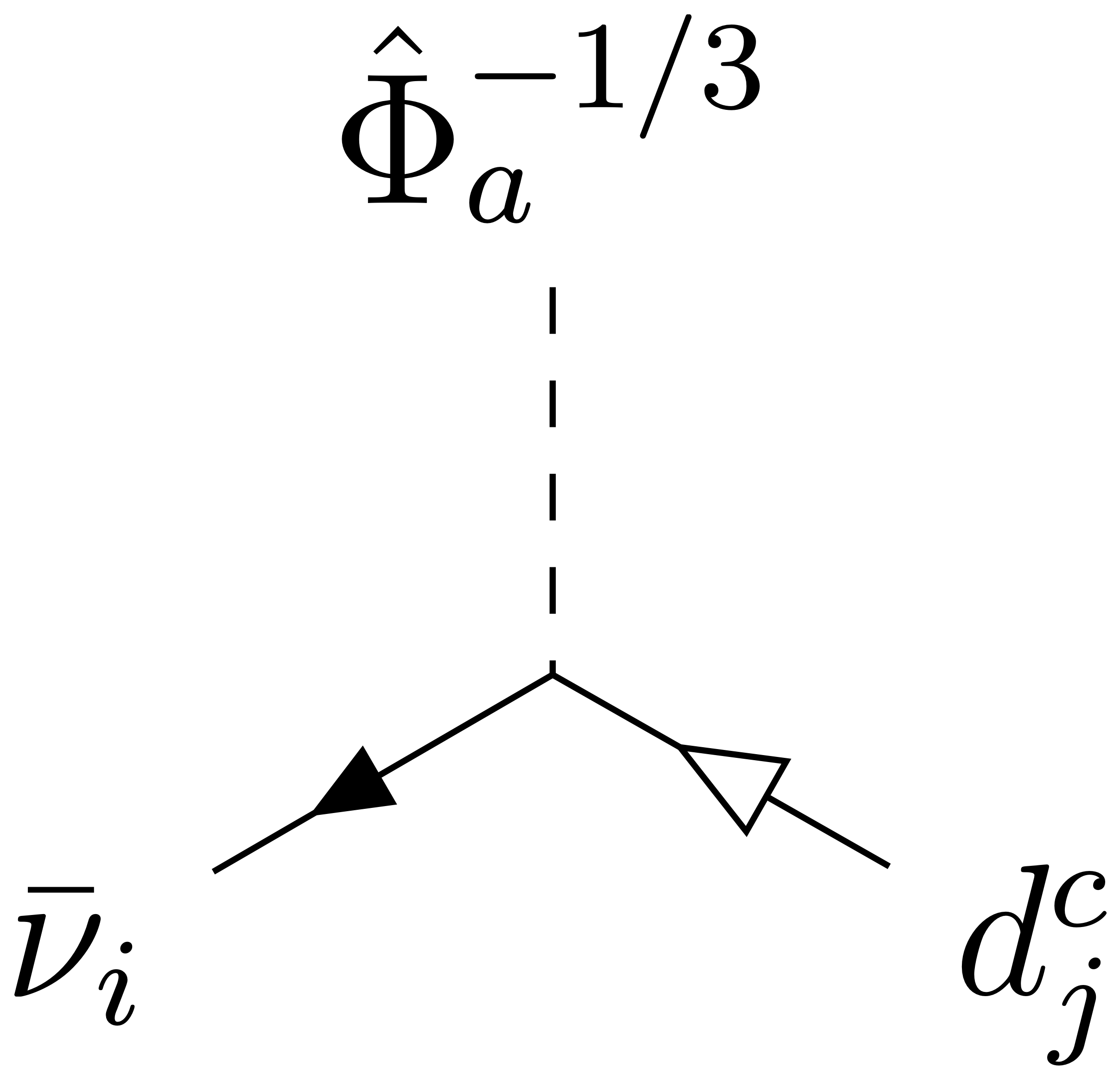}&
        $\begin{aligned}
        &- \ iV^*_{kj}  \ Y_{1, ki}^{\text{LL}*} \ W^{-1/3*}_{n_a1} \ \text{P}_{\text{R}} \\
        &- \ i V^*_{kj} \ Y_{3, ki}^{\text{LL}*} \ W^{-1/3*}_{n_a3} \ \text{P}_{\text{R}} 
        \end{aligned}$  \newline  \\
        \hline
    \end{tabular}
\end{table}
\end{center}

\begin{center}
\begin{table}[!ht]
    \centering\begin{tabular}{>{\centering\arraybackslash}m{4cm} m{8.5cm}} 
    \multicolumn{2}{c}{Charge 1/3} \\
    \hline
        Diagram & Feynman Rule \\
        \hline
        \vspace{5px}
        \includegraphics[height=2.25cm]{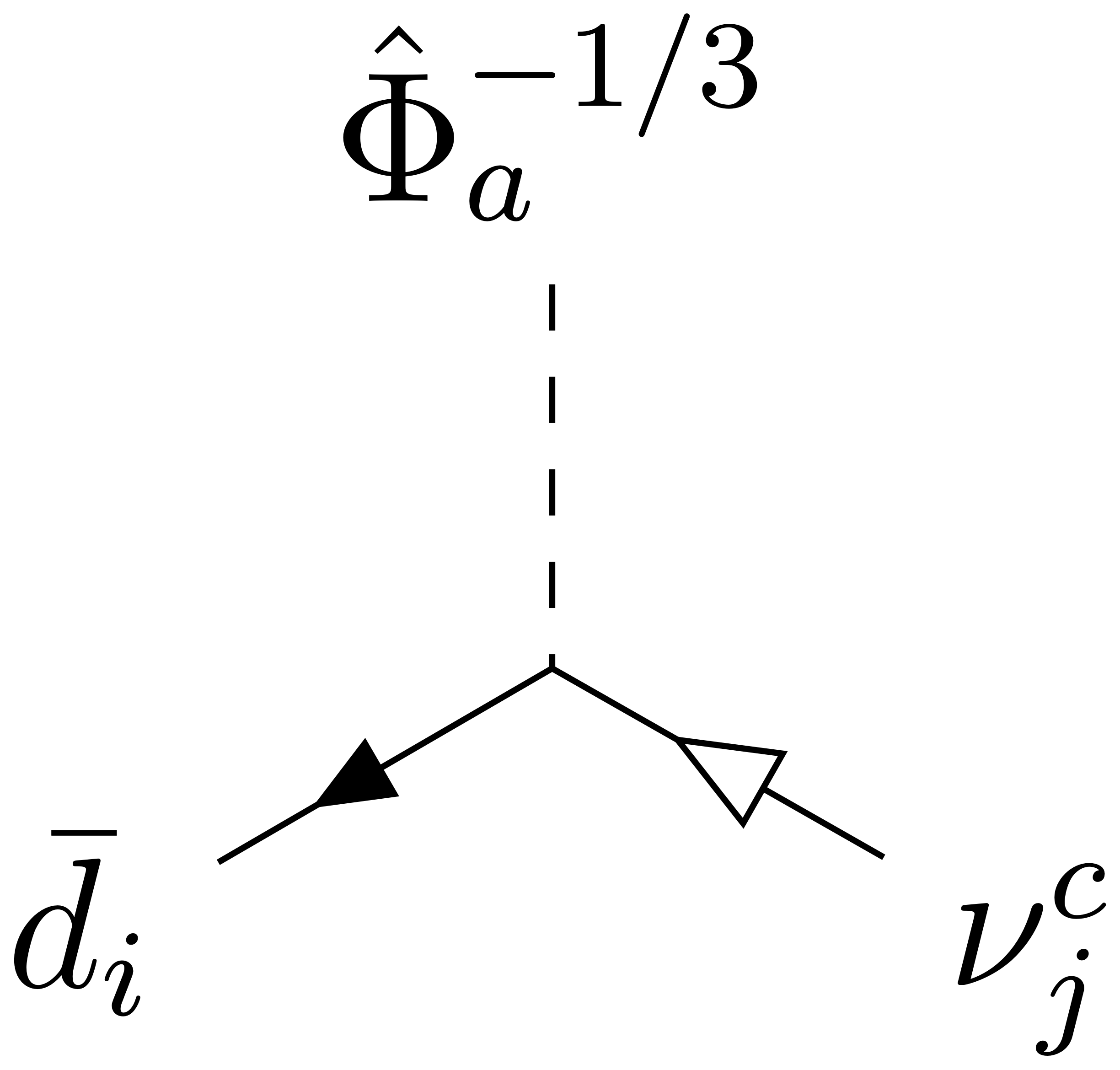}&
        $\begin{aligned}
        &- \ iV^*_{ki}  \ Y_{1, kj}^{\text{LL}*} \ W^{-1/3*}_{n_a1} \ \text{P}_{\text{R}} \\
        &- \ i V^*_{ki} \ Y_{3, kj}^{\text{LL}*} \ W^{-1/3*}_{n_a3} \ \text{P}_{\text{R}} 
        \end{aligned}$  \newline  \\
        \vspace{5px}
        \includegraphics[height=2.25cm]{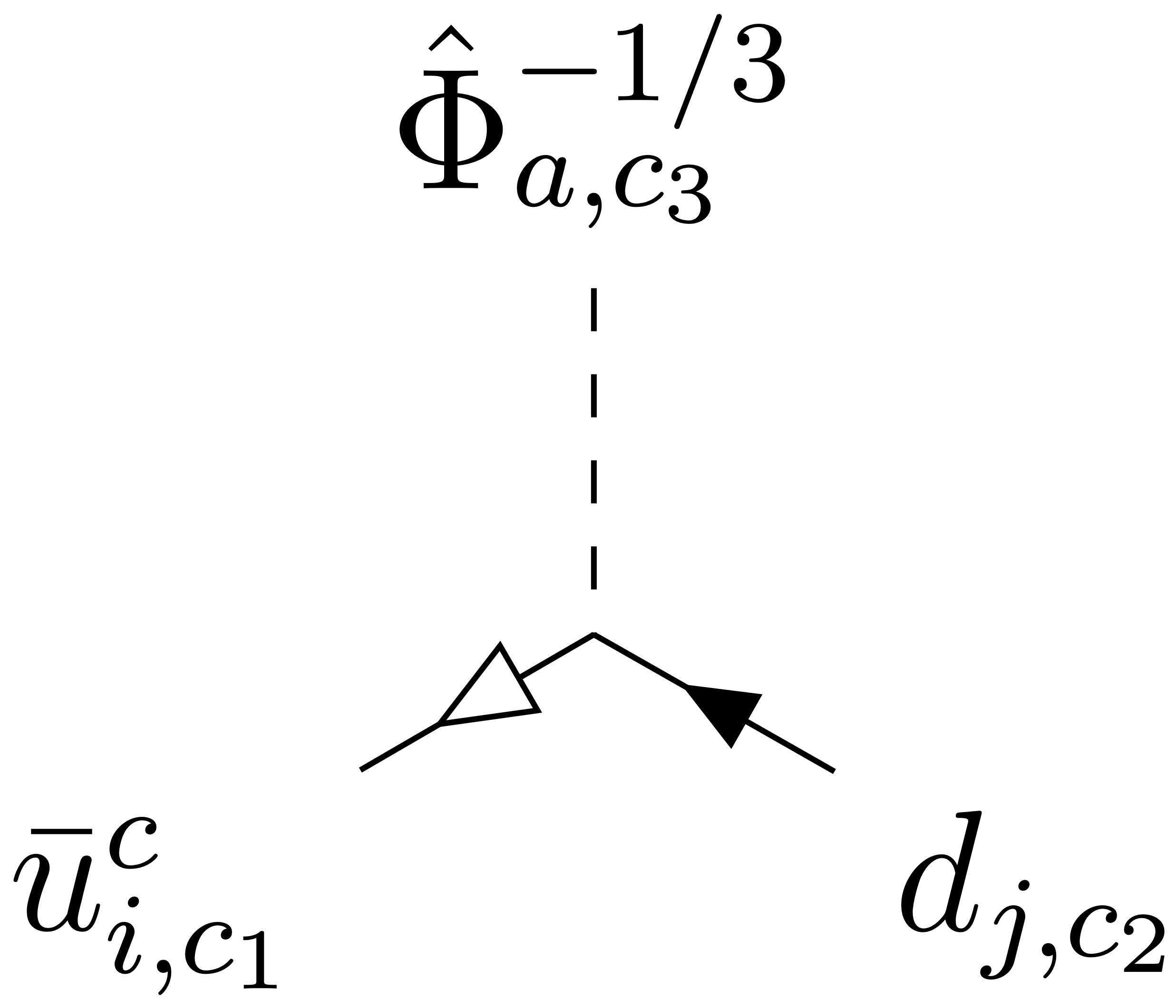}&
        \vspace{10px}
        $\begin{aligned}
        &+\ 2i V_{kj} \ Y_{1, ik}^{\text{Q, LL}} \ W^{-1/3*}_{n_a1} \epsilon^{c_1 c_2 c_3}\ \text{P}_\text{L} \\
        &+\ i Y_{1, ij}^{\text{Q, RR}} \ W^{-1/3*}_{n_a1} \epsilon^{c_1 c_2 c_3}\ \text{P}_\text{R} \\
        &-\ 2i V_{kj} \ Y_{3, ik}^{\text{Q, LL}} \ W^{-1/3*}_{n_a3} \epsilon^{c_1 c_2 c_3}\ \text{P}_\text{L}
        \end{aligned}$  \newline  \\
        \vspace{5px}
        \includegraphics[height=2.25cm]{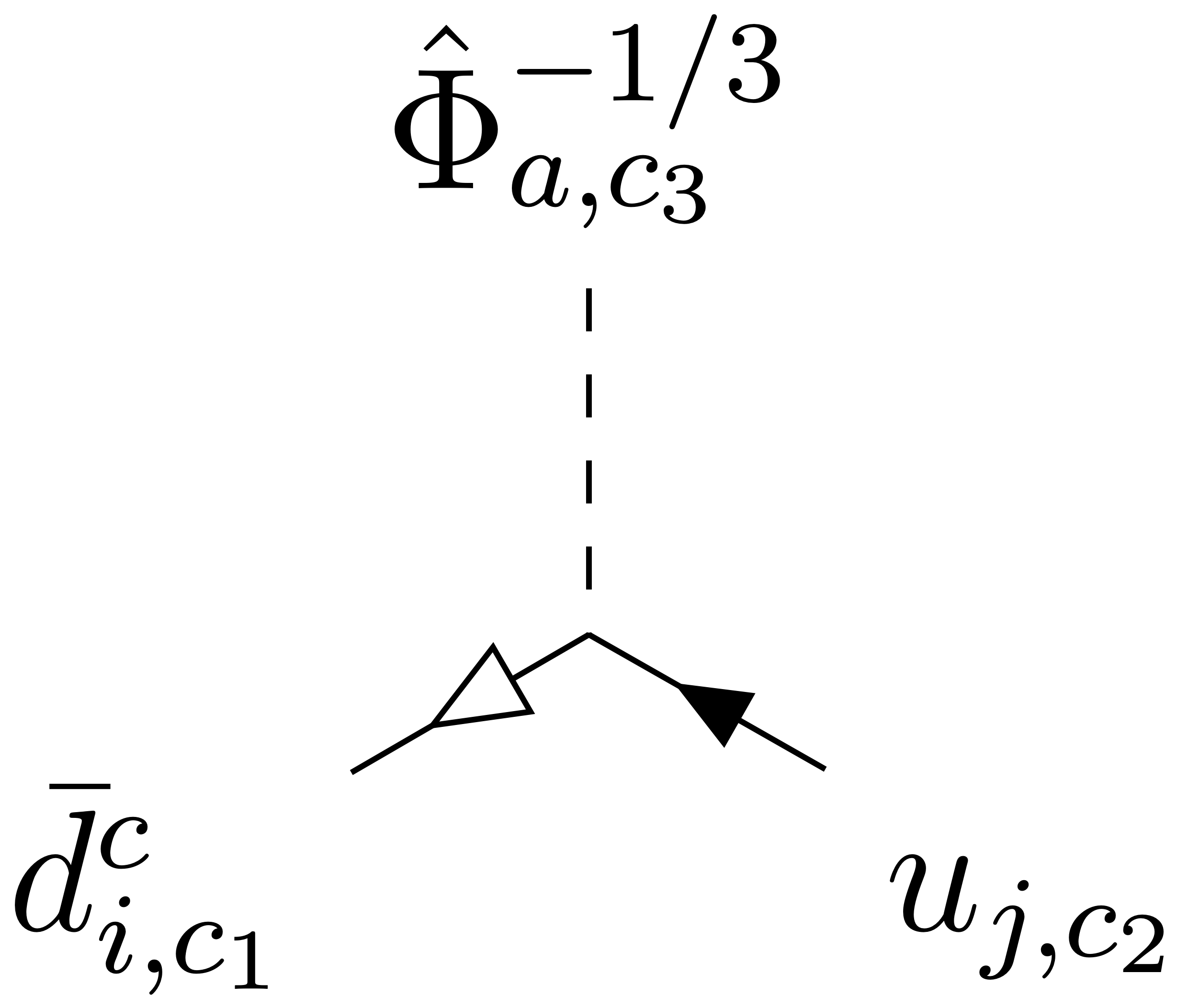}&
        \vspace{10px}
        $\begin{aligned}
        &-\ 2i V_{ki} \ Y_{1, jk}^{\text{Q, LL}} \ W^{-1/3*}_{n_a1} \epsilon^{c_1 c_2 c_3}\ \text{P}_\text{L} \\
        &-\ i Y_{1, ji}^{\text{Q, RR}} \ W^{-1/3*}_{n_a1} \epsilon^{c_1 c_2 c_3}\ \text{P}_\text{R} \\
        &+\ 2i V_{ki} Y_{3, jk}^{\text{Q, LL}} \ W^{-1/3*}_{n_a3} \epsilon^{c_1 c_2 c_3}\ \text{P}_\text{L}
        \end{aligned}$  \newline  \\
        \hline
    \end{tabular}
\end{table}
\end{center}
%
%
\subsubsection{Charge 2/3}

\begin{center}
\begin{table}[!ht]
    \centering\begin{tabular}{>{\centering\arraybackslash}m{4cm} m{8.5cm}} 
        \multicolumn{2}{c}{Charge 2/3} \\
    \hline
        Diagram & Feynman Rule \\
        \hline
        \vspace{5px}
        \includegraphics[height=2.25cm]{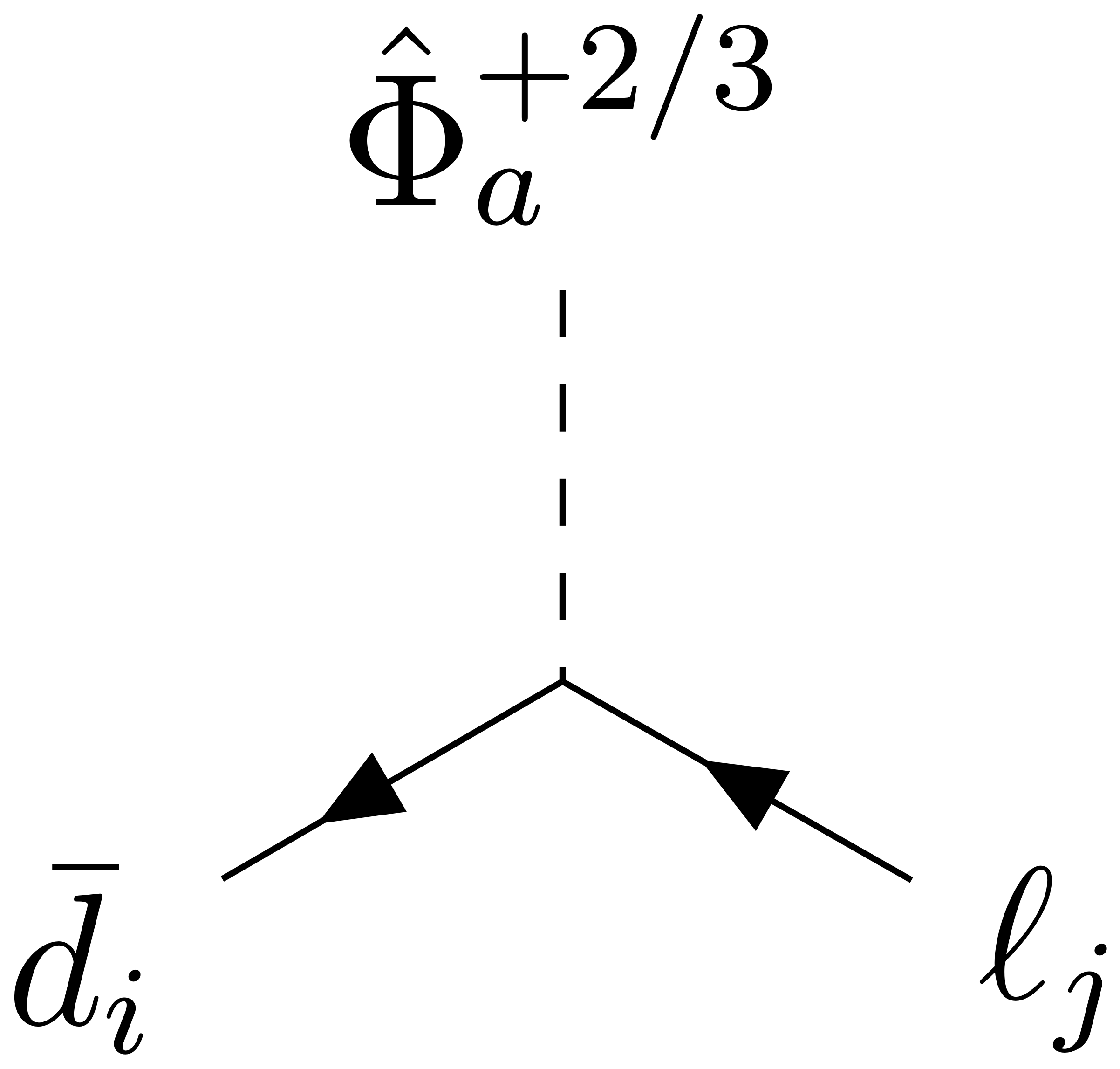}&
        \vspace{5px}
        $\begin{aligned}
        &+ \ i V_{ki}^* \ Y_{2, kj}^{\text{LR}} \ W^{+2/3*}_{n_a1} \ \text{P}_{\text{R}} \\
        &+ \ i  Y_{\tilde 2, ij}^{\text{RL}} \ W^{+2/3*}_{n_a2} \ \text{P}_{\text{L}}
        \end{aligned}$  \newline  \\

        \hline
    \end{tabular}
\end{table}
\end{center}

\begin{center}
\begin{table}[!ht]
    \centering\begin{tabular}{>{\centering\arraybackslash}m{4cm} m{8.5cm}} 
        \multicolumn{2}{c}{Charge 2/3} \\
    \hline
        Diagram & Feynman Rule \\
        \hline
        \vspace{5px}
        \includegraphics[height=2.25cm]{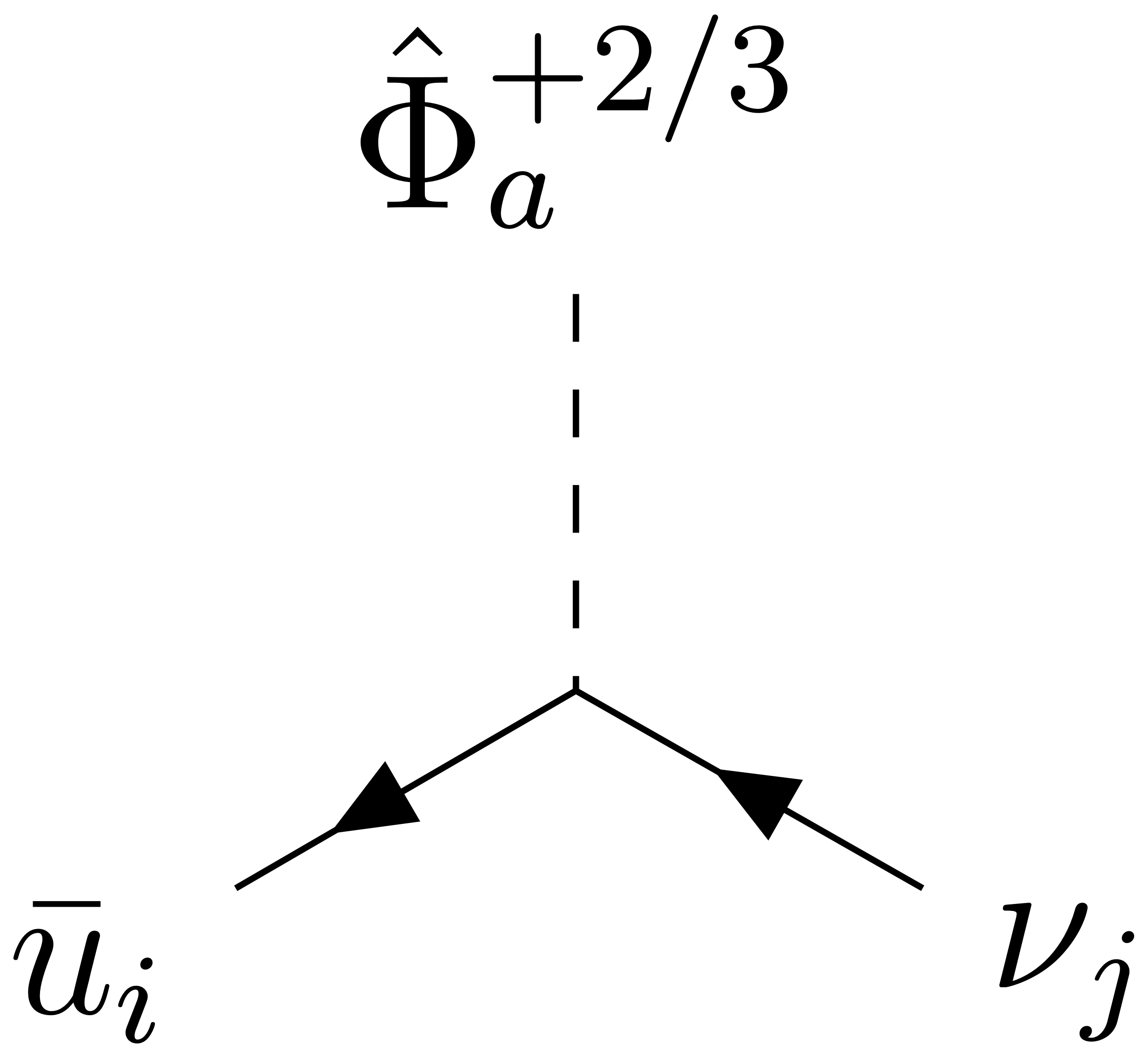}&
        $- \ i  Y_{2, ij}^{\text{RL}} \ W^{+2/3*}_{n_a1} \ \text{P}_{\text{L}}  $  \newline  \\
        \vspace{5px}
        \includegraphics[height=2.25cm]{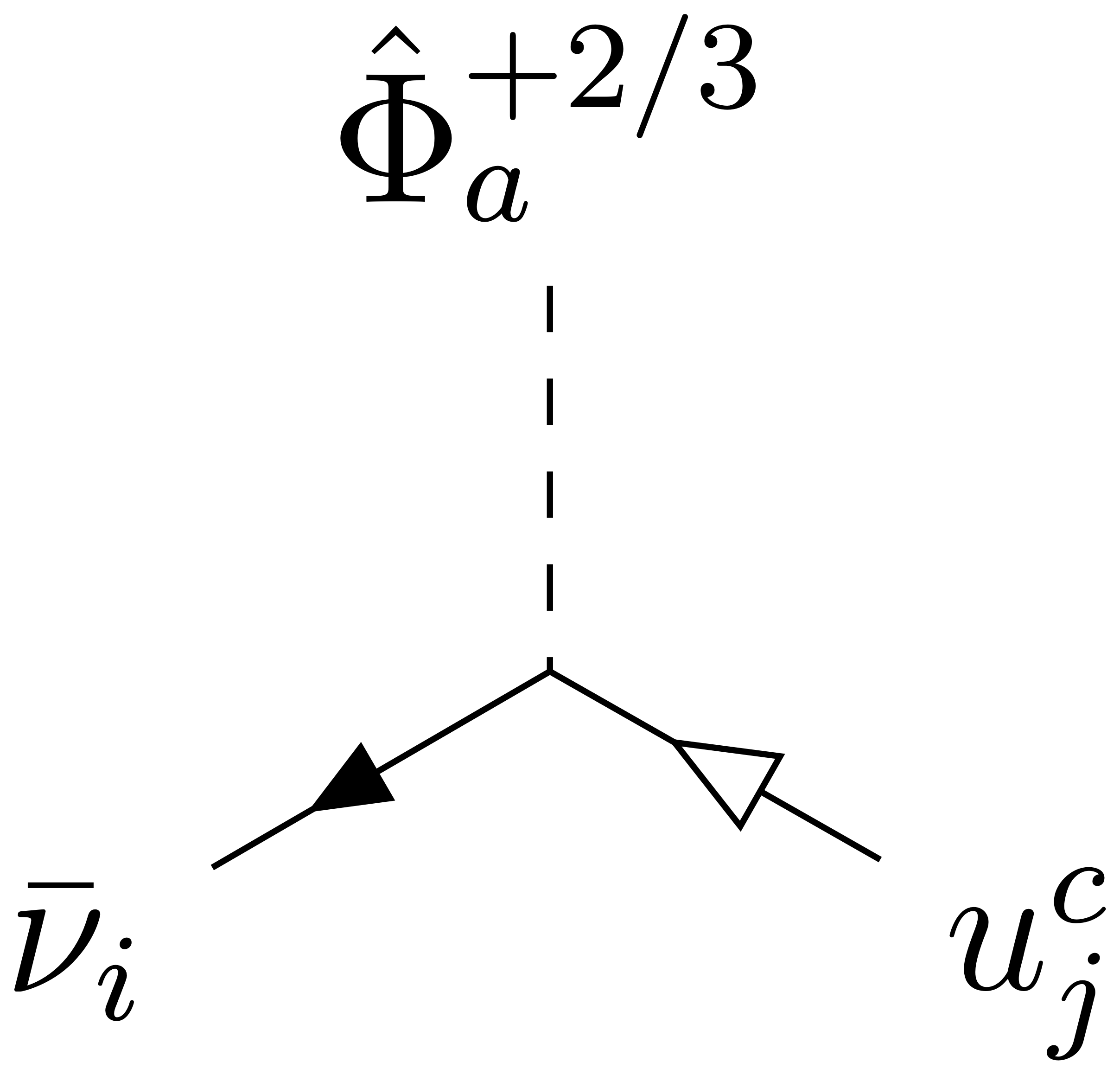}&
        $+ \ \sqrt{2} i  Y_{3, ji}^{\text{LL}*} \ W^{+2/3*}_{n_a3} \ \text{P}_{\text{R}}$   \newline  \\
        \vspace{5px}
        \includegraphics[height=2.25cm]{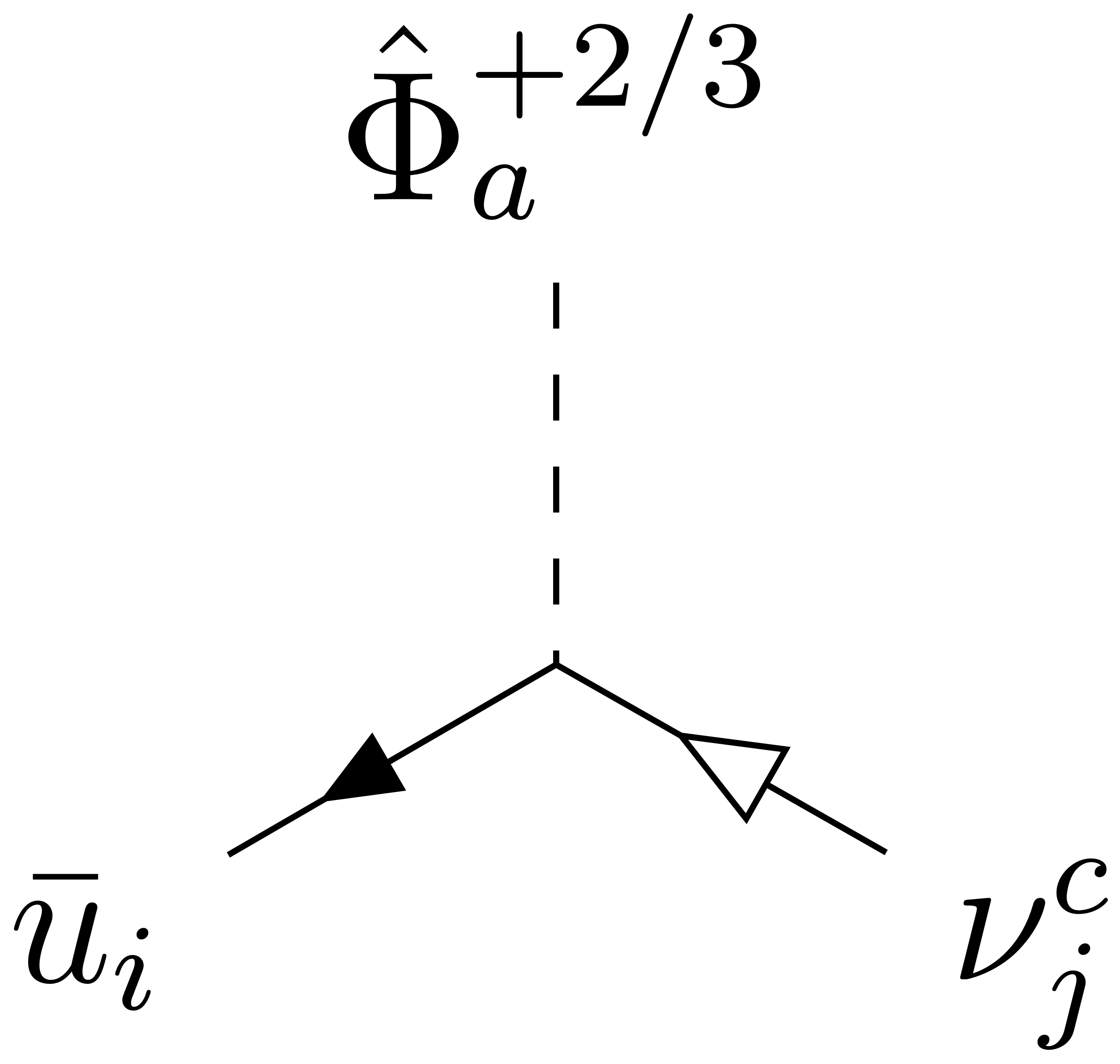}&
        $+ \ \sqrt{2} i  Y_{3, ij}^{\text{LL}*} \ W^{+2/3*}_{n_a3} \ \text{P}_{\text{R}}$   \newline  \\
        \vspace{5px}
        \includegraphics[height=2.25cm]{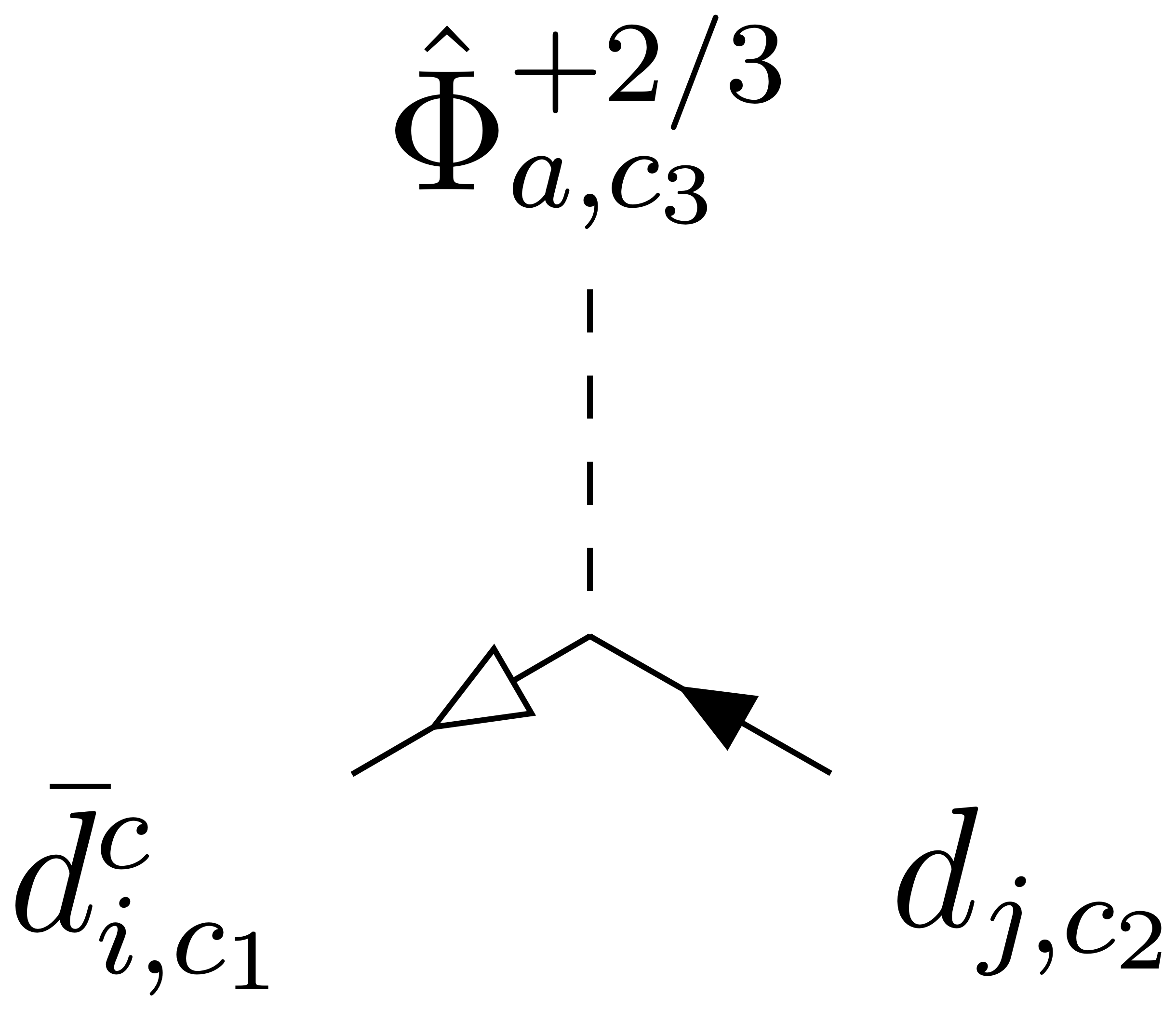}&
        $- \ 2\sqrt{2} i V_{ki} V_{lj}  \ Y_{3, kl}^{\text{Q, LL}}  \ W^{+2/3*}_{n_a3} \epsilon^{c_1 c_2 c_3} \ \text{P}_{\text{L}}$  \newline   \\
        \hline
    \end{tabular}
\end{table}
\end{center}

\clearpage
\subsubsection{Charge 4/3}


\begin{center}
\begin{table}[!ht]
    \centering\begin{tabular}{>{\centering\arraybackslash}m{4cm} m{8.5cm}} 
    \multicolumn{2}{c}{Charge 4/3} \\
    \hline
        Diagram& Feynman Rule \\
        \hline
        \vspace{5px}
        \includegraphics[height=2.25cm]{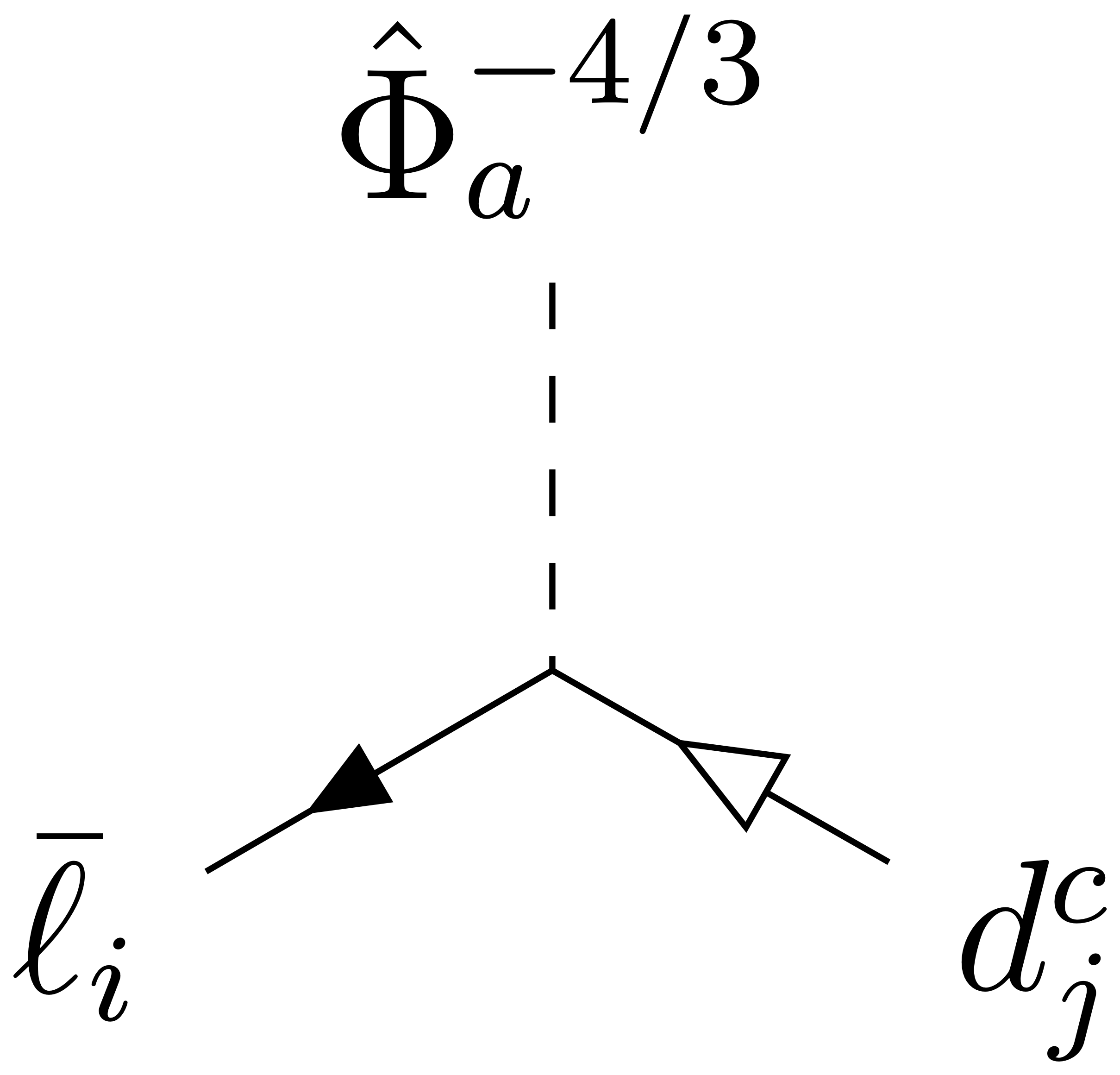}&
        \vspace{5px}
        $\begin{aligned}
        &+ \ i  Y_{\tilde 1, ji}^{\text{RR}*} \ W^{-4/3*}_{n_a1}  \vspace{5px} \ \text{P}_{\text{L}} \\
        &- \ \sqrt{2} i  V^*_{kj} \ Y_{3, ki}^{\text{LL}*} \ W^{-4/3*}_{n_a2}  \ \text{P}_{\text{R}}
        \end{aligned}$ \newline   \\
        \vspace{5px}
        \includegraphics[height=2.25cm]{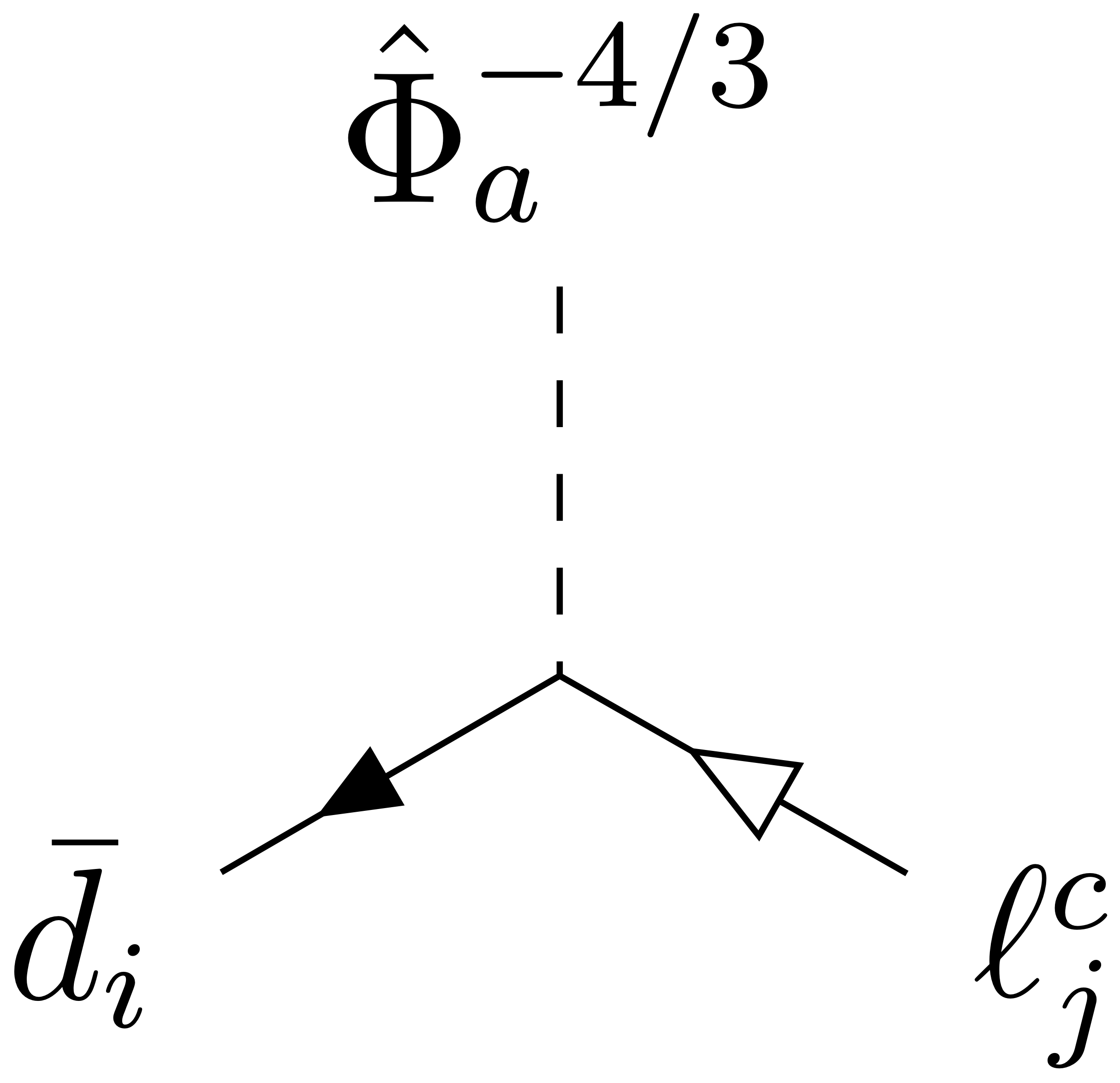}&
        \vspace{5px}
        $\begin{aligned}
        &+ \ i  Y_{\tilde 1, ij}^{\text{RR}*} \ W^{-4/3*}_{n_a1}  \vspace{5px} \ \text{P}_{\text{L}} \\
        &- \ \sqrt{2} i  V^*_{ki} \ Y_{3, kj}^{\text{LL}*} \ W^{-4/3*}_{n_a2}  \ \text{P}_{\text{R}}
        \end{aligned}$ \newline   \\
        \vspace{5px}
        \includegraphics[height=2.25cm]{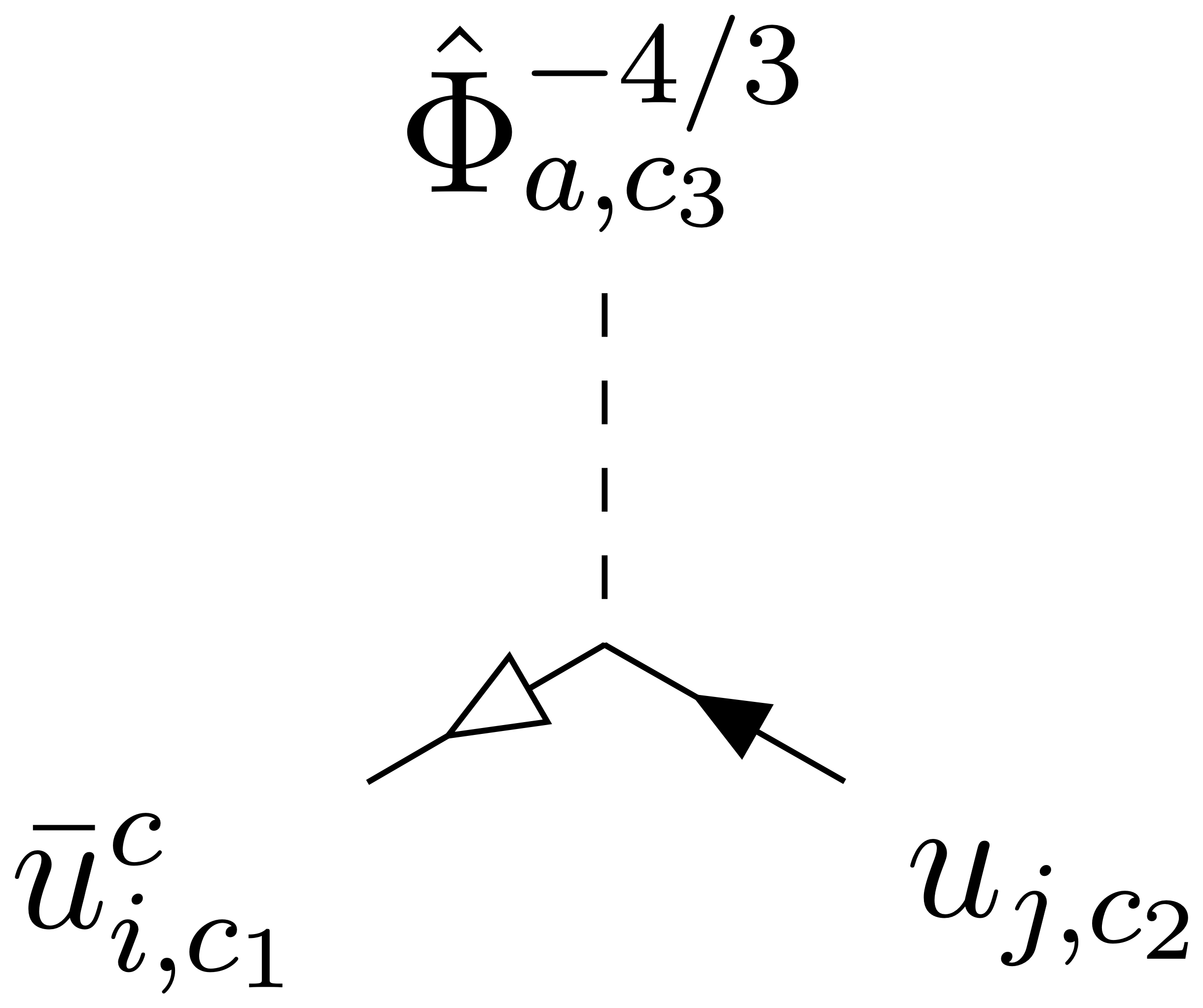}&
        \vspace{5px}
        $\begin{aligned}
        &+ \ 2i  Y_{\tilde{1}, ij}^{\text{Q, RR}} \ W^{-4/3*}_{n_a1}  \vspace{5px} \epsilon^{c_1 c_2 c_3}\ \text{P}_{\text{R}} \\
        &+ \ 2\sqrt{2}i  Y_{3, ij}^{\text{Q, LL}} \ W^{-4/3*}_{n_a2}  \vspace{5px} \epsilon^{c_1 c_2 c_3}\ \text{P}_{\text{L}}
        \end{aligned}$ \newline  \\

        \hline
    \end{tabular}
\end{table}
\end{center}

\subsubsection{Charge 5/3}

\begin{center}
\begin{table}[!ht]
    \centering\begin{tabular}{>{\centering\arraybackslash}m{4cm} m{8.5cm}} 
    \multicolumn{2}{c}{Charge 5/3} \\
    \hline
        Diagram & Feynman Rule \\
        \hline
        \vspace{5px}
        \includegraphics[height=2.25cm]{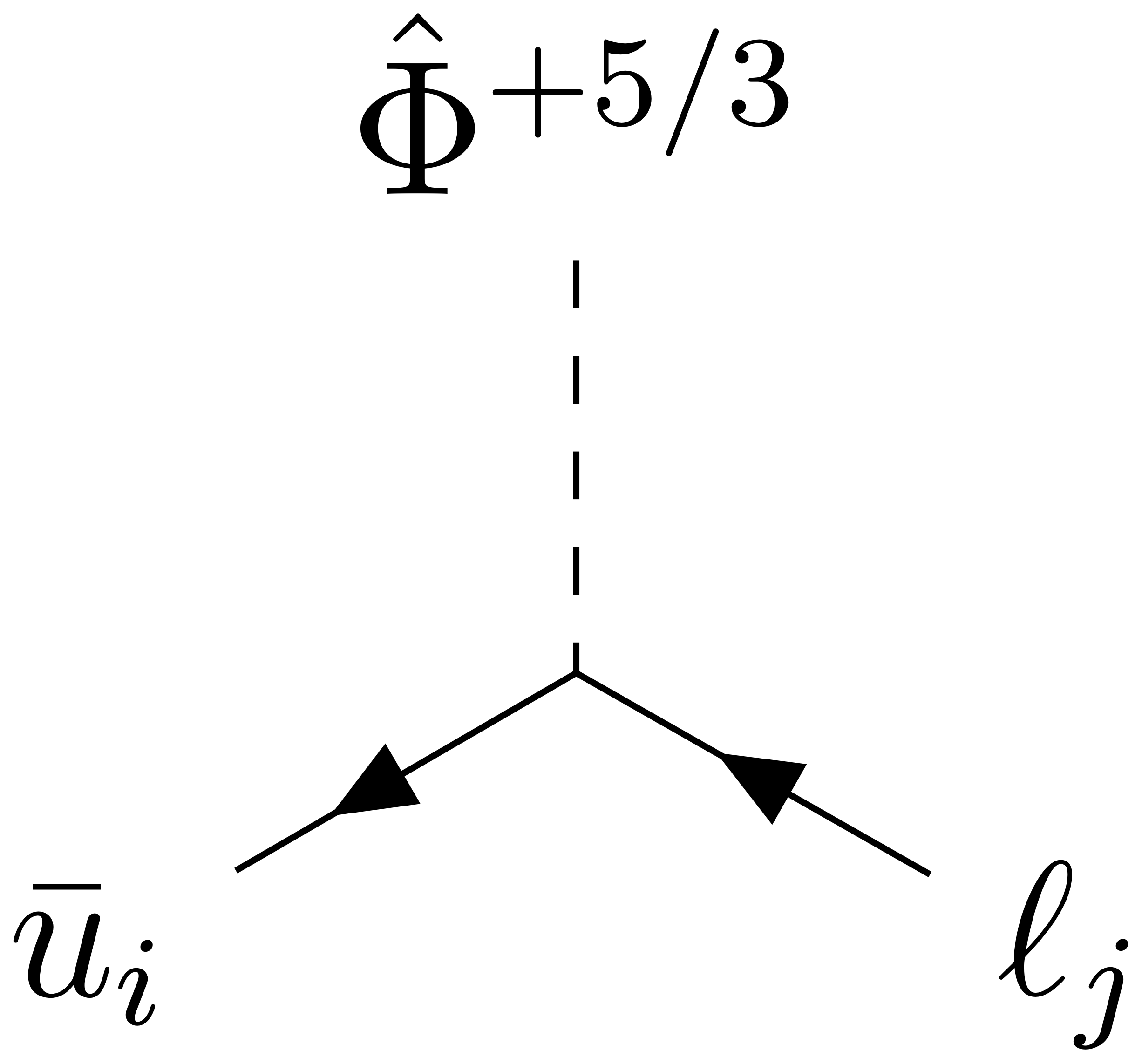}&
        \vspace{5px}
        $\begin{aligned}
        &+ \ i Y_{2, ij}^{\text{LR}} \vspace{5px} \ \text{P}_{\text{R}} \\
        &+ \ i Y_{2, ij}^{\text{RL}} \ \text{P}_{\text{L}}
        \end{aligned}$  \newline  \\
        \hline
        \hline
    \end{tabular}
\end{table}
\end{center}

\clearpage


\subsection{Interactions with SM Gauge Bosons}
The Feynman rules for the LQ interactions with the SM gauge bosons are given below.
\label{sec:LQGaugeBosons}

\subsubsection{Charge 1/3}



\begin{center}
\begin{table}[!ht]
    \centering
\begin{tabular}{>{\centering\arraybackslash}m{5cm} m{7.5cm}} 
        \multicolumn{2}{c}{Charge 1/3, Interactions with Photons} \\
    \hline
        Diagram & Feynman Rule \\
        \hline
        \vspace{5px}
        \includegraphics[height=2.25cm]{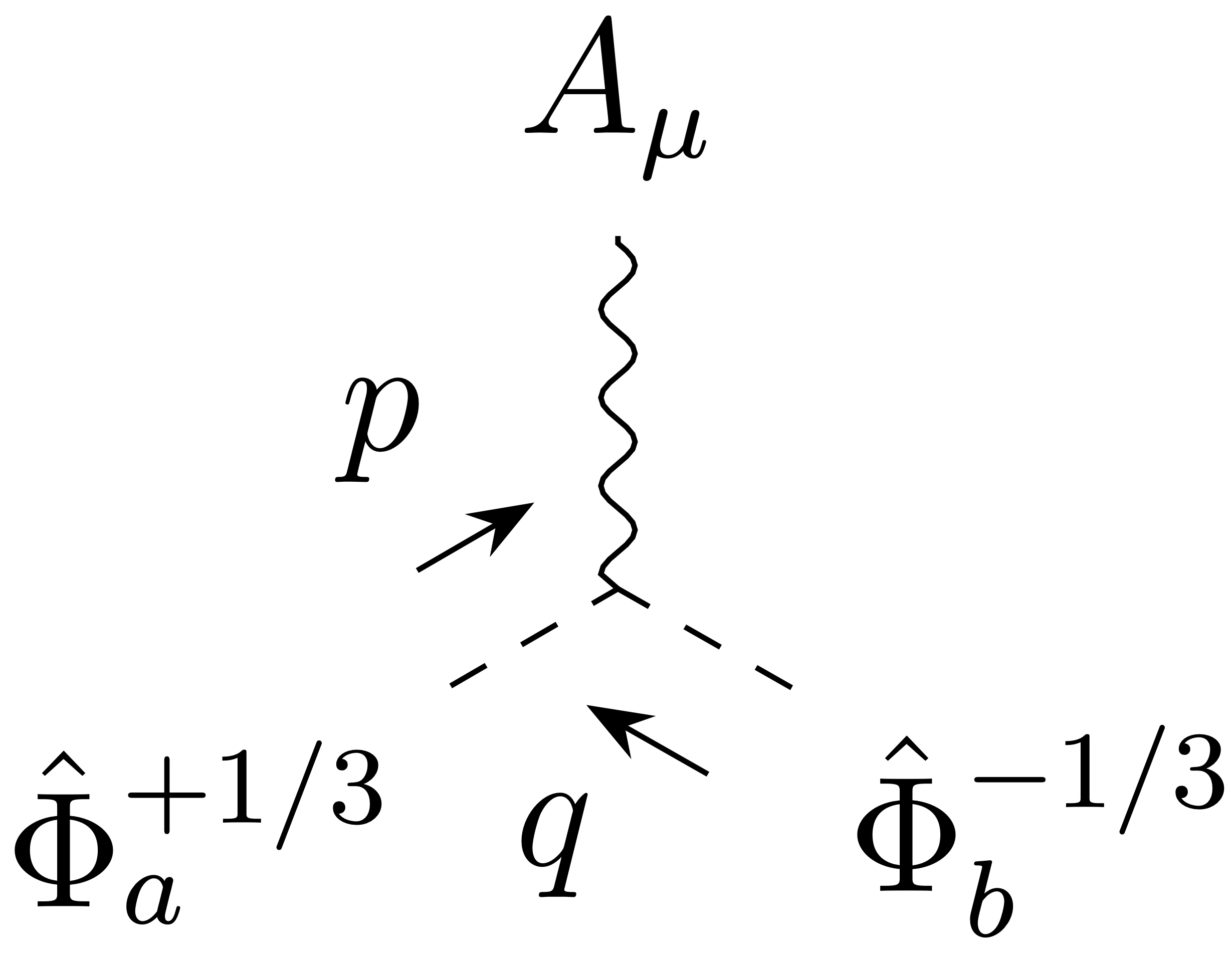}&
        $- \ \dfrac{i}{3} e \left(q^\mu - p^\mu \right) \delta_{ab}$\\
        \vspace{5px}
        \includegraphics[height=3.5cm]{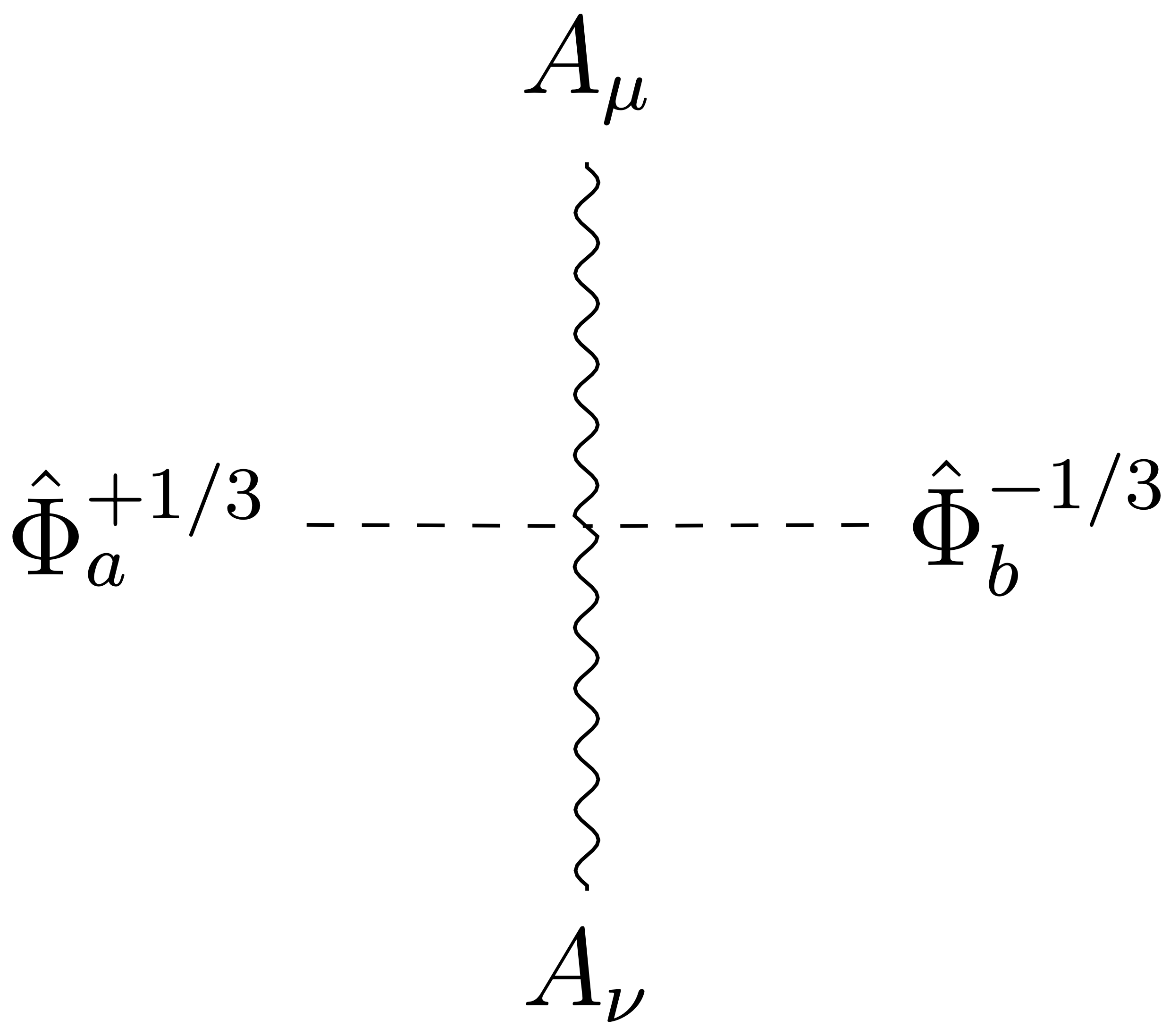}&
        $+ \ \dfrac{2i}{9} e^2 \eta^{\mu \nu} \delta_{ab}  $\\
        \hline
    \end{tabular}
\end{table}
\end{center}

\begin{center}
\begin{table}[!ht]
    \centering\begin{tabular}{>{\centering\arraybackslash}m{5cm} m{7.5cm}} 
    \multicolumn{2}{c}{Charge 1/3, Interactions with $Z$ Bosons} \\
    \hline
        Diagram & Feynman Rule \\
        \hline
        \vspace{5px}
        \includegraphics[height=2.25cm]{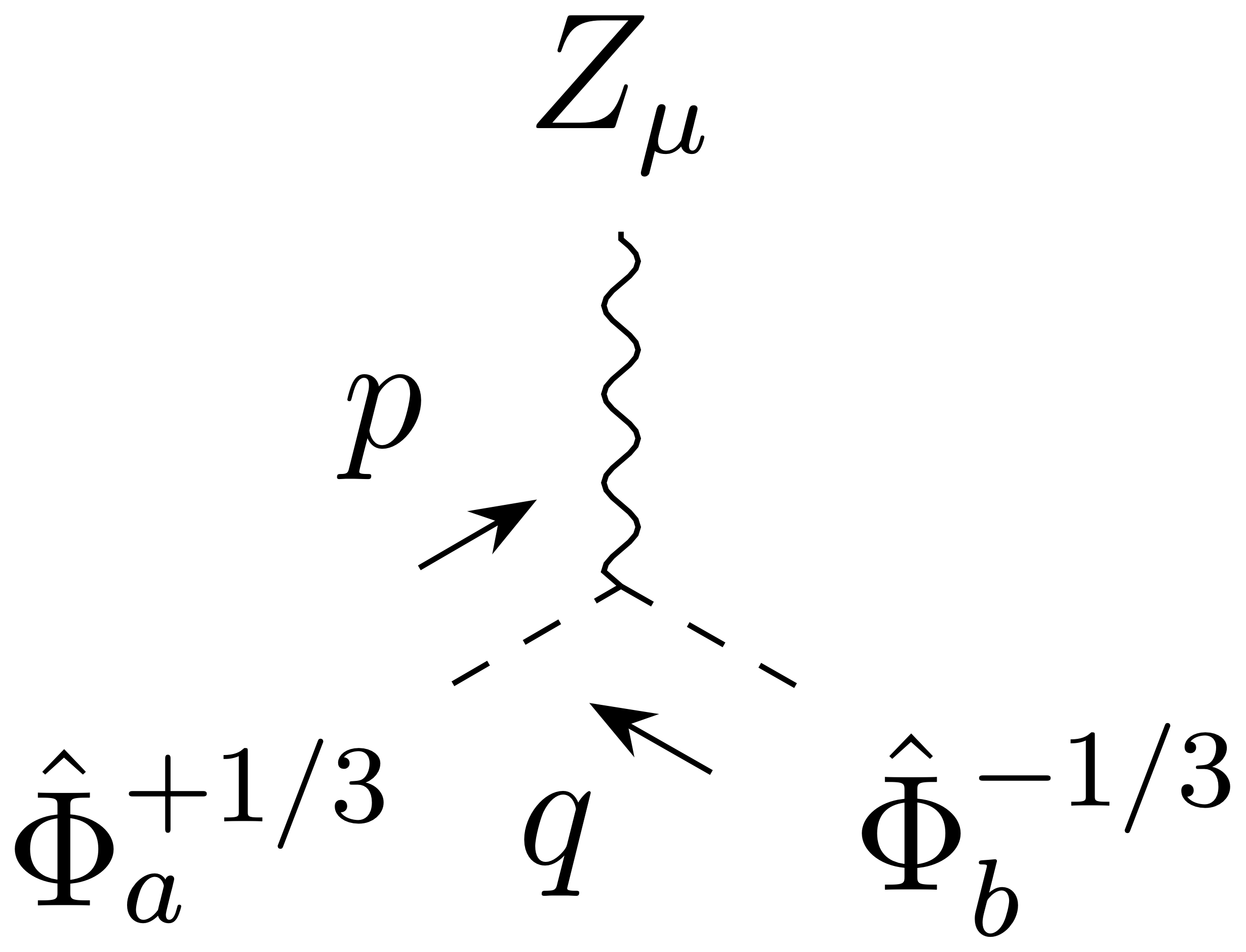}&
        $+ \ i \dfrac{e}{s_w c_w} \left(q^\mu - p^\mu \right) 
        \hat{\mathbbm{Z}}^{-1/3}
       _{n_an_b}$\\
        \vspace{5px}
        \includegraphics[height=3.5cm]{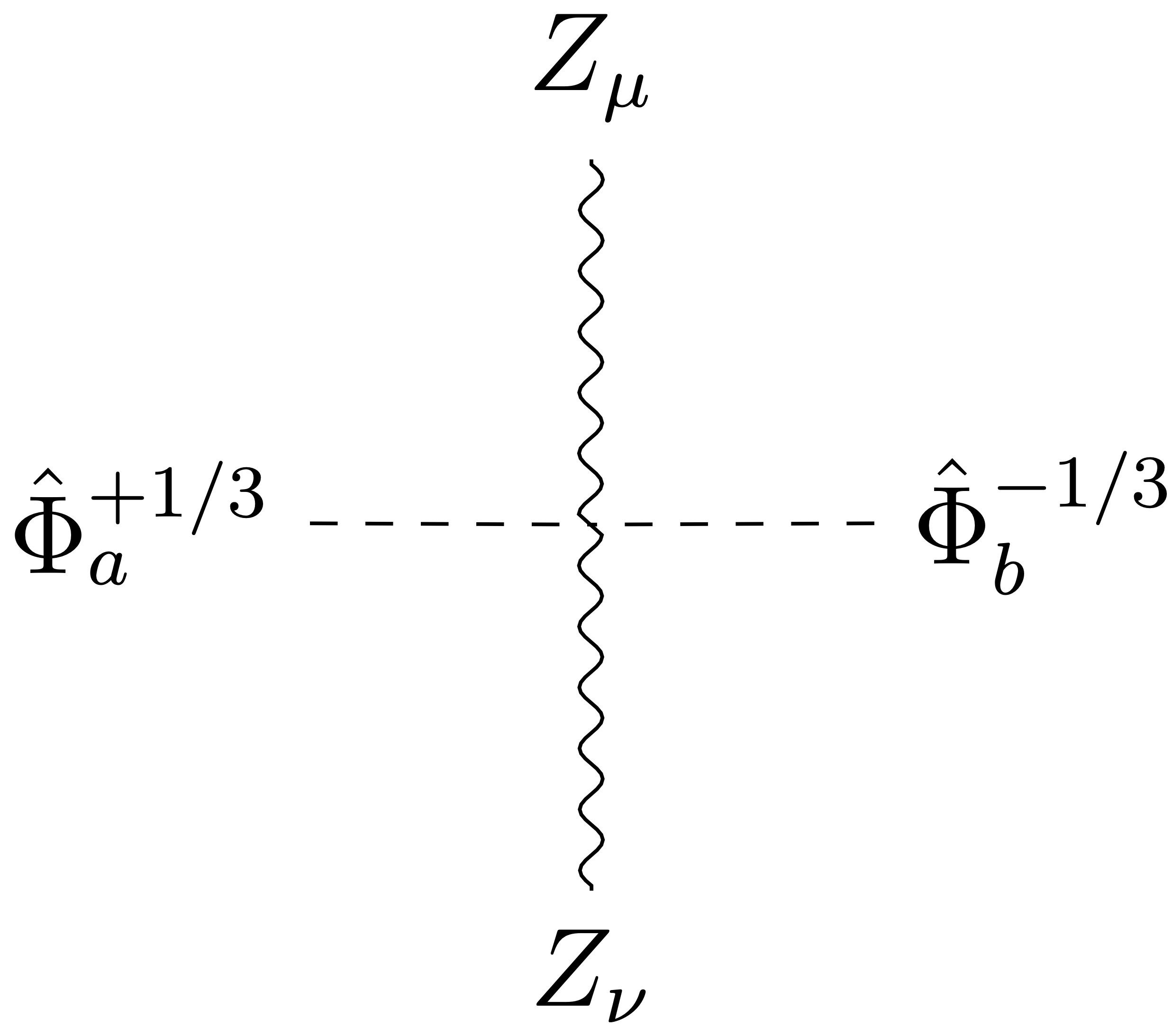}&
        $+ \ 2 i \left(\dfrac{e}{s_w c_w}\right)^2 \eta^{\mu \nu}  \left(\hat{\mathbbm{Z}}^{-1/3\dagger} \ \hat{\mathbbm{Z}}^{-1/3} \right)_{n_an_b}  $\\
        \hline
    \end{tabular}
\end{table}
\end{center}

\begin{center}
\begin{table}[!ht]
    \centering\begin{tabular}{>{\centering\arraybackslash}m{5cm} m{7.5cm}} 
    \multicolumn{2}{c}{Charge 1/3, Interactions with $W$ Bosons} \\
    \hline
        Diagram & Feynman Rule \\
        \hline
        \vspace{5px}
        \includegraphics[height=3.5cm]{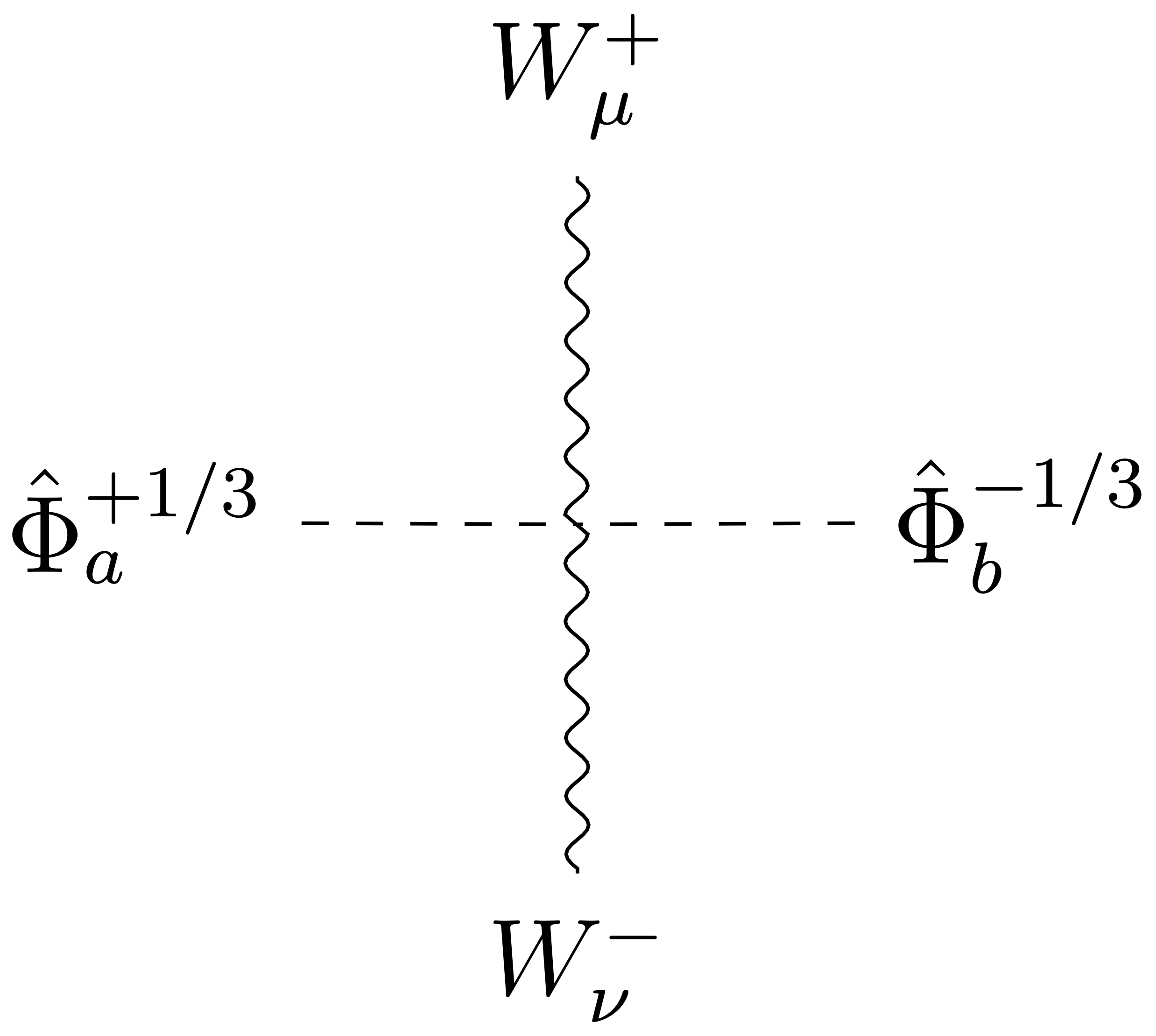}&
        $+ \ i \left(\dfrac{e}{\sqrt{2}s_w}\right)^2 \eta^{\mu \nu}  
        \left( 
        \hat{\mathbbm{W}}^{+2/3 \dagger}_{-1/3} \ \hat{\mathbbm{W}}^{+2/3}_{-1/3} \right)_{n_an_b} \vspace{5px} \newline 
         + \ i \left(\dfrac{e}{\sqrt{2} s_w}\right)^2 \eta^{\mu \nu}  
        \left(\hat{\mathbbm{W}}^{-1/3}_{-4/3} \ 
         \hat{\mathbbm{W}}^{-1/3 \dagger}_{-4/3}
        \right)_{n_an_b}$ \newline \\
        \hline
    \end{tabular}
\end{table}
\end{center}

\begin{center}
\begin{table}[!ht]
    \centering\begin{tabular}{>{\centering\arraybackslash}m{5cm} m{7.5cm}} 
    \multicolumn{2}{c}{Charge 1/3, Electroweak Interactions} \\
    \hline
        Diagram & Feynman Rule \\
        \hline
        \vspace{5px}
        \includegraphics[height=3.5cm]{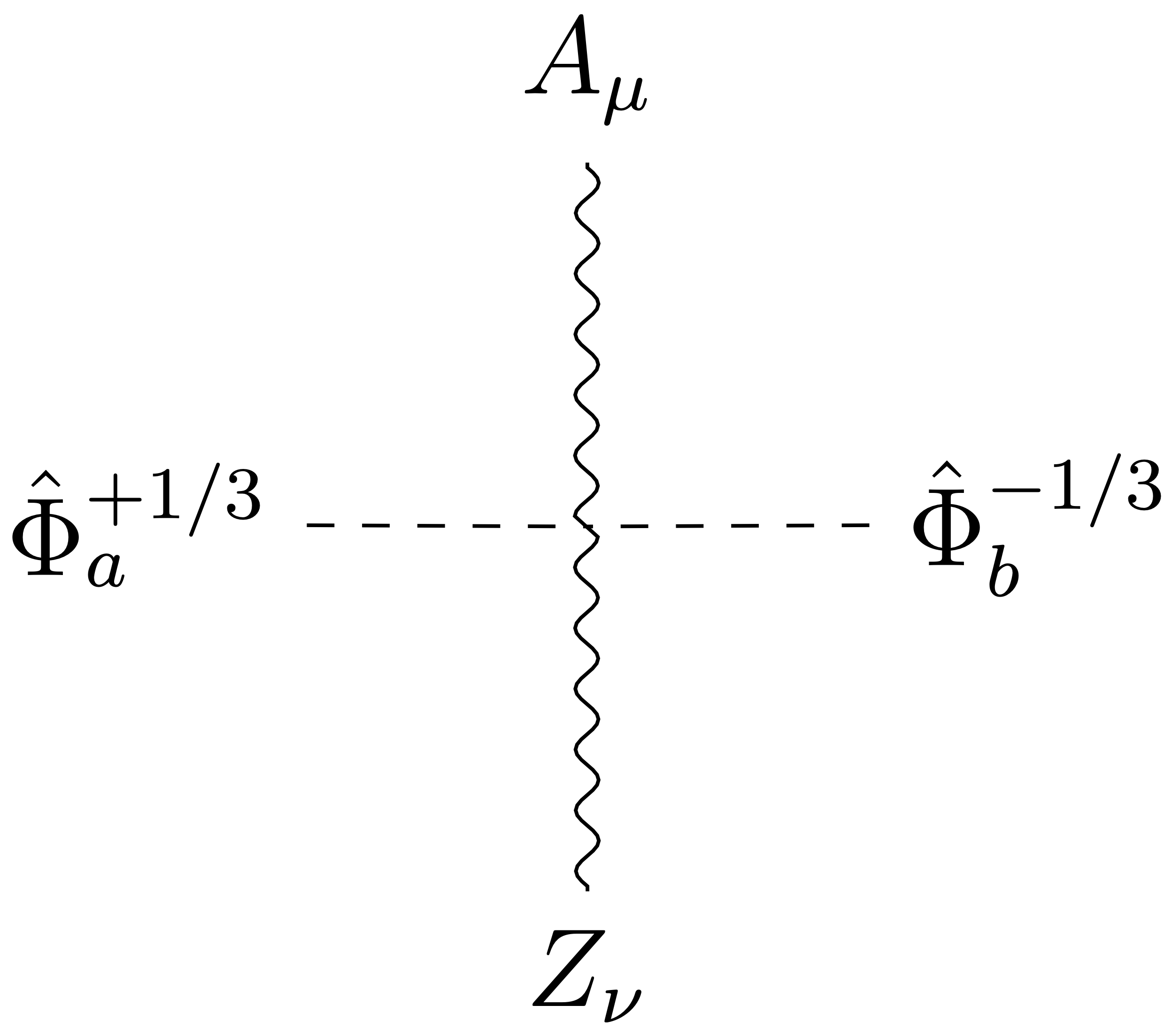}&
        $- \ \dfrac{2i}{3} \dfrac{e^2}{s_w c_w} \eta^{\mu \nu}  
                \ \hat{\mathbbm{Z}}^{-1/3}
               _{n_an_b}$ \newline \\
        \hline
    \end{tabular}
\end{table}
\end{center}

\begin{center}
\begin{table}[!ht]
    \centering\begin{tabular}{>{\centering\arraybackslash}m{5cm} m{7.5cm}} 
    \multicolumn{2}{c}{Charge 1/3, Strong Interactions} \\
    \hline
        Diagram & Feynman Rule \\
        \hline
        \vspace{5px}
        \includegraphics[height=2.25cm]{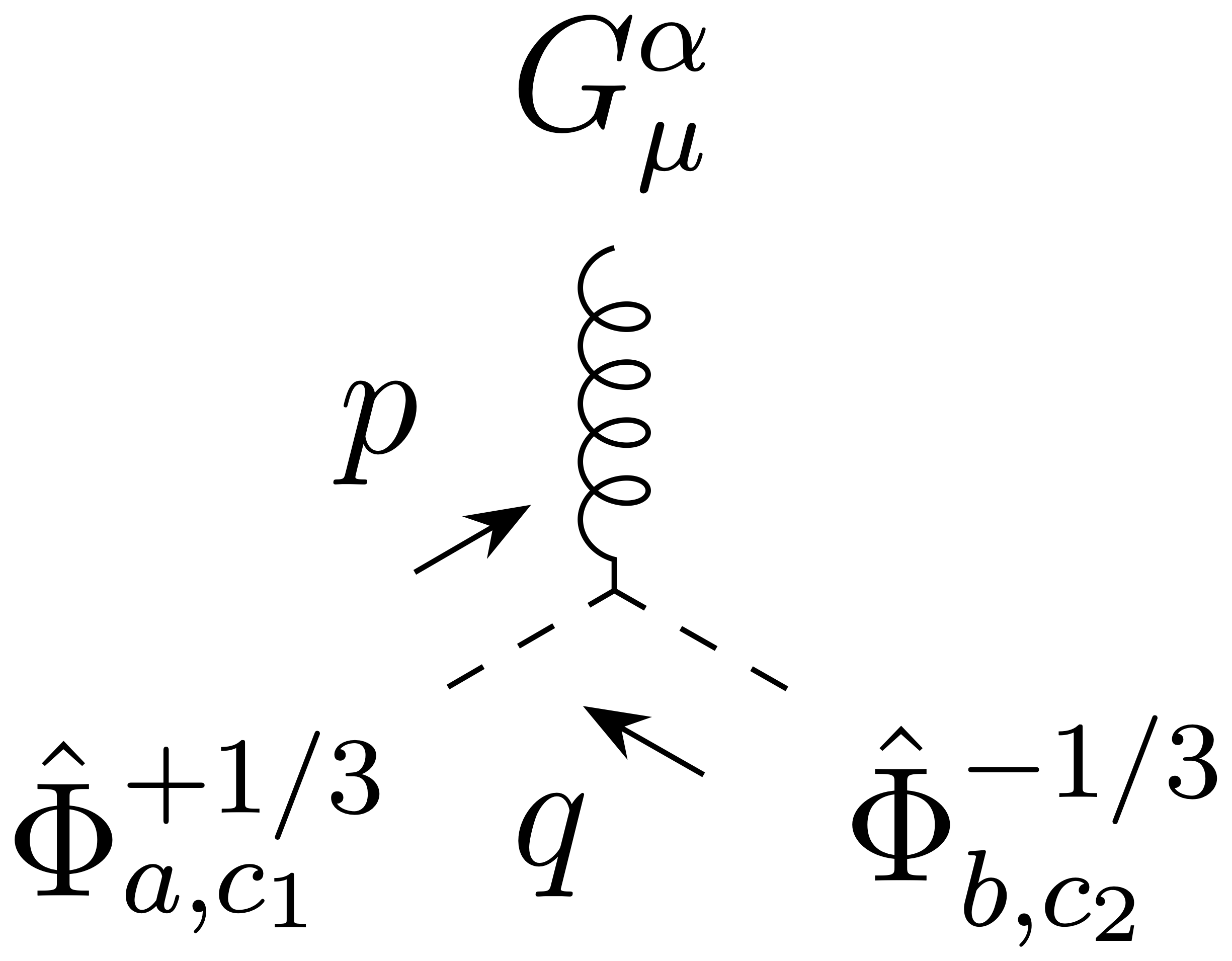}&
        $+ \ i g_s \left(q^\mu - p^\mu \right) \delta_{ab} \ T^\alpha_{c_1 c_2}$\\
        \vspace{5px}
        \includegraphics[height=3.5cm]{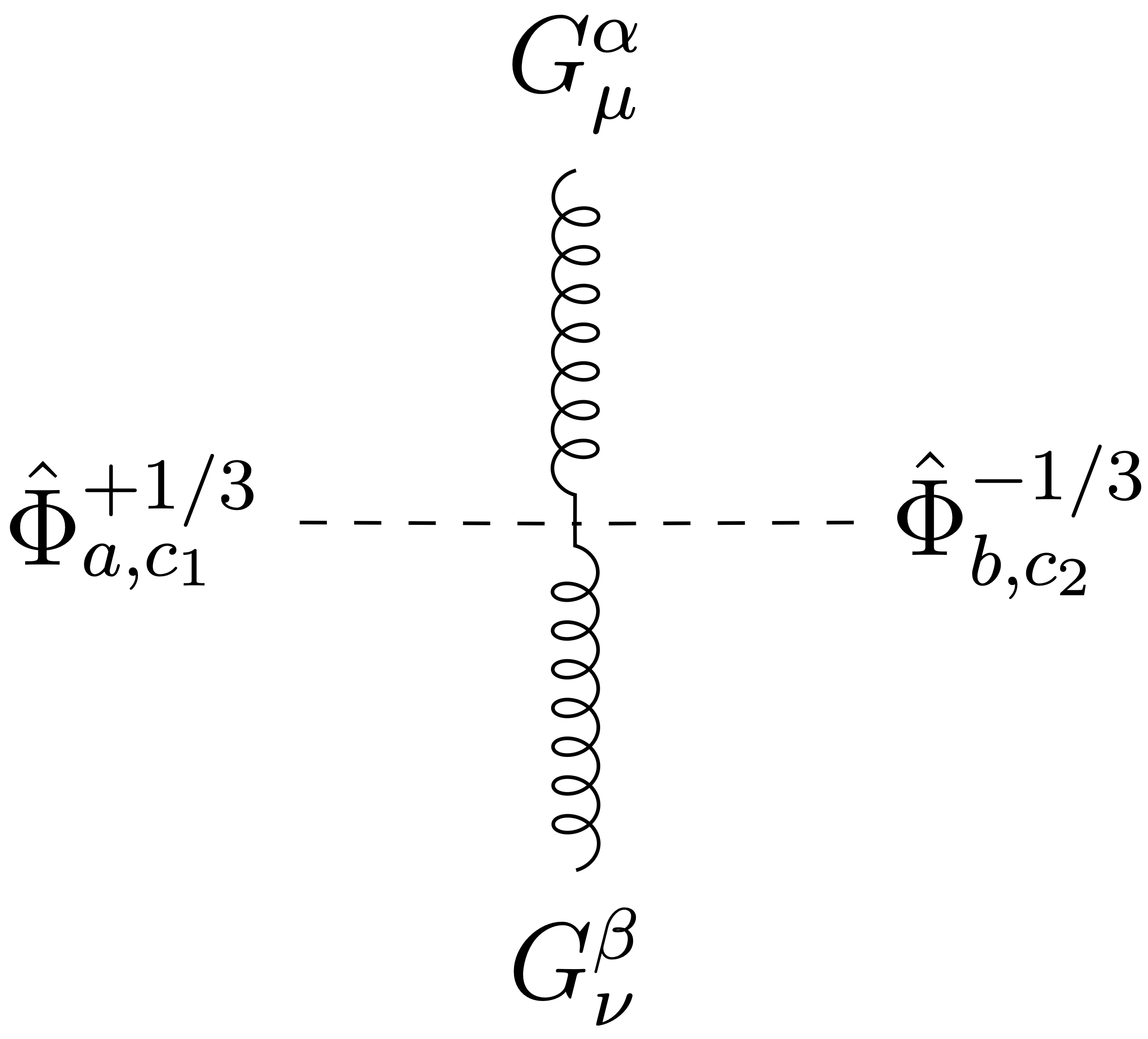}&
        $+\ i g_s^2 \eta^{\mu \nu} \delta_{ab} \left(T^\alpha T^\beta + T^\beta T^\alpha \right)_{c_1 c_2}$\\
        \vspace{5px}
        \includegraphics[height=3.5cm]{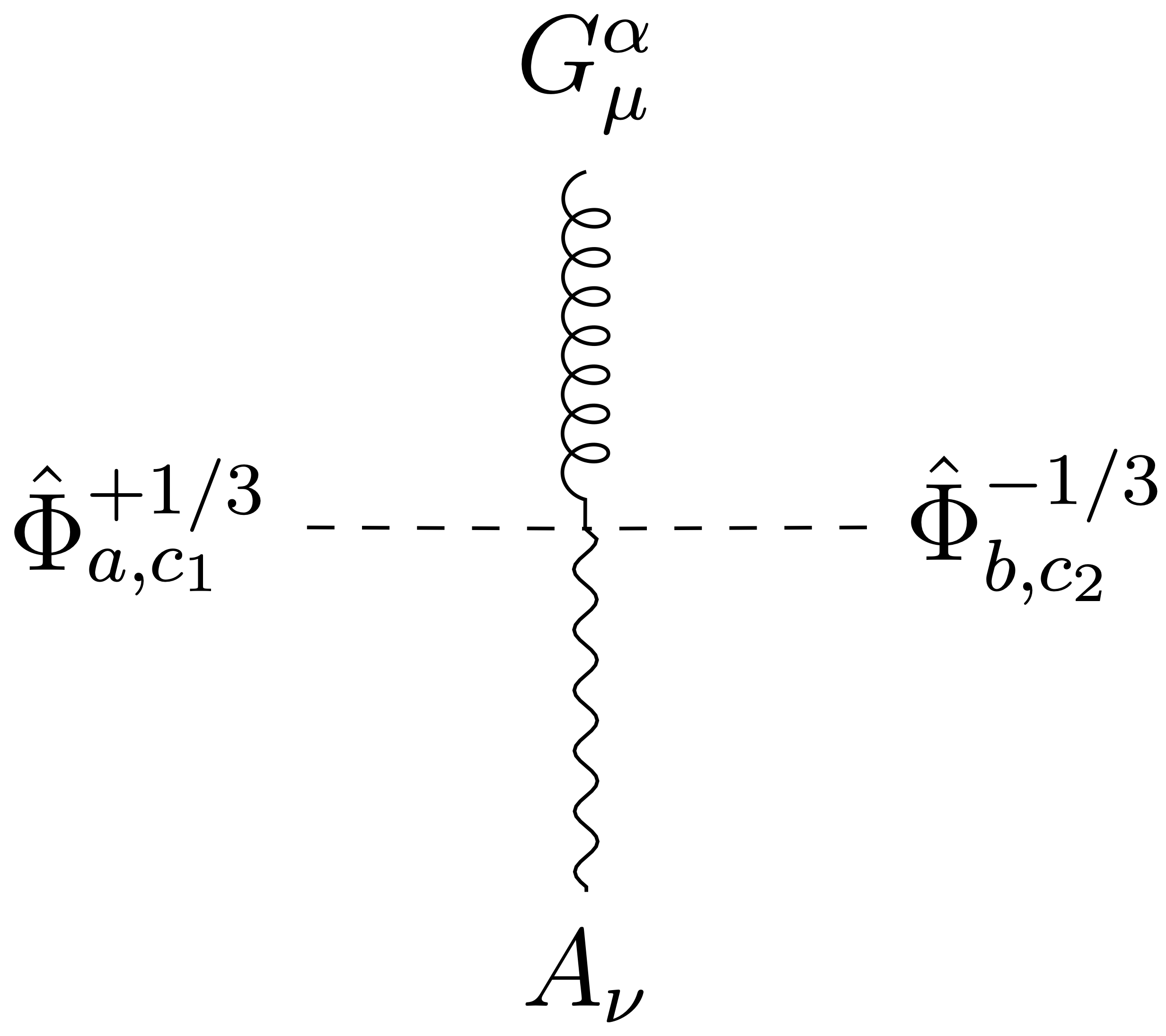}&
        $-\ \dfrac{2i}{3} g_s e \ \eta^{\mu \nu} \delta_{ab} \ T^\alpha_{c_1 c_2} $\\
        \vspace{5px}
        \includegraphics[height=3.5cm]{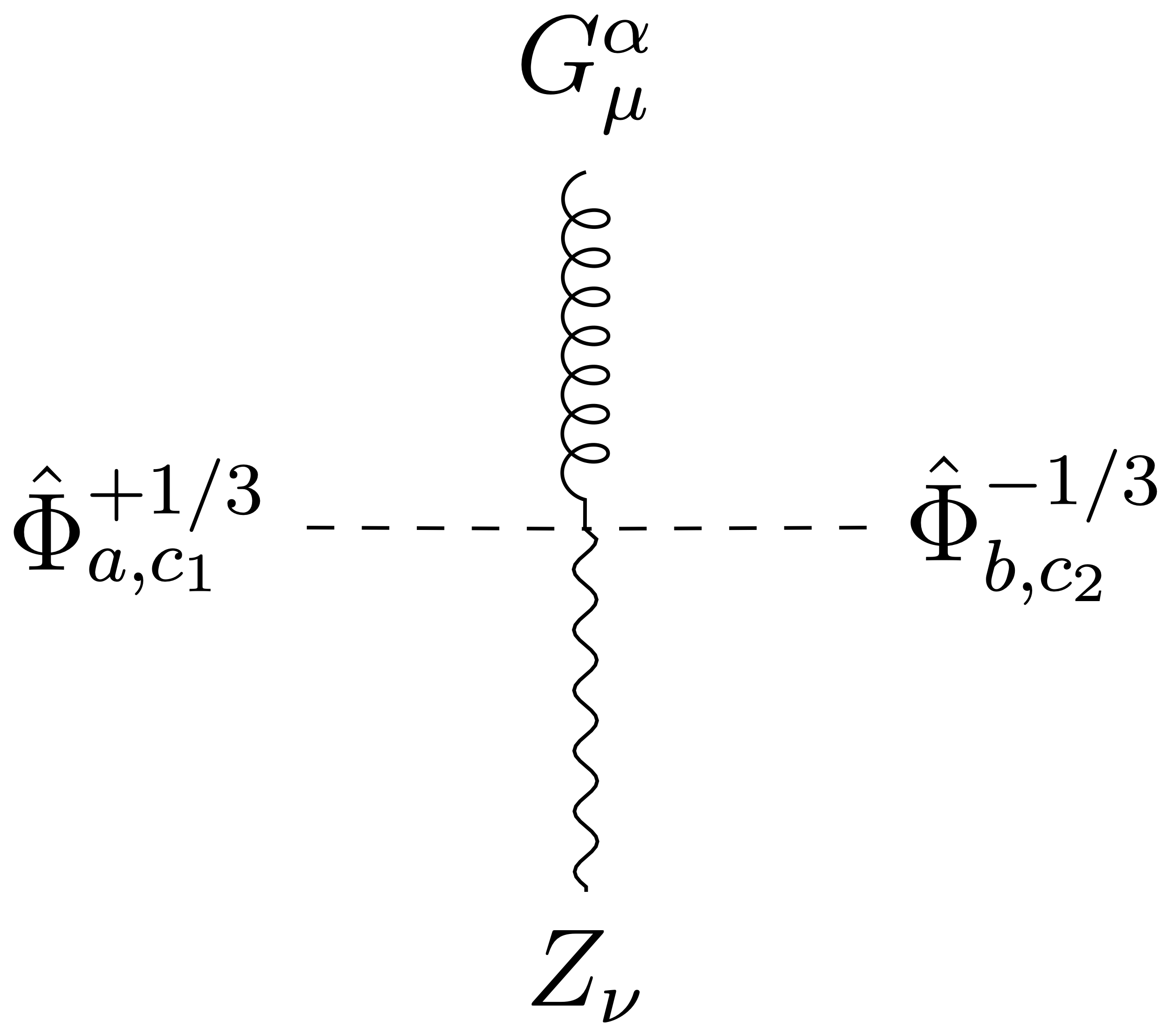}&
        $+\ 2i \dfrac{g_s e}{c_w s_w} \eta^{\mu \nu} \ T^\alpha_{c_1 c_2} 
        \ \hat{\mathbbm{Z}}^{-1/3}
       _{n_an_b} $\\
        \hline
    \end{tabular}
\end{table}
\end{center}

\clearpage
\subsubsection{Charge 2/3}

\begin{center}
\begin{table}[!ht]
    \centering\begin{tabular}{>{\centering\arraybackslash}m{5cm} m{7.5cm}} 
    \multicolumn{2}{c}{Charge 2/3, Interactions with Photons} \\
    \hline
        Diagram & Feynman Rule \\
        \hline
         \vspace{5px}
        \includegraphics[height=2.25cm]{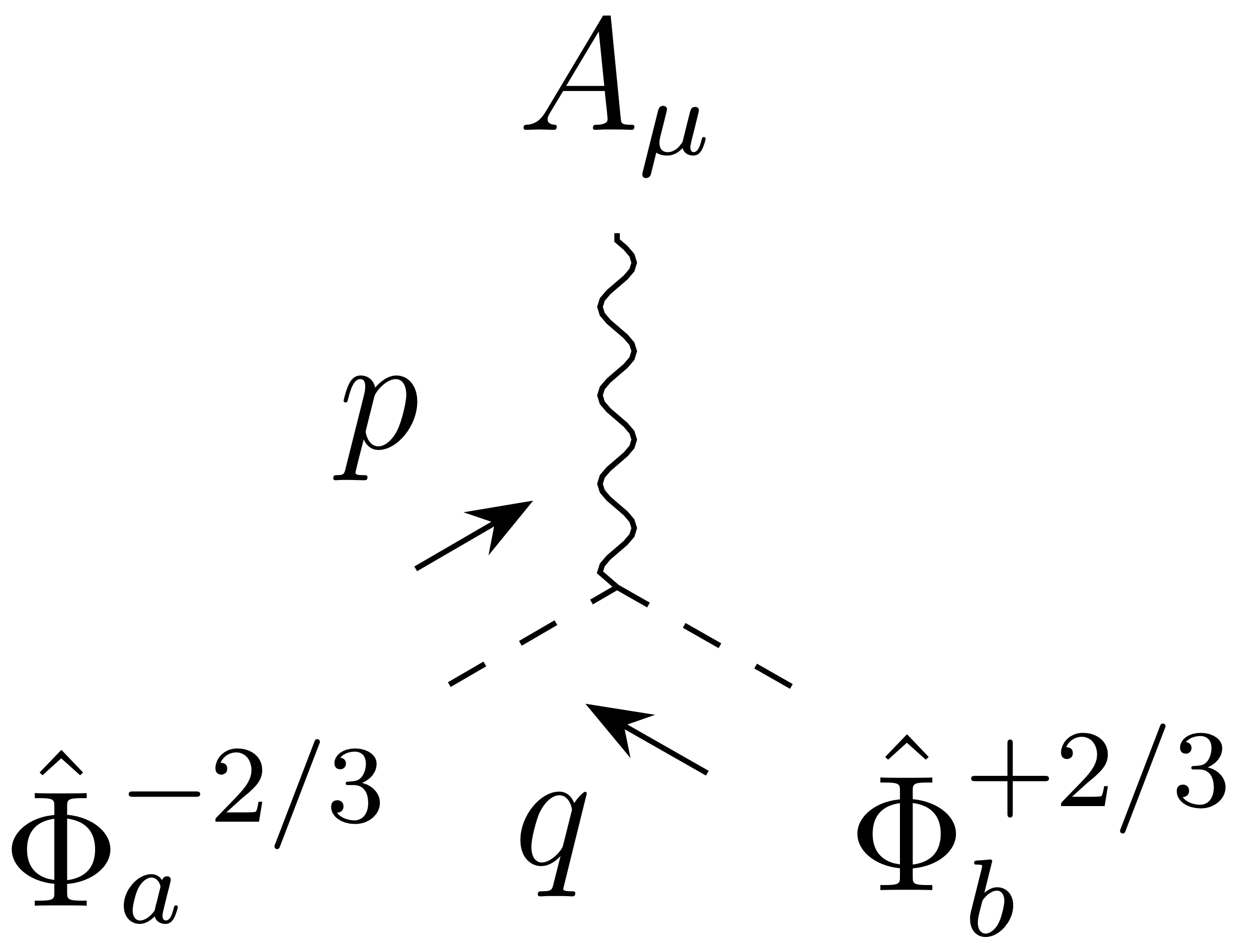}&
        $+ \ \dfrac{2i}{3} e \left(q^\mu - p^\mu \right) \delta_{ab}$\\
        \vspace{5px}
        \includegraphics[height=3.5cm]{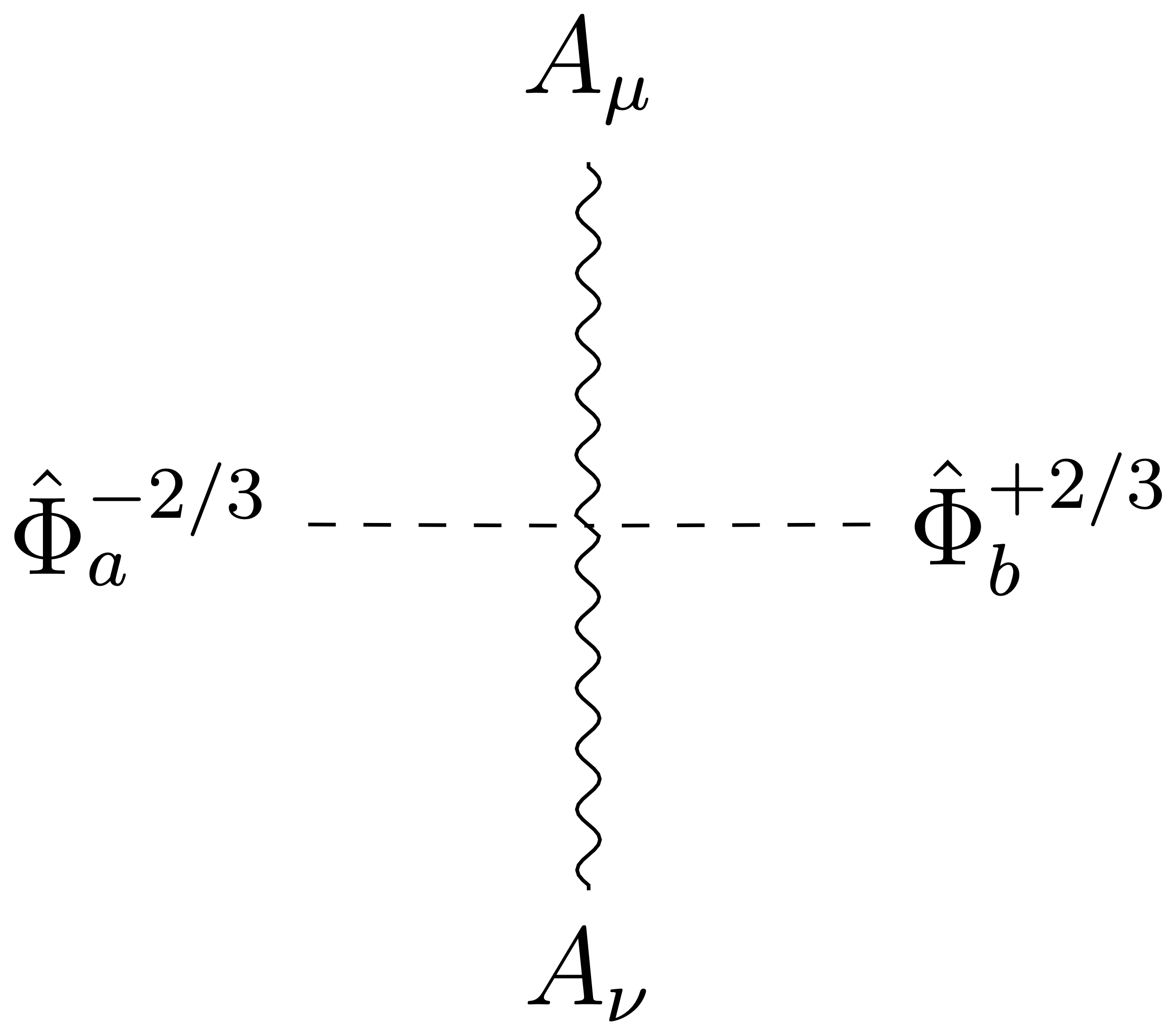}&
        $+ \ \dfrac{8i}{9} e^2 \eta^{\mu \nu} \delta_{ab}  $\\
        \hline
    \end{tabular}
\end{table}
\end{center}

\begin{center}
\begin{table}[!ht]
    \centering\begin{tabular}{>{\centering\arraybackslash}m{5cm} m{7.5cm}} 
    \multicolumn{2}{c}{Charge 2/3, Interactions with $Z$ Bosons} \\
    \hline
        Diagram & Feynman Rule \\
        \hline
        \vspace{5px}
        \includegraphics[height=2.25cm]{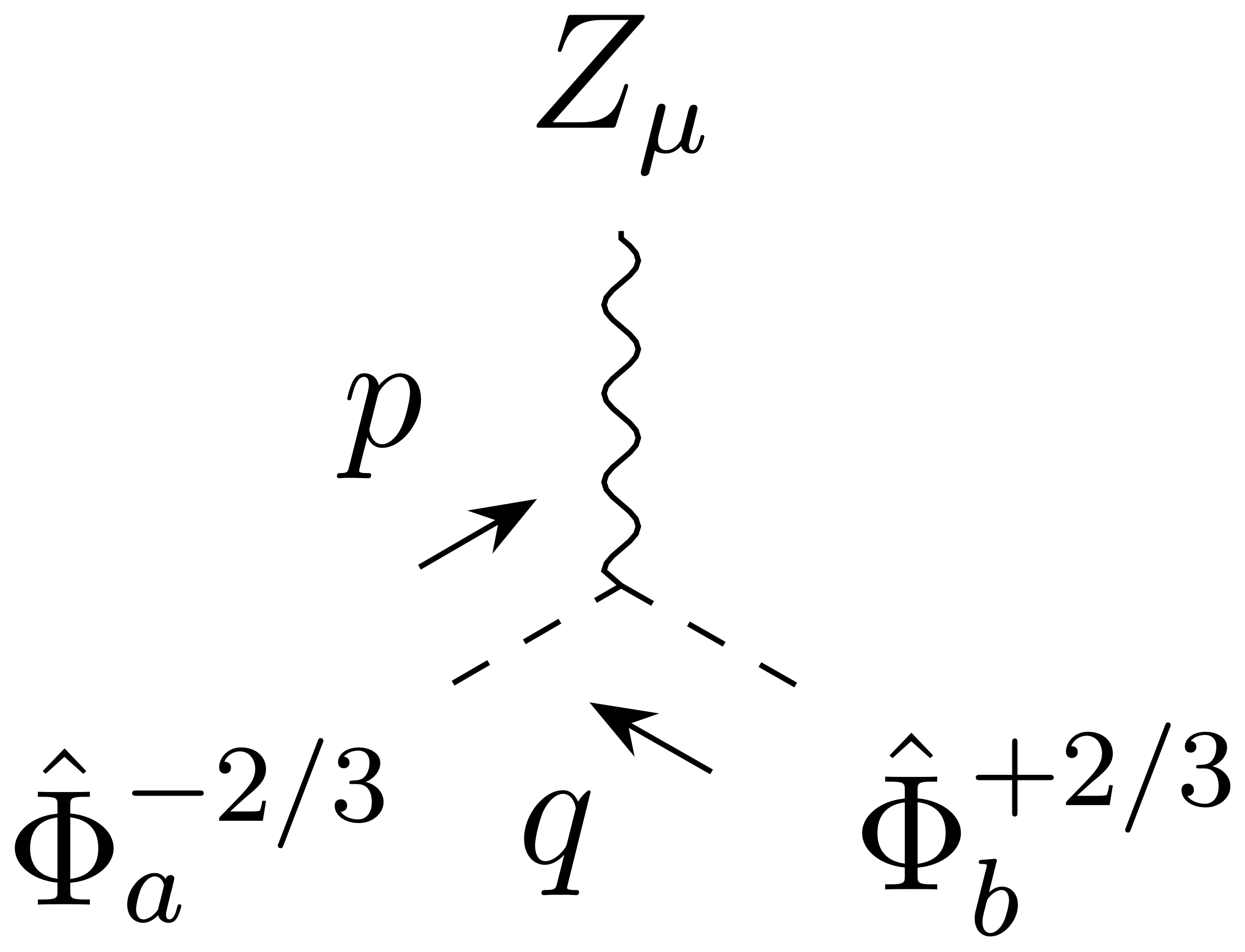}&
        $+ \ i \dfrac{e}{s_w c_w} \left(q^\mu - p^\mu \right) 
        \hat{\mathbbm{Z}}^{+2/3}
       _{n_an_b}$\\
        \hline
    \end{tabular}
\end{table}
\end{center}

\begin{center}
\begin{table}[!ht]
    \centering\begin{tabular}{>{\centering\arraybackslash}m{5cm} m{7.5cm}} 
    \multicolumn{2}{c}{Charge 2/3, Interactions with $Z$ Bosons} \\
    \hline
        Diagram & Feynman Rule \\
        \hline
        \vspace{5px}
        \includegraphics[height=3.5cm]{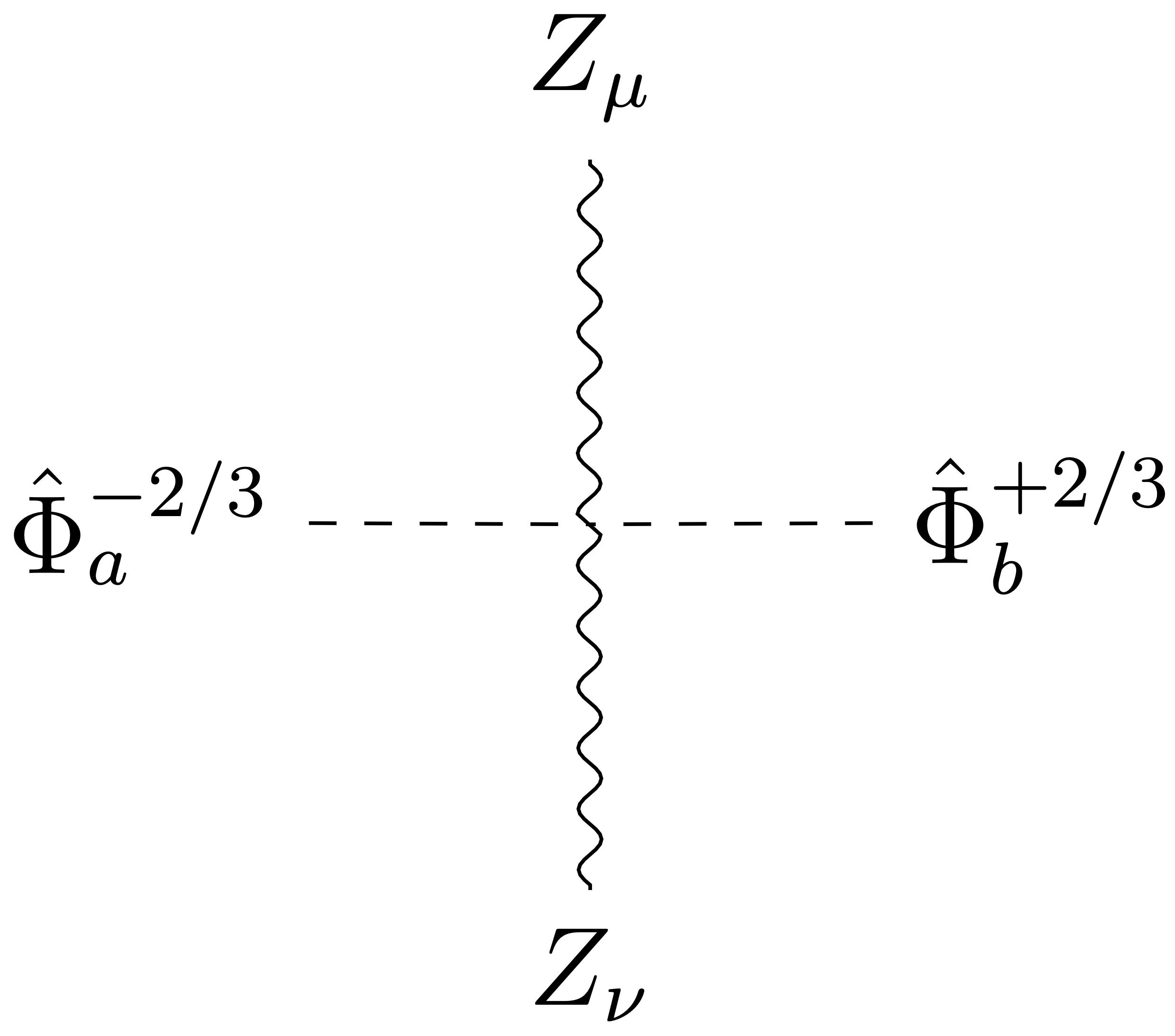}&
        $+ \ 2 i \left(\dfrac{e}{s_w c_w}\right)^2 \eta^{\mu \nu}  \left(\hat{\mathbbm{Z}}^{+2/3\dagger} \  \hat{\mathbbm{Z}}^{+2/3} \right)_{n_an_b}  $\\
        \hline
    \end{tabular}
\end{table}
\end{center}

\begin{center}
\begin{table}[!ht]
    \centering\begin{tabular}{>{\centering\arraybackslash}m{5cm} m{7.5cm}} 
    \multicolumn{2}{c}{Charge 2/3, Interactions with $W$ Bosons} \\
    \hline
        Diagram & Feynman Rule \\
        \hline
        \vspace{5px}
        \includegraphics[height=3.5cm]{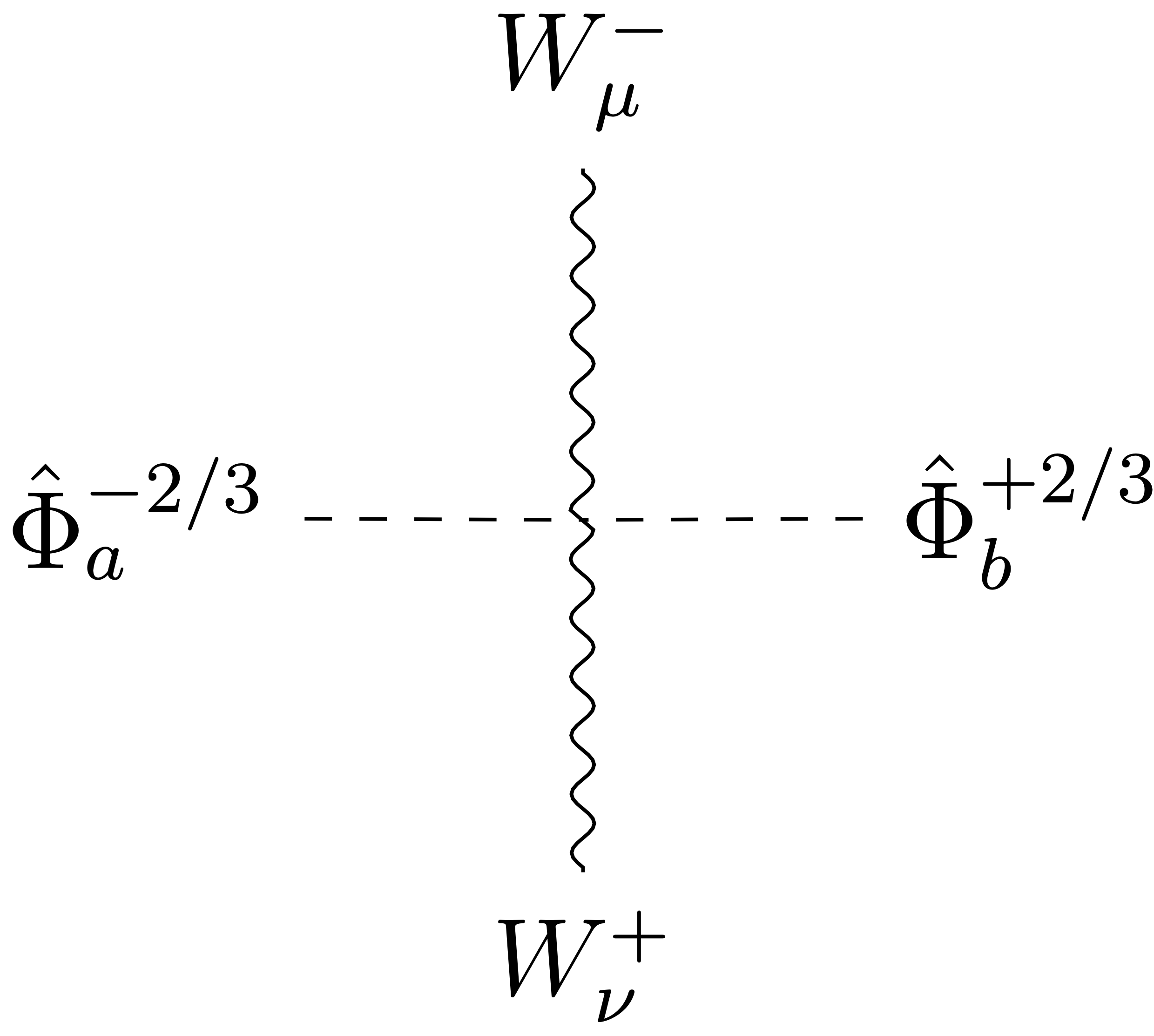}&
        $+ \ i \left(\dfrac{e}{\sqrt{2}s_w}\right)^2 \eta^{\mu \nu}  
                \left( 
                \hat{\mathbbm{W}}^{+2/3}_{-1/3} \  \hat{\mathbbm{W}}^{+2/3 \dagger}_{-1/3} \right)_{n_an_b} \vspace{5px} \newline
                + \ i \left(\dfrac{e}{\sqrt{2}s_w}\right)^2 \eta^{\mu \nu}  
                \left(
                \hat{\mathbbm{W}}^{+5/3 \dagger}_{+2/3} \ 
                \hat{\mathbbm{W}}^{+5/3}_{+2/3}
                \right)_{n_an_b}$ \\
        \hline
    \end{tabular}
\end{table}
\end{center}

\begin{center}
\begin{table}[!ht]
    \centering\begin{tabular}{>{\centering\arraybackslash}m{5cm} m{7.5cm}} 
    \multicolumn{2}{c}{Charge 2/3, Electroweak Interactions} \\
    \hline
        Diagram & Feynman Rule \\
        \hline
        \vspace{5px}
        \includegraphics[height=3.5cm]{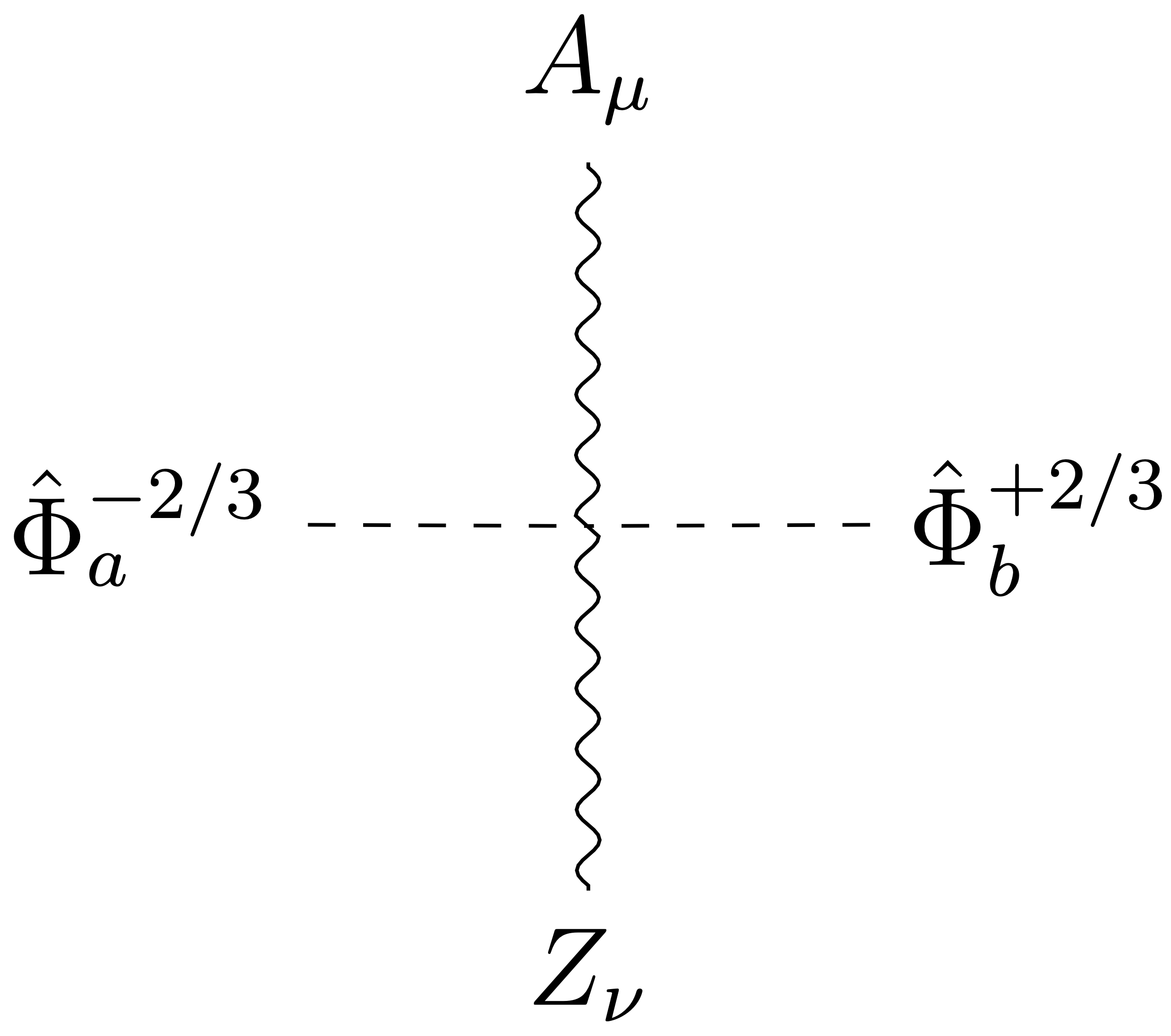}&
        $+ \ \dfrac{4i}{3} \dfrac{e^2}{s_w c_w} \ \eta^{\mu \nu}  
                \hat{\mathbbm{Z}}^{+2/3}_{n_an_b}
                \vspace{5 px}$ \newline \\
        \hline
    \end{tabular}
\end{table}
\end{center}

\begin{center}
\begin{table}[!ht]
    \centering\begin{tabular}{>{\centering\arraybackslash}m{5cm} m{7.5cm}} 
    \multicolumn{2}{c}{Charge 2/3, Strong Interactions} \\
    \hline
        Diagram & Feynman Rule \\
        \hline
        \vspace{5px}
        \includegraphics[height=2.25cm]{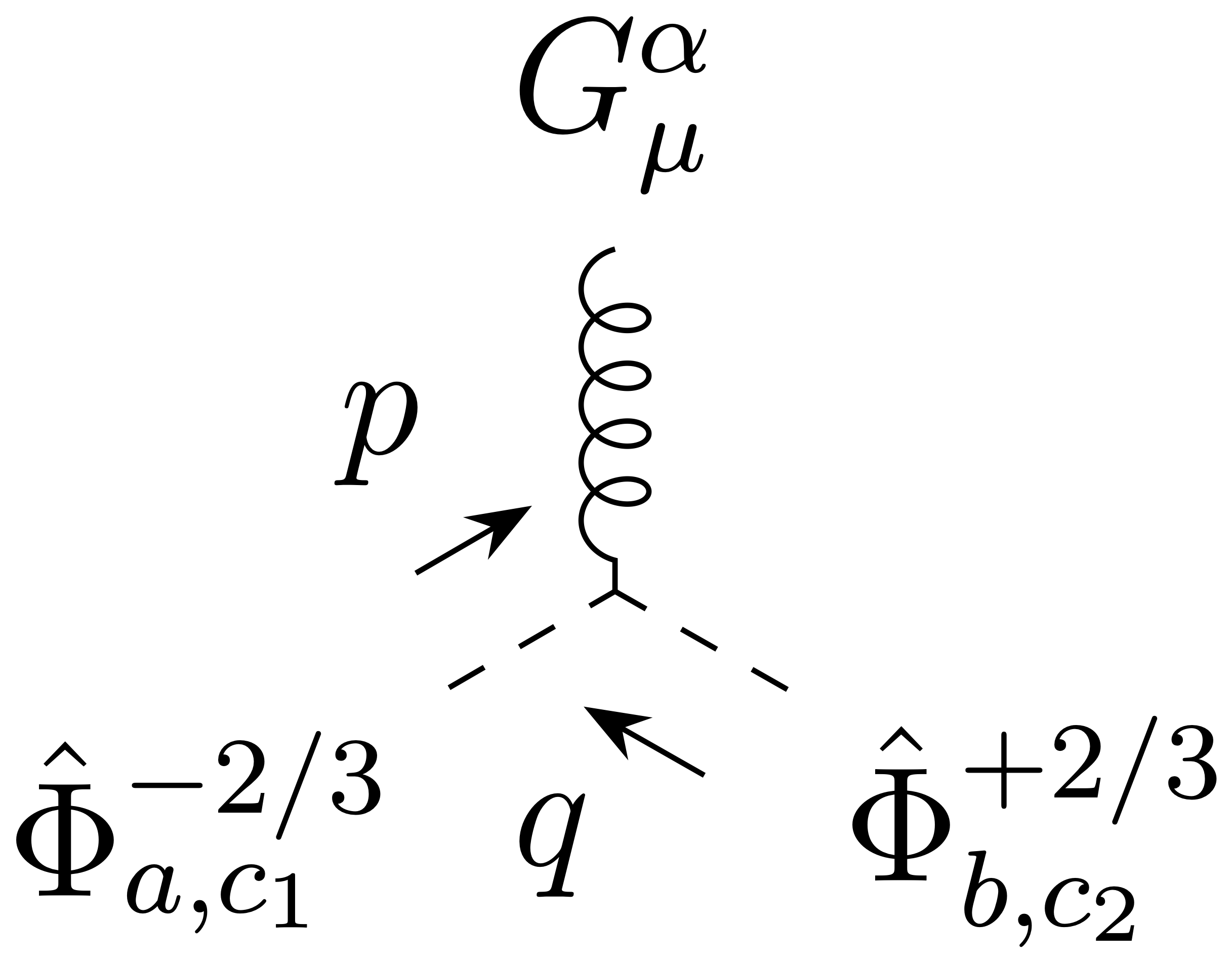}&
        $+ \ i g_s \left(q^\mu - p^\mu \right) \delta_{ab} \  T^\alpha_{c_1 c_2}$\\
        \vspace{5px}
        \includegraphics[height=3.5cm]{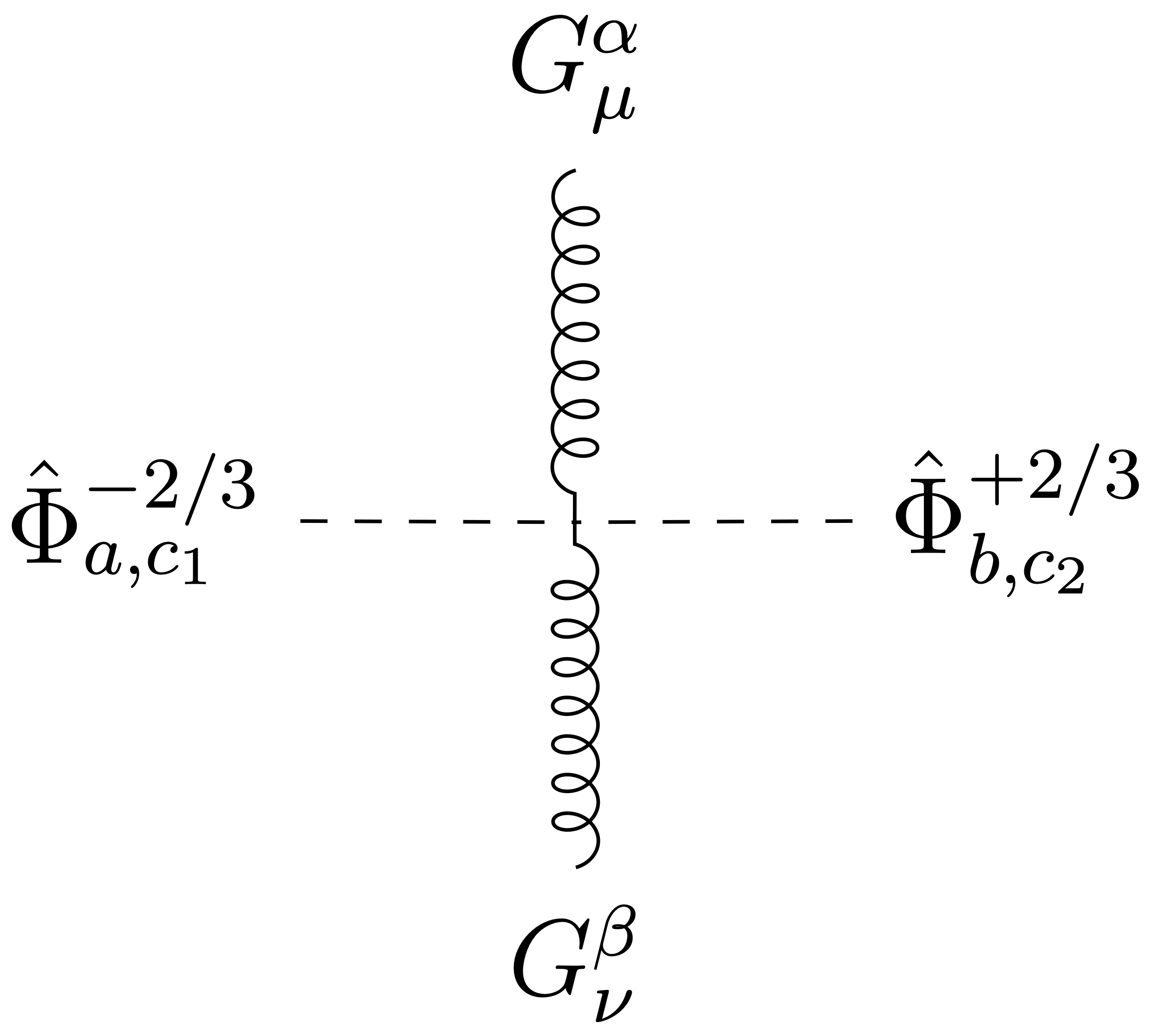}&
        $+\ i g_s^2 \eta^{\mu \nu} \delta_{ab} \left(T^\alpha T^\beta + T^\beta T^\alpha \right)_{c_1 c_2}$\\
        \vspace{5px}
        \includegraphics[height=3.5cm]{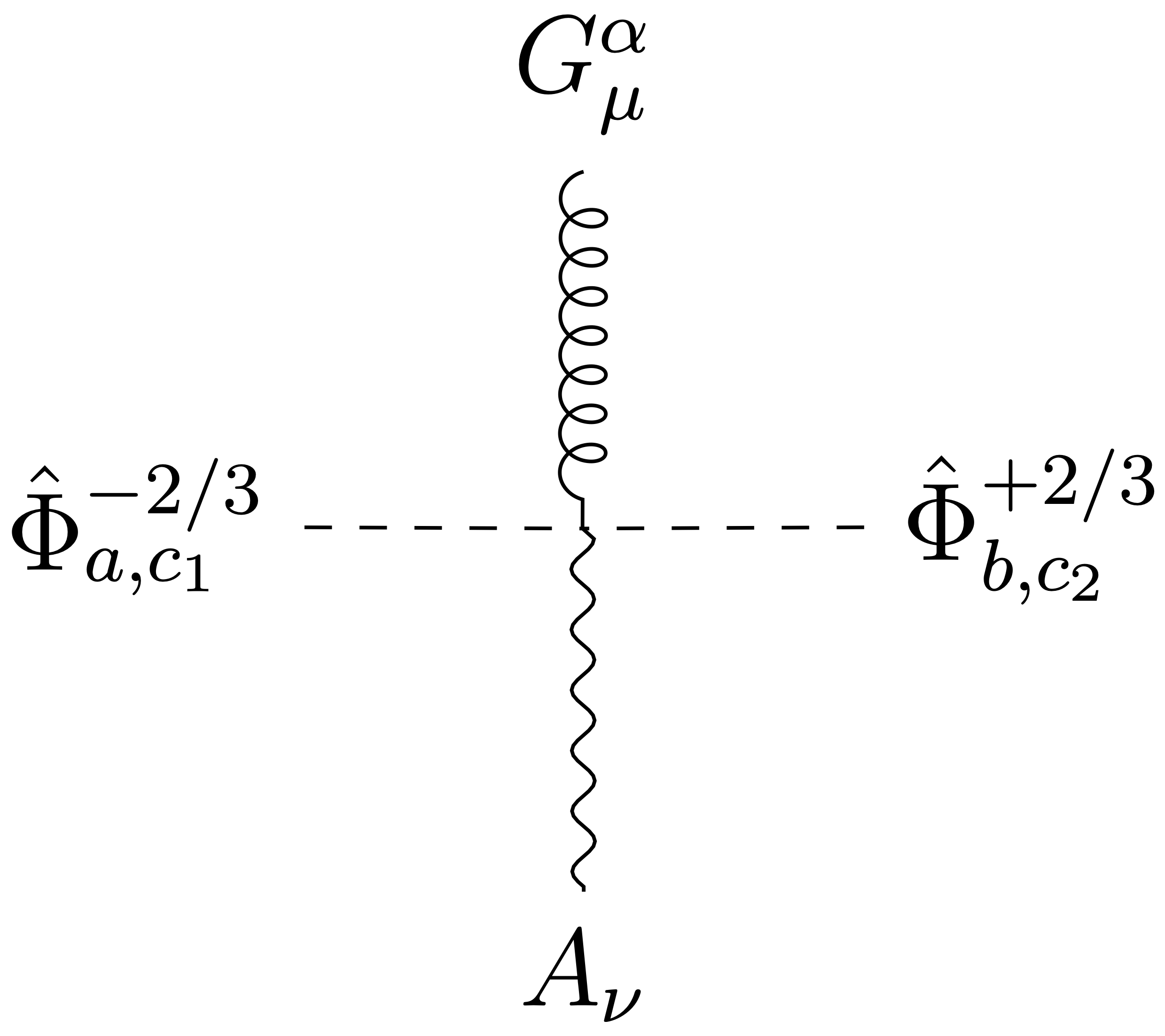}&
        $+\ \dfrac{4i}{3} g_s e \ \eta^{\mu \nu} \delta_{ab} \  T^\alpha_{c_1 c_2} $\\
        \vspace{5px}
        \includegraphics[height=3.5cm]{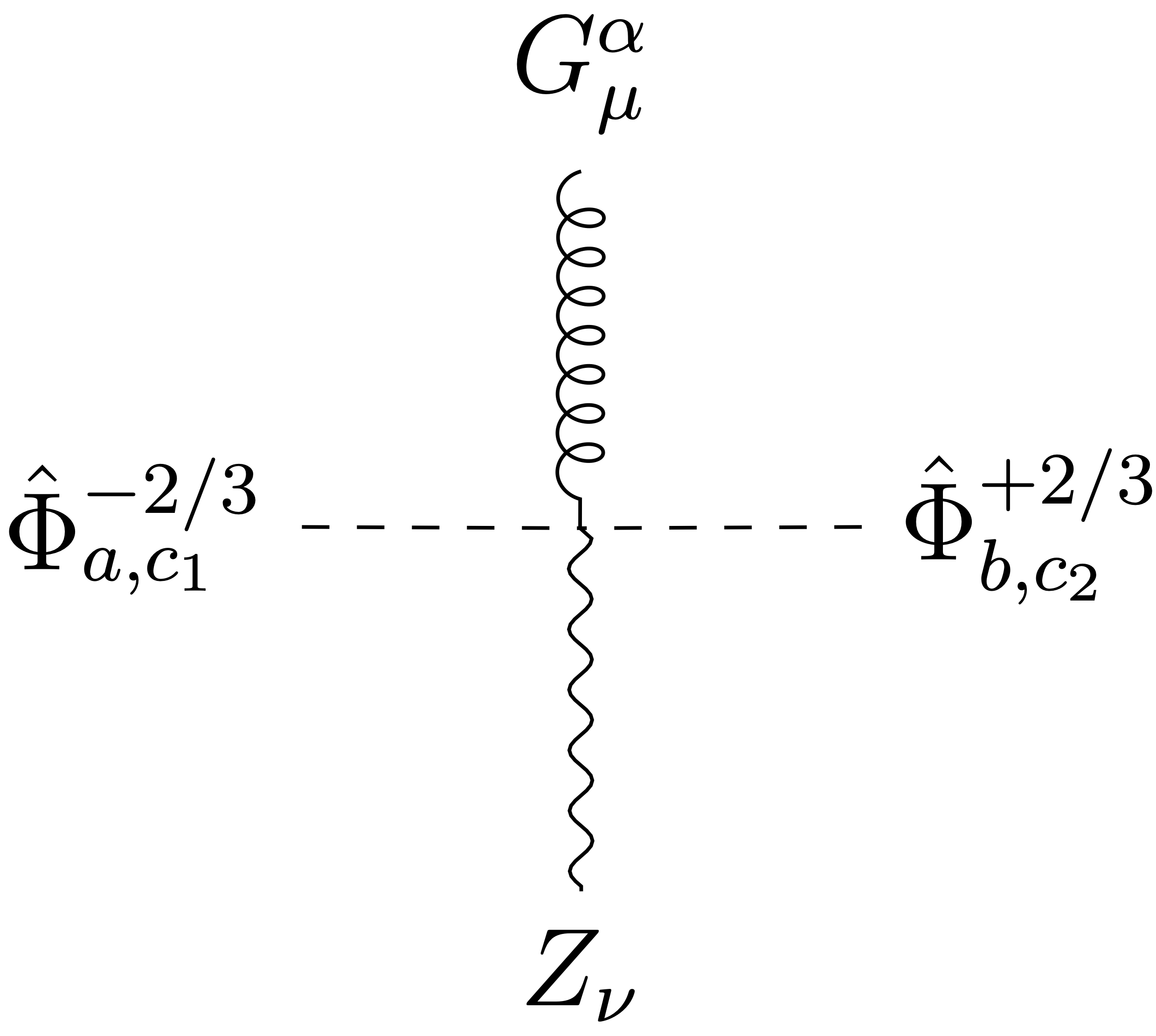}&
        $+\ 2i \dfrac{g_s e}{c_w s_w} \eta^{\mu \nu} \ T^\alpha_{c_1 c_2} 
        \ \hat{\mathbbm{Z}}^{+2/3}
       _{n_an_b} $\\
        \hline
    \end{tabular}
\end{table}
\end{center}


\clearpage
\subsubsection{Charge 4/3}

\begin{center}
\begin{table}[!ht]
    \centering\begin{tabular}{>{\centering\arraybackslash}m{5cm} m{7.5cm}} 
    \multicolumn{2}{c}{Charge 4/3, Interactions with Photons} \\
    \hline
        Diagram & Feynman Rule \\
        \hline
        \vspace{5px}
        \includegraphics[height=2.25cm]{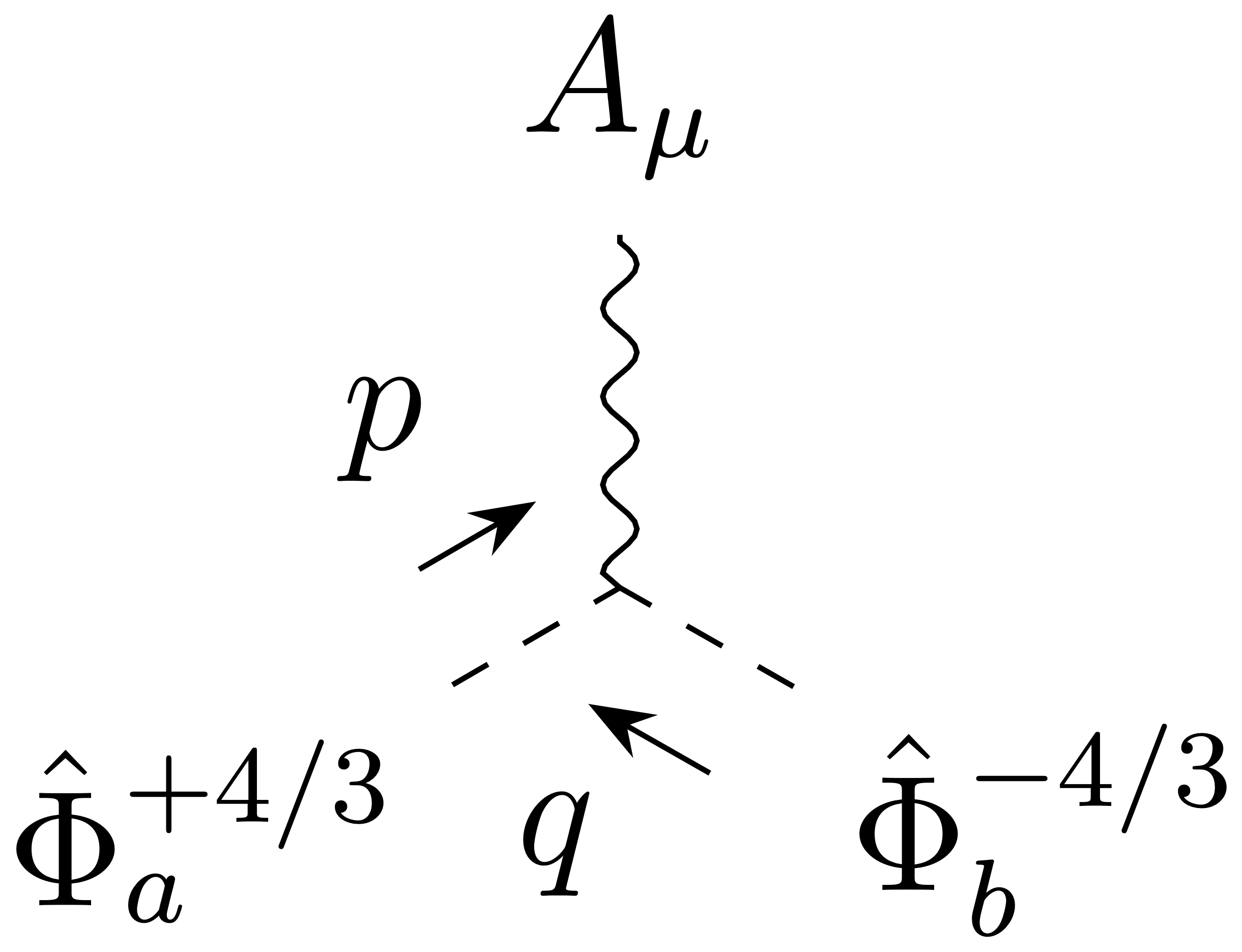}&
        $- \ \dfrac{4i}{3} e \left(q^\mu - p^\mu \right) \delta_{ab}$\\
        \vspace{5px}
        \includegraphics[height=3.5cm]{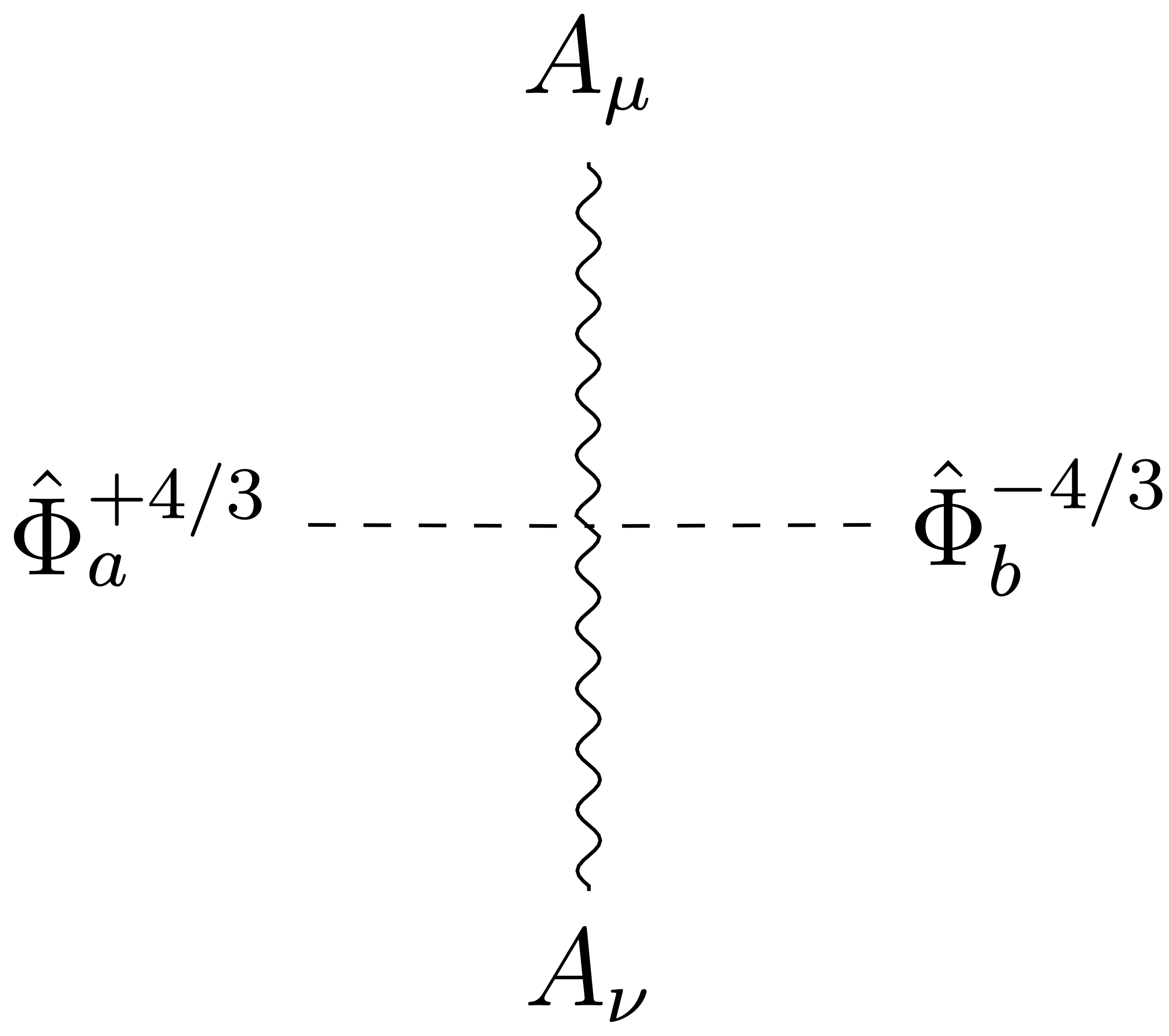}&
        $+ \ \dfrac{32i}{9} e^2 \eta^{\mu \nu} \delta_{ab}  $\\
        \hline
    \end{tabular}
\end{table}
\end{center}

\begin{center}
\begin{table}[!ht]
    \centering\begin{tabular}{>{\centering\arraybackslash}m{5cm} m{7.5cm}} 
    \multicolumn{2}{c}{Charge 4/3, Interactions with $Z$ Bosons} \\
    \hline
        Diagram & Feynman Rule \\
        \hline
        \vspace{5px}
        \includegraphics[height=2.25cm]{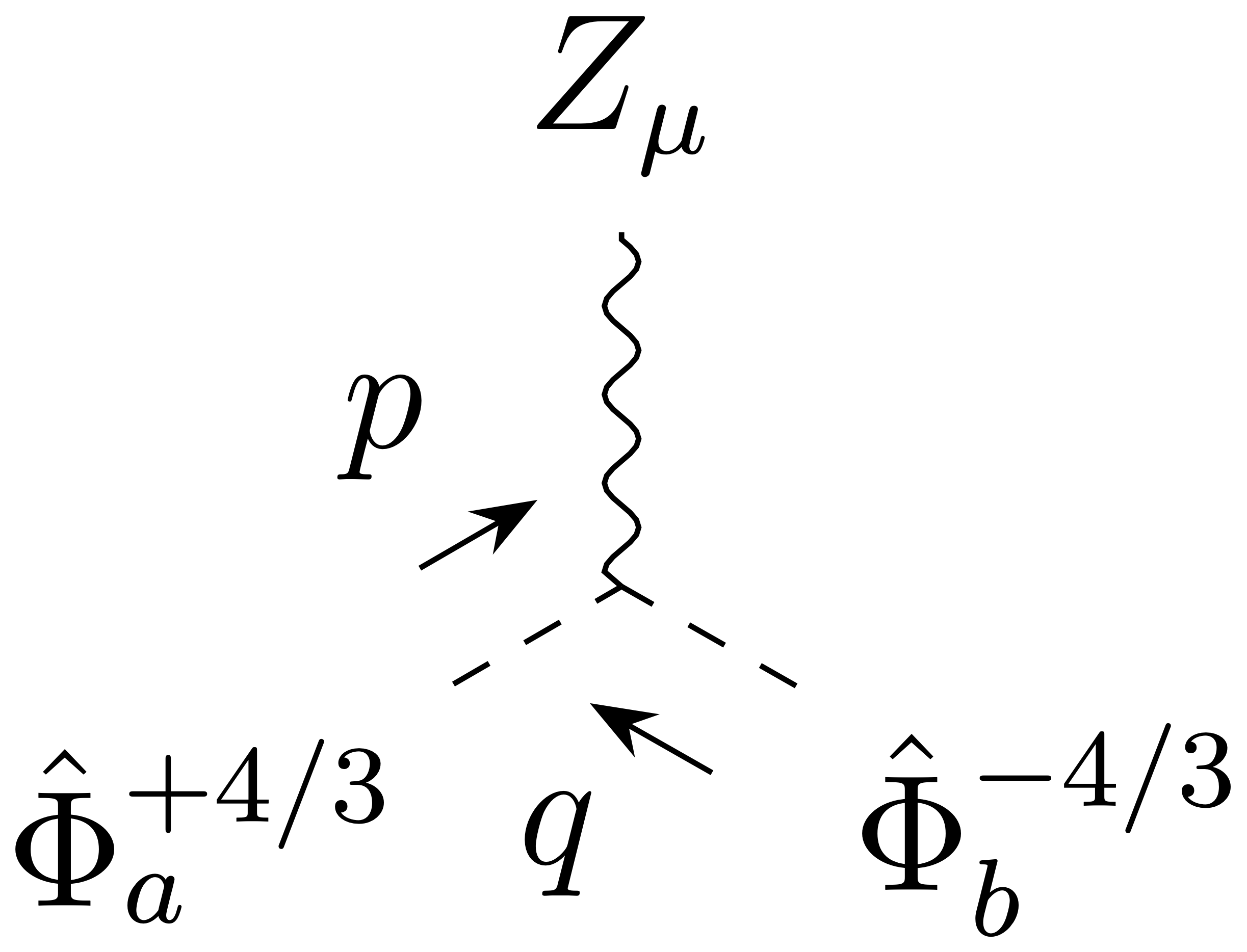}&
        $+ \ i \dfrac{e}{s_w c_w} \left(q^\mu - p^\mu \right) 
        \hat{\mathbbm{Z}}^{-4/3}
       _{n_an_b}$\\
        \hline
    \end{tabular}
\end{table}
\end{center}

\begin{center}
\begin{table}[!ht]
    \centering\begin{tabular}{>{\centering\arraybackslash}m{5cm} m{7.5cm}} 
    \multicolumn{2}{c}{Charge 4/3, Interactions with $Z$ Bosons} \\
    \hline
        Diagram & Feynman Rule \\
        \hline
        \vspace{5px}
        \includegraphics[height=3.5cm]{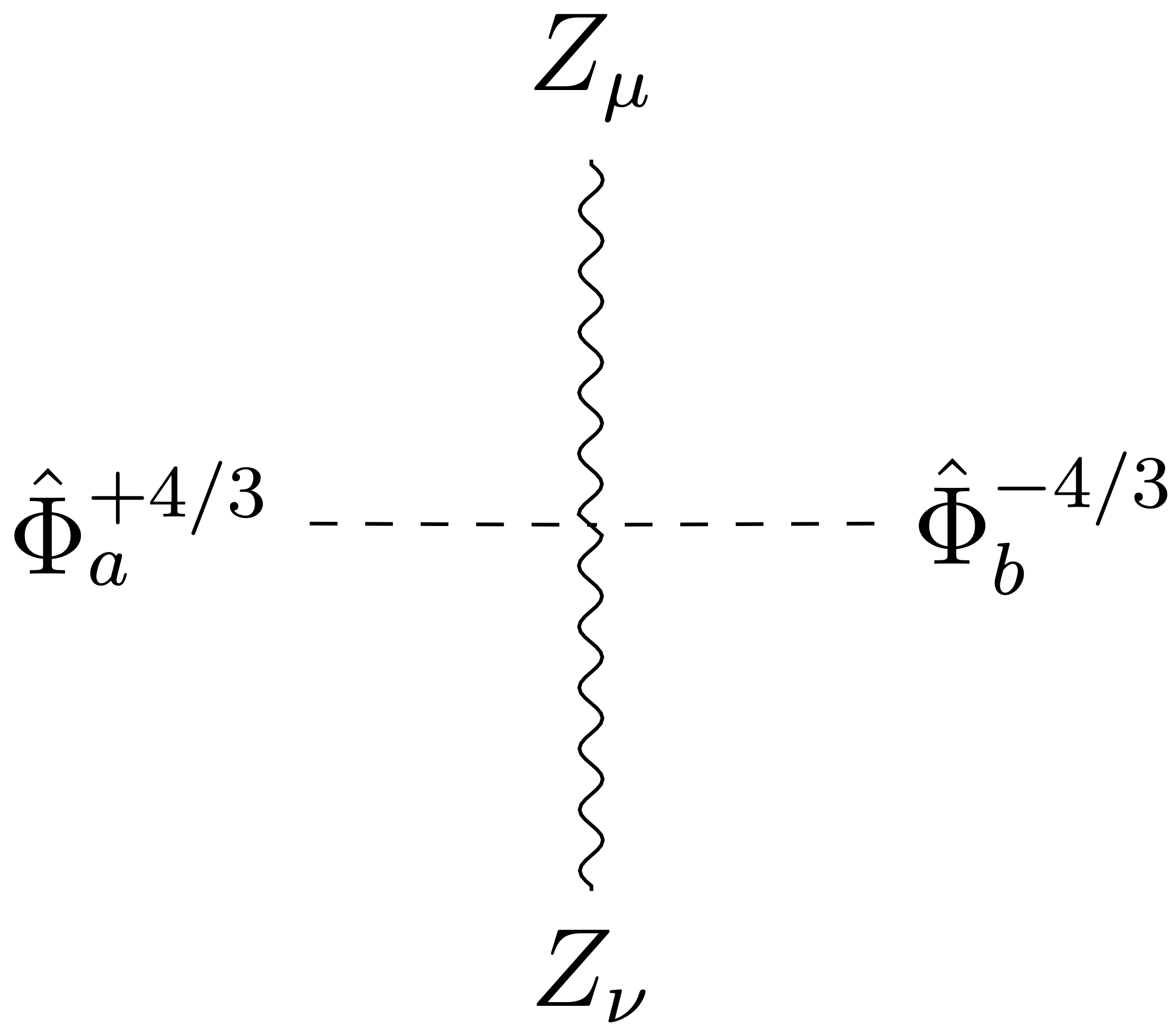}&
        $+ \ 2i \left(\dfrac{e}{s_w c_w}\right)^2 \eta^{\mu \nu}  \left(\hat{\mathbbm{Z}}^{-4/3 \dagger} \ \hat{\mathbbm{Z}}^{-4/3}\right)_{n_an_b}  $\\
        \hline
    \end{tabular}
\end{table}
\end{center}

\begin{center}
\begin{table}[!ht]
    \centering\begin{tabular}{>{\centering\arraybackslash}m{5cm} m{7.5cm}} 
    \multicolumn{2}{c}{Charge 4/3, Interactions with $W$ Bosons} \\
    \hline
        Diagram & Feynman Rule \\
        \hline
        \vspace{5px}
        \includegraphics[height=3.5cm]{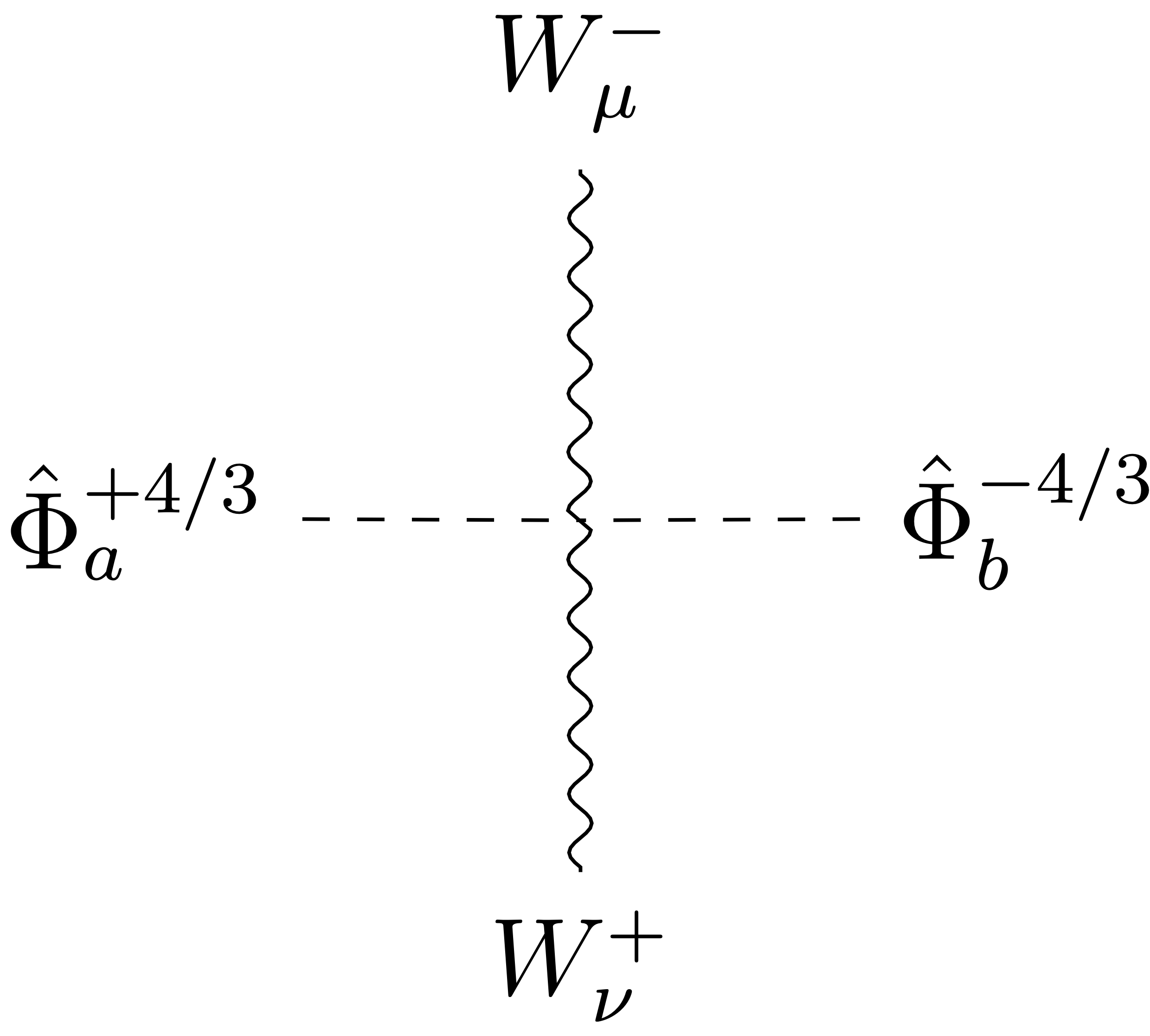}&
        $+ \ i \left(\dfrac{e}{\sqrt{2} s_w}\right)^2 \eta^{\mu \nu}  
                \left(\hat{\mathbbm{W}}^{-1/3 \dagger}_{-4/3} \ 
                \hat{\mathbbm{W}}^{-1/3}_{-4/3}\right)_{n_an_b}$ \newline \\
        \hline
    \end{tabular}
\end{table}
\end{center}

\begin{center}
\begin{table}[!ht]
    \centering\begin{tabular}{>{\centering\arraybackslash}m{5cm} m{7.5cm}} 
    \multicolumn{2}{c}{Charge 4/3, Electroweak Interactions} \\
    \hline
        Diagram & Feynman Rule \\
        \hline
        \vspace{5px}
        \includegraphics[height=3.5cm]{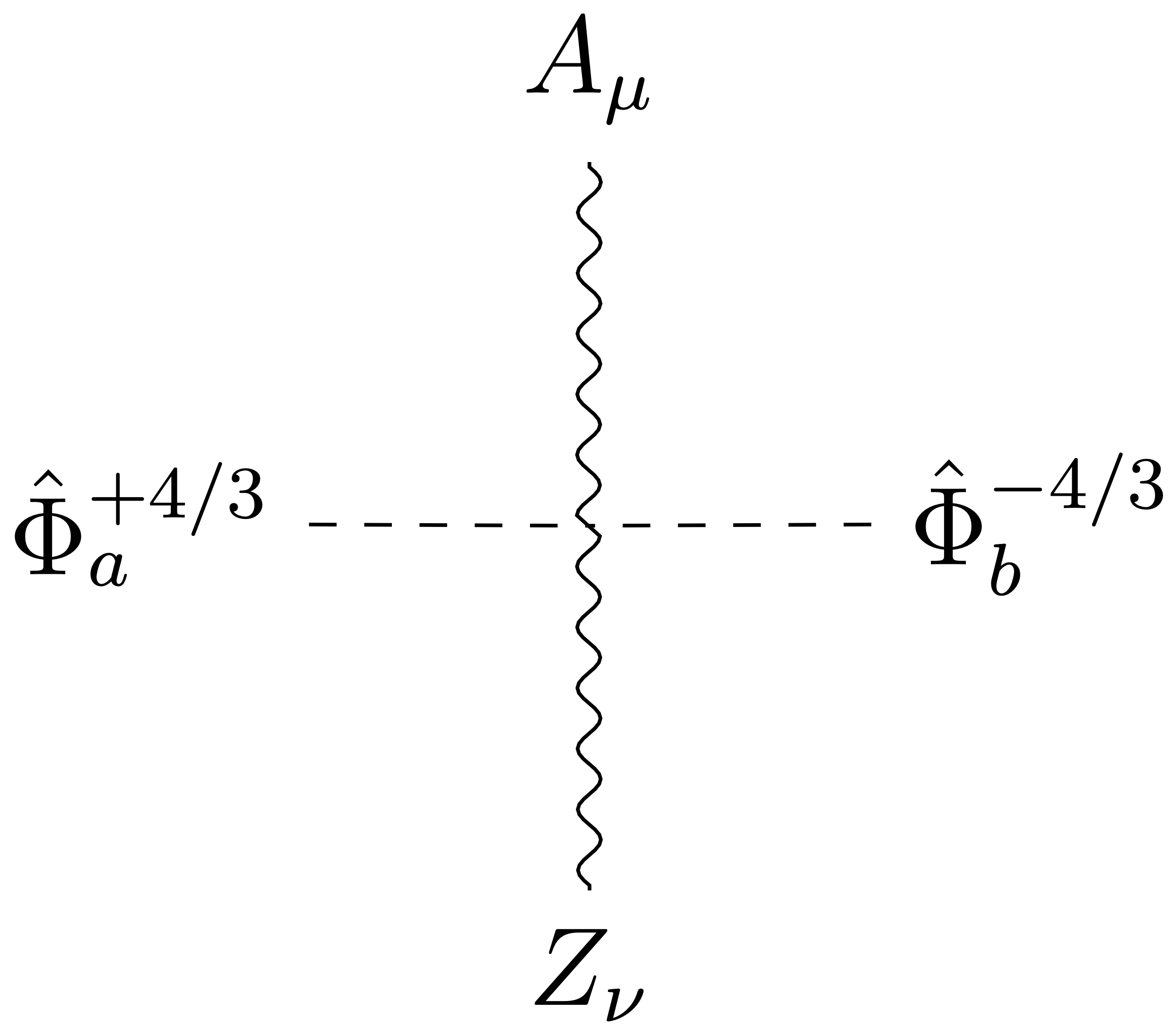}&
        $- \ \dfrac{8i}{3} \dfrac{e^2}{s_w c_w} \eta^{\mu \nu}  
                \ \hat{\mathbbm{Z}}^{-4/3}_{n_an_b}$ \newline \\
        \hline
    \end{tabular}%
\end{table}
\end{center}


\clearpage
\subsubsection{Charge 5/3}


\begin{center}
\begin{table}[!ht]
    \centering\begin{tabular}{>{\centering\arraybackslash}m{5cm} m{7.5cm}} 
    \multicolumn{2}{c}{Charge 5/3, Interactions with Photons} \\
    \hline
        Diagram & Feynman Rule \\
        \hline
        \vspace{5px}
        \includegraphics[height=2.25cm]{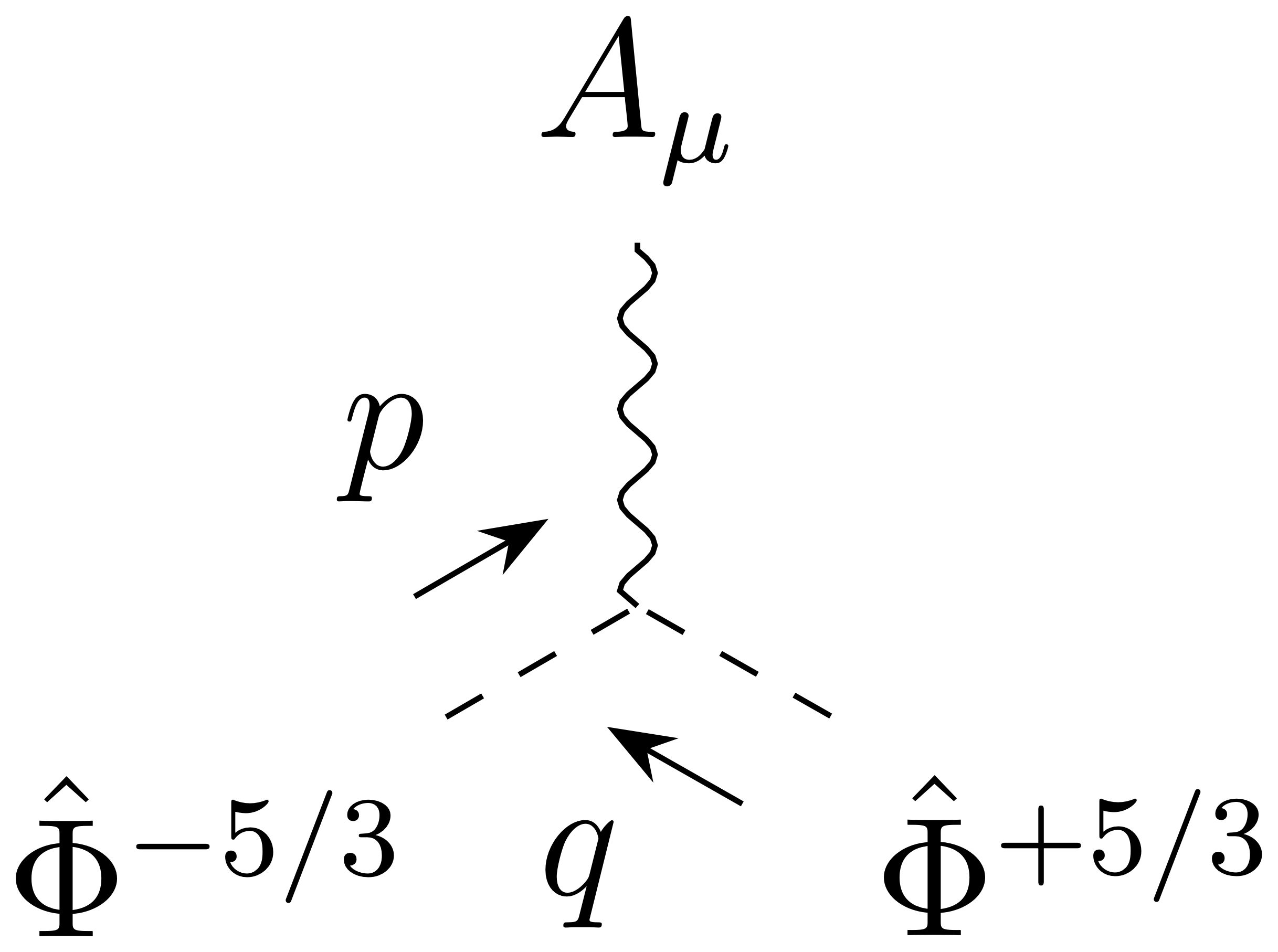}&
        $+ \ \dfrac{5i}{3} e \left(q^\mu - p^\mu \right)$\\
        \vspace{5px}
        \includegraphics[height=3.5cm]{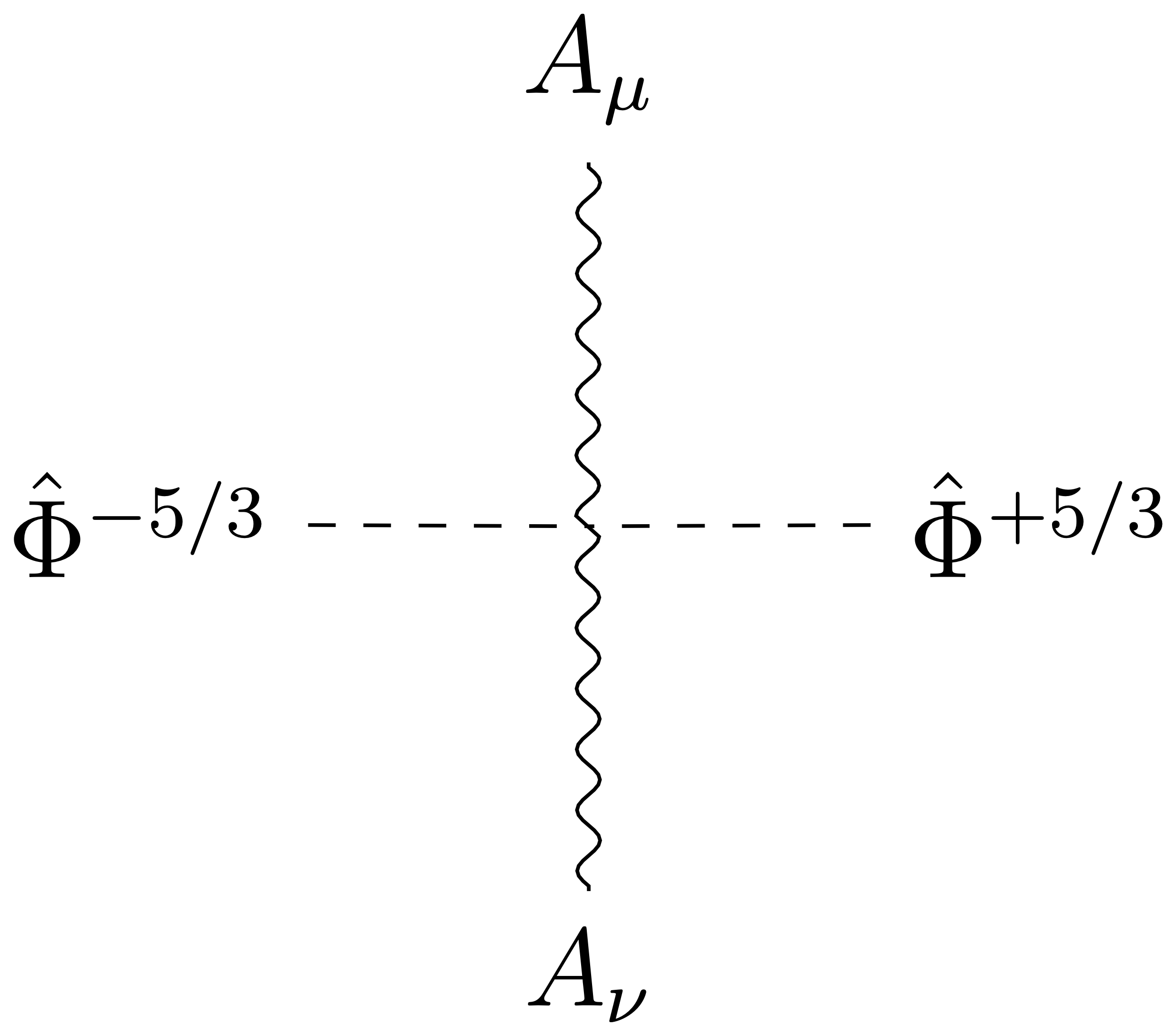}&
        $+ \ \dfrac{50i}{9} e^2 \eta^{\mu \nu} $\\
        \hline
    \end{tabular}%
\end{table}
\end{center}


\clearpage
\subsubsection{Charge 1/3 and 2/3}


\begin{center}
\begin{table}[!ht]
    \centering%
\end{table}
\end{center}


\clearpage
\subsubsection{Charge 1/3 and 4/3}


\begin{center}
\begin{table}[!ht]
    \centering%
\end{table}
\end{center}


\clearpage
\subsubsection{Charge 2/3 and 4/3}


\begin{center}
\begin{table}[!ht]
    \centering%
\end{table}
\end{center}

\subsubsection{Charge 2/3 and 5/3}


\begin{center}
\begin{table}[!ht]
    \centering%
\end{table}
\end{center}

\clearpage
\subsection{Higgs Interactions}
The Feynman rules for LQ interactions with the SM Higgs field are given below. 
\label{sec:LQHiggs}



\subsubsection{Charge 1/3}

\begin{center}
\begin{table}[!ht]
    \centering

\end{table}
\end{center}

\clearpage
 \subsubsection{Charge 2/3}

\begin{center}
\begin{table}[!ht]
    \centering%
\end{table}
\end{center}


\subsubsection{Charge 4/3}

\begin{center}
\begin{table}[!ht]
    \centering%
\end{table}
\end{center}

\subsubsection{Charge 5/3}
         
\begin{center}
\begin{table}[!ht]
    \centering%
\end{table}
\end{center}

\clearpage
\subsection{Goldstone Interactions}
The Feynman rules for LQ interactions with the SM Goldstone fields are given below. 


\subsubsection{Charge 1/3}

\begin{center}
\begin{table}[!ht]
    \centering%
\end{table}
\end{center}

 \clearpage        
\subsubsection{Charge 2/3}

\begin{center}
\begin{table}[!ht]
    \centering%
\end{table}
\end{center}

\clearpage
\subsubsection{Charge 4/3}

\begin{center}
\begin{table}[!ht]
    \centering%
\end{table}
\end{center}

\clearpage
\subsubsection{Charge 5/3}
         
\begin{center}
\begin{table}[!ht]
    \centering%
\end{table}
\end{center}

\clearpage
\subsubsection{Charge 1/3 and 2/3}

\begin{center}
\begin{table}[!ht]
    \centering%
\end{table}
\end{center}

\clearpage
\subsubsection{Charge 1/3 and 4/3}

\begin{center}
\begin{table}[!ht]
    \centering%
\end{table}
\end{center}
        
\clearpage

\subsubsection{Charge 1/3 and 5/3}

\begin{center}
\begin{table}[!ht]
    \centering%
\end{table}
\end{center}

\subsubsection{Charge 2/3 and 4/3}

\begin{center}
\begin{table}[!ht]
    \centering%
\end{table}
\end{center}

\clearpage
\subsubsection{Charge 2/3 and 5/3}

\begin{center}
\begin{table}[!ht]
    \centering%
\end{table}
\end{center}

\clearpage
\subsection{LQ Triple Interactions}
The LQ self-interactions involving three LQs are given below. 

\begin{center}
\begin{table}[!ht]
    \centering%
\end{table}
\end{center}

\clearpage
\subsection{LQ Quartic Interactions}
\label{sec:LQQuartic}
In the following, the Feynman rules for the LQ self-interactions involving four LQs are listed. If $n_a$ or $n_b$ are equal to zero, $W^q_{n_a n_b}$ should be set to zero (i.e.~the corresponding term neglected). 

\begin{center}
\begin{table}[!ht]
    \centering\begin{tabular}{>{\centering\arraybackslash}m{4.5cm} m{8.5cm}} 
    \multicolumn{2}{c}{Quartic Interactions} \\
    \hline
        Diagram  & Feynman Rule \\
\hline
        \end{tabular}%
\end{table}
\end{center}

\clearpage
\section{Conclusions}
\label{sec:Conclusions}
In this article we provided the complete scalar LQ Lagrangian, including all interactions with SM fermions, gauge bosons and the Higgs. After EW symmetry breaking the mass terms of the Lagrangian are diagonalized, leading to mixing among different LQ representations. We present all Feynman rules, including for the first time the LQ-LQ-LQ(-Higgs) and LQ-LQ-LQ-LQ interactions. On the computational side, we provided a FeynRules model file containing the complete scalar LQ Lagrangian, as well as a Mathematica notebook to access it. The export to MadGraph and FeynArts is possible and the corresponding files are provided, rendering this a powerful tool for the automatization of studies of scalar LQ phenomenology.

\appendix
\clearpage
\section{Charge-Conjugate SM Fermions}
\label{sec:appendix}

\subsection{Treatment of Charge-Conjugate SM Fermions}
Our treatment of interactions involving charge-conjugate SM fermions is based on the method of Ref.~\cite{Denner:1992vza}. It uses the properties of the charge-conjugation matrix $C$ (introduced in Eq.~(\ref{eq:Cmatrix}))~\cite{Denner:1992vza, Nieves:2003in}
\begin{equation}
\begin{aligned}
	C^\dagger &= C^{-1} \,, \\
	C^\intercal &= -C \,, \\
	C \Gamma_i^\intercal C^{-1} &= \eta_i \Gamma_i \equiv \Gamma_i^\prime \,,
\end{aligned}
\label{eq:Cequations}
\end{equation}
with $\Gamma_i = 1, i\gamma^5, \gamma^\mu, \gamma^\mu \gamma^5, \sigma^{\mu \nu}$ and 
\begin{equation}
\eta_i =     \begin{cases}
      +1, & \text{ for } \Gamma_i = 1, i\gamma^5, \gamma^\mu \gamma^5 \\
      -1, & \text{ for } \Gamma_i = \gamma^\mu, \sigma^{\mu\nu}
    \end{cases}
\end{equation}
to derive two equivalent sets of Feynman rules for internal propagators, external lines and vertices involving (anti-)fermions. The first set corresponds to the expressions 
$\bar{\psi}_1 \Gamma_i \psi_2$ that appear in the Lagrangian, the second set to their charge-conjugate analogues $\bar{\psi}_2^c \Gamma^\prime_i \psi_1^c$. The two are equivalent 
\begin{equation}
\bar{\psi}_1 \Gamma_i \psi_2 = \bar{\psi}_2^c \Gamma^\prime_i \psi_1^c \,.
\end{equation}
This double allocation allows for an intuitive treatment of interactions involving charge-conjugate fermion fields. While the underlying principle is the same, our notation differs from Ref.~\cite{Denner:1992vza}. We use two types of arrows on the fermion lines (black and white ones) instead of additional arrows next to them. This avoids potential confusion with momentum arrows as well as having clashing fermion arrows. We proceed by stating the recipe for calculating Feynman amplitudes for processes involving charge-conjugate fermions, which we demonstrate using an example. The Feynman rules for diagrams featuring charge-conjugate SM fermions only are given in \ref{sec:CCFeynman}. 

\clearpage
\subsubsection*{Prescription for the Evaluation of Feynman diagrams} 
\label{sec:CCprescription}
The amplitudes for specific processes are determined according to the following prescription:

\begin{itemize}
\item Draw all Feynman diagrams for a given process. Two fermion lines with clashing arrows (i.e.~one line ending in a fermion and the other in its charge-conjugate) may be connected by inverting the direction and the colors of the arrows for one of the two lines. It does not matter which of the two lines are inverted. 

\item For each internal propagator, external line and vertex write down the appropriate analytic expressions proceeding opposite to the fermion flow. The Feynman rules for normal fermions (black arrows) are given in Sec.~\ref{sec:SMInteractions}, the rules for charge-conjugated fermions (white arrows) in \ref{sec:CCFeynman} and the rules involving both in Sec.~\ref{sec:LQfermions}. 

\item Multiply by a factor $(-1)$ for every closed loop.  

\item Multiply by the permutation parity of the spinors in the obtained analytical expression with respect to some reference order.
\end{itemize}
The evaluation of the bosonic parts of a diagram proceed as usual. 

\subsubsection*{Example} 
\label{sec:CCexample}
In order to exemplify our Feynman rules, we calculate the Feynman amplitude for the process $e\nu_\mu \to \Phi_1^{-1/3} \Phi_2^{-2/3}$ with an up quark $t$-channel contribution. A similar example is given in Sec.~4 of Ref.~\cite{Denner:1992vza}, which allows for a direct comparison of our notation to theirs. For simplicity, we neglect the mixing among the LQs and take the SM fermions to be massless. \\

\begin{figure}[ht!]
     \centering
     \begin{subfigure}[b]{0.3\textwidth}
         \centering
         \includegraphics[width=\textwidth]{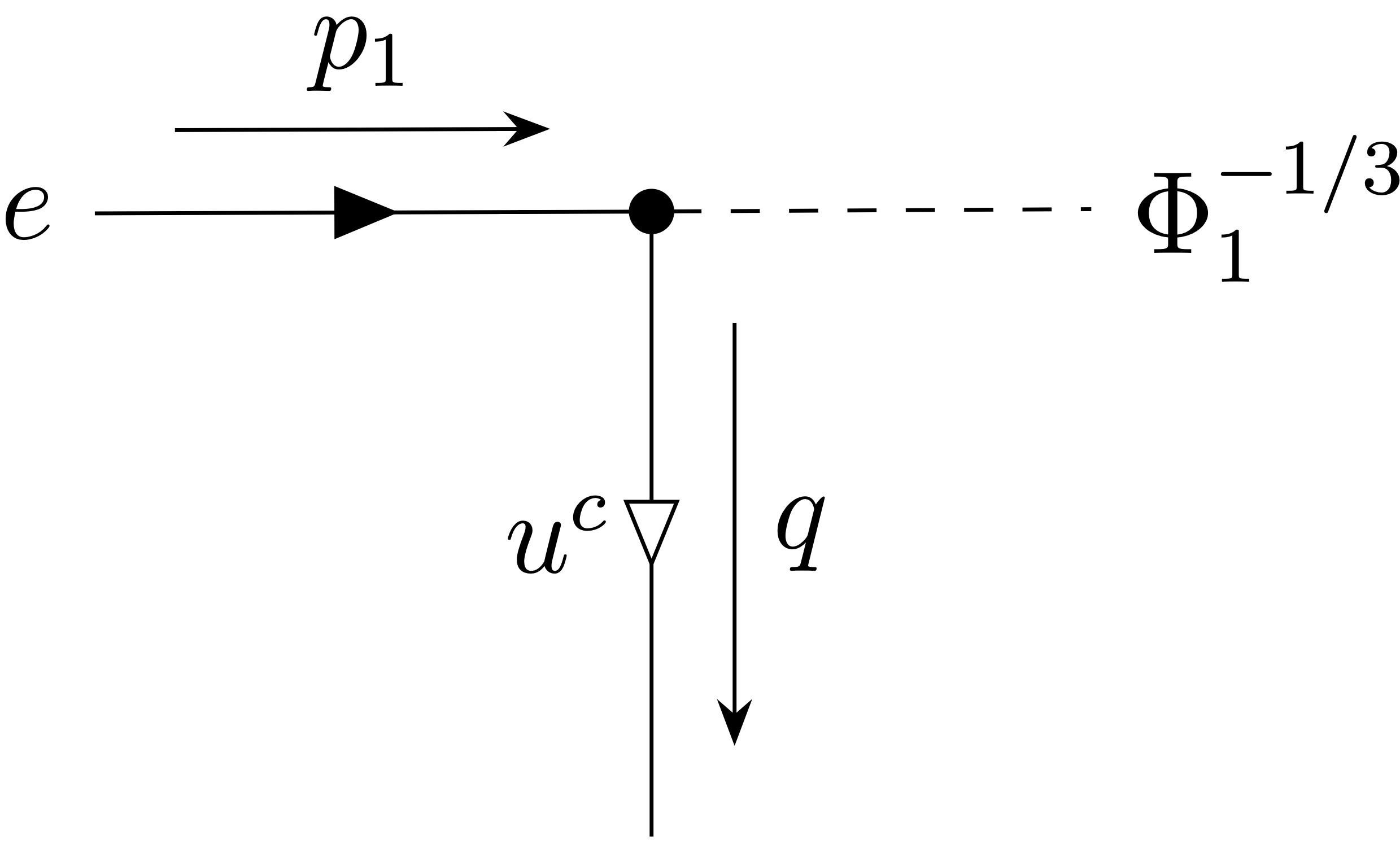} \newline
         \includegraphics[width=\textwidth]{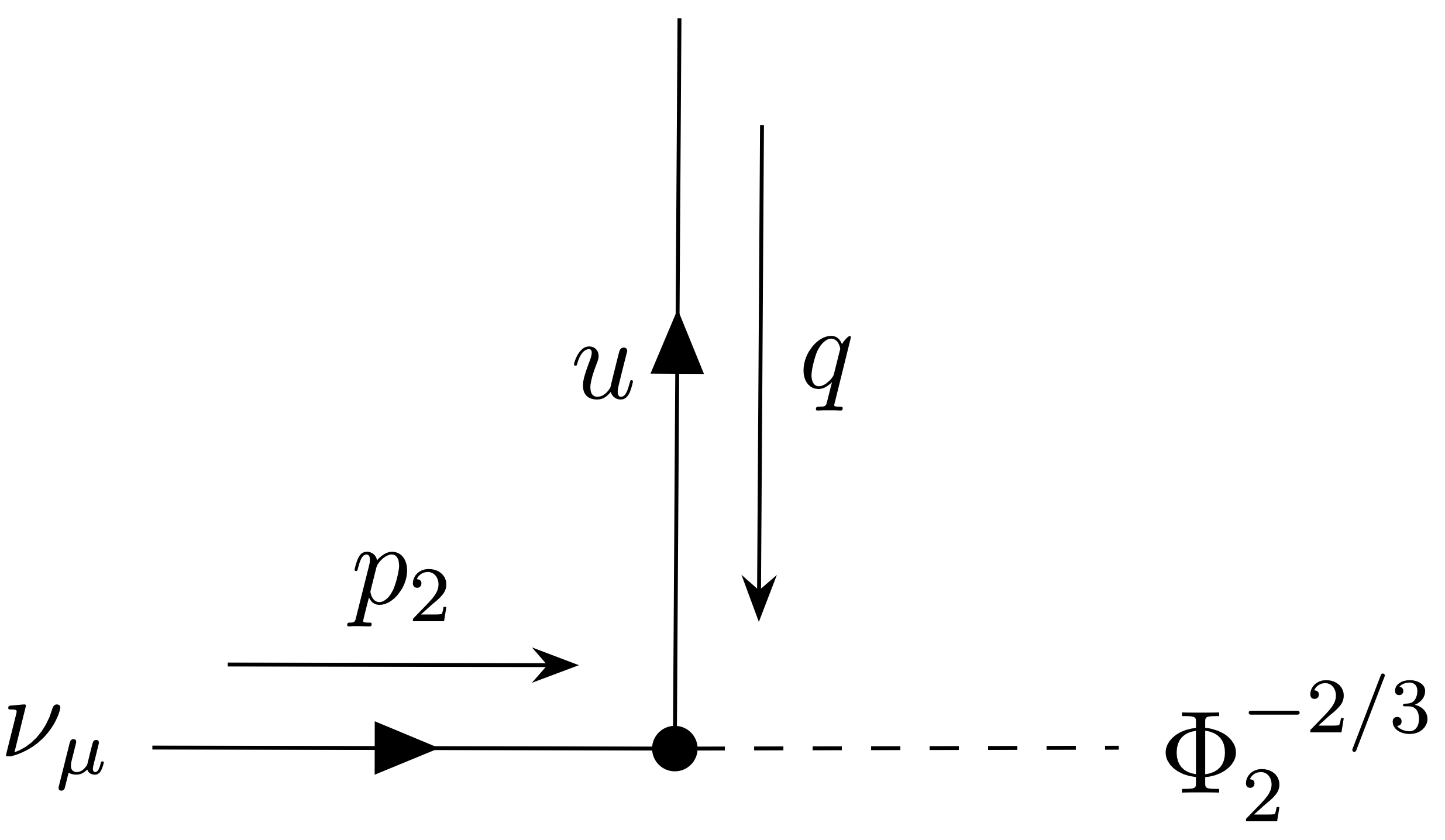}
         \caption{}
         \label{fig:Example1a}
     \end{subfigure}
     \hfill
     \begin{subfigure}[b]{0.3\textwidth}
         \centering
         \includegraphics[width=\textwidth]{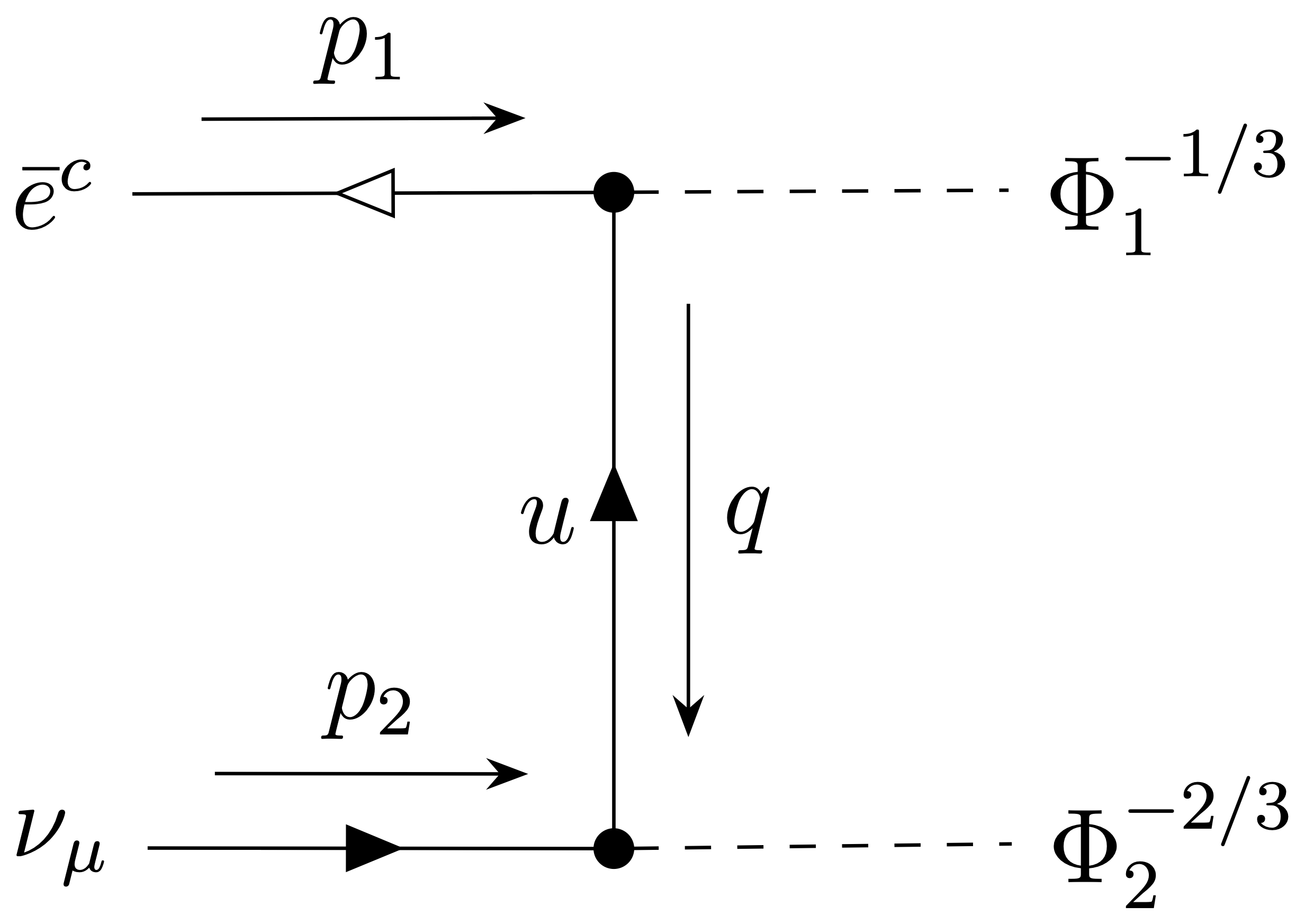}
         \caption{}
         \label{fig:Example1b}
     \end{subfigure}
     \hfill
     \begin{subfigure}[b]{0.3\textwidth}
         \centering
         \includegraphics[width=\textwidth]{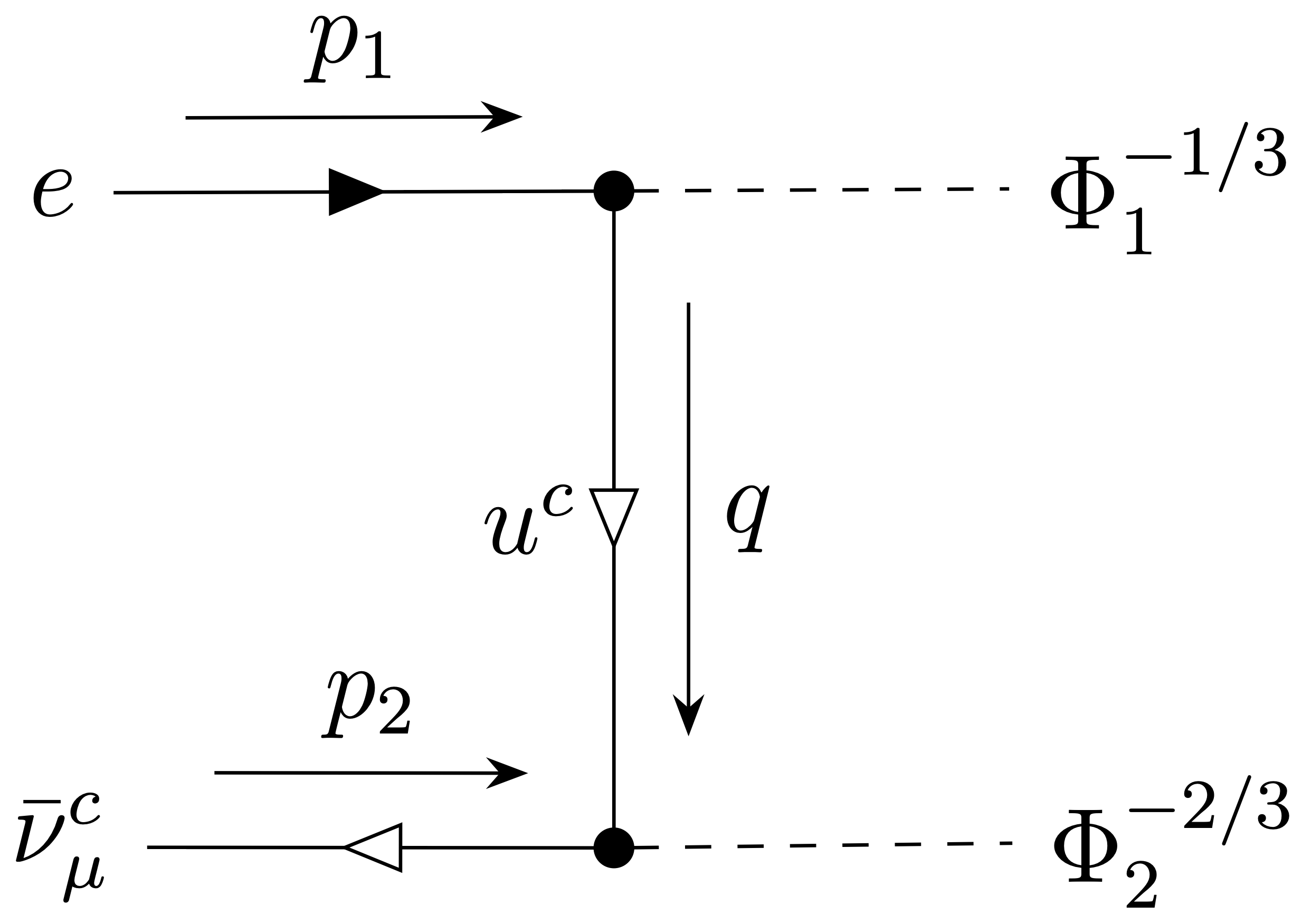}
         \caption{}
         \label{fig:Example1c}
     \end{subfigure}
        \caption{Feynman diagrams for the process $e\nu_\mu \to \Phi_1^{-1/3} \Phi_2^{-2/3}$ involving a $t$-channel up quark. The up quark fermion lines with clashing arrows shown in the diagram on the left can be combined by inverting the fermion arrows (color and direction) of one of the two subdiagrams. This yields the two equivalent diagrams on the right.  }
        \label{fig:Example1}
\end{figure}

The Feynman diagram is shown in Fig.~\ref{fig:Example1}. Fig.~\ref{fig:Example1a} shows two parts of the diagram with clashing fermion line arrows for the up quark. In order to combine the two parts, the fermion arrows of one of them have to be inverted. In Fig.~\ref{fig:Example1b}, the arrows of the upper part are inverted, in Fig.~\ref{fig:Example1c} the ones of the lower part. In the following, we will derive the analytic expressions for both of them, showing their equivalence. Choosing the reference order $(1,2)$ for the external states, the Feynman amplitude for the diagram in Fig.~\ref{fig:Example1b} reads
\begin{equation}
\begin{aligned}
i \mathcal{M} &= \bar{v}^{(s_1)}(\vec{p}_1) \left(+i Y_{1, 11}^{LL} \text{P}_\text{L} +i Y_{1, 11}^{RR} \text{P}_\text{R}   \right) \frac{-i\slashed{q}}{q^2 + i\epsilon} \left(-i Y_{2, 12}^{RL} \text{P}_\text{L} \right) u^{(s_2)}(\vec{p}_2) \\
&= -i \frac{Y_{1, 11}^{RR} Y_{2, 12}^{RL}}{q^2 + i\epsilon} \bar{v}^{(s_1)}(\vec{p}_1) \ \slashed{q} \text{P}_\text{L} \  u^{(s_2)}(\vec{p}_2)\,.
\end{aligned}
\end{equation}
The amplitude for the diagram in Fig.~\ref{fig:Example1c} is
\begin{equation}
\begin{aligned}
i \mathcal{M}^\prime &= (-1) \bar{v}^{(s_2)}(\vec{p}_2) \left(-i Y_{2, 12}^{RL} \text{P}_\text{L} \right) \frac{i\slashed{q}}{q^2 + i\epsilon} \left(+i Y_{1, 11}^{LL} \text{P}_\text{L} +i Y_{1, 11}^{RR} \text{P}_\text{R}   \right) u^{(s_1)}(\vec{p}_1) \\
&= -i \frac{Y_{1, 11}^{RR} Y_{2, 12}^{RL} }{q^2 + i\epsilon}  \bar{v}^{(s_2)}(\vec{p}_2) \ \slashed{q} \text{P}_\text{R} \  u^{(s_1)}(\vec{p}_1)\,,
\end{aligned}
\end{equation}
where the factor $(-1)$ is added since the order of the external states is now reversed. The two amplitudes are equivalent, since 
\begin{equation}
\begin{aligned}
i\mathcal{M} &\propto \bar{v}^{(s_1)}(\vec{p}_1) \ \slashed{q} \text{P}_\text{L} \  u^{(s_2)}(\vec{p}_2) = q_\mu \bar{v}^{(s_1)}(\vec{p}_1) \ \frac{\gamma^\mu - \gamma^\mu \gamma^5}{2} \  u^{(s_2)}(\vec{p}_2) \\
&= q_\mu  u^{(s_2)}(\vec{p}_2)^\intercal \ \frac{\gamma^{\mu \intercal} - \left(\gamma^{\mu} \gamma^{5} \right)^\intercal }{2} \  \bar{v}^{(s_1)}(\vec{p}_1)^\intercal \\
&=  q_\mu  u^{(s_2)}(\vec{p}_2)^\intercal \ C^{-1} \frac{-\gamma^{\mu} -  \gamma^{\mu}\gamma^{5}  }{2}C \  \bar{v}^{(s_1)}(\vec{p}_1)^\intercal \\
&= q_\mu \bar{v}^{(s_2)}(\vec{p}_2) \  \frac{\gamma^{\mu} +  \gamma^{\mu}\gamma^{5}  }{2} \ u^{(s_1)}(\vec{p}_1) =  \bar{v}^{(s_2)}(\vec{p}_2) \ \slashed{q} \text{P}_\text{R} \ u^{(s_1)}(\vec{p}_1) \propto i\mathcal{M}^\prime\,,
\end{aligned}
\end{equation}
where we have used Eq.~(\ref{eq:Cequations}) and
\begin{equation}
\begin{aligned}
u^{(s)}(\vec{p}) &= C \bar{v}^{(s)}(\vec{p})^\intercal\,, \\
v^{(s)}(\vec{p}) &= C \bar{u}^{(s)}(\vec{p})^\intercal
\end{aligned}
\end{equation}
for spinors $u, v$, which can be derived from Eq.~(\ref{eq:Cmatrix}). This equivalence is true for all Feynman diagrams involving charge-conjugate fermions, as argued in Ref.~\cite{Denner:1992vza}.

\clearpage
\subsection{Charge-Conjugate SM Fermion Feynman Rules}
\label{sec:CCFeynman}
\subsubsection*{External Fields}
The Feynman rules for the external charge-conjugate fermions are listed in this section. See Ref.~\cite{Denner:1992vza} for details. 

\begin{center}
\begin{table}[!h]
    \centering\begin{tabular}{>{\centering\arraybackslash}m{3cm} m{9.5cm}}
    \multicolumn{2}{c}{External Fermions} \\
    \hline
        Diagram & Feynman Rule \\
        \hline
        \vspace{10 px}
        \includegraphics[scale=0.025]{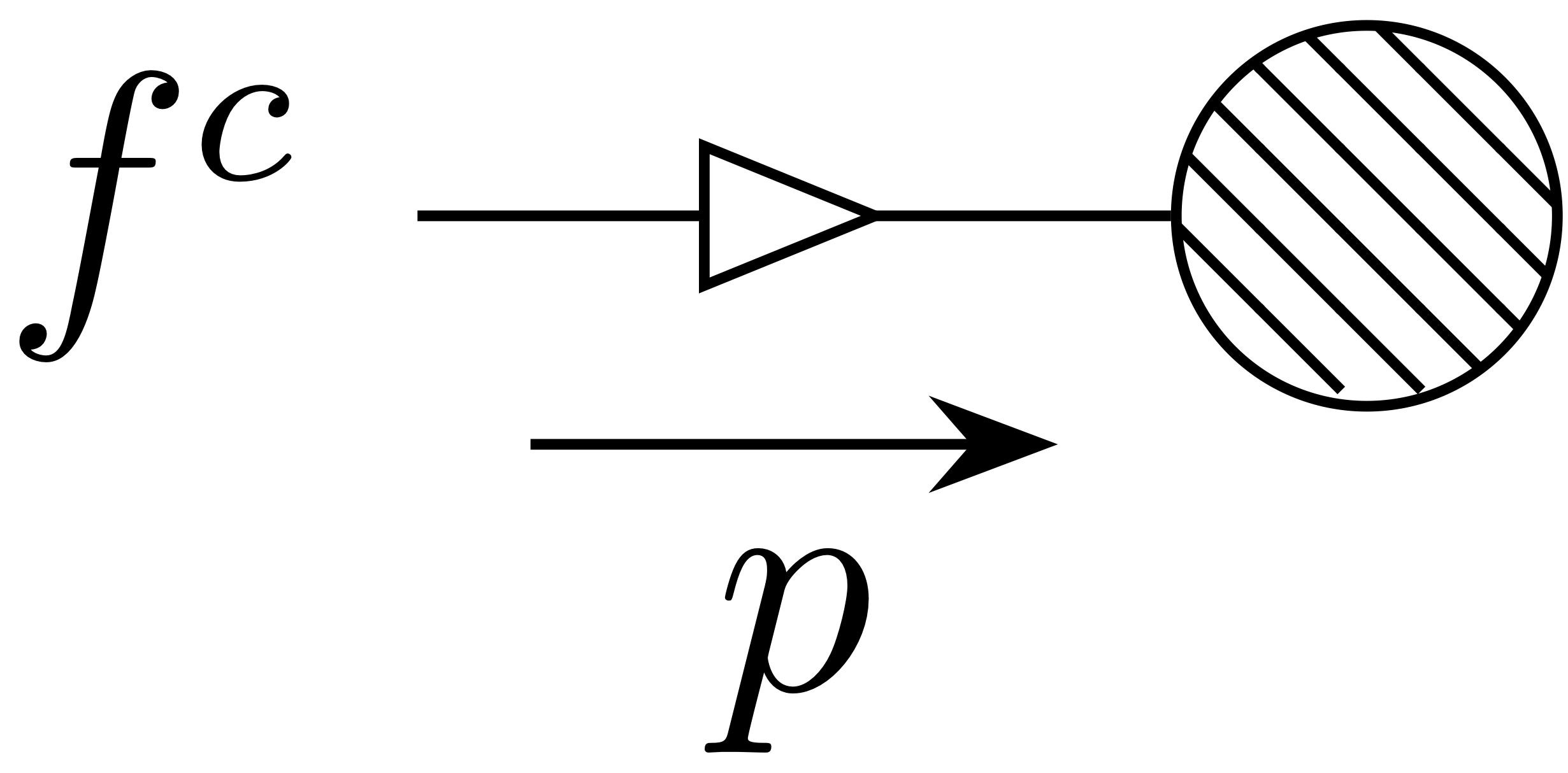} &
        \vspace{5px}
        $u^{(s)}(\vec{p})$ \newline \\ 
        \vspace{10 px}
        \includegraphics[scale=0.025]{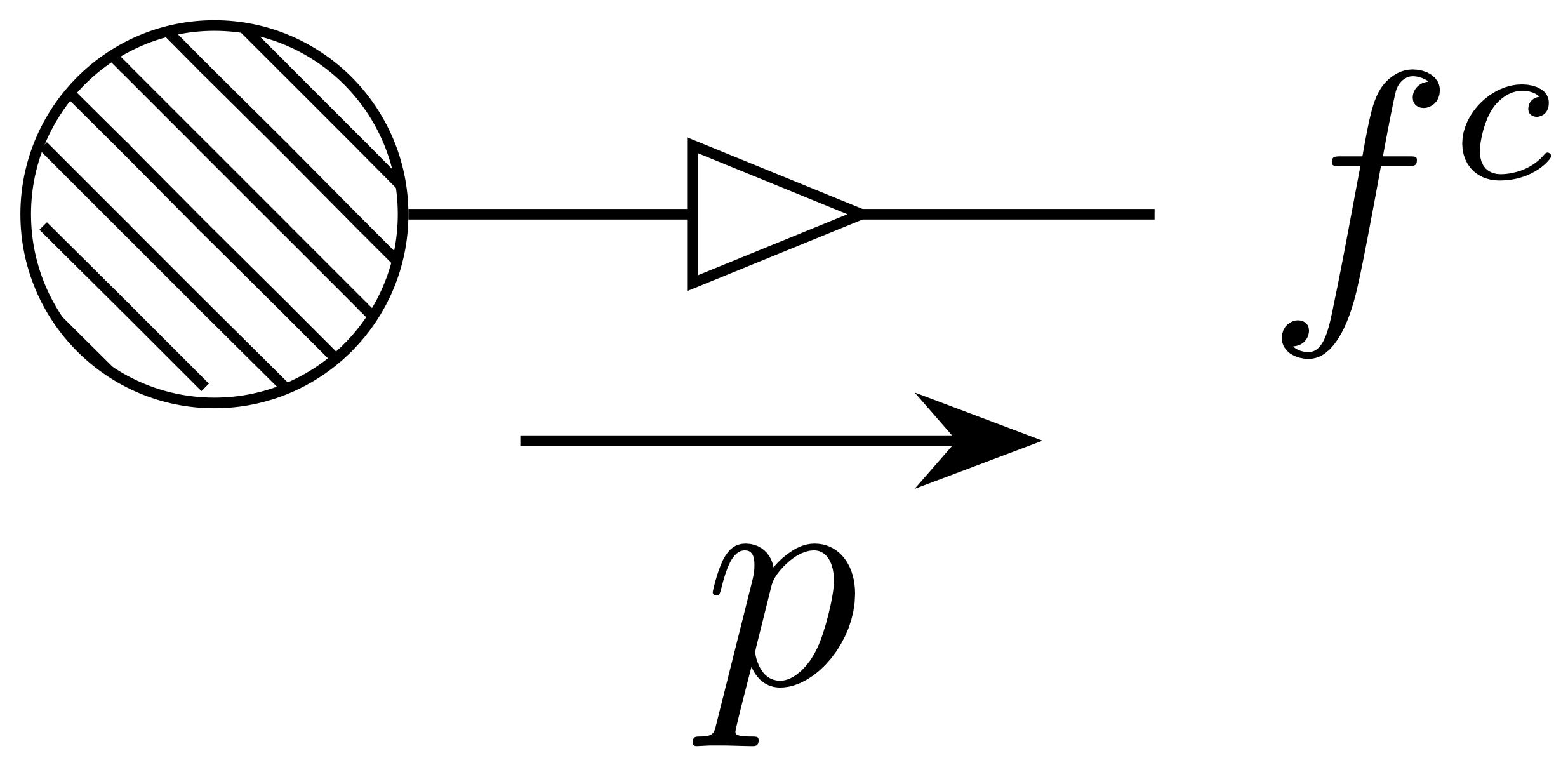} &
        \vspace{5px}
        $\bar{u}^{(s)}(\vec{p})$ \newline \\
        \vspace{10 px}
        \includegraphics[scale=0.025]{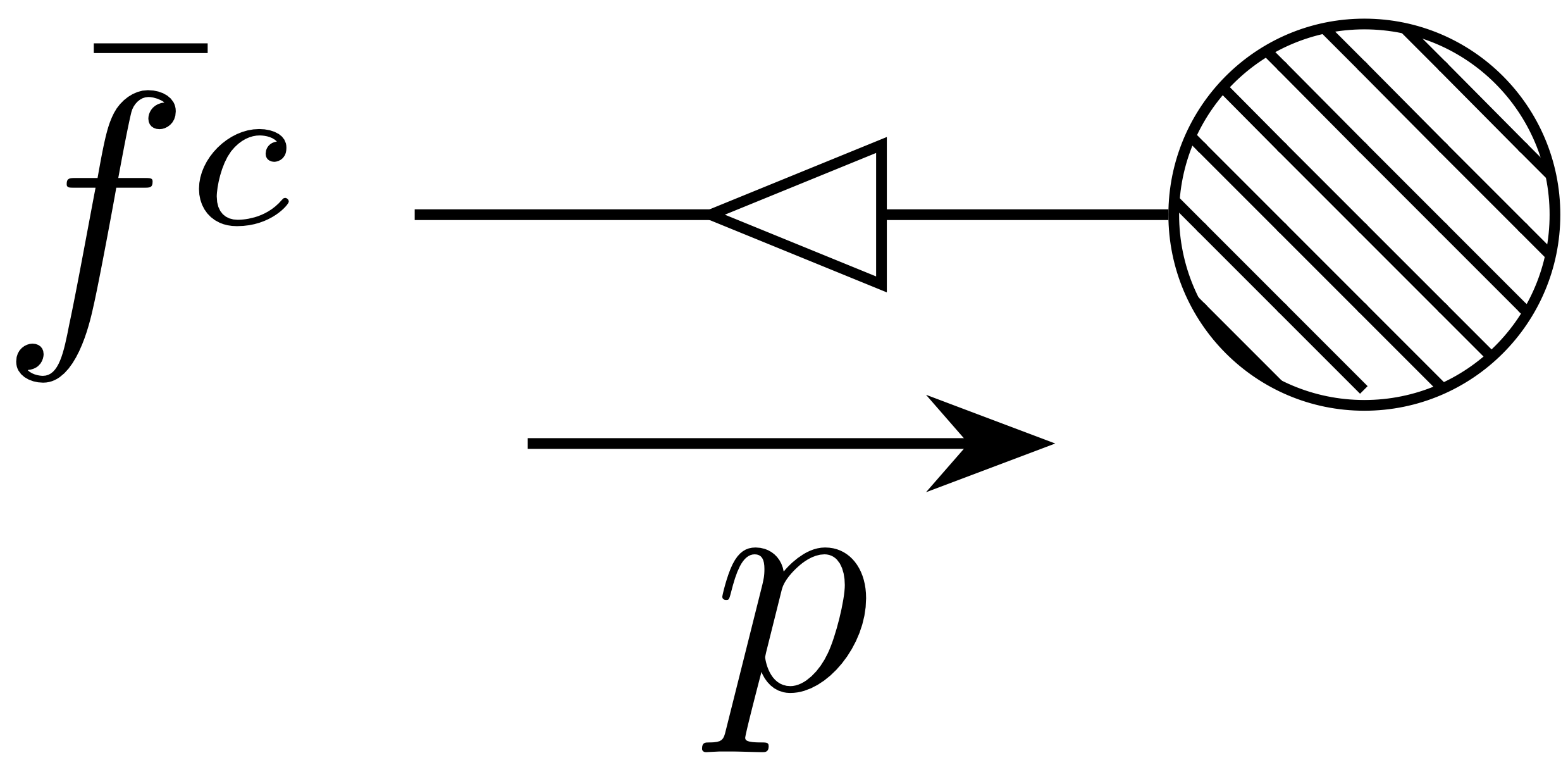} &
        \vspace{5px}
        $\bar{v}^{(s)}(\vec{p})$ \newline \\
        \vspace{10 px}
        \includegraphics[scale=0.025]{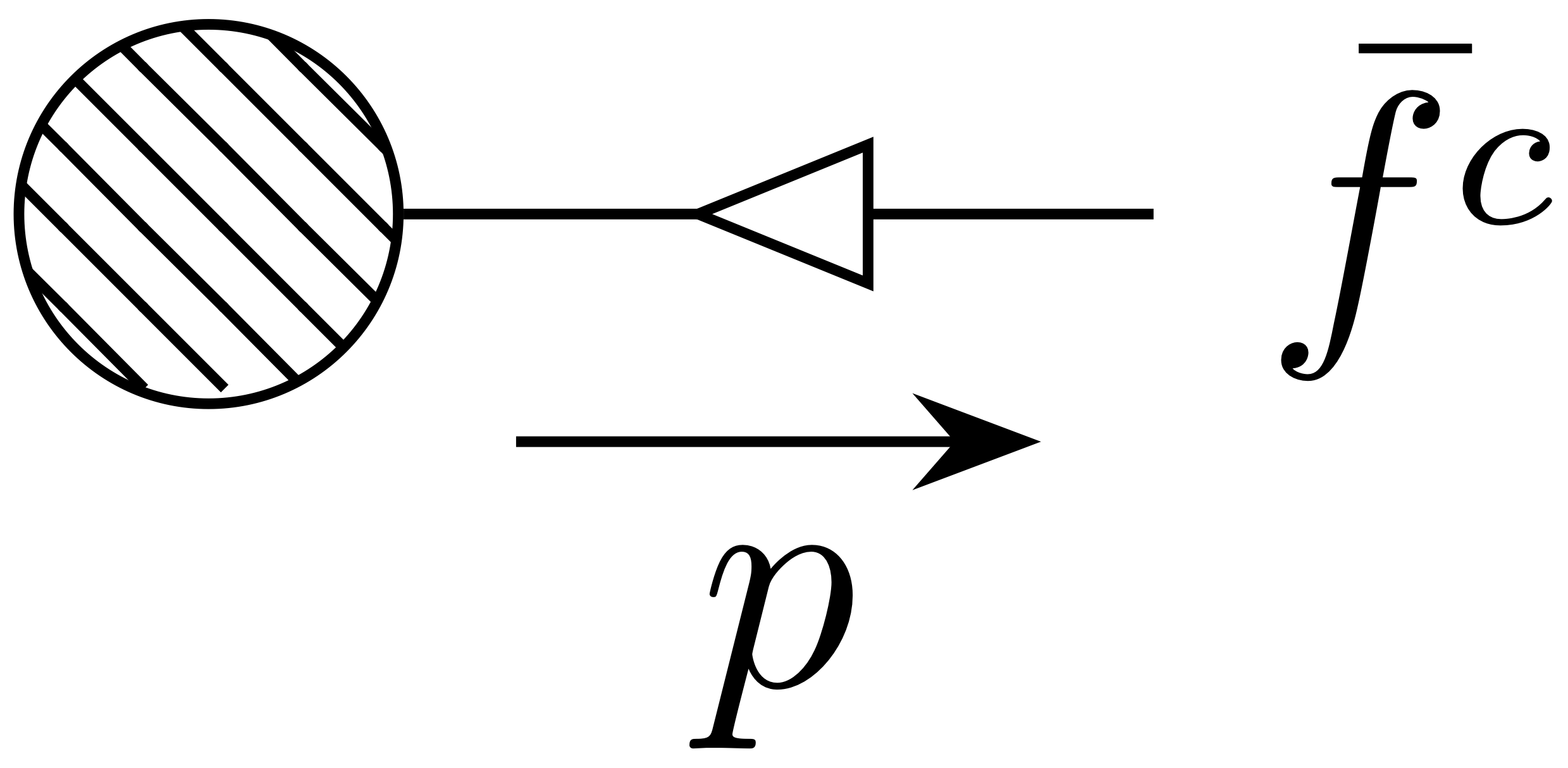} &
        \vspace{5px}
        $v^{(s)}(\vec{p})$ \newline \\
        \hline
        \hline
    \end{tabular}
\end{table}
\end{center}

\clearpage
\subsubsection*{Propagators}
The propagators for the charge-conjugate SM fermions are given below. The fermion masses $m_{\ell_i}, m_{u_i}, m_{d_i}$ are defined in Eq.~(\ref{eq:fermionMasses}).

\begin{center}
\begin{table}[!h]
    \centering\begin{tabular}{>{\centering\arraybackslash}m{3.5cm} m{9cm}} 
        \multicolumn{2}{c}{Fermion Propagators} \\
    \hline
        Diagram & Feynman Rule \\
        \hline
        \vspace{10 px}
        \includegraphics[scale=0.035]{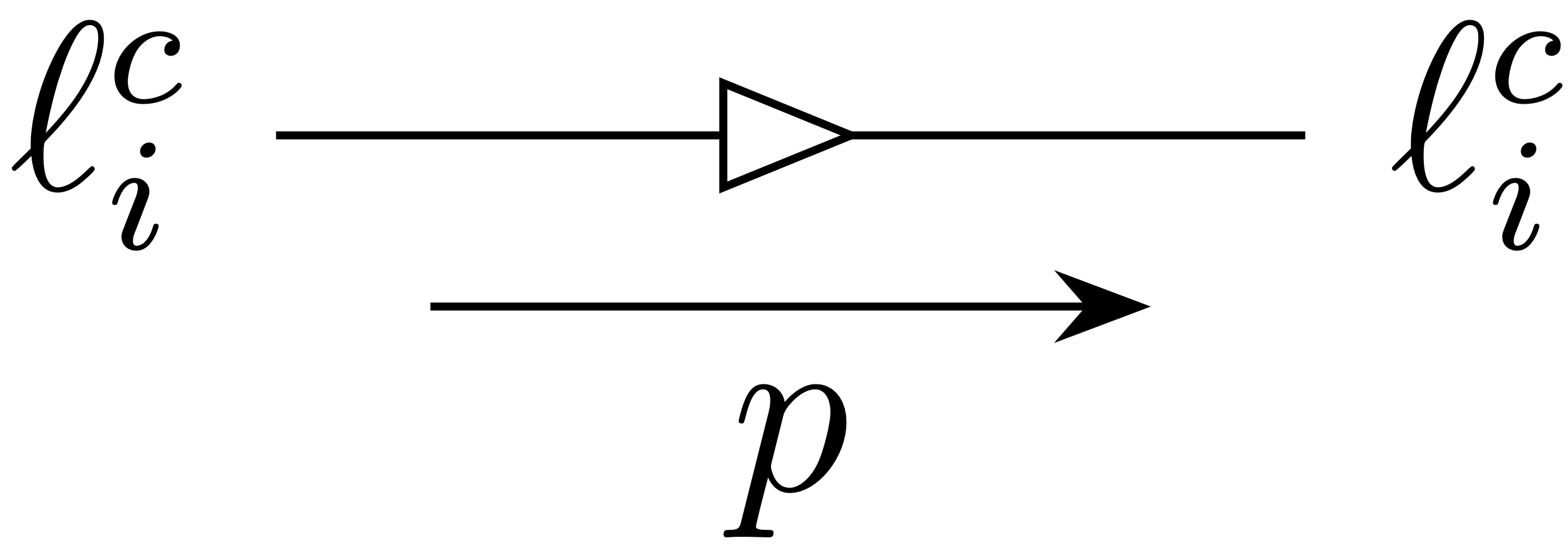}&
        \vspace{10 px}
        $\dfrac{i \left(\slashed{p} + m_{\ell_i}\right)}{p^2 - m^2_{\ell_i} +i\epsilon}$ \newline \\
        \vspace{10 px}
        \includegraphics[scale=0.035]{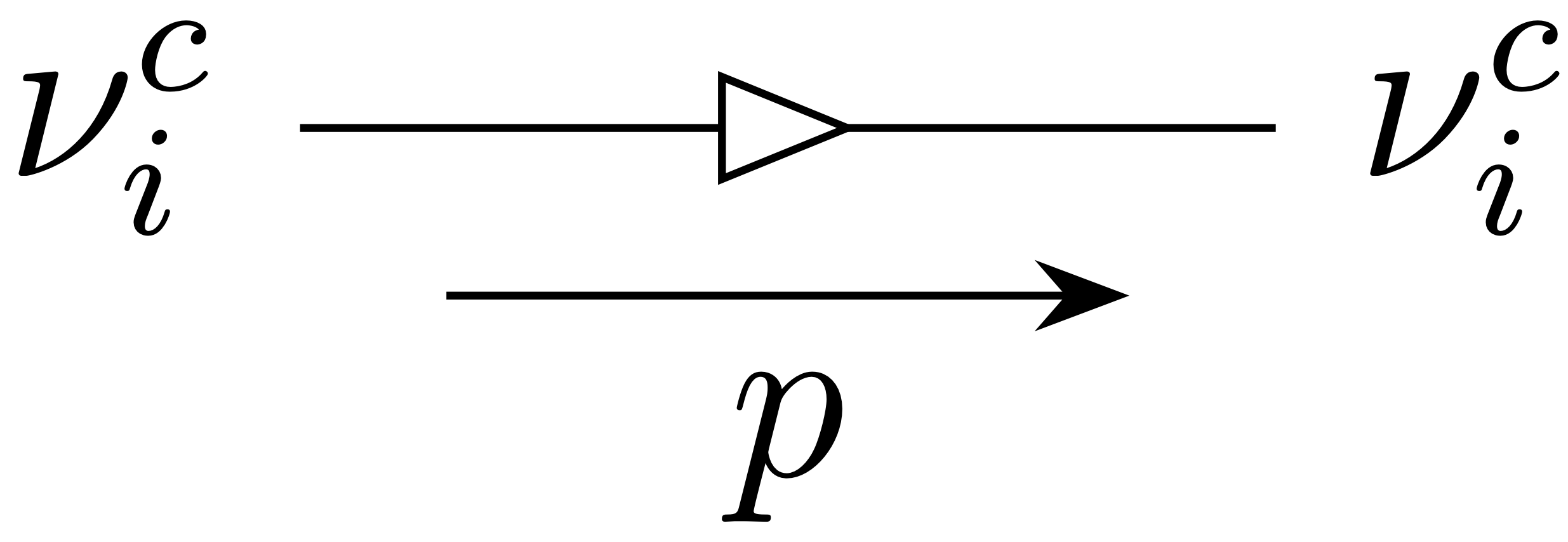}&
        \vspace{10 px}
         $\dfrac{i \slashed{p} }{p^2 +i\epsilon}$ \newline \\
        \vspace{10 px}
        \includegraphics[scale=0.035]{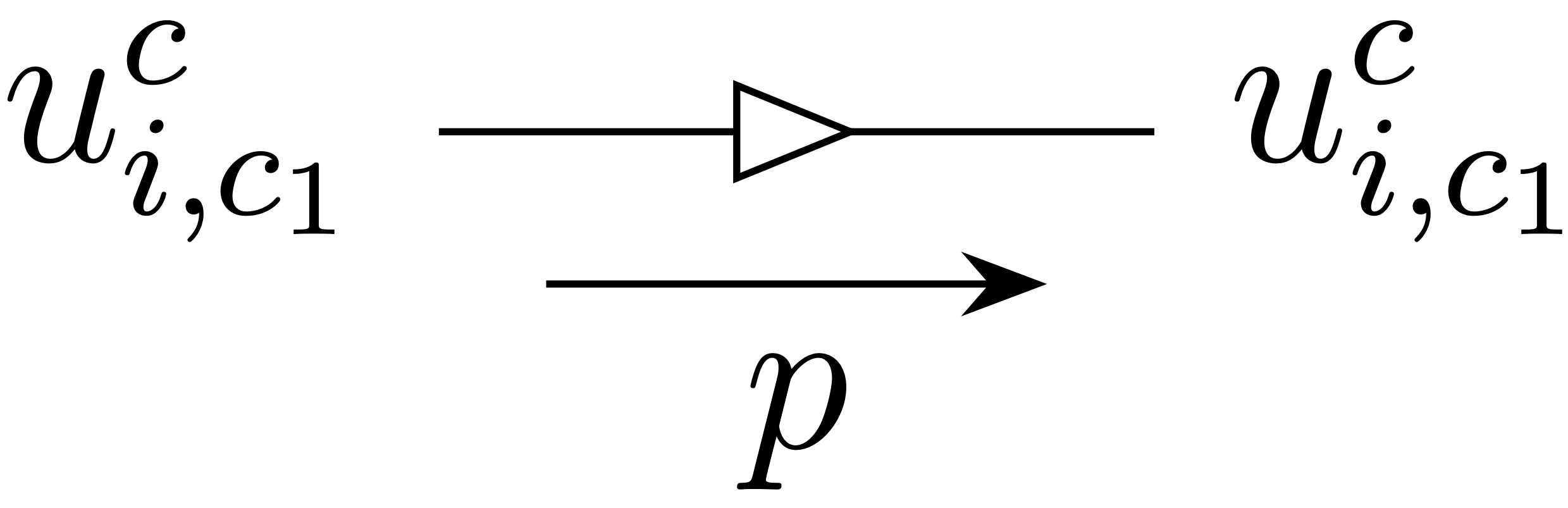}&
        \vspace{10 px}
         $\dfrac{i \left(\slashed{p} + m_{u_i}\right)}{p^2 - m^2_{u_i} +i\epsilon}$ \newline \\
        \vspace{10 px}
        \includegraphics[scale=0.035]{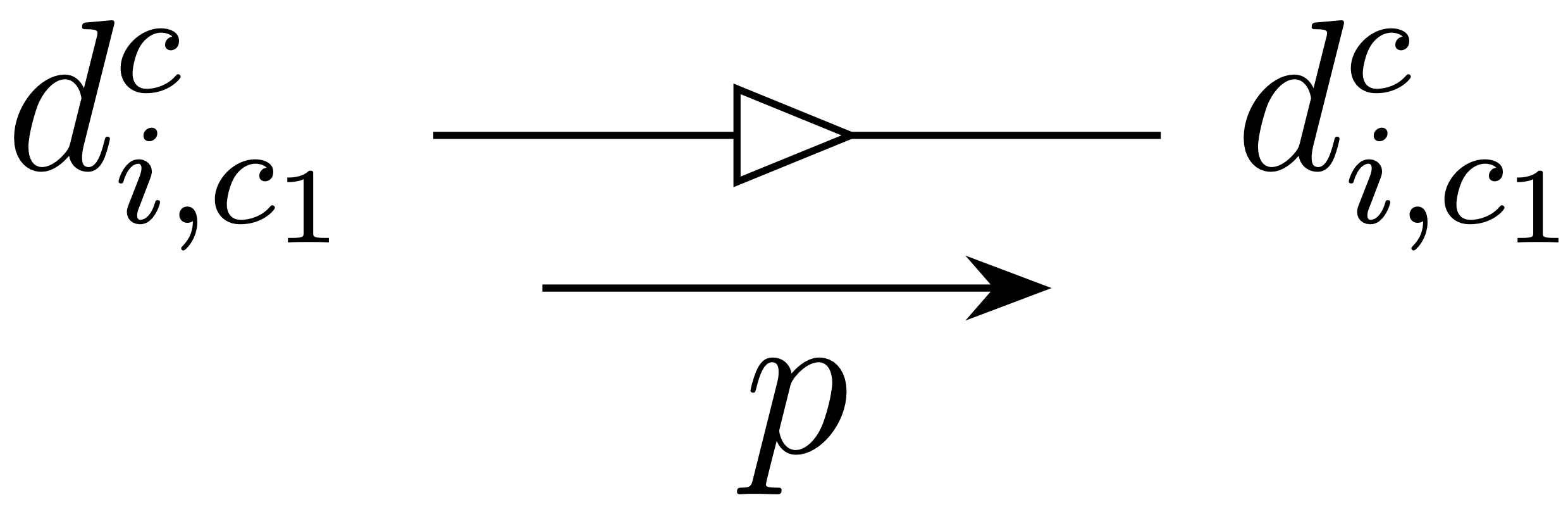}&
        \vspace{10 px}
         $\dfrac{i \left(\slashed{p} + m_{d_i}\right)}{p^2 - m^2_{d_i} +i\epsilon}$ \newline \\
        \hline
        \hline
    \end{tabular}
\end{table}
\end{center}

\clearpage
\subsubsection*{Interactions with SM Gauge Bosons}
The Feynman rules for charge-conjugate SM fermion fields interacting with SM gauge fields are given below. 

\begin{center}
\begin{table}[!ht]
    \centering\begin{tabular}{>{\centering\arraybackslash}m{4cm} m{9cm}} 
    \multicolumn{2}{c}{Interactions with Photons} \\
    \hline
        Diagrams & Feynman Rules \\
        \hline
        \vspace{5 px}
        \includegraphics[height=2.25cm]{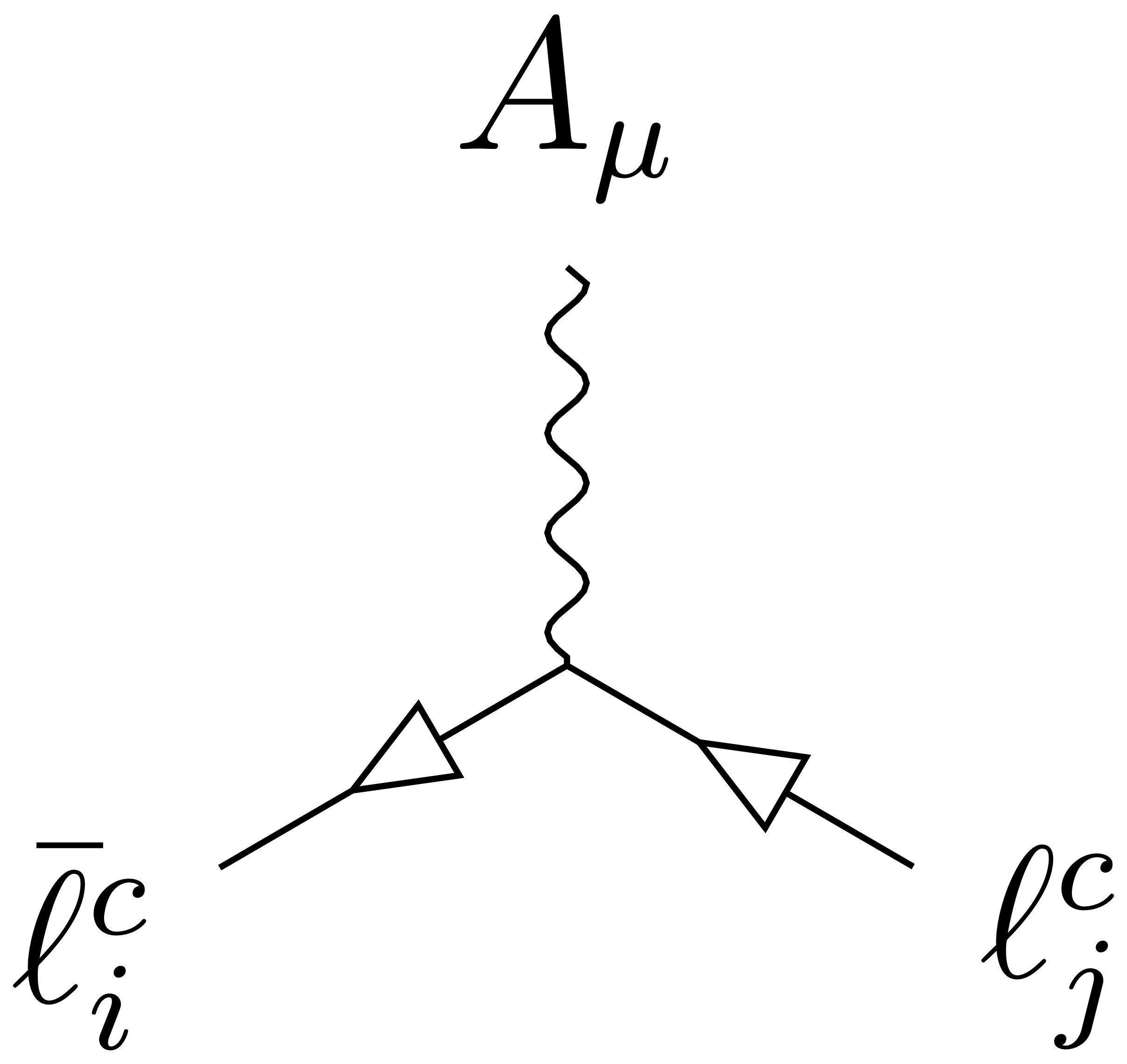}&
        $+ \ i e \delta_{ij} \gamma^{\mu}$\\
        \vspace{5 px}
        \includegraphics[height=2.25cm]{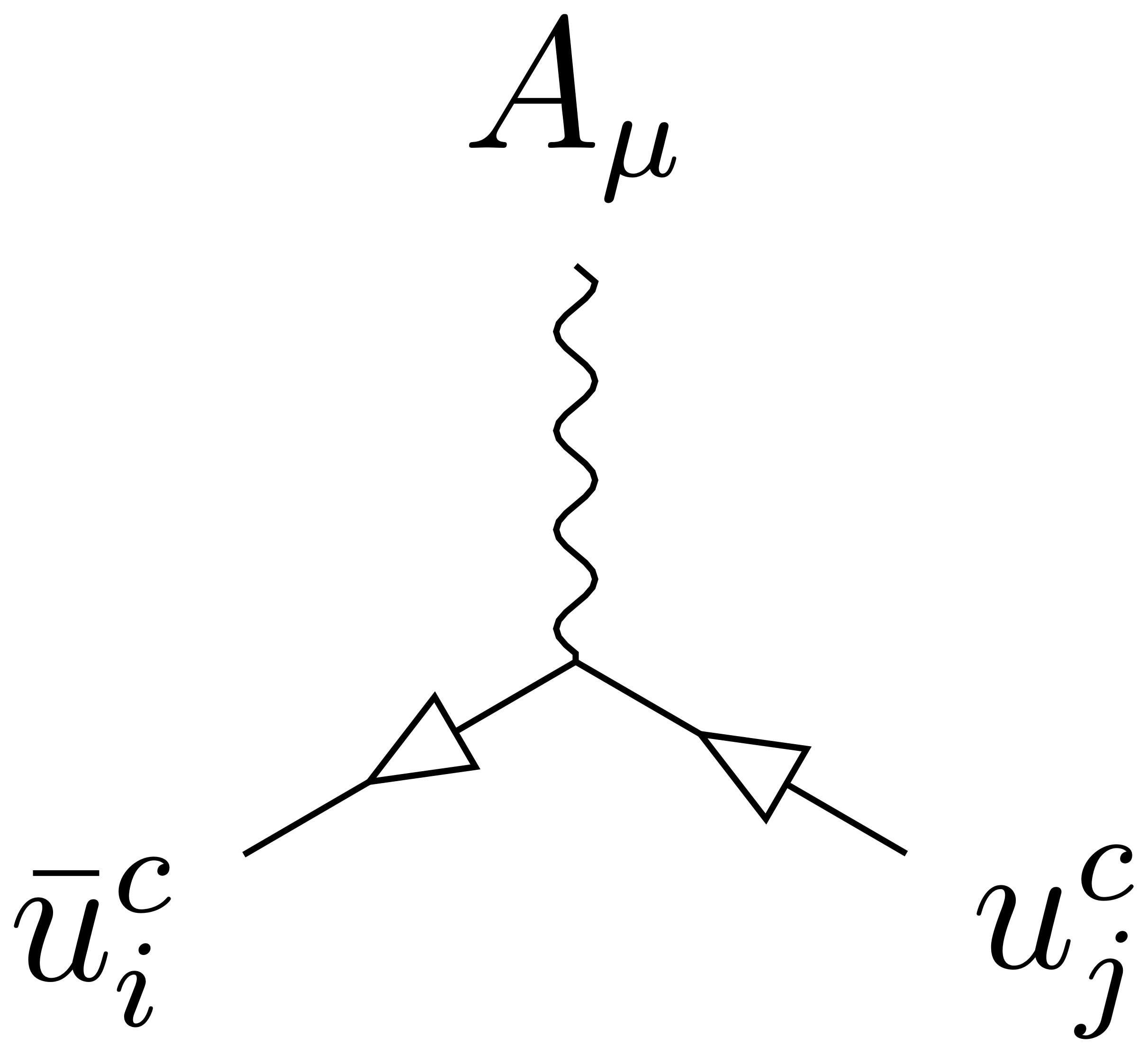}&
        $- \ \dfrac{2i}{3} e \delta_{ij} \gamma^{\mu}$\\
        \vspace{5 px}
        \includegraphics[height=2.25cm]{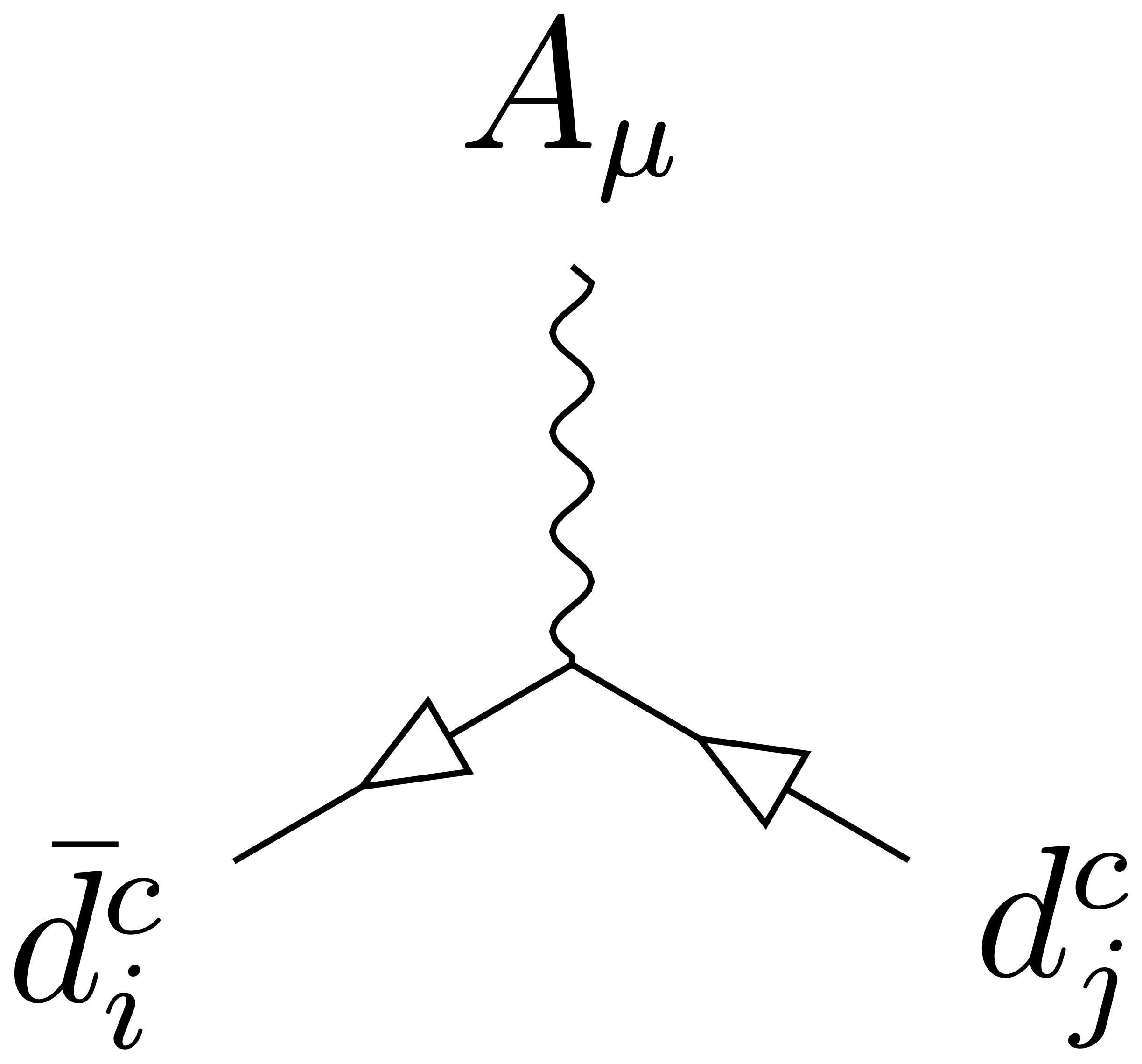}&
        $+ \ \dfrac{i}{3} e \delta_{ij} \gamma^{\mu}$\\
        \hline
    \end{tabular}
\end{table}
\end{center}

\pagebreak

\begin{center}
\begin{table}[!ht]
    \centering\begin{tabular}{>{\centering\arraybackslash}m{4cm} m{9cm}} 
    \multicolumn{2}{c}{Interactions with $Z$ Bosons} \\
    \hline
        Diagrams & Feynman Rules \\
        \hline
        \vspace{5 px}
        \includegraphics[height=2.25cm]{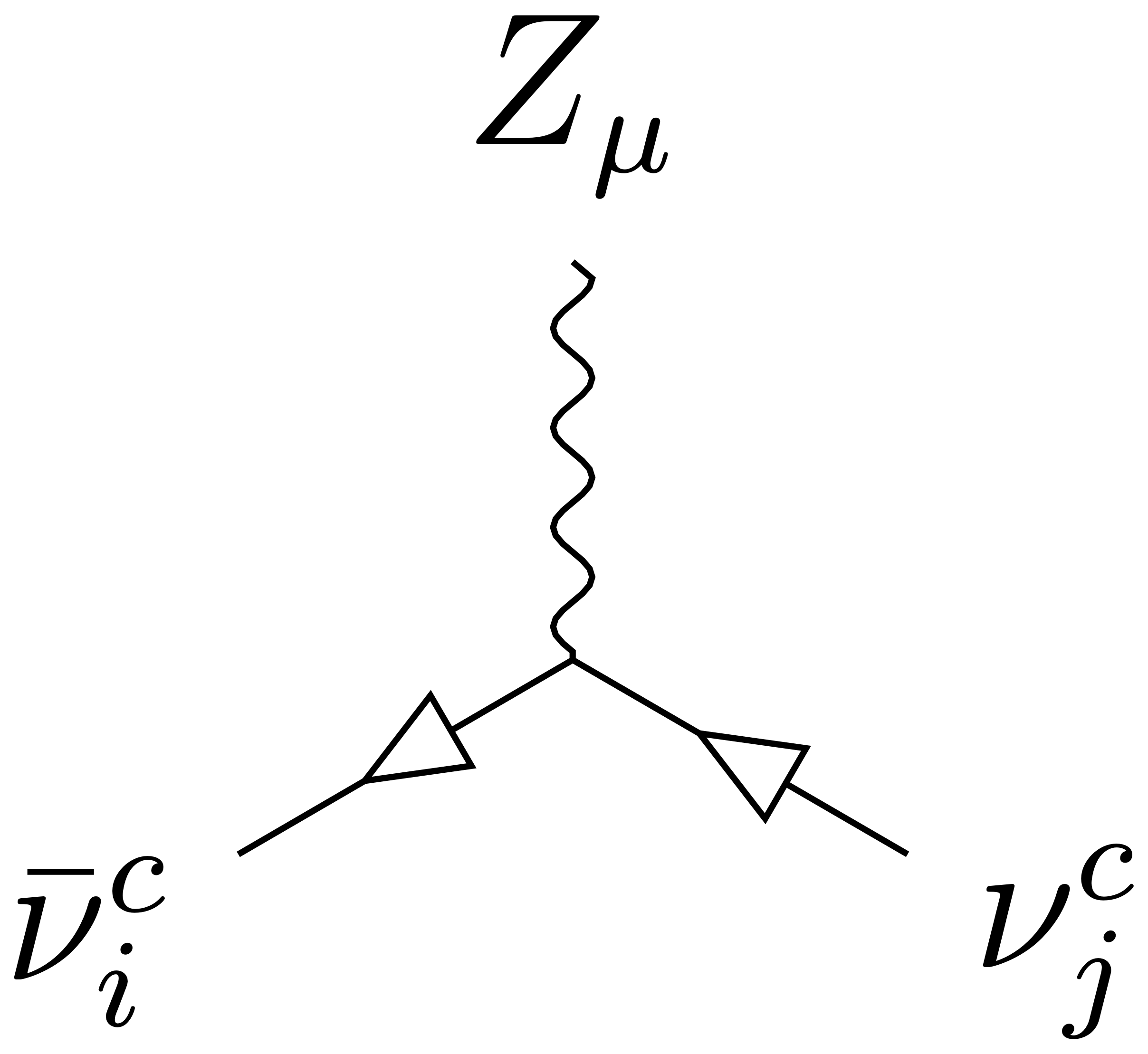}&
        $- \ \dfrac{i}{2} \dfrac{e}{s_w c_w} \delta_{ij} \gamma^{\mu} \ \text{P}_{\text{R}}$\\
        \vspace{5 px}
        \includegraphics[height=2.25cm]{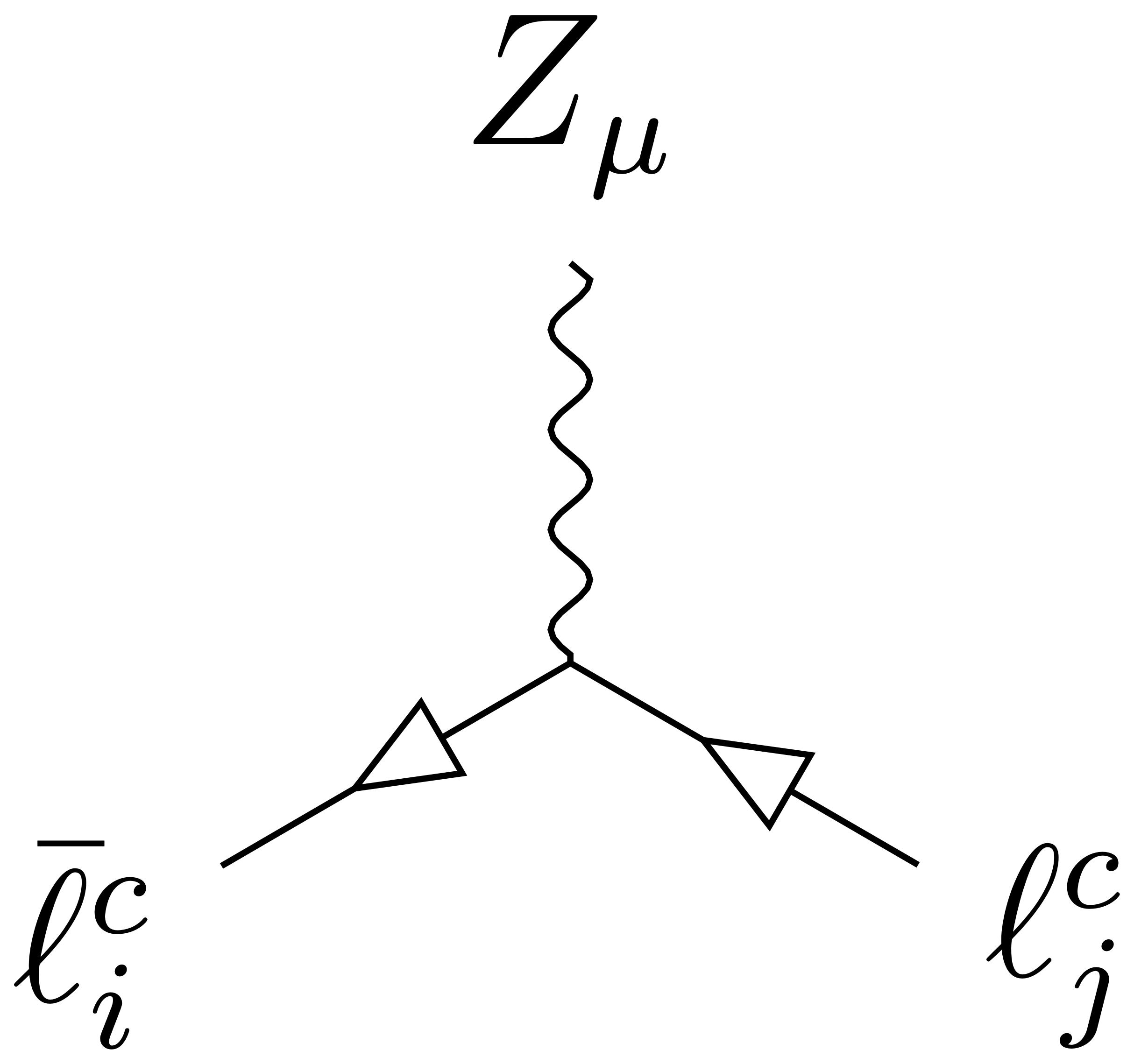}&
        $+ \ i \dfrac{e}{s_w c_w} \delta_{ij} \gamma^{\mu} \left(+\dfrac{1}{2} - s_w^2 \right) \ \text{P}_{\text{R}} \vspace{5px} \newline
        - i e \dfrac{s_w}{c_w} \delta_{ij} \gamma^{\mu} \  \text{P}_{\text{L}}$\\
        \vspace{5 px}
        \includegraphics[height=2.25cm]{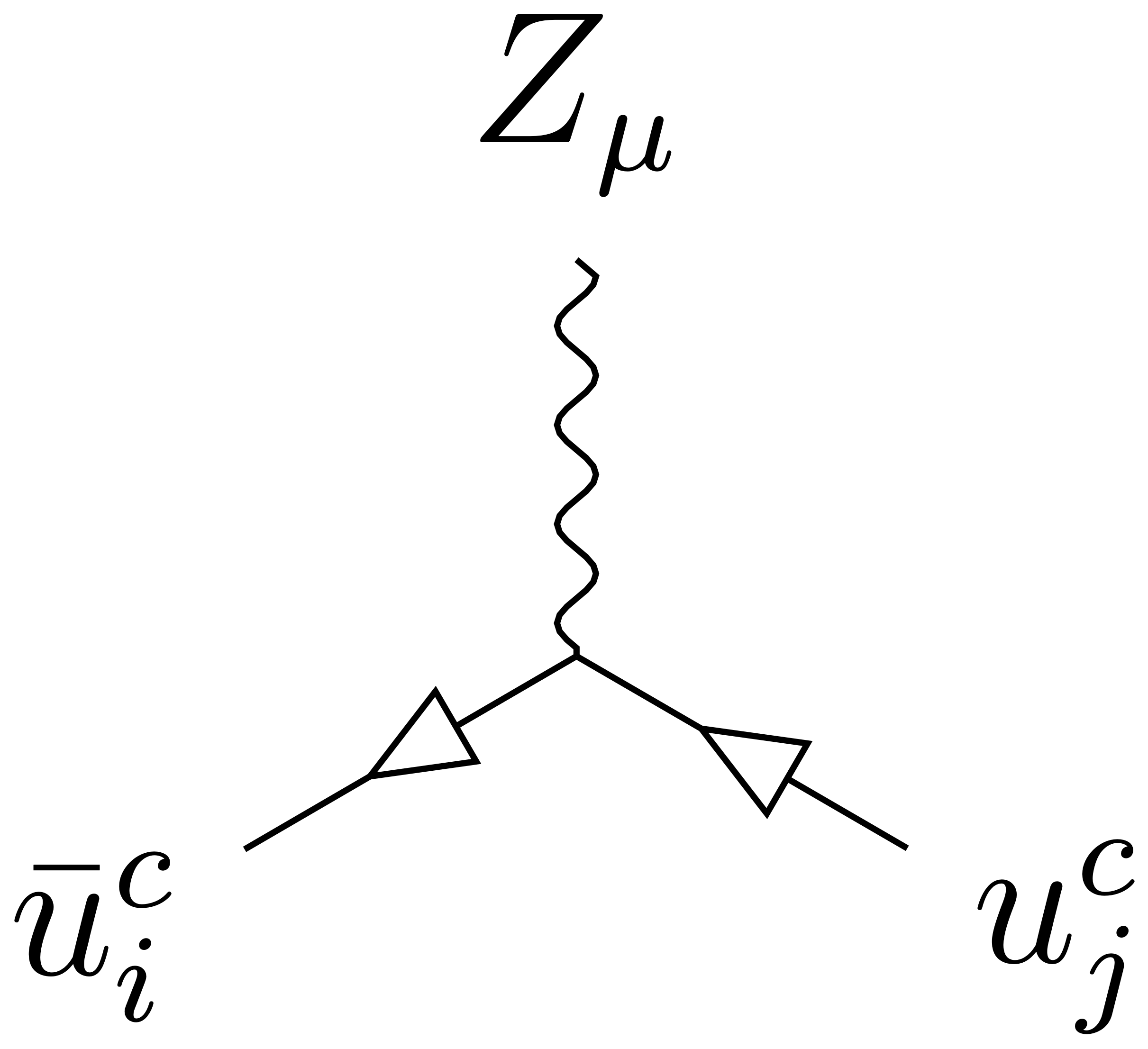}&
        $+ \ i \dfrac{e}{s_w c_w} \delta_{ij} \gamma^{\mu} \left(-\dfrac{1}{2} + \dfrac{2}{3}s_w^2 \right) \ \text{P}_{\text{R}} \vspace{5px} \newline
        + \dfrac{2i}{3} e \dfrac{s_w}{c_w} \delta_{ij} \gamma^{\mu} \  \text{P}_{\text{L}}$\\
        \vspace{5 px}
        \includegraphics[height=2.25cm]{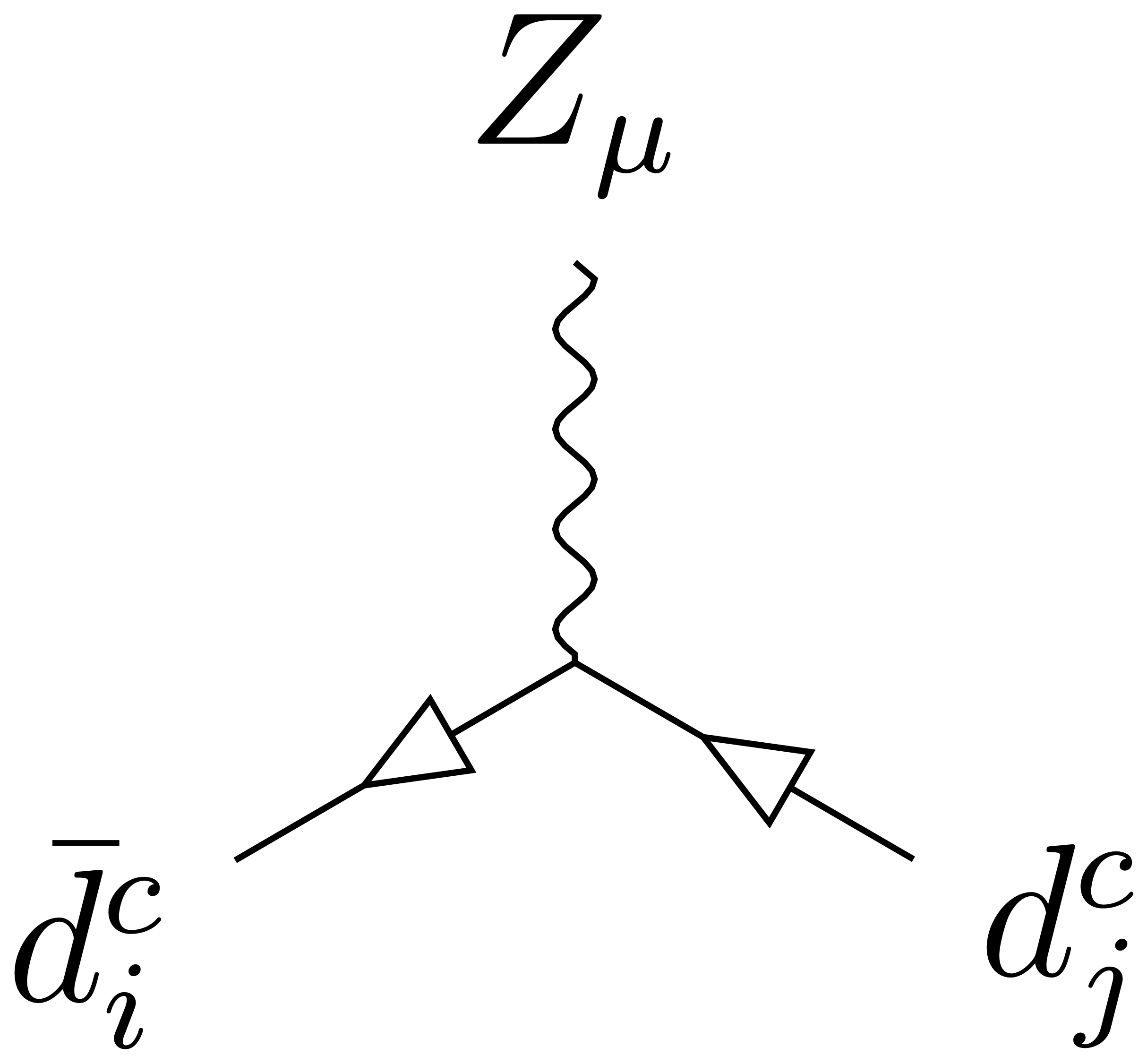}&
        $+ \ i \dfrac{e}{s_w c_w} \delta_{ij} \gamma^{\mu} \left(+\dfrac{1}{2} - \dfrac{1}{3}s_w^2 \right) \ \text{P}_{\text{R}}  \vspace{5px} \newline
        - \dfrac{i}{3} e\dfrac{s_w}{c_w} \delta_{ij} \gamma^{\mu} \ \text{P}_{\text{L}}$\\
        \hline
    \end{tabular}
\end{table}
\end{center}

\pagebreak

\begin{center}
\begin{table}[!ht]
    \centering\begin{tabular}{>{\centering\arraybackslash}m{4cm} m{9cm}} 
    \multicolumn{2}{c}{Interactions with $W$ Bosons} \\
    \hline
        Diagrams & Feynman Rules \\
        \hline
        \vspace{5 px}
        \includegraphics[height=2.25cm]{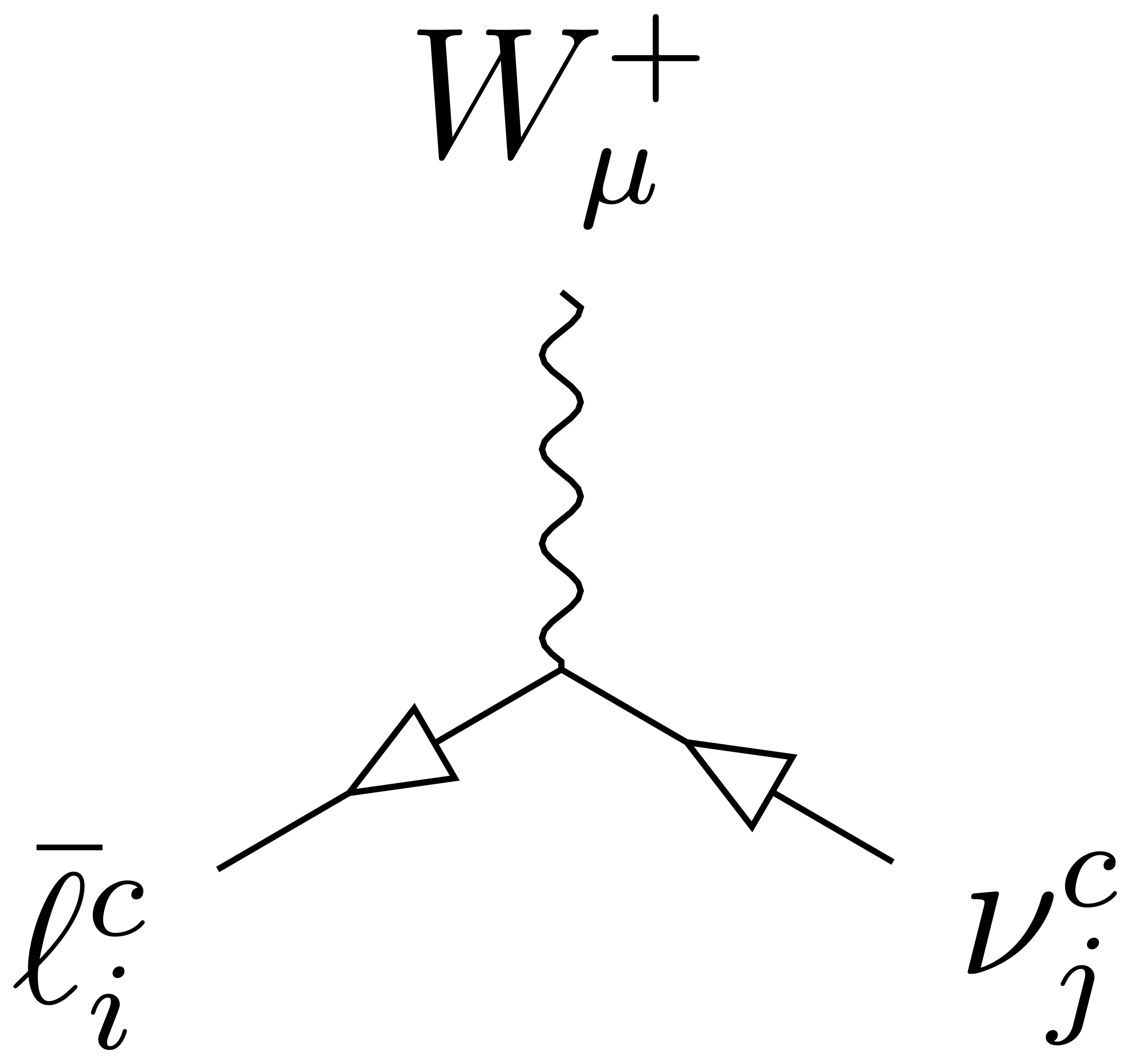}&
        $- \ \dfrac{i}{\sqrt{2}} \dfrac{e}{s_w} \delta_{ij} \gamma^{\mu} \  \text{P}_{\text{R}}$\\
        \vspace{5 px}
        \includegraphics[height=2.25cm]{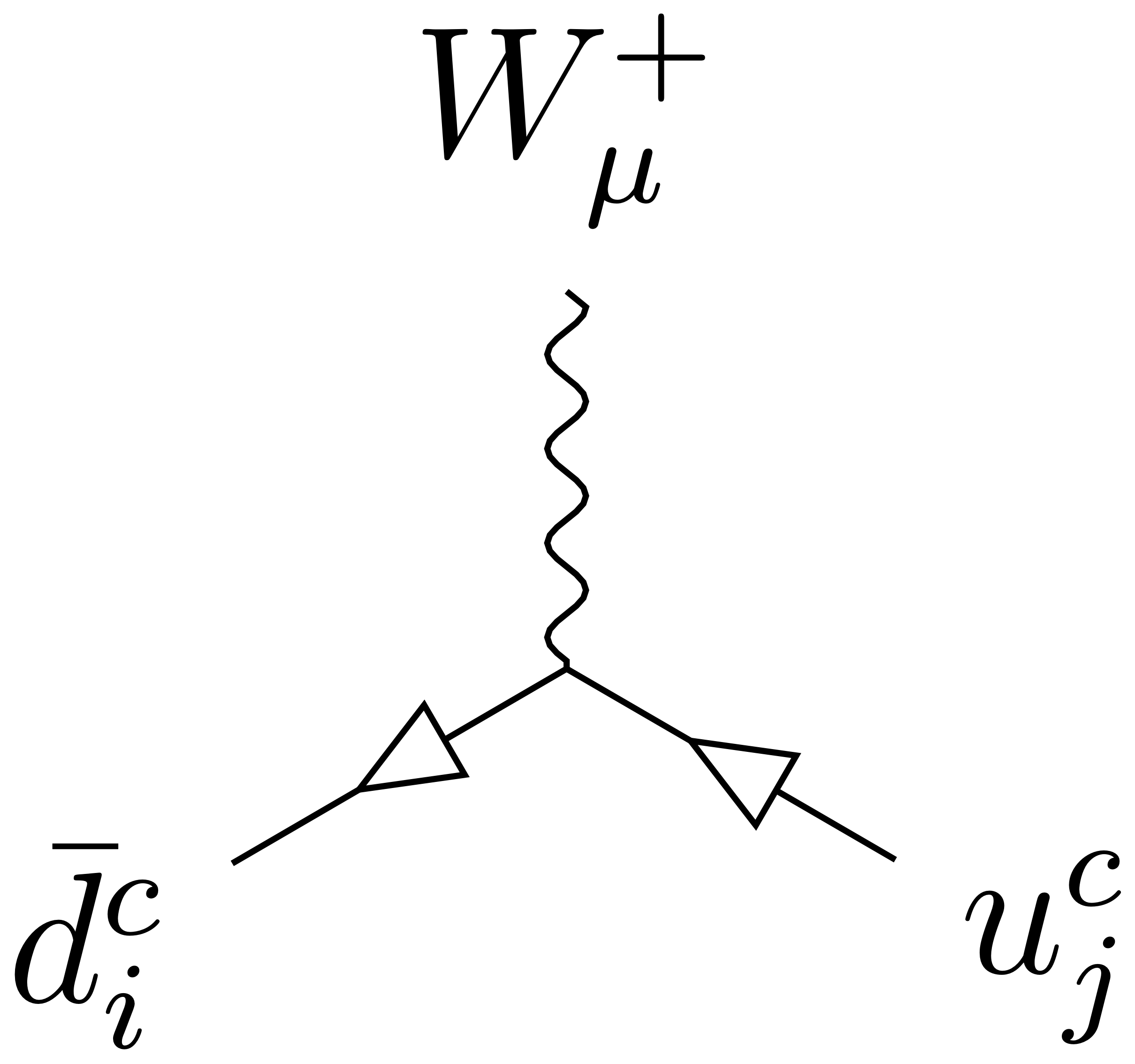}&
        $- \ \dfrac{i}{\sqrt{2}} \dfrac{e}{s_w} V_{ji} \gamma^{\mu} \ \text{P}_{\text{R}}$\\
        \hline
    \end{tabular}
\end{table}
\end{center}

\begin{center}
\begin{table}[!ht]
    \centering\begin{tabular}{>{\centering\arraybackslash}m{4cm} m{9cm}} 
    \multicolumn{2}{c}{Interactions with Gluons} \\
    \hline
        Diagrams & Feynman Rules \\
        \hline
        \vspace{5 px}
        \includegraphics[height=2.25cm]{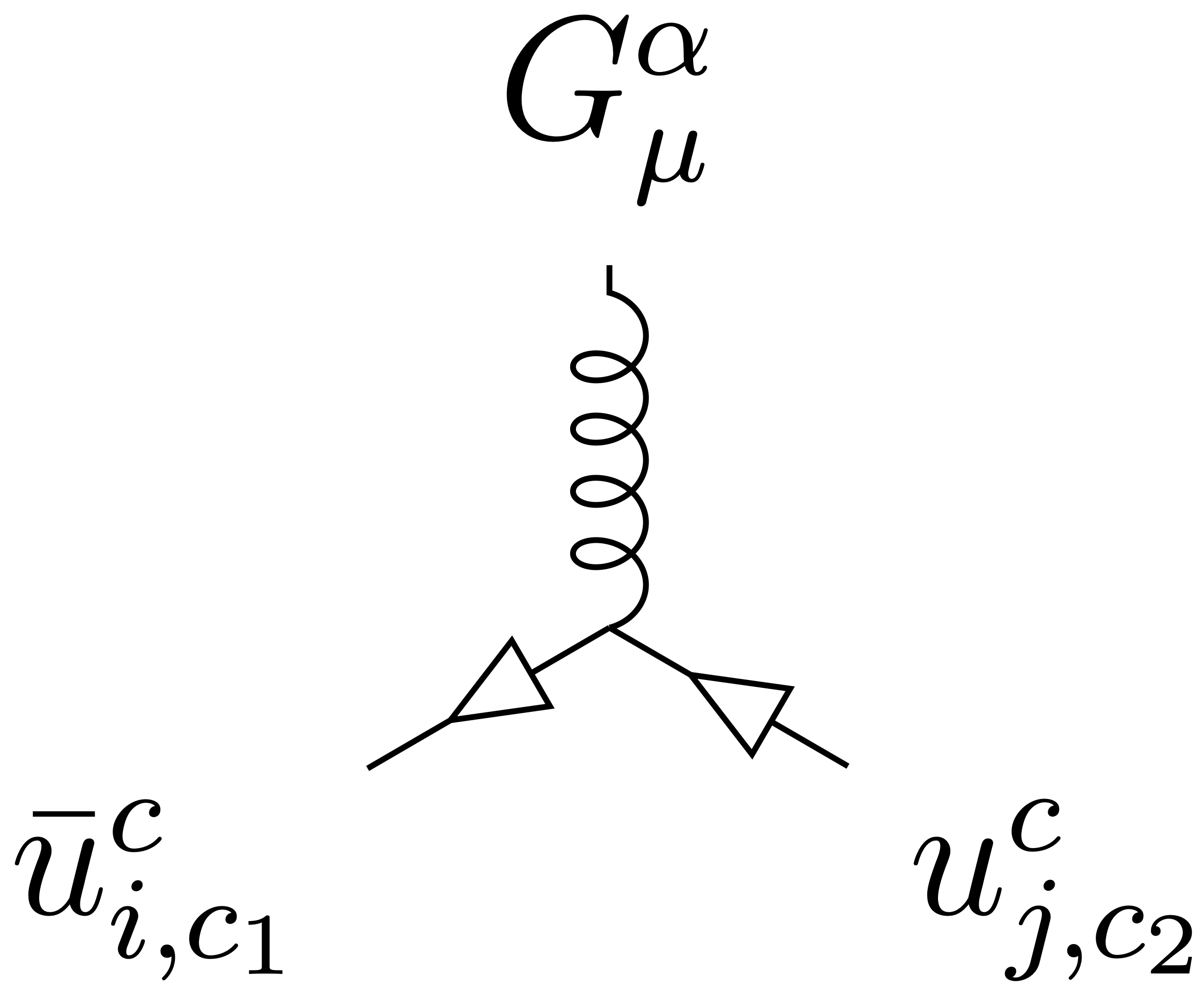}&
        $- \ i g_s \delta_{ij}  T^{\alpha}_{c_2 c_1} \gamma^{\mu}$\\
        \vspace{5 px}
        \includegraphics[height=2.25cm]{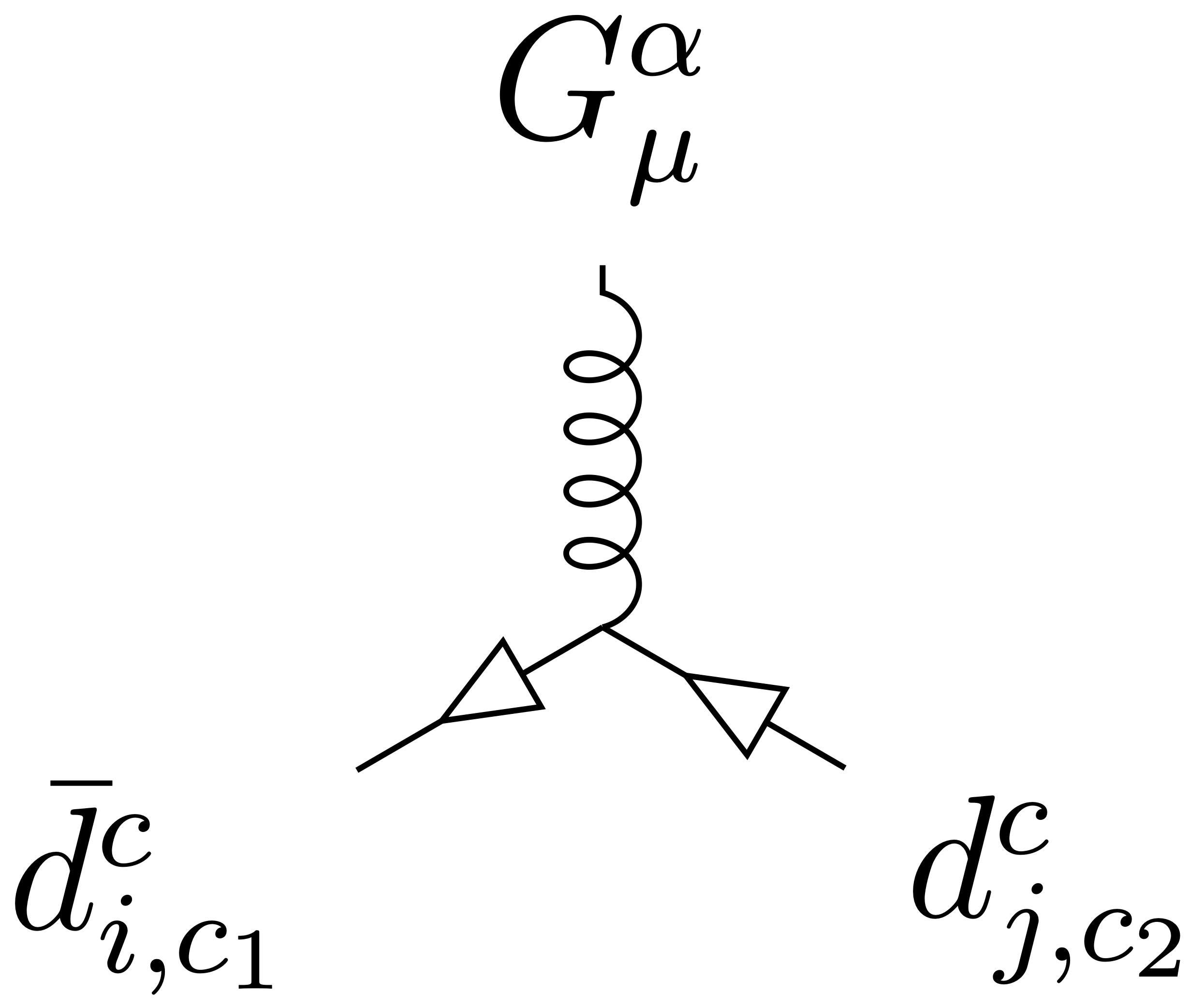}&
        $- \ i g_s \delta_{ij}  T^{\alpha}_{c_2 c_1} \gamma^{\mu}$\\
        \hline
        \hline
    \end{tabular}
\end{table}
\end{center}

\clearpage
\subsubsection*{Interactions with the Higgs Field}
The Feynman rules for charge-conjugate SM fermion fields interacting with Higgs field are given below. 

\begin{center}
\begin{table}[!ht]
    \centering\begin{tabular}{>{\centering\arraybackslash}m{4cm} m{9cm}} 
    \multicolumn{2}{c}{Interactions with Higgs Field} \\
    \hline
        Diagrams & Feynman Rules \\
        \hline
        \vspace{5 px}
        \includegraphics[height=2.25cm]{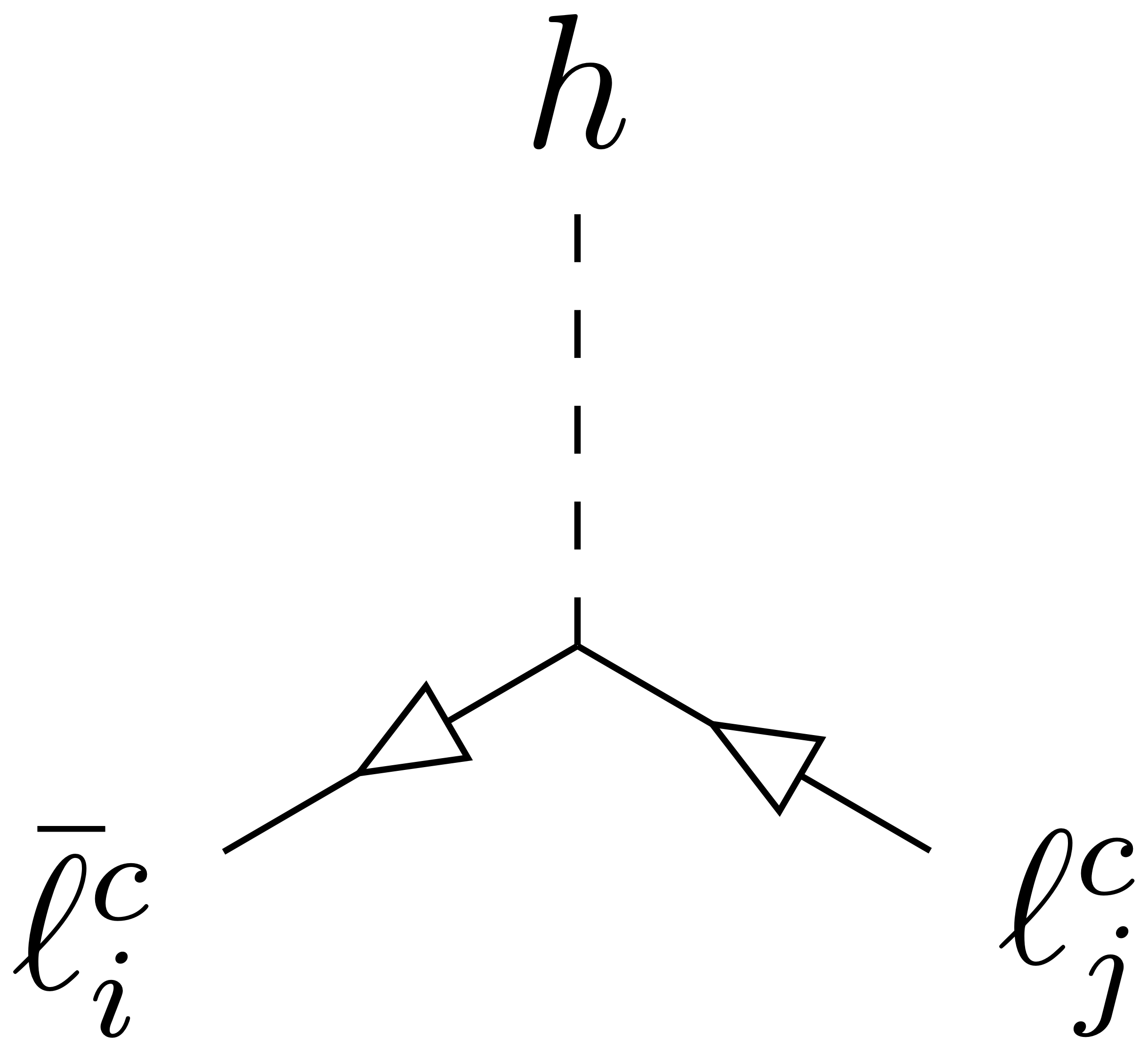}&
        $- \ i \dfrac{m_{\ell_i}}{v} \delta_{ij}$\\
        \vspace{5 px}
        \includegraphics[height=2.25cm]{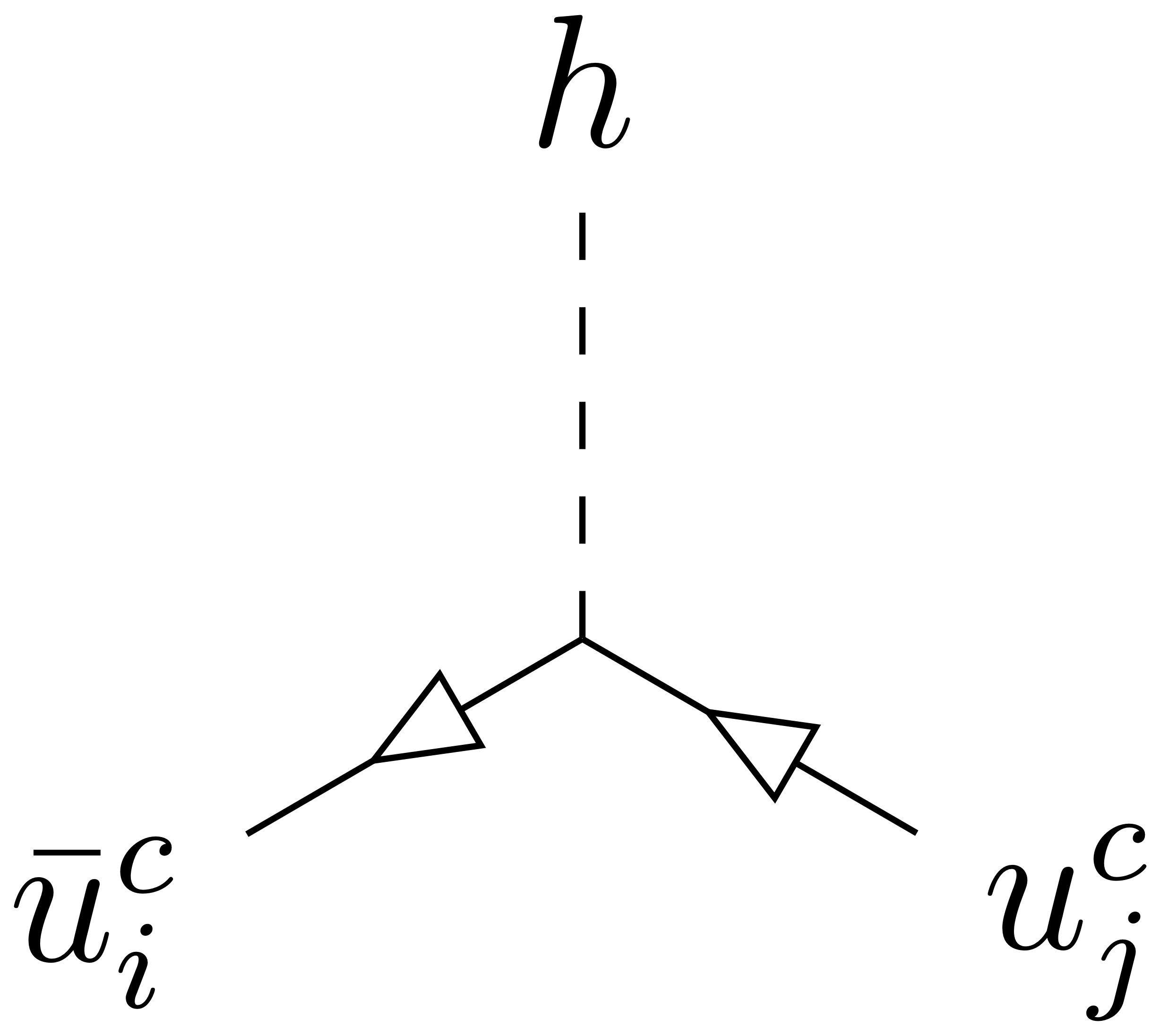}&
        $- \ i \dfrac{m_{u_i}}{v} \delta_{ij}$\\
        \vspace{5 px}
        \includegraphics[height=2.25cm]{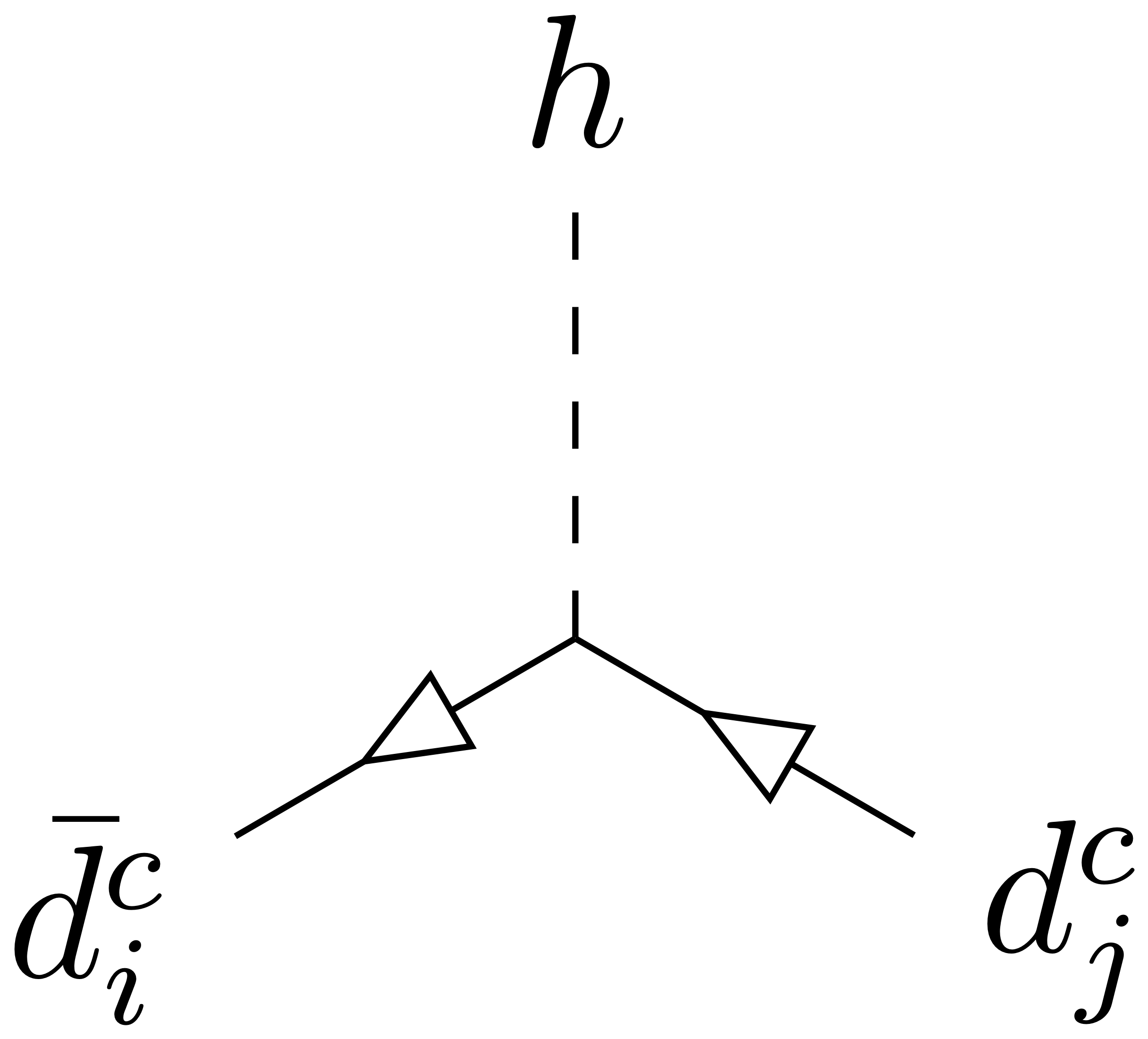}&
        $- \ i \dfrac{m_{d_i}}{v} \delta_{ij}$\\
        \hline
        \hline
    \end{tabular}
\end{table}
\end{center}

\clearpage
\subsubsection*{Interactions with the Goldstone Bosons}
The Feynman rules for charge-conjugate SM fermion fields interacting with SM Goldstone bosons are given below. 

\begin{center}
\begin{table}[!ht]
    \centering\begin{tabular}{>{\centering\arraybackslash}m{4cm} m{9cm}} 
    \multicolumn{2}{c}{Interactions with Goldstone Bosons} \\
    \hline
        Diagrams & Feynman Rules \\
        \hline
        \vspace{5 px}
        \includegraphics[height=2.25cm]{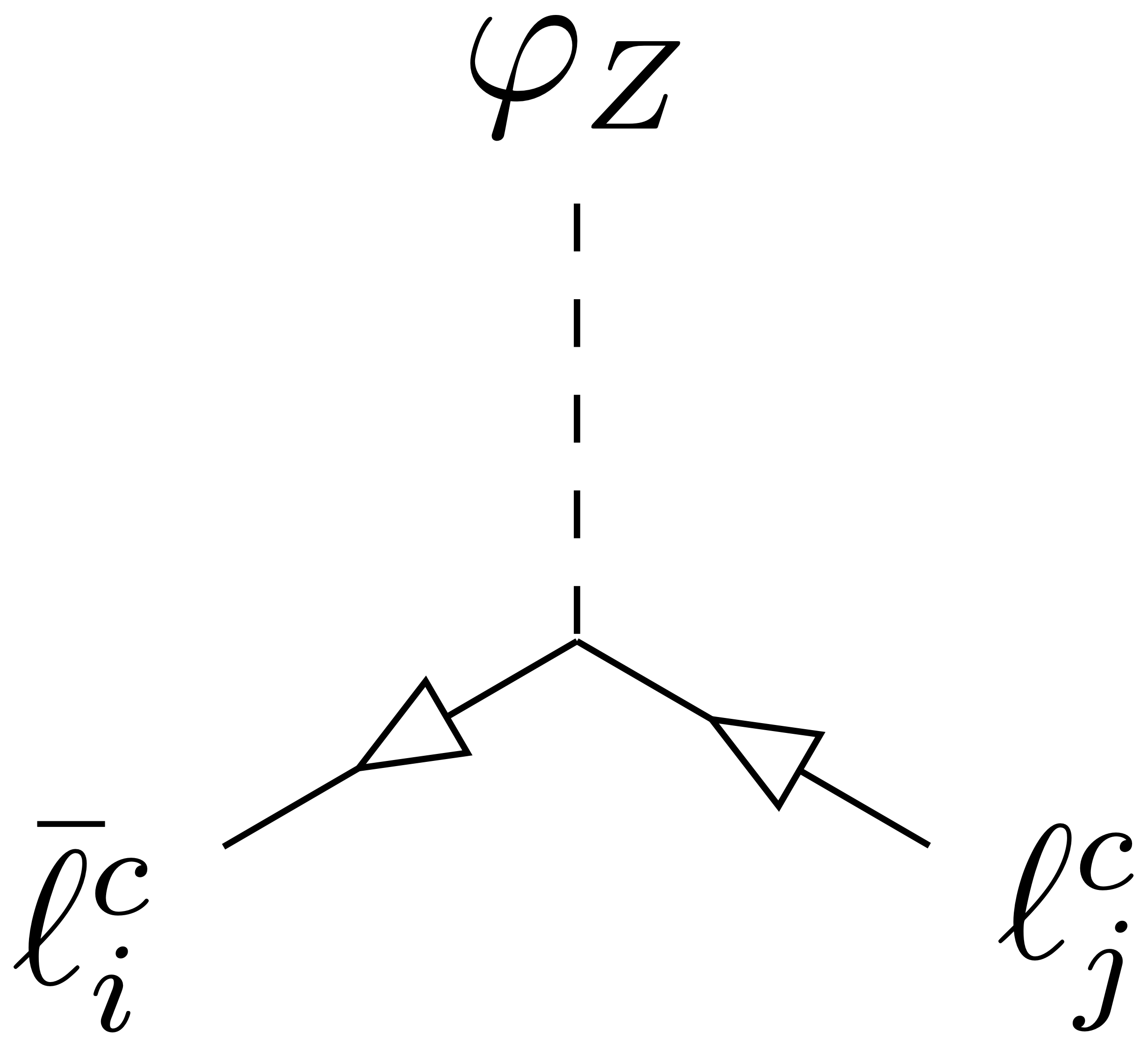}&
        $+ \dfrac{m_{\ell_i}}{v} \gamma^5 \delta_{ij}$\\
        \vspace{5 px}
        \includegraphics[height=2.25cm]{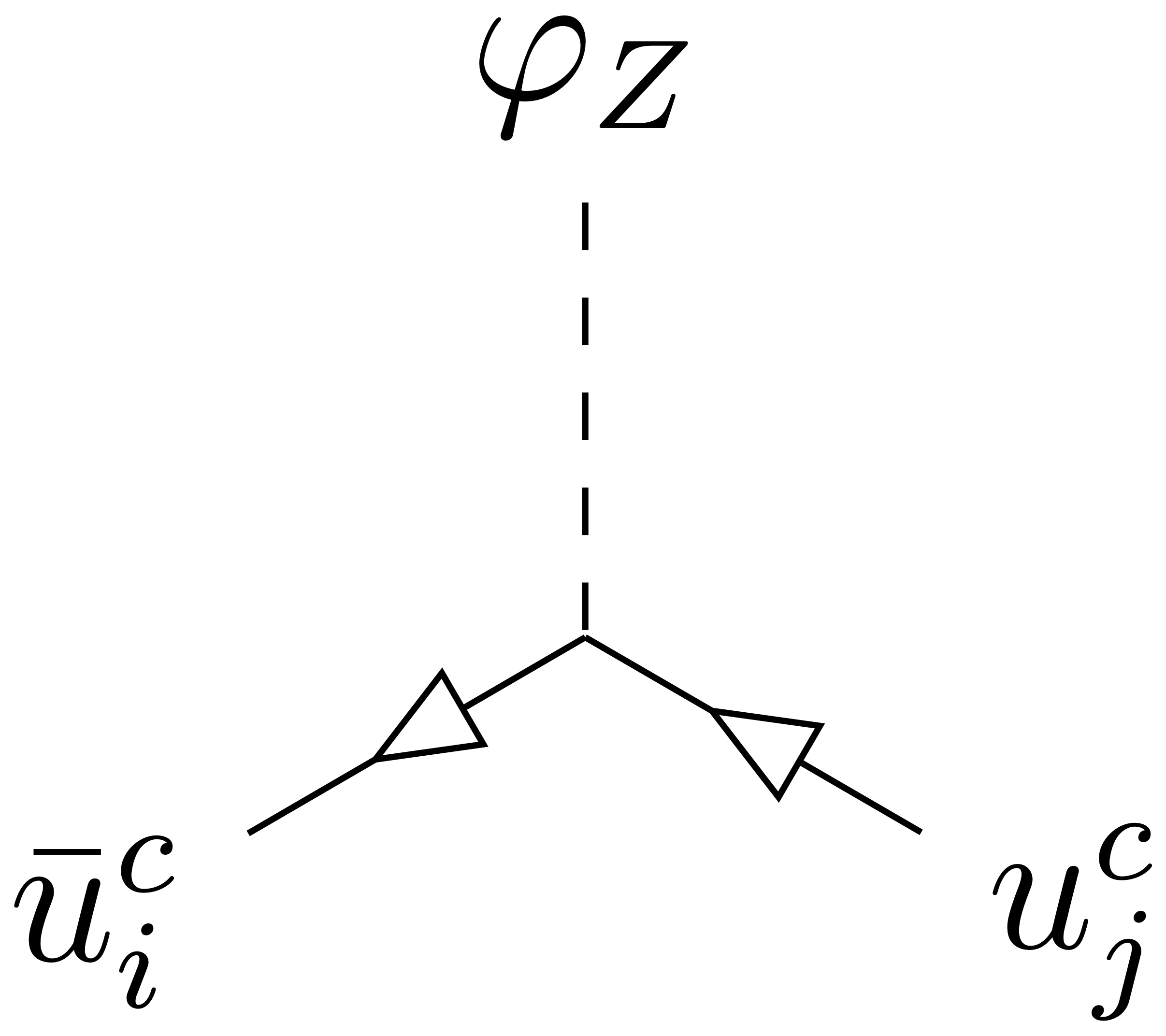}&
        $- \dfrac{m_{u_i}}{v} \gamma^5 \delta_{ij}
        $\\
        \vspace{5 px}
        \includegraphics[height=2.25cm]{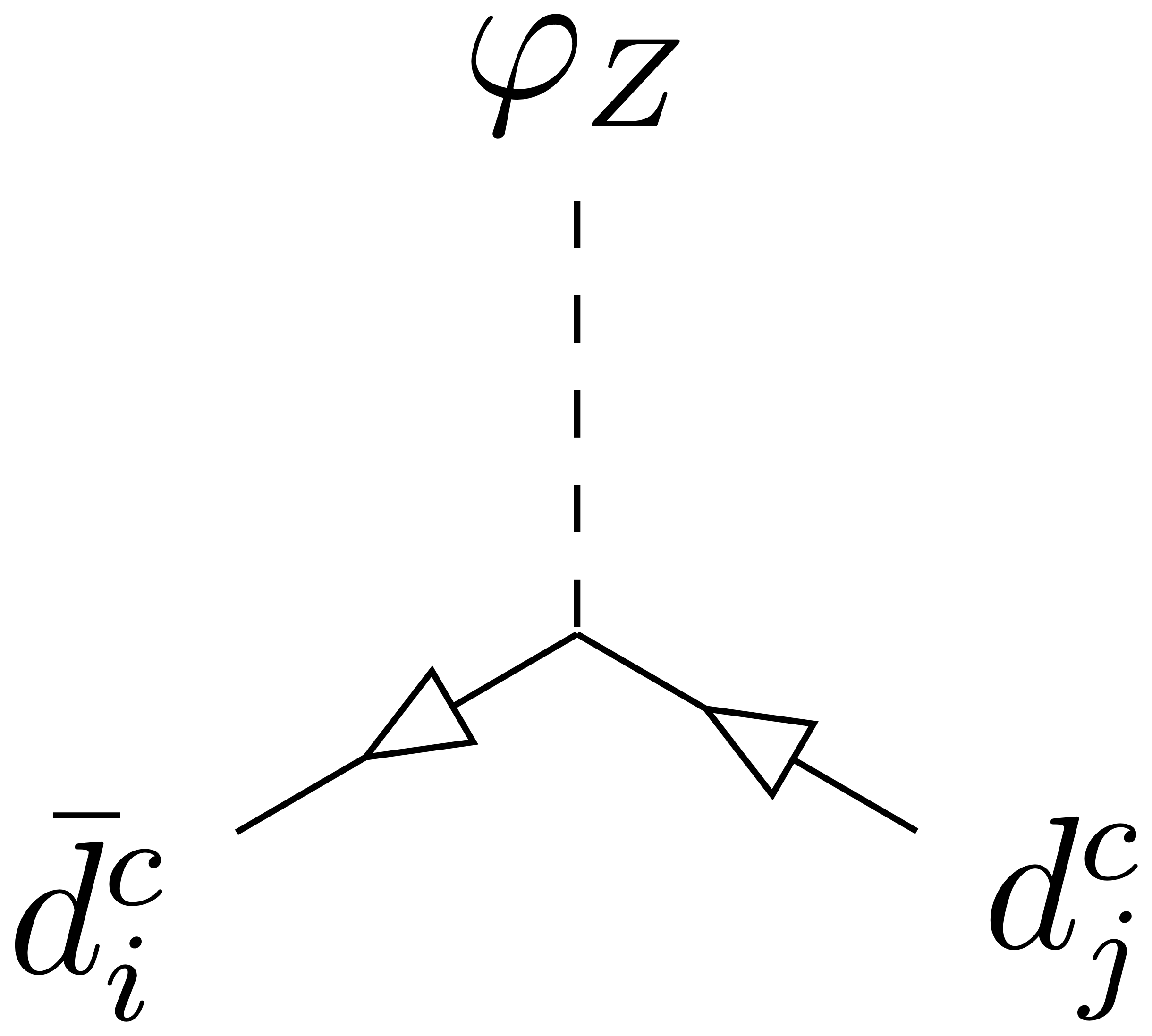}&
        $+ \dfrac{m_{d_i}}{v} \gamma^5 \delta_{ij}
        $\\
        \hline
    \end{tabular}
\end{table}
\end{center}

\begin{center}
\begin{table}[!ht]
    \centering\begin{tabular}{>{\centering\arraybackslash}m{4cm} m{9cm}} 
    \multicolumn{2}{c}{Interactions with Goldstone Bosons} \\
    \hline
        Diagrams & Feynman Rules \\
        \hline
        \vspace{5 px}
        \includegraphics[height=2.25cm]{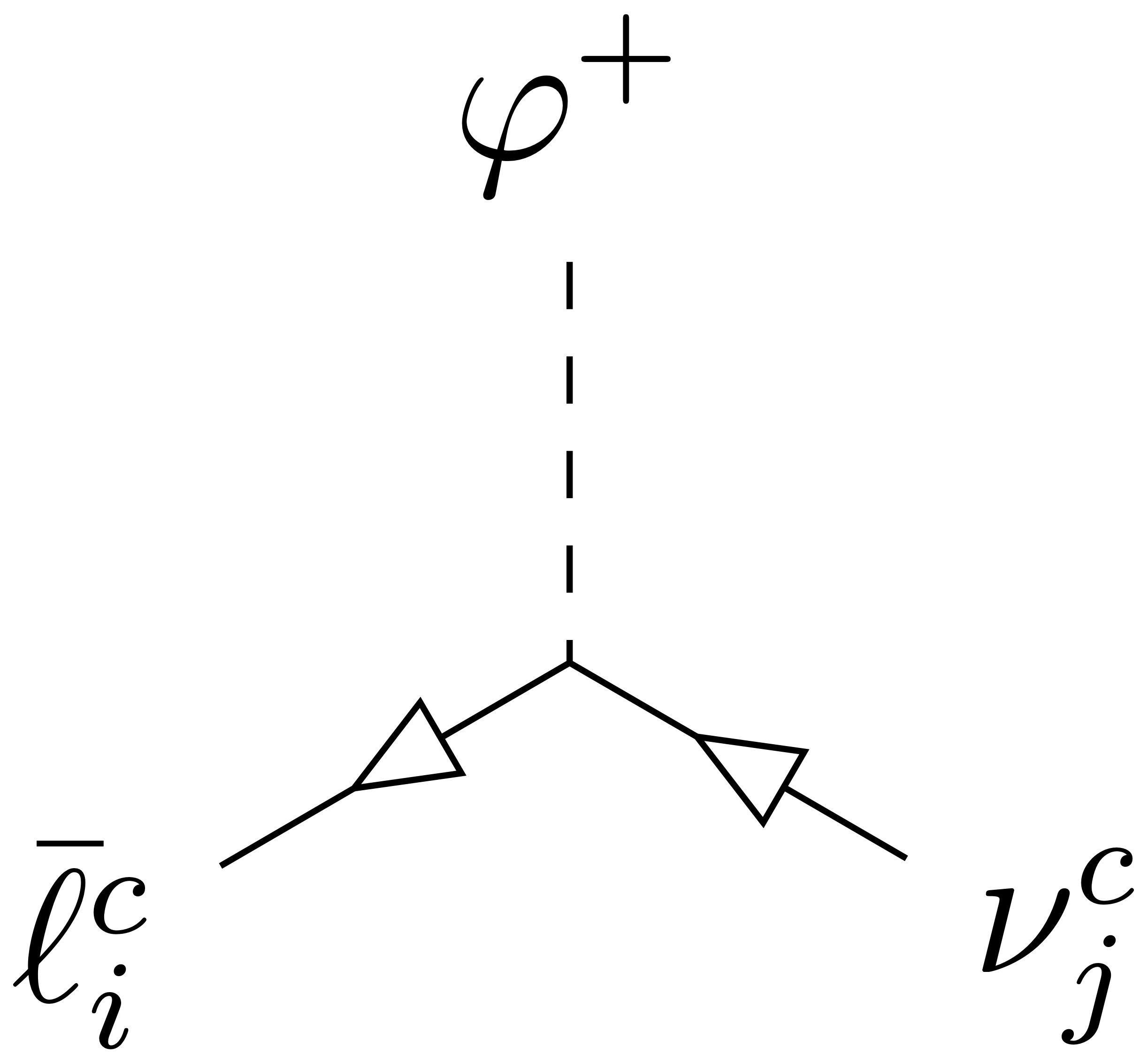}&
        $- \ \dfrac{\sqrt{2}}{v} m_{\ell_i} \delta_{ij}\  \text{P}_\text{R}
        $\\
        \vspace{5 px}
        \includegraphics[height=2.25cm]{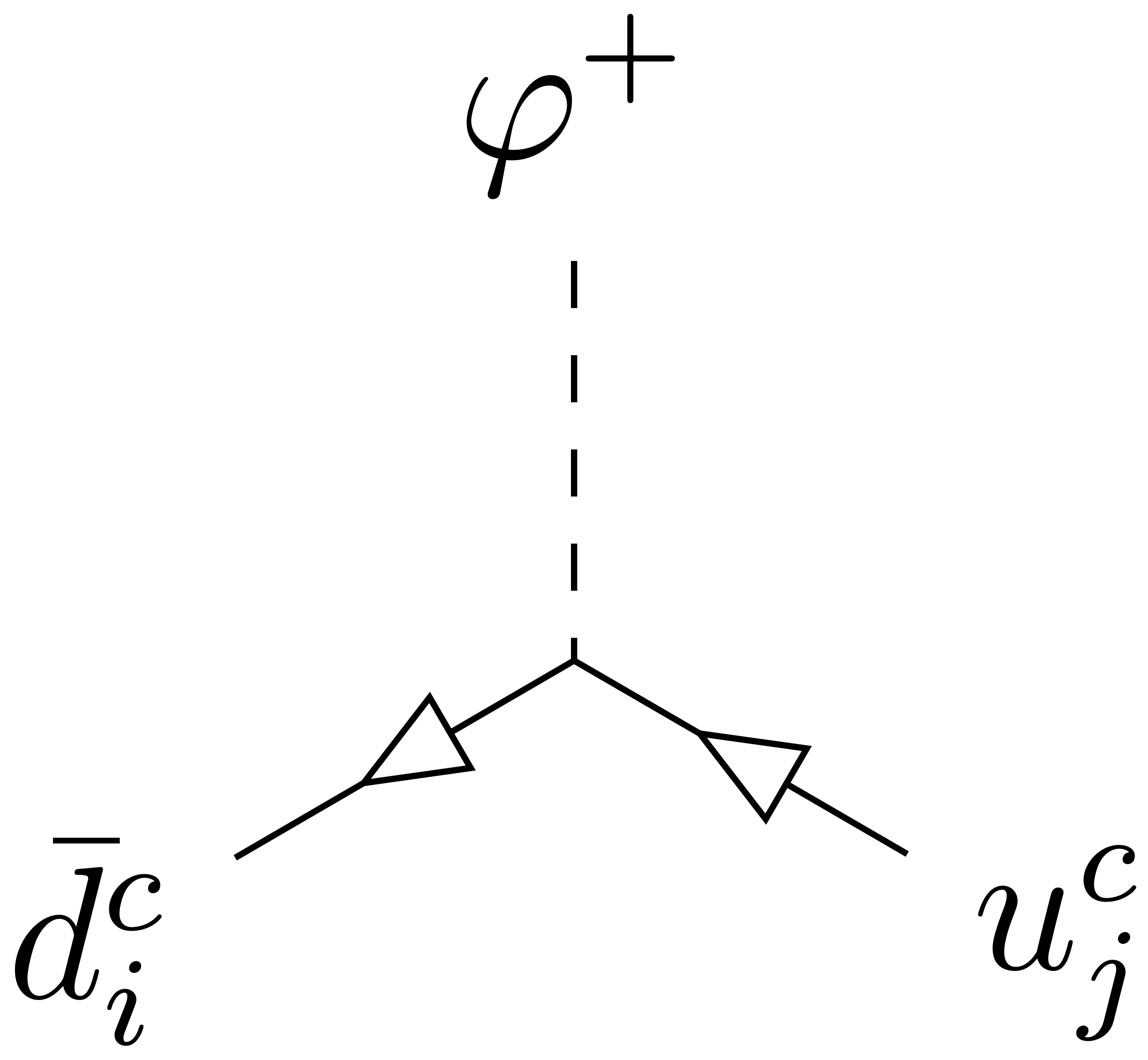}&
        $+ \ \dfrac{\sqrt{2}}{v} V_{ji}  \left(m_{u_i}\text{P}_\text{L} - m_{d_i} \text{P}_\text{R} \right)
        $\\
        \hline
        \hline
    \end{tabular}
\end{table}
\end{center}

\clearpage
\subsubsection*{Interactions with LQs}
The Feynman rules for LQ interactions with the charge-conjugate SM fermions are given below. 


\begin{center}
\begin{table}[!ht]
    \centering\begin{tabular}{>{\centering\arraybackslash}m{4cm} m{8.5cm}} 
    \multicolumn{2}{c}{Charge 1/3} \\
    \hline
        Diagram & Feynman Rule \\
        \hline
        \vspace{5px}
        \includegraphics[height=2.25cm]{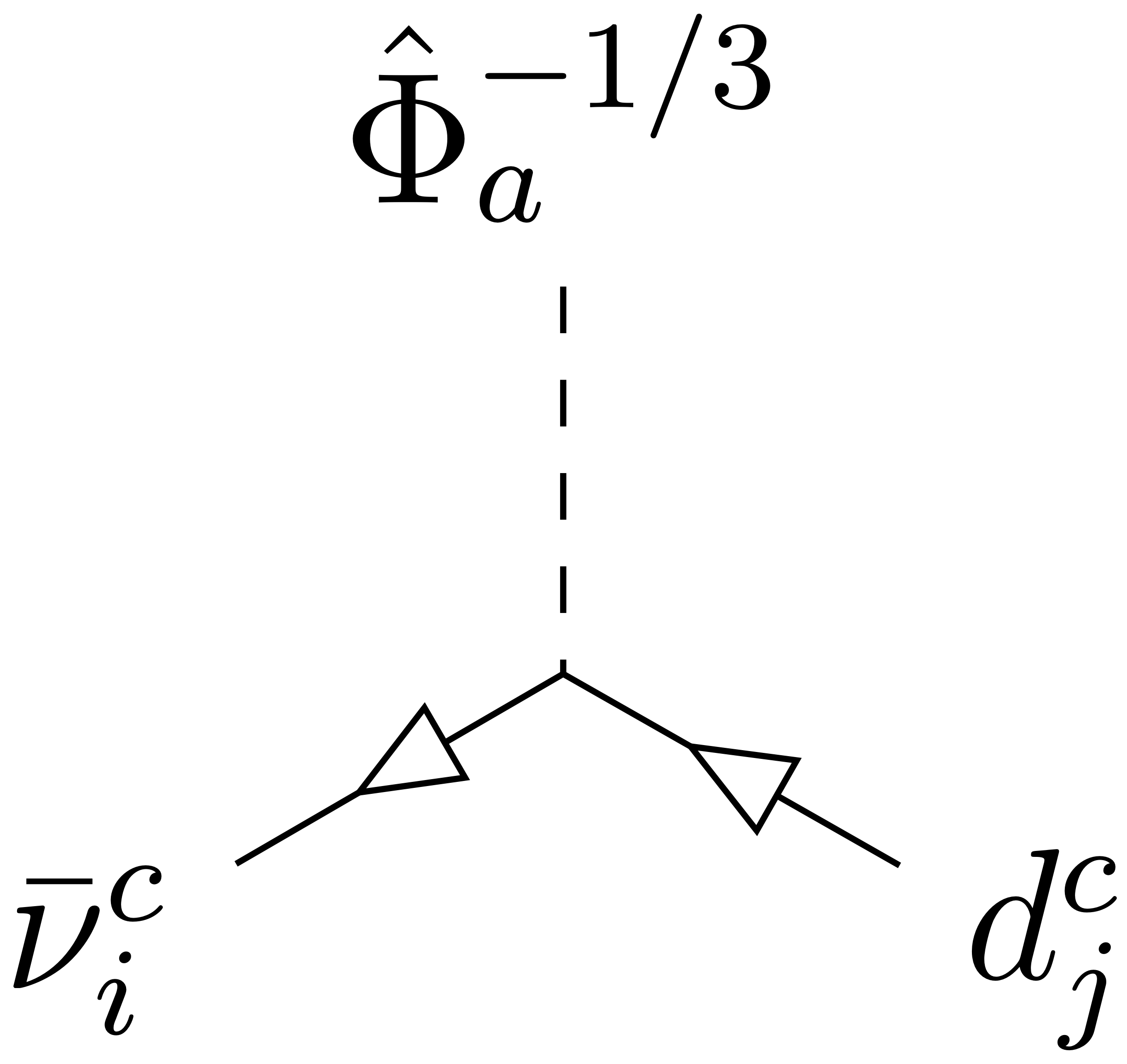}&
        $- \ i Y_{\tilde 2, ji}^{\text{RL}} \ W^{-1/3*}_{n_a2} \ \text{P}_{\text{L}} $ \\
        \hline
    \end{tabular}
\end{table}
\end{center}

\begin{center}
\begin{table}[!ht]
    \centering\begin{tabular}{>{\centering\arraybackslash}m{4cm} m{8.5cm}} 
        \multicolumn{2}{c}{Charge 2/3} \\
    \hline
        Diagram & Feynman Rule \\
        \hline
        \vspace{5px}
        \includegraphics[height=2.25cm]{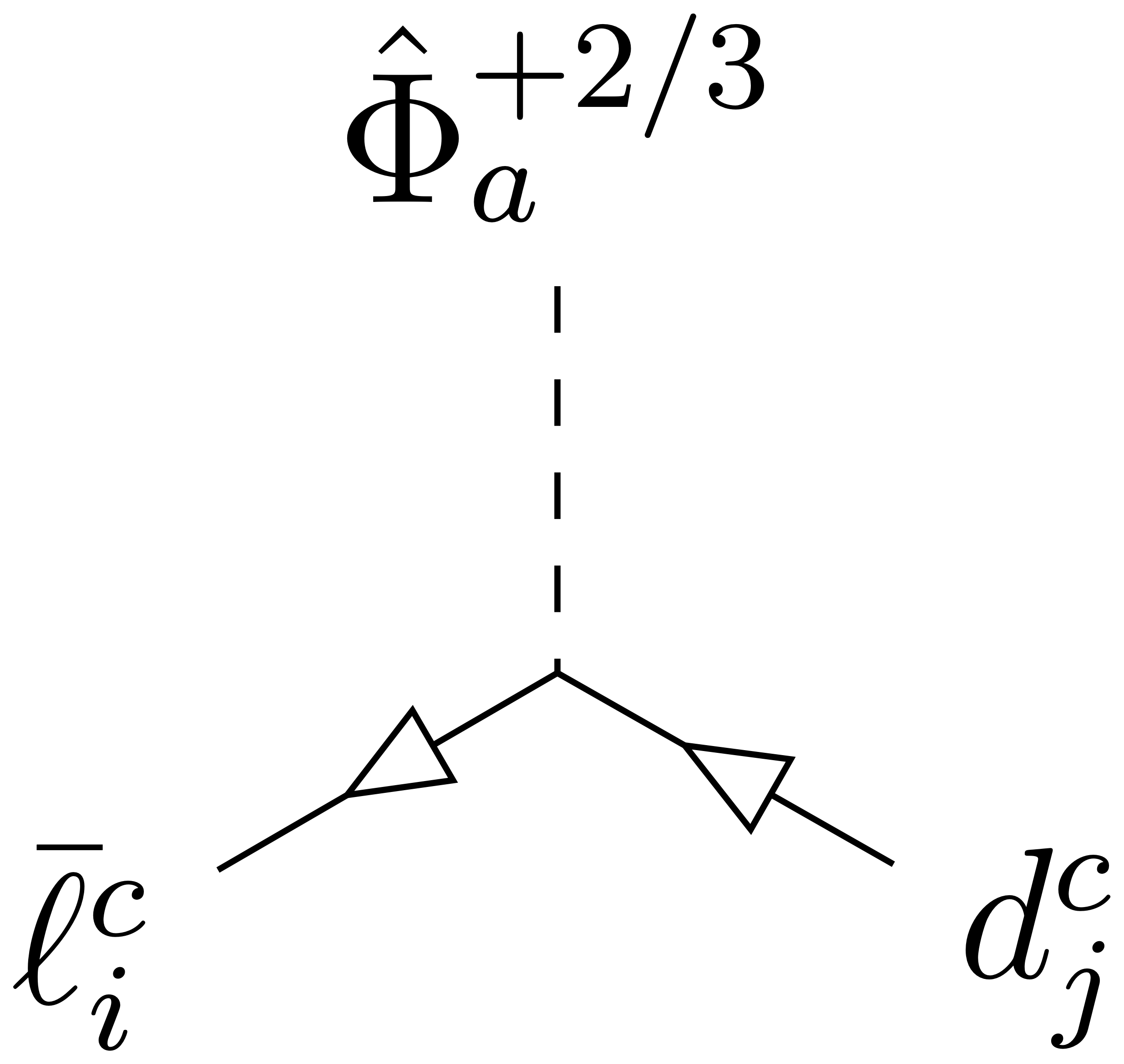}&
        $+ \ i V_{kj}^* Y_{2, ki}^{\text{LR}} \ W^{+2/3*}_{n_a1} \ \text{P}_{\text{R}}  \newline
        + i  Y_{\tilde 2, ji}^{\text{RL}} \ W^{+2/3*}_{n_a2} \ \text{P}_{\text{L}} $\\
        \vspace{5px}
        \includegraphics[height=2.25cm]{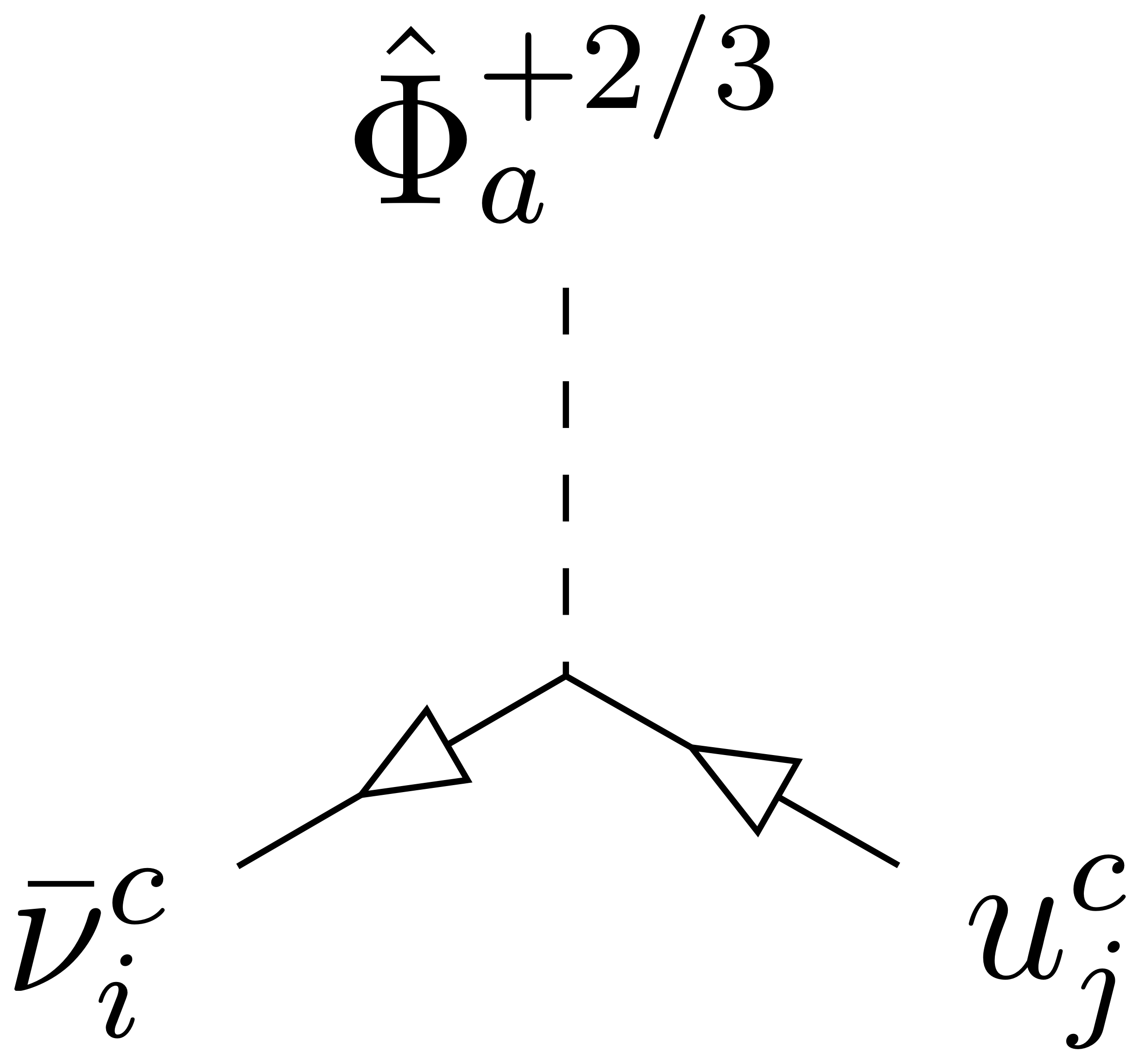}&
        $- \ i  Y_{2, ji}^{\text{RL}} \ W^{+2/3*}_{n_a1} \ \text{P}_{\text{L}}  $\\
        \hline
    \end{tabular}
\end{table}
\end{center}

\begin{center}
\begin{table}[!ht]
    \centering\begin{tabular}{>{\centering\arraybackslash}m{4cm} m{8.5cm}} 
    \multicolumn{2}{c}{Charge 5/3} \\
    \hline
        Diagram & Feynman Rule \\
        \hline
        \vspace{5px}
        \includegraphics[height=2.25cm]{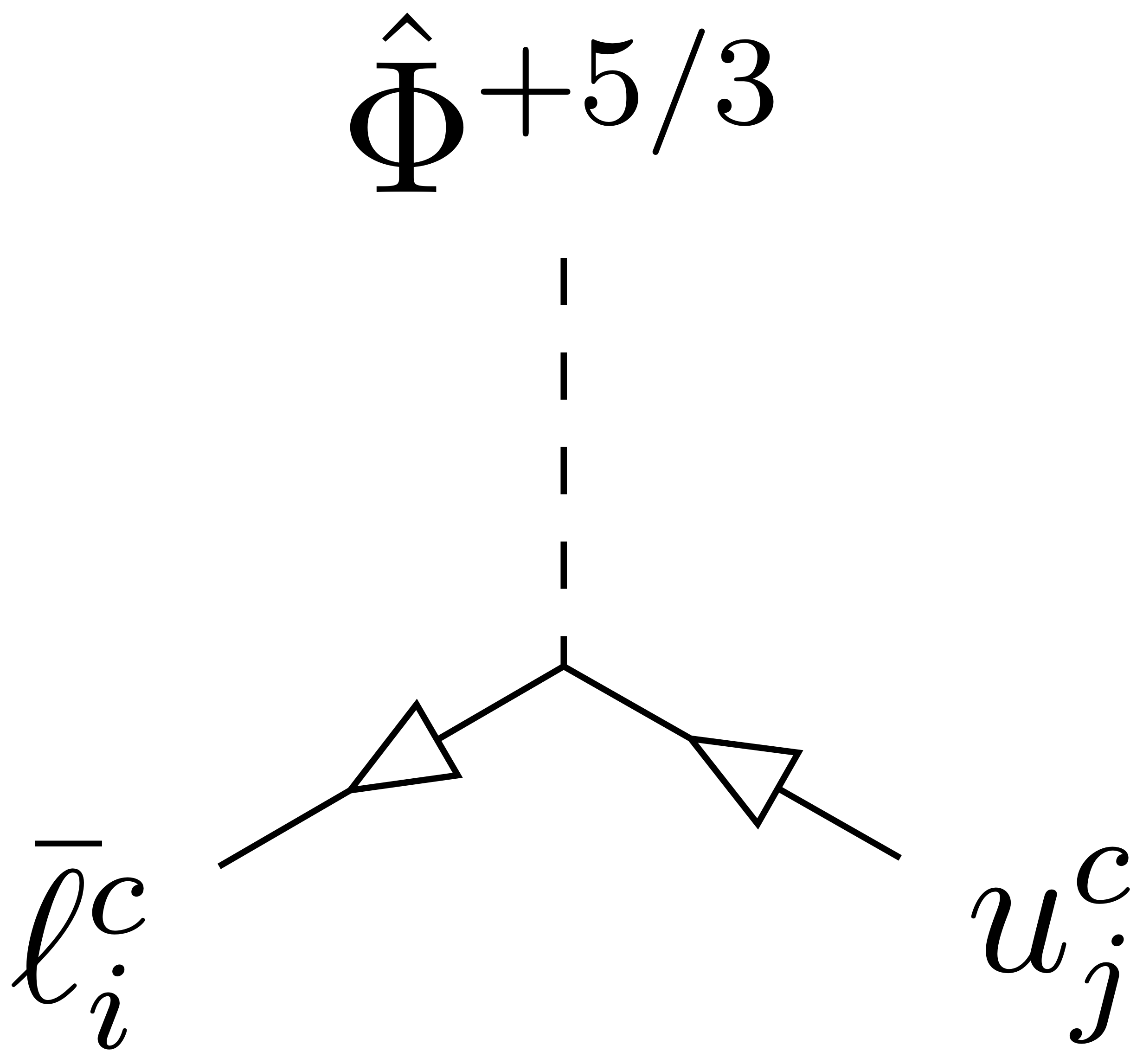}&
        $+ \ i Y_{2, ji}^{\text{LR}} \vspace{5px} \ \text{P}_{\text{R}} \newline
        + i \ Y_{2, ji}^{\text{RL}} \ \text{P}_{\text{L}} $\\
        \hline
        \hline
    \end{tabular}
\end{table}
\end{center}





\clearpage
\section*{Acknowledgements}
We thank Dario Müller and Francesco Saturnino for useful discussions. The work of A.C. is supported by a Professorship Grant (PP00P2\_176884) of the Swiss National Science Foundation. L.S. thanks the Restaurant Veranda Bern for the use of their meeting room during the Covid-19 outbreak. The work of L.S. is supported by the ``Excellence Scholarship \& Opportunity Programme'' of the ETH Z\"urich Foundation.

\bibliographystyle{elsarticle-num}
\biboptions{sort&compress}
\bibliography{main.bib}







\end{document}